\documentclass[12pt]{report}
\usepackage[letterpaper]{geometry}
 \usepackage[doublespacing]{setspace}
 \usepackage{tocloft}
\usepackage[rm, tiny,center, compact]{titlesec}
 \usepackage{indentfirst}
 \usepackage{etoolbox}
\usepackage{tocvsec2}
 \usepackage[titletoc]{appendix}
 \usepackage{appendix}
 \usepackage{tamuconfig}
\usepackage{rotating}
\usepackage{graphicx}
\usepackage{caption}
\usepackage{picture}
\usepackage{subcaption}
\usepackage{amsmath,amssymb}
\usepackage{multirow}
\usepackage{float}

\begin{document}

\renewcommand{\tamumanuscripttitle}{Partially-averaged Navier-Stokes (PANS) Method for Turbulence Simulations: Near-wall Modeling and Smooth-surface Separation Computations}
\renewcommand{\tamupapertype}{Dissertation}
\renewcommand{\tamufullname}{Pooyan Razi}
\renewcommand{\tamudegree}{Doctor of Philosophy}
\renewcommand{\tamuchairone}{Sharath Girimaji}
\renewcommand{\tamumemberone}{N.K. Anand}
\newcommand{\tamumembertwo}{Kalyan Annamalai}
\newcommand{\tamumemberthree}{Robert Handler}
\renewcommand{\tamudepthead}{Andreas A. Polycarpou}
\renewcommand{\tamugradmonth}{Aug}
\renewcommand{\tamugradyear}{2015}
\renewcommand{\tamudepartment}{Mechanical Engineering}

%
%
%


\providecommand{\tabularnewline}{\\}

\begin{titlepage}
\begin{center}
\MakeUppercase{\tamumanuscripttitle}
\vspace{4em}

A \tamupapertype

by

\MakeUppercase{\tamufullname}

\vspace{4em}

\begin{singlespace}

Submitted to the Office of Graduate and Professional Studies of \\
Texas A\&M University \\

in partial fulfillment of the requirements for the degree of \\
\end{singlespace}

\MakeUppercase{\tamudegree}
\par\end{center}
\vspace{2em}
\begin{singlespace}
\begin{tabular}{ll}
 & \tabularnewline
& \cr
Chair of Committee, & \tamuchairone\tabularnewline
Committee Members, & \tamumemberone\tabularnewline
 & \tamumembertwo\tabularnewline
 & \tamumemberthree\tabularnewline
Head of Department, & \tamudepthead\tabularnewline

\end{tabular}
\end{singlespace}
\vspace{3em}

\begin{center}
\tamugradmonth \hspace{2pt} \tamugradyear

\vspace{3em}

Major Subject: \tamudepartment \par
\vspace{3em}
Copyright \tamugradyear \hspace{.5em}\tamufullname 
\par\end{center}
\end{titlepage}
\pagebreak{}

%
%
%

\chapter*{ABSTRACT}
\addcontentsline{toc}{chapter}{ABSTRACT} 

\pagestyle{plain} 
\pagenumbering{roman} 
\setcounter{page}{2}

\indent In the context of computational fluid dynamics (CFD), accurate simulation of turbulent flows remains as a challenging field of research. Although direct numerical simulation (DNS) and large-eddy simulation (LES) are able to capture the turbulent flow features to a great extent, they are not viable for complex engineering flows. On the other hand, Reynolds-averaged Navier-Stokes (RANS) models involve too many simplifying assumptions making them inadequate to capture complex flow features. Variable resolution (VR) bridging methods such as the Partially-averaged Navier-Stokes (PANS) model fill the gap between these two limits by allowing a tunable degree of resolution from RANS to DNS.   

The goal of this dissertation is to investigate the the PANS model capabilities in providing significant improvement over RANS predictions at slightly higher computational expense and producing LES quality results at significantly lower computational cost. The PANS model will be implemented in OpenFoam, and is then assessed over several separated flows. This research work is divided into three main studies. The objective of each study is now given: (i) investigate the model fidelity at a fixed level of scale resolution (Generation1-PANS/G1-PANS) for smooth surface separation, (ii) Derive the PANS closure model in regions of resolution variation (Generation2-PANS/G2-PANS), and (iii) Validate G2-PANS model for attached and separated flows. The separated flows considered in this study have been designated as critical benchmark flows by NASA CFD study group.  

The key contributions of this dissertation are summarized as follows. The turbulence closure model of varying resolution, G2-PANS, is developed by deriving mathematically-consistent commutation residues and using energy conservation principles. The log-layer recovery and accurate computation of Reynolds stress anisotropy is accomplished by transitioning from steady RANS to scaled resolved simulations using the G2-PANS model. This represents a major advantage of PANS as most other hybrid approaches encounter significant errors in the log-layer region. Finally, several smooth-separation flows on the NASA turbulence website have been computed with high degree of accuracy at a significantly reduced computational effort over LES using the G1-PANS and G2-PANS models. These results along with strong theoretical foundation demonstrates that PANS has the potential to become a transformative CFD approach for scale-resolving turbulence simulations.

\pagebreak{}

%
%
%

\chapter*{DEDICATION}
\addcontentsline{toc}{chapter}{DEDICATION}  

\indent To my Mother, Father, and Brother for all your love and support

\pagebreak{}

%
%
%

\chapter*{ACKNOWLEDGEMENTS}
\addcontentsline{toc}{chapter}{ACKNOWLEDGEMENTS}  

\indent First, I would like to thank my advisor, Dr. Sharath Girimaji for his many years of guidance and mentoring during my graduate career. I would also like to acknowledge my committee members, Dr. N.K. Anand, Dr. Kalyan Annamalai, and Dr. Robert Handler for participating on my committee and for the feedback they have provided me which has greatly enhanced my dissertation. I truly appreciate my colleagues for their discussions and assistance throughout the years. I would like to especially thank Vishnu Venugopal for his invaluable help and support during my research work.

I would like to thank my very dear Sepeedeh for all her love and support and for sharing experiences, frustrations, and friendship which helped me a lot to grow as a better person and researcher. To all my friends, Mohsen Mahdavi, Behnam Moghadassian, Masood Darabi, Azadeh Haghshenas, Eisa Rahmani, Mohammad Reza Keshavarz, thanks for all your support and encouragement throughout the years. Graduate school was definitely more enjoyable and adventurous with you. Finally, I would like to express my gratitude to my family in U.S. and Iran for their great support and cordial manner during my stay at Texas A$\&$M University.

\pagebreak{}


%
%
%

\phantomsection
\addcontentsline{toc}{chapter}{TABLE OF CONTENTS}  

\begin{singlespace}
\renewcommand\contentsname{\normalfont} {\centerline{TABLE OF CONTENTS}}


\setlength{\cftaftertoctitleskip}{1em}
\renewcommand{\cftaftertoctitle}{%
\hfill{\normalfont {Page}\par}}

\tableofcontents

\end{singlespace}

\pagebreak{}


\phantomsection
\addcontentsline{toc}{chapter}{LIST OF FIGURES}  

\renewcommand{\cftloftitlefont}{\center\normalfont\MakeUppercase}

\setlength{\cftbeforeloftitleskip}{-12pt} 
\renewcommand{\cftafterloftitleskip}{12pt}

\renewcommand{\cftafterloftitle}{%
\\[4em]\mbox{}\hspace{2pt}FIGURE\hfill{\normalfont Page}\vskip\baselineskip}

\begingroup

\begin{center}
\begin{singlespace}
\setlength{\cftbeforechapskip}{0.4cm}
\setlength{\cftbeforesecskip}{0.30cm}
\setlength{\cftbeforesubsecskip}{0.30cm}
\setlength{\cftbeforefigskip}{0.4cm}
\setlength{\cftbeforetabskip}{0.4cm} 

\listoffigures

\end{singlespace}
\end{center}

\pagebreak{}

%
\phantomsection
\addcontentsline{toc}{chapter}{LIST OF TABLES}  

\renewcommand{\cftlottitlefont}{\center\normalfont\MakeUppercase}

\setlength{\cftbeforelottitleskip}{-12pt} 

\renewcommand{\cftafterlottitleskip}{12pt}

\renewcommand{\cftafterlottitle}{%
\\[4em]\mbox{}\hspace{4pt}TABLE\hfill{\normalfont Page}\vskip\baselineskip}

\begin{center}
\begin{singlespace}

\setlength{\cftbeforechapskip}{0.4cm}
\setlength{\cftbeforesecskip}{0.30cm}
\setlength{\cftbeforesubsecskip}{0.30cm}
\setlength{\cftbeforefigskip}{0.4cm}
\setlength{\cftbeforetabskip}{0.4cm}

\listoftables 

\end{singlespace}
\end{center}
\endgroup
\pagebreak{}  

%
%
%


\pagestyle{plain} 
\pagenumbering{arabic} 
\setcounter{page}{1}

\chapter{\uppercase {Introduction}}
\label{Intro}

Turbulent flow is characterized by rapid and chaotic variation of flow properties (i.e velocity, pressure and etc.) in space and time. Turbulence is triggered by instabilities inherent in many flows. Turbulence consists of a continuous spectrum of scales ranging from largest to smallest. The wide range of scales inherent in a turbulent flow stem from interactions between fluctuations of different wavelengths and directions. This interaction is very complex as it is rotational, fully three dimensional and time-dependent. Nearly, all flows of practical engineering interest are turbulent.

Turbulent flows exhibit irregularity, enhanced mixing, rotationality, and rapid energy dissipation. Due to the inherent irregularity, turbulent flows are often treated statistically rather than deterministically. Further, turbulent flows are influenced by a strong three-dimensional vortex generation mechanism known as vortex stretching which is the main reason behind the energy cascade (transfer of energy from larger flow structures to smaller structures). Turbulence dissipation occurs due to the conversion of kinetic energy to internal energy via intensified viscous action.  

\section{Historical perspective on turbulence}

The problem of turbulence has been an intriguing topic of research among the greatest physicists and engineers of the 19th and 20th centuries. This phenomenon was first recognized as a distinct flow behavior by the great artist in the 15th century, Leonardo da Vinci. He sketched the following artwork to illustrate the turbulent flow and described it with a remarkably modern note: 
\begin{quote}
. . . the smallest eddies are almost innumerate, and large
eddies are rotated only by large eddies and not by small ones,
and small eddies are turned by small eddies and large \cite{davinci}.
\end{quote}

\begin{figure}[!htb]
\begin{center}
\includegraphics[width=.55\textwidth]{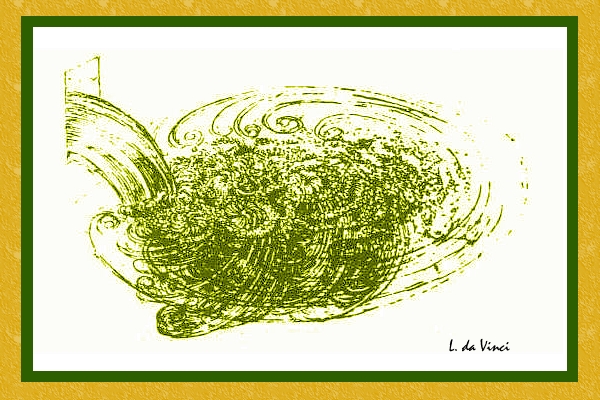}
\caption{da Vinci sketch of turbulent flow}
\end{center}
\end{figure}

This phenomenon in the fluid flow was named “turbolenza” by da Vinci, and hence is the origin of the current name, turbulence. 

Navier and Stokes introduced equations which are believed to embody physics of all fluid flow including the turbulent ones in the early to mid 19th century. These equations are non-linear and difficult to solve. There have been a few attempts to solve Navier-Stokes equations analytically which are all accompanied by some sort of simplification and often unrealistic assumptions. Little progress toward understanding turbulence has been made by analytical solution of Navier-Stokes equations and the early studies on turbulent flows are mainly focused on experimental analysis until the emergence of computational tools.   

As noted previously, da Vinci realized turbulence as a distinct physical process in fluid flow around 500 years ago. However, there was no substantial progress in understanding turbulent flow until the late 19th century. Osborne Reynolds was among the first scholars to experimentally investigate the transition from laminar to turbulent flow \cite{reynolds}. He injected a dye streak into flow passing through a pipe with smooth transparent walls. During these experiments, he identified a single dimensionless parameter, later called the Reynolds number, which was responsible for the observed flow behavior and transition to turbulence. In addition, Reynolds concluded that a detailed understanding of turbulence is difficult due to its randomness and introduced the concept of decomposing the flow variables into mean and fluctuating parts. This viewpoint of Reynolds is still followed by majority of people in turbulence community to come up with feasible methods to predict turbulent flow behavior. Boussinesq \cite{boussinesq} postulated that the turbulent stresses are linearly proportional to mean strain rate. This hypothesis is still the keystone in developing most of the turbulence models.   

The next major contribution was the "mixing length theory" of Prandtl \cite{prandtl} which proposed a functional form for the eddy viscosity (introduced by Boussinesq \cite{boussinesq}) for some simple flows. Based on an analogy between turbulent eddied and molecules/atoms of a gas, this theory proposed a way to construct eddy viscosity from a length and velocity (time) scale determined from kinetic theory. 

The next big step in analysing turbulence was taken by a British mathematician and physicist, G. I. Taylor, during the 1930s. He employed advanced mathematical and statistical methods to the turbulence literature by introducing correlations, Fourier transforms and power spectra. In his paper \cite{taylor}, he defined turbulence as a random phenomenon and established statistical approaches to investigate homogeneous, isotropic turbulence. He also conducted a wind tunnel experiment to show success of his analytical methods in predicting the turbulent flow behavior. Besides, he developed "Taylor hypothesis" which is specifically valuable for analysing the experiments and converting temporal behavior to spatial behavior. Other important works in that period are those of \cite{von1}, \cite{von2}. In addition, in 1941, A. N. Kolmogorov \cite{kolmogorov} developed the most important and well cited theories of turbulence.

During 1970s and 1980s, most studies focused on numerical computation rather than experimental/analytical investigations. The first direct numerical simulation (DNS) of turbulent flow was performed by Orszag and Patterson \cite{orzag}. However, DNS simulation of turbulent flows has been always challenging, especially with increasing Reynolds number, due to the presence of wide range of length and time scales and the necessity to resolve them. Therefore, majority of  scientific and engineering calculations of turbulent flows, at high Reynolds numbers, are based on some degree of turbulence modeling. 


\section{Turbulence modeling}

The complexity of a turbulence calculation is usually related to the the information we seek for a specified application. In some applications, we may only want to inquire about the friction coefficient, separation size or heat transfer coefficient. In these cases, a simple mathematical model of turbulence might suffice and provide the required information. Following this idea, it was in 1972 that Reynolds-averaged Navier Stokes (RANS) approach was started to develop \cite{launder} and \cite{launder2}. However, if we desire a complete time history of every aspect of a turbulent flow, only a direct simulation of Navier-Stokes equation \cite{orzag} and/or a large eddy simulation (LES) of a turbulent flow \cite{LES} are needed which require vast computer resources and accurate numerical schemes. Most engineering applications fall within these two limits, and thus a model which suggests the simplicity of the first category and some accuracy level of the second category will be an ideal and practical way to approach turbulent flow simulations. 

In other words, computationally-intensive approaches like DNS and LES provide the most accurate solutions for the turbulent flow. However, application of these methods is limited for high Reynolds number flows due to the extraordinary computational resource required which is beyond the current capabilities. RANS simulations, on the other hand, are computationally inexpensive but resolve only the mean velocity field, while the fluctuating fields are all modeled. Since the accuracy of RANS simulations for many engineering applications is inadequate and DNS/LES calculations are not viable, it has been suggested that variable resolution methods would be computationally more suitable as they can provide any intermediate level of resolution.

Partially-averaged Navier-Stokes (PANS) model \cite{PANS1,PANS2} is a hybrid method that is intended to bridge smoothly between RANS and LES/DNS. In PANS, the accuracy of results can be optimized based on available computational resources. In the RANS method, only the unsteady mean flow i.e. scales that are comparable to the geometry of the flow are resolved, whereas all other scales are modeled. In LES, all the large scale motions or energy carrying eddies are computed exactly, and the small scale motions that are more universal in nature, are modeled \cite{LES2}. 

\begin{figure}[!htb]
\begin{center}
\includegraphics[width=.55\textwidth]{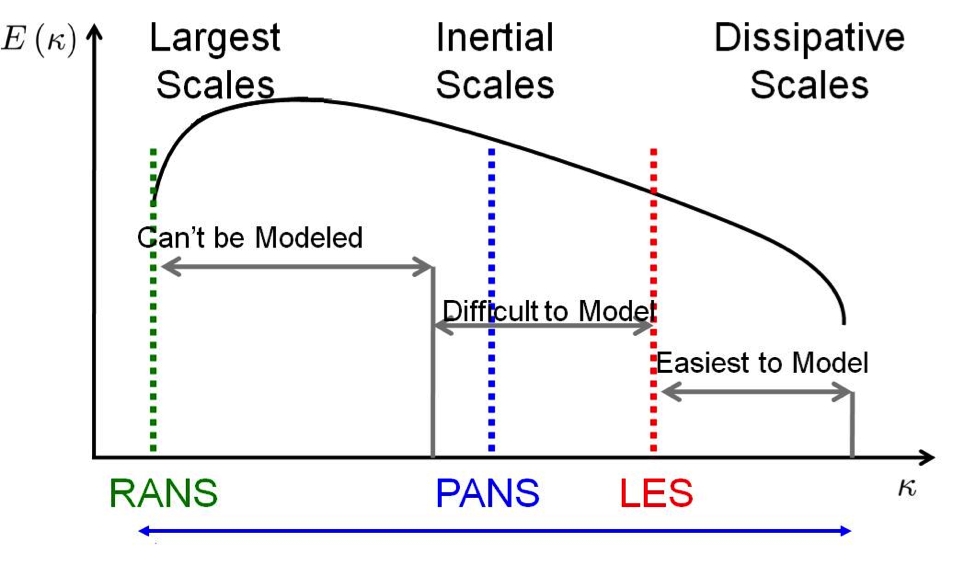}
\caption{Energy spectrum of turbulent flow with relative URANS, PANS, and LES spectrum cut-off \label{spectrum}}
\end{center}
\end{figure}

To illustrate the operative regions of PANS, a typical spectrum of energy as a function of wavelength for turbulent flow is shown in Fig. \ref{spectrum}. The relative cut-off for unresolved flow scales is shown for RANS, PANS, and LES. The cut off parameters $f_k$ and $f_\epsilon$ for PANS are defined as the ratio of the unresolved to resolved turbulent kinetic energy and dissipation, respectively. Value of $f_k$ close to zero indicates DNS and the value of unity is essentially a RANS simulation.

PANS is based on the premise that physical phenomena contributing to accurate predictive computations reside in flow scales that are not resolved in RANS but are significantly larger than the smallest LES scales. PANS seeks to resolve only up to the scales that critically contribute to the desired objective function, excluding the computationally expensive small scales. Thus, PANS does not seek to combine two models (RANS and LES) in different regions, it rather provides a closure model for any intermediate degree of scale resolution.

\section{Dissertation Objectives}
The first generation of the PANS model (G1-PANS) has been tested for several benchmark and complex flow geometries \cite{lak10,jeong10,murthithesis,basara2008pans,basara2010pans,frendi2006}, and promising results have been obtained. However, comprehensive assessment studies over a broader range of flow physics needs to be performed for the G1-PANS model. Therefore, the objective of the first study is to further evaluate model capabilities in predicting the turbulent flow features associated with flow separation over smooth curved surfaces. The simulations are carried out with fixed $f_k$ throughout the entire computational domain. 

In the next step of PANS progression, closure modeling in region of varying $f_k$ has been investigated. The objective of this study is to develop the second generation of the PANS model known as G2-PANS for variable resolution calculations based on sound physical concepts. This study is aimed to demonstrate the ability of the G2-PANS model to accurately capture the boundary layer physics in wall-bounded turbulent flows and separation process from smooth curved surfaces. 

To achieve these objective, this dissertation is divided into three distinct studies. Specification and main objectives of each study are presented in the following sub-sections.

\subsection{Simulation of Smooth Surface Separation Using G1-PANS Model}
\label{study2}

The objective of this study is to investigate the flow separation from smooth curved surfaces using G1-PANS Model. Flow over periodic hill and wall-mounted hump are considered for this study. Flow over periodic hill geometry exhibits complex flow features and was introduced in ERCOFTAC/IAHR workshop \cite{workshop} as a benchmark test-case for turbulence modeling validation. Wall-mounted hump configuration is also a challenge flow in the NASA turbulence resource website, and is generally considered a very difficult case to predict.

The following goals are investigated for the simulation of smooth-surface separation by G1-PANS method:

\begin{enumerate}

   \item
Establish the model fidelity in high Reynolds number flows at a fixed level of scale resolution throughout the computational domain-that is, $f_k$ = const. $(<
1)$ and $f_\epsilon$ = 1;
   \item
      Assess the PANS method in predicting the turbulent flow features of the separation process associated with smooth curved surfaces
   \item
study the effect of cut-off length scale and grid size on flow statistics and structures          
   \item
Recovery of the PANS filter parameter is sought as to check the validity of calculations (known as internal consistency criteria)      
   \item
Anisotropy of the flow will be investigated by constructing the anisotropy-invariant maps.
   \item
Outline the need for near-wall modeling of the G1-PANS model for high $Re$ wall-bounded turbulent flow calculations

\end{enumerate}

\subsection{Near Wall Modeling of the PANS Method}
\label{study1}

This represents one of the most important challenges in modern-day turbulence research. Available computational resources demand spatio-temporal variation in resolution for calculating complex engineering flows. However, implementing a variable resolution model is not a trivial task due to the interaction between resolved and unresolved fields. The objective of this study is to derive the PANS model for the bridging region based on energy conservation principles. Then, the validity of assumptions and mathematical derivations are tested for turbulent channel flow simulation. The simplicity of the channel flow allows a thorough examination of several aspects of the proposed closure which are discussed in detail in section \ref{wall-model}. 

The objectives of the this study can be summarized as:

\begin{enumerate}
   \item
Develop a bridging region closure model in the PANS framework for variable resolution calculations near the wall
   \item
Validate the model against DNS data for turbulent channel flow simulations at different Reynolds numbers and compute the mean flow properties as well as determining the instantaneous unsteady features 
   \item
Study the effect of the energy scale transfer terms on the flow domain and accuracy of the results

   \item
Investigate the influence of resolution variation location on the resolved scales near the wall, and consequently on mean flow statistics

   \item
Evaluate the ability of the model at high Reynolds numbers flow calculations where DNS and LES could be extremely costly and not feasible to perform
\end{enumerate}

\subsection{Simulation of Smooth Surface Separation Using G2-PANS model}
\label{study3}
Results from wall-modeled PANS computations of the flow over mounted hump are investigated using the second generation of the PANS model. In this work, the details of G2-PANS simulations have been presented. The primary objectives of this work can be summarized as

\begin{enumerate}
   \item
Asses the ability of the G2-PANS model in predicting flow separation from a smooth body, and the subsequent reattachment and flow recovery
   \item
Investigate the reattachment and separation point at different grid resolutions and filter parameters 
   \item
Visualization of unsteady flow structures and interpreting the flow behavior 

\end{enumerate}

\section{Outline}

The remainder of the dissertation is arranged as follows. Each of the studies detailed in subsections \ref{study2}-\ref{study3}, are presented in Secs. \ref{hill}, \ref{wall-model}, and \ref{hump}, respectively. This work concludes with a summary in Sec. \ref{conclusion}.
%
%
%

\chapter{\uppercase{Simulation of Smooth Surface Separation Using G1-PANS}}
\label{hill}

\section{Introduction} 
\label{sec2-intro}

Flow separation over smooth curved surfaces occurs in many engineering applications such as flow over wings and airfoils, turbine blades, ships, automobile bodies and curved obstructions in pipes. Reliable and accurate modeling of this  phenomenon is important for effective and safe design of the aforementioned industrial components. The prediction of flow separation over curved and continuous surface is challenging due to the spatial and temporal fluctuations of the separation line and failure of the law of the wall assumption for separated shear layer regions. Separation from curved surfaces differs from the one from sharp edges in that the point or line of separation is not fixed in space and is very sensitive to external flow properties, turbulence level and development of streamwise pressure gradient \cite{Basara}. 

For practical engineering applications involving flow separation, there are several design parameters which are essential to be accurately estimated. These parameters are mainly associated with the separation point, size of the recirculation zone, and reattachment point. Accurate prediction of the separation process depends on several factors given a particular flow geometry and simulation procedure. Figure \ref{sep} shows a sketch of a flow geometry which involves smooth-surface separation and identifies the main important physics regarding that. As seen in this figure, the flow separation occurs between a curved wall and a thin shear layer. There are mainly two sources of unsteadiness inherent in this flow geometry which are the Kelvin-Helmholtz (K.H.) and Tollmein-Schlichting (T.S.) instabilities. The former one occurs in a flow due to existence of inflection point in the velocity field, and the latter one happens in a developing boundary layer. Figure \ref{sep} illustrates the spots where these instabilities are developed and affects the separation and reattachment locations. Based on this figure, it can be inferred that prediction of the important locations in a separated flow depends on:

\begin{enumerate}
\item
Separation-point
\begin{enumerate}
\item
flow behavior unlike the sharp-edge separation
\item
Inflow 
\item
Accurate representation of BL coherent structures (e.g., T.S. waves)
\end{enumerate}
\item
Recirculation-zone
\begin{enumerate}
\item
How well the thin shear layer is resolved
\item
Recovering the K.H. instability 
\end{enumerate}
\item
Reattachment-point
\begin{enumerate}
\item
Resolving both BL and shear layer structures
\item
Accurately capturing the Kelvin-Helmholtz and Tollmein-Schlichting instabilities
\end{enumerate}
\end{enumerate}

\begin{figure}[H]
\centering
  \includegraphics[trim=0cm 0cm 0cm 0.3cm, clip=true, scale=0.55]{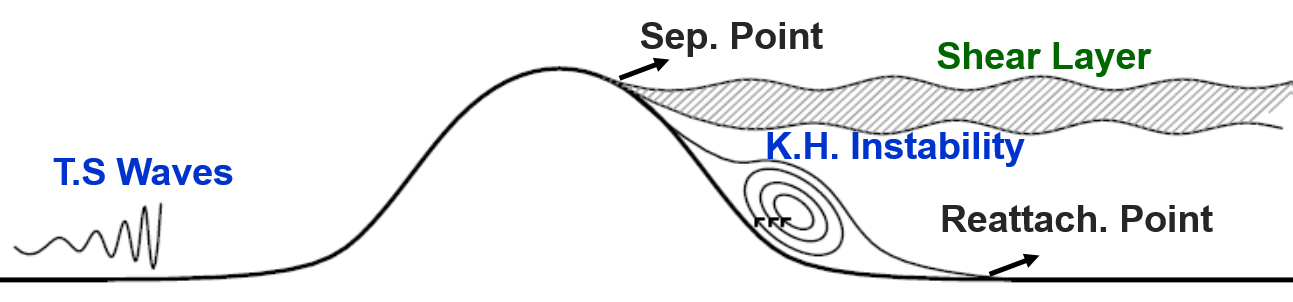}
\vspace{-6pt}

\caption{Separation physics}
\label{sep}
\end{figure}

These features associated with smooth-surface separation imply that a great extent of physical and computational sophistication is required for modelling flow separation over curved surfaces. To address the fidelity of different turbulence closure models in this matter, flow over periodic hill \cite{workshop} and flow over wall-mounted hump \cite{rumsey2013} are considered in this research work.

\section{Governing equations for the PANS model- generation 1}

The first generation of the PANS model (G1-PANS) is derived for a fixed level of scale resolution. A brief description of the G1-PANS model is presented in this section.

The development of all the hybrid models commences from the incompressible Navier-Stokes equations and continuity equation:
\begin{eqnarray}
\label{eq:NS}
\frac{\partial U_i}{\partial t}+U_j\frac{\partial U_i}{\partial x_j}&=&-\frac{\partial p}{\partial x_i}+\nu \frac{\partial^2 U_i}{\partial x_j \partial x_j},\\
\label{cont}
\frac{\partial U_i}{\partial x_i}&=&0.
\end{eqnarray}

The difference between URANS and PANS/LES lies in how the averaged or filtered velocity equations are obtained from Equation \ref{eq:NS}. URANS uses an averaging operator leading to equations that describe the mean velocity field. On the other hand, PANS/LES uses a generalized homogeneous filter to decompose the velocity into resolved and unresolved part \cite{PANS3}:

\begin{eqnarray}
\label{eq:PANS_filt}
V_i = U_i + u_i, \: U_i = \langle V_i \rangle, \: \langle u_i \rangle &\neq 0
\end{eqnarray}
By applying the filter which is not variable in time/space to the Navier-Stokes equations, the momentum equations for the resolved field are given as:
\begin{eqnarray}
\label{eq:NSfiltered}
\frac{\partial U_i}{\partial t} + U_j \frac{\partial U_i}{\partial x_j} &=& -\frac{\partial \tau (U_i,U_j)}{\partial x_j} - \frac{\partial \langle p \rangle}{\partial x_i} + \nu \frac{\partial^2 U_i}{\partial x_j \partial x_j}, \\
\label{eq:secondmom}
\tau(U_i,U_j) &=& \langle U_i U_j \rangle - \langle U_i \rangle \langle U_j \rangle.
\end{eqnarray}

The term $\tau(U_i,U_j)$ in eqn. \ref{eq:NSfiltered} represents the ``sub-filter stress", given by eqn. \ref{eq:secondmom} \cite{germano}. The sub filter stress term (SFS) is modeled differently in various turbulence models. 

\textbf{LES:} The generalized sub-filter stress in LES is modeled via Boussinesq-type approximation \cite{germano,murthi,Lakshmipathy}:
\begin{eqnarray}
\label{eq:Bousstype}
\tau_{ij} = \tau(V_i,V_j) &=& \frac{2}{3} k \delta_{ij} - \nu_T S_{ij}
\end{eqnarray}
Where $\nu_T,$ and $k$ are the ``unresolved eddy viscosity" and the ``unresolved kinetic energy'' respectively. $S_{ij}$  is the resolved strain-rate tensor defined as: 
\begin{eqnarray}
S_{ij} = \frac{1}{2}\left( \frac{\partial U_i}{\partial x_j} + \frac{\partial U_j}{\partial x_i} \right)
\end{eqnarray}
In LES Smagorinsky model, the unresolved eddy viscosity is modeled algebraically by assuming that the energy production and dissipation are in equilibrium, resulting in the following relationship \cite{germano1991}:
\begin{eqnarray}
\label{eq:LES_Smag}
\nu_T = C\Delta^2 \lvert S \rvert, \; where \; \lvert S \rvert = (2 S_{ij} S_{ij})^{1/2}
\end{eqnarray}
Here, $\Delta$ is the grid spacing and $C$ is a subgrid scale constant determined from decaying isotropic turbulence. If the filter cut-off is in the inertial range, then the values of the Smagorinsky constant ($C_s = \sqrt{C}$) usually lie between 0.18 to 0.23. In the dynamic Smagorinsky model, the model parameter C is not constant, rather it is calculated from the energy content of the smallest resolved scales \cite{ghosal}. In order to make the model self-contained, an additional test filter ($\widehat{\Delta} > \Delta$ ) is introduced and $C = C_d (x, y, z, t)$ is dynamically adjusted based on the following identity: 
\begin{eqnarray}
\label{eq:LES_dynSmag}
L_{ij} = T_{ij} - \widehat{\tau_{ij}}
\end{eqnarray}
Where $L_{ij} = \widehat{\overline{u_i} \; \overline{u_j}} - \widehat{\overline{u_i}} \; \widehat{\overline{u_j}} $ is the Loenard stress, and $T_{ij} = \widehat{\overline{u_i u_j}} - \widehat{\overline{u_i}} \; \widehat{\overline{u_j}} $ is the `test-level' subgrid scale stress or subtest stress. It is assumed that the subtest stress can also be expressed with eddy viscosity model:
\begin{eqnarray}
\label{eq:LES_dynSmag1}
T_{ij} -\frac{1}{3} \delta_{ij}T_{kk} = - 2C \widehat{\Delta}^2  \lvert \widehat{\overline{S}} \rvert \widehat{\overline{S_{ij}}}
\end{eqnarray}
Incorporating equations \ref{eq:Bousstype}, \ref{eq:LES_Smag} and \ref{eq:LES_dynSmag1} into \ref{eq:LES_dynSmag}, we obtain an equation for determing C \cite{germano1991}:
\begin{eqnarray}
\label{eq:LES_dynSmag2}
L_{ij} -\frac{1}{3} \delta_{ij}L_{kk} = \alpha_{ij} C - \widehat{\beta_{ij} C}
\end{eqnarray}
where, 
\begin{eqnarray}
\label{eq:LES_dynSmag2}
\alpha_{ij}  =  - 2 \widehat{\Delta}^2  \lvert \widehat{\overline{S}} \rvert \widehat{\overline{S_{ij}}} \\
\beta_{ij}   =    - 2 {\Delta}^2  \lvert {\overline{S}} \rvert {\overline{S_{ij}}}
\end{eqnarray}

\textbf{G1-PANS:} The filtering procedure in PANS modeling is similar to LES, which separates the flow into resolved and unresolved features. However, in PANS, rather than cut-off length scale or grid
size, the unresolved-to-total ratios of kinetic energy and dissipation are the resolution control parameters and their value must be in commensurate with grid size \cite{girimaji2005}.

\begin{eqnarray}
\label{eq:filter}
f_k=\frac{k_u}{k};~f_{\epsilon}=\frac{\epsilon_u}{\epsilon};~ f_{\omega}=\frac{\omega_u}{\omega};~\frac{f_k^{3/2}}{f_{\epsilon}}=\frac{1}{C_{\mu}^{1/3}}\left(\frac{\Delta}{L}\right)
\end{eqnarray} 

where, $\Delta=(\Delta_x\times\Delta_y\times\Delta_z)^{1/3}$ is the grid dimension and $L \equiv k^{1.5}/\epsilon$ is the integral scale of turbulence.  

In PANS, the sub-filter stress term in eqn. \ref{eq:secondmom} is also modeled with Boussinesq approximation:
\begin{eqnarray}
\label{eq:Bousstype1}
\tau_{ij} = \tau(U_i,U_j) &=& \frac{2}{3} k_u \delta_{ij} - \nu_u \left( \frac{\partial U_i}{\partial x_j} + \frac{\partial U_j}{\partial x_i} \right).
\end{eqnarray}
Where $\nu_u$ is the ``unresolved eddy viscosity" defined as $\nu_u = k_u / \omega_u = C_\mu k_u^2/ \varepsilon_u$. The final model equations for the unresolved kinetic energy, $k_u$ and the unresolved dissipation, $\epsilon_u$ can be obtained from spectral or fixed-point analysis \cite{sch05,PANS1}:

\begin{eqnarray}
\label{eq:k}
\frac{Dk_u}{Dt}=(P_u-\epsilon_u)+\frac{\partial}{\partial{x_j}}\left[\left(\nu+\frac{\nu_u}{\sigma_{ku}}\right)\frac{\partial{k_u}}{\partial{x_j}}\right]
\vspace{-0.2cm}
\end{eqnarray} 
\vspace{-0.2cm}
\begin{eqnarray}
\label{eq:e}
\frac{D\epsilon_u}{Dt}=C_{\epsilon1}P_u\frac{\epsilon_u}{k_u}-C_{\epsilon2}^*\frac{\epsilon_u^2}{k_u}+\frac{\partial}{\partial{x_j}}\left[\left(\nu+\frac{\nu_u}{\sigma_{\epsilon u}}\right)\frac{\partial{\epsilon_u}}{\partial{x_j}}\right]
\vspace{-0.2cm}
\end{eqnarray}

\begin{figure}[!htb]
\begin{center}
\includegraphics[width=.75\textwidth]{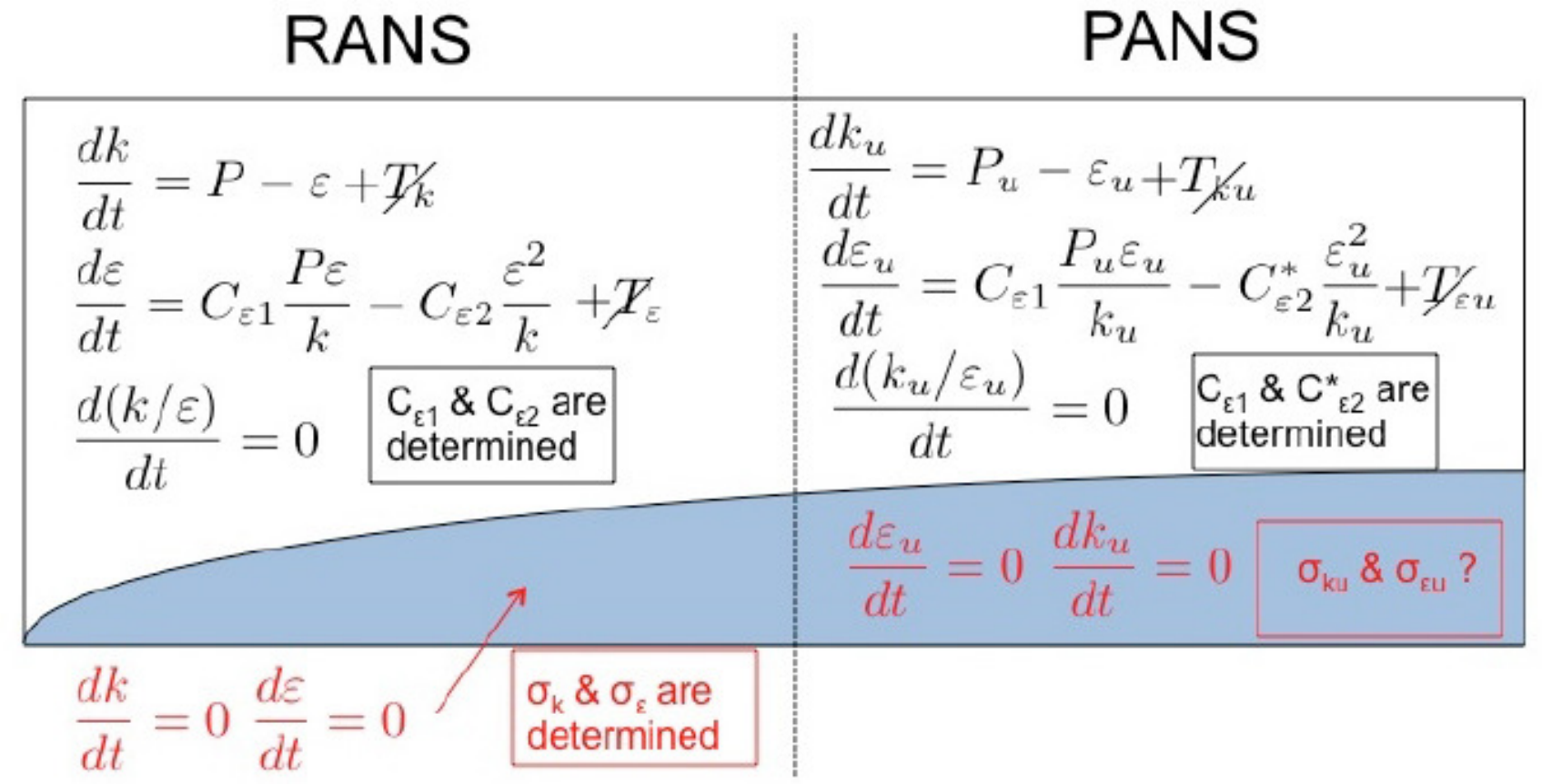}
\caption{Boundary layer analysis}
\label{rey1}
\end{center}
\end{figure}

$C_{\epsilon1}$ and $C_{\epsilon2}^*$ are obtained by investigating homogeneous shear turbulence (HST) and decaying isotropic turbulence (DIT). In the above equations, $\sigma_{k u}$ and $\sigma_{\epsilon u}$ are the transport coefficients for the PANS model which need to be determined. Two proposals are developed to obtain these coefficients \cite{PANS1,PANS2}. The first approach given the name "zero transport model (ZTM)" assumes that the resolved fluctuations do not contribute to the net transport of unresolved kinetic energy. Based on this assumption, the following equations are obtained for the turbulent Prandtl numbers for the G1-PANS ZTM model:
\begin{eqnarray}
\label{eq:sigma1}
	\sigma_{k u} = \sigma_k \frac{f_k^2}{f_\epsilon}, \: \sigma_{\epsilon u} = \sigma_\epsilon  \frac{f_k^2}{f_\epsilon}
\end{eqnarray}
In the second approach, the transport of unresolved kinetic energy is assumed to be proportional to the eddy viscosity of the resolved fluctuations which is referred as maximum transport model (MTM). In this case, the transport coefficients are given by
\begin{eqnarray}
\label{eq:sigma2}
	\sigma_{k u} = \sigma_k, \: 
	\sigma_{\omega u} = \sigma_\omega
\end{eqnarray}
Derivation of $\sigma_{ku}$ and $\sigma_{\epsilon u}$ from equilibrium boundary layer analysis shows consistency with the assumptions made for the ZTM model. Therefore, G1-PANS ZTM model is selected for the majority of calculations in this report. Figure \ref{rey1} summarizes the procedure to obtain the model coefficients \cite{rey13thesis}. The scale dependent model coefficients are thus given as
\begin{eqnarray}
\label{eq:coeff}
C_{\epsilon2}^*=C_{\epsilon1}+\frac{f_k}{f_{\epsilon}}(C_{\epsilon2}-C_{\epsilon1});~~\sigma_{k_u, \epsilon_u}=\sigma_{k, \epsilon}\frac{f_k^2}{f_\epsilon}
\vspace{-0.2cm}
\end{eqnarray} 
The standard RANS values are used for other coefficients:
\vspace{-0.2cm}
\begin{eqnarray}
\label{eq:coeff}
C_{\mu}=0.09;~C_{\epsilon1}=1.44;~C_{\epsilon2}=1.92;~\sigma_k=1~
\sigma_{\epsilon}=1
\vspace{-0.2cm}
\end{eqnarray} 
Similarly, the G1-PANS $k_u-\omega_u$ equations can be derived as \cite{PANS3}:
\begin{subequations}
\label{eq:PANS1}
\begin{align}
	\frac{\partial k_u}{\partial t} + U_j\frac{\partial k_u}{\partial x_j} &= P_u - \beta^* k_u \omega_u + \frac{\partial}{\partial x_j} \left[ \left(\nu + \nu_{u} / \sigma_{ku} \right) \frac{\partial k_u}{\partial x_j} \right] ,\\
	\frac{\partial \omega_u}{\partial t} + U_j\frac{\partial \omega_u}{\partial x_j} &= \alpha \frac{\omega_u}{k_u} P_u - \beta' \omega_u^2 + \frac{\partial}{\partial x_j} \left[ \left(\nu + \nu_{u} / \sigma_{\omega_u} \right) \frac{\partial \omega_u}{\partial x_j} \right]
\end{align}
\end{subequations}
\begin{eqnarray}
\label{eq:coeff2}
\beta' &= \alpha \beta^* - \alpha \frac{\beta^*}{f_{\omega}} + \frac{\beta}{f_{\omega}}.;~~\sigma_{k_u, \omega_u}=\sigma_{k, \omega}\frac{f_k}{f_\omega};~~f_{\omega}=\frac{f_{\epsilon}}{f_k}
\vspace{-0.2cm}
\end{eqnarray} 
Where, the values of the RANS closure coefficients are: $\beta^* = 0.09$, $\alpha = 5/9$, $\beta = 0.075$, $\sigma_{k} = 2.0$, and $\sigma_{\omega} = 2.0$.  

\textbf{URANS:} In URANS, eqn. \ref{eq:secondmom} becomes the familiar Reynolds stress term, where  the filter $\langle \rangle$ denotes an average. With two-equation URANS models, the Reynolds stress term is given by:
\begin{eqnarray}
\label{eq:Boussinesq}
\overline{u_i u_j} &=& \frac{2}{3} k \delta_{ij} - \nu_T \left( \frac{\partial U_i}{\partial x_j} + \frac{\partial U_j}{\partial x_i} \right)
\end{eqnarray}
where $\nu_T$ is the eddy viscosity, defined as $\nu_T = C_\mu{k^2}/{\varepsilon} = {k}/{\omega} $. When $f_k = f_\varepsilon = 1.0$ in PANS formulation, we recover the evolution equations for URANS.

\section{Flow over periodic hill}
\label{G1-hill}
Rapp and Manhart \cite{Rapp} carried out an experimental study for the flow over periodic hills at Reynolds numbers in the range of 5600 to 37000 using particle image velocimetry and laser doppler anemometry measurements. The experiment was performed in a water channel with 10 hills in the streamwise direction to ensure periodicity for the range of Reynolds numbers under investigation. In order to minimize the effect of side walls on the flow statistics, the spanwise direction was extended to 18 hill heights. They observed formation of secondary vortex structures due to the unstable curvature of streamlines on the windward side of the hills particularly at low Reynolds numbers. They also noticed frequent break up of the separation bubble and strong intermittency in the location of the reattachment line. It was also found that the reattachment length decreases with increasing Reynolds number.  

Computations of flow over periodic hills have also been performed by several researchers. Frohlich et al. \cite{Frohlich} provided extensive statistical and structural features of the flow obtained from a highly resolved large eddy simulation (LES) at Re=10590. They indicated a high level of spanwise turbulence intensity resulted by splatting of large-scale eddies originating from the shear layer above the recirculation zone. They reported a strongly time-dependent separation line, unsteady reattachment and large scale structures for the flow over periodic hill. They also identified spanwise rollers originating from Kelvin-Helmholtz instability in the shear layer. In another study on the current flow geometry, Breuer et al. \cite{Breuer} presented a complementary numerical and experimental investigation for different Reynolds numbers in the range of 100-10590. They performed LES and direct numerical simulations (DNS) on very fine grids and established an experimental set-up to examine the flow behavior and place confidence in their numerical results. They observed existence of a very small recirculation region at the crest of the hill for Re=10590 which did not exist for the lower Reynolds numbers. They also found a decreasing trend of the reattachment length with increasing Reynolds number. Additionally, they noticed steady and two-dimensional flow for Re=100 and three dimensional and chaotic flow for Re$\geq$200. 

It is noteworthy that the LES and DNS simulations \cite{Breuer} at Re=5600 were performed on 13.1 and 231 million grid nodes, respectively. For practical applications, it is important to develop and validate more affordable 'hybrid' computational techniques. The subject of this study is the so-called bridging techniques which includes partially integrated transport modeling (PITM) and partially-averaged Navier-Stokes (PANS) method.     

Chaouat and Schiestel \cite{PITM} simulated the flow over periodic hills using PITM for Re=37000 and compared their results against experimental data \cite{Rapp}. Simulations were performed on coarse and medium grid sizes and aimed to achieve a reasonable agreement with experimental data \cite {Rapp} at low computational cost. At the finest grid size of 0.9 million cells, the mean turbulence quantities were well predicted. Chaouat and Schiestel \cite{PITM} also demonstrated the failure of Reynolds stress transport model to predict the correct behavior of mean velocity and stress components at different streamwise locations. It was also observed that PITM method may not be accurate if the grid is too coarse especially in the spanwise direction.   
%

The objective of this work is to perform a comprehensive investigation of flow over periodic hills using G1-PANS to examine the effect of cut-off wave number and grid resolution at two different Reynolds numbers. Various statistics and flow features from G1-PANS will be compared against available experimental data, LES \cite{Frohlich} and partially integrated transport modeling (PITM) \cite{PITM} results. 

\subsection{Simulation procedure}

\begin{figure}[H]
\centering
  \includegraphics[trim=4cm 2.5cm 1cm 19cm, clip=true, scale=0.3]{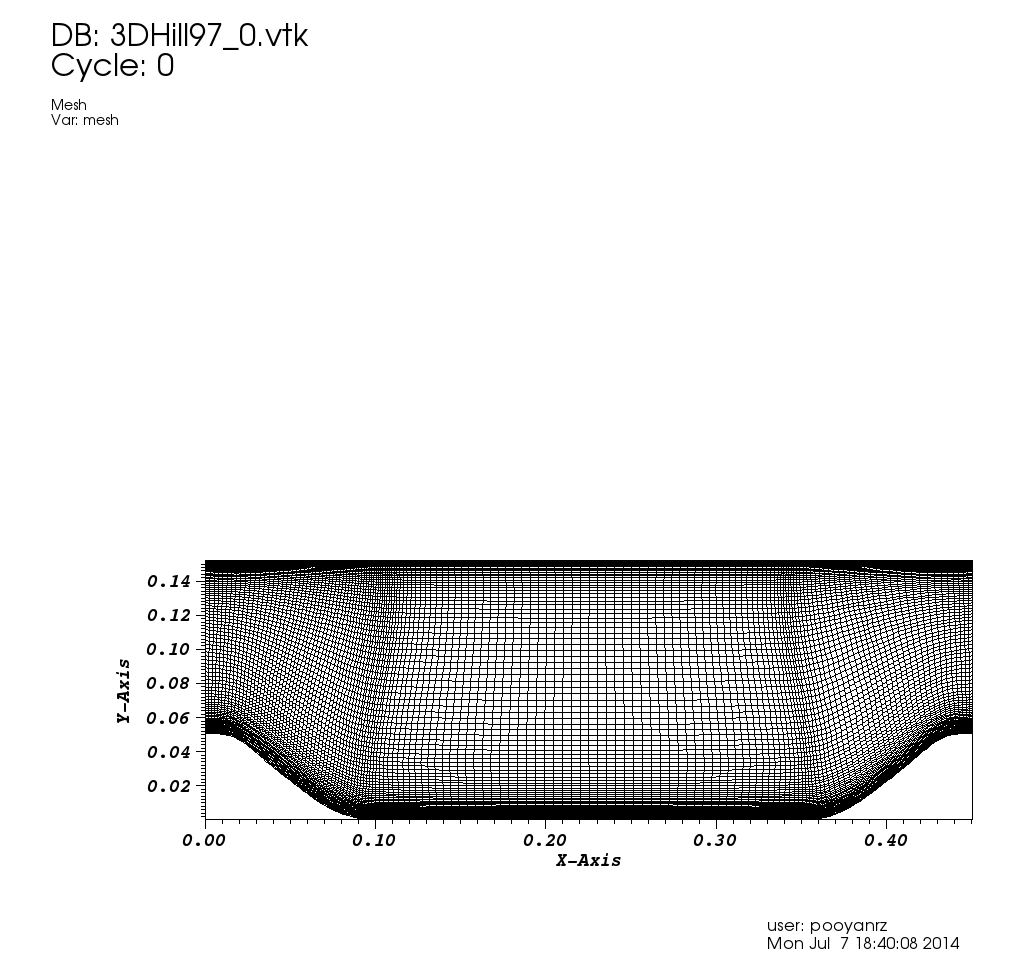}
\vspace{-6pt}

\caption{Flow Configuration}
\label{grid}
\end{figure}

The computational domain and flow configuration are summarized in Fig. \ref{grid}. The geometry and domain dimensions are consistent with \cite{Frohlich, PITM}. A body fitted, curvilinear grid very similar to the one studied in \cite{Frohlich} is generated for the flow geometry. The hill height ($h$) is one-third of the total channel height. In the streamwise direction, the domain extends from one crest to the next for a total of 9h. In the spanwise direction, the domain size is 4.5h. The flow is periodic in both streamwise and spanwise directions and no slip boundary condition is used at the bottom and top walls. The special features associated with this flow geometry makes it suitable for turbulence modeling validation. Frohlich et al. \cite{Frohlich} discussed that streamwise periodicity removes uncertainties posed by the inlet and outlet boundary conditions. Besides, they argued that extending domain size in streamwise and spanwise directions has minor effects on the flow structure and statistics. Flow is driven by pressure gradient which is added as a source term to the momentum equation. The flow Reynolds number is calculated based on the following equation

\begin{equation}
Re=\frac{U_bh}{\nu}
\end{equation}
Where, $U_b$ is the bulk velocity. 

PANS simulations are performed over a range of $f_k$ values. It is important to investigate the mean flow properties after flow has reached statistically steady state condition. For LES calculation \cite{Frohlich}, mean quantities were collected after 23 flow-through times and over a time period of 55T. The time to start averaging and the averaging period for PANS calculations depend on the cut-off length scale; the lower the cut-off ratio, the higher averaging period is required to obtain the steady flow statistics. It is also important to note that the mean flow properties are also averaged in the spanwise direction. A summary of the various test cases simulated in this study are given in Table \ref{case} which details the range of $f_k$ values along with various grid resolutions investigated in this study.

OpenFoam \cite{openfoam}, an open source finite volume code written in C++, is used to solve the equations. Out of the many available solvers in OpenFOAM, an incompressible transient solver, ChannelFOAM, is used with a second order accurate spatial discretization scheme. The Backward time scheme, which is second order accurate, is applied for time integration.

\begin{table}[H]
\centering
\caption{Details of the test cases simulated}
\begin{tabular}{ l l l l l}
\hline\noalign{\smallskip}
\textbf{Study} & \textbf{$f_k$} & \textbf{$f_\epsilon$} & \textbf{Grid} & \textbf{Averaging Period}\\ \hline\noalign{\smallskip}
\textbf{Re=10590} \\ \hline\noalign{\smallskip}
$f_k$ study & 0.35 & 1 & $150\times100\times60$ & 10T-15T \\ \hline\noalign{\smallskip}
$f_k$ study & 0.25 & 1 & $150\times100\times60$ & 10T-24T \\ \hline\noalign{\smallskip}
$f_k$ study & 0.15 & 1 & $150\times100\times60$ & 18T-36T \\ \hline\noalign{\smallskip}
LES & - & - & $196\times128\times186$ & 23T-55T \\ \hline\noalign{\smallskip}

\textbf{Re=37000} \\ \hline\noalign{\smallskip}
$f_k$ study & 1 & 1 & $150\times100\times60$ & 10T-15T \\ \hline\noalign{\smallskip}
$f_k$ study & 0.35 & 1 & $150\times100\times60$ & 10T-15T \\ \hline\noalign{\smallskip}
$f_k$ study & 0.25 & 1 & $150\times100\times60$ & 10T-24T \\ \hline\noalign{\smallskip}
$f_k$ study & 0.15 & 1 & $150\times100\times60$ & 18T-36T \\ \hline\noalign{\smallskip}
Resolution study & 0.35 & 1 & $150\times100\times60$ & 10T-15T \\ \hline\noalign{\smallskip}
Resolution study & 0.35 & 1 & $100\times100\times30$ & 10T-15T \\ \hline\noalign{\smallskip}
Resolution study & 0.15 & 1 & $150\times100\times60$ & 18T-36T \\ \hline\noalign{\smallskip}
Resolution study & 0.15 & 1 & $100\times100\times30$ & 18T-36T \\ \hline\noalign{\smallskip}
PITM (Fine Grid) & - & - & $160\times100\times60$ & - \\ \hline\noalign{\smallskip}
PITM (Coarse Grid) & - & - & $80\times100\times30$ & - \\ \hline\noalign{\smallskip}
\end{tabular}
\label{case}
\end{table}

\subsection{Results}

The simulations are performed at two different Reynolds numbers of 10590 and 37000. From literature, LES results are only available for Reynolds number up to 10590. Therefore, the first part of this section is aimed at comparing PANS against LES and experimental data at Re=10590. For Re=37000, the study has been carried out to address the model performance at different physical and computational resolutions. For this case, PANS results are compared against experimental data \cite{Rapp} and PITM simulations \cite{PITM}. Results for RANS is also presented whenever possible. 

\subsubsection{Internal consistency criteria}

In the PANS approach, the degree of resolution is determined by $f_k$ which is the purported function of the modeled kinetic energy. Simulations are performed for a specific $f_k$ field. After the simulation, it is possible to compute the actual fraction of the modeled eddy viscosity:

\begin{equation}
{\left(f_{k}^{R}\right)}^2=\frac{\nu_u}{\nu_t}
\label{eqn:fkr}
\end{equation}
where $\nu_u$ is the sub-filter eddy viscosity and $\nu_t$ is the total eddy viscosity. 

The "a posteriori" fraction, $\frac{\nu_u}{\nu}$ is given the name, recovered eddy viscosity fraction. If the closure model performance is consistent with flow physics, then the recovered value must be close to $f_k^2$. Comparison of specified $f_k^2$ and recovered $\frac{\nu_u}{\nu}$ serves as an important internal consistency test of the model performance. We will perform this consistency test whenever possible.

\begin{figure}[H]
\centering
\begin{subfigure}{.5\textwidth}
  \centering
  \includegraphics[scale=0.28]{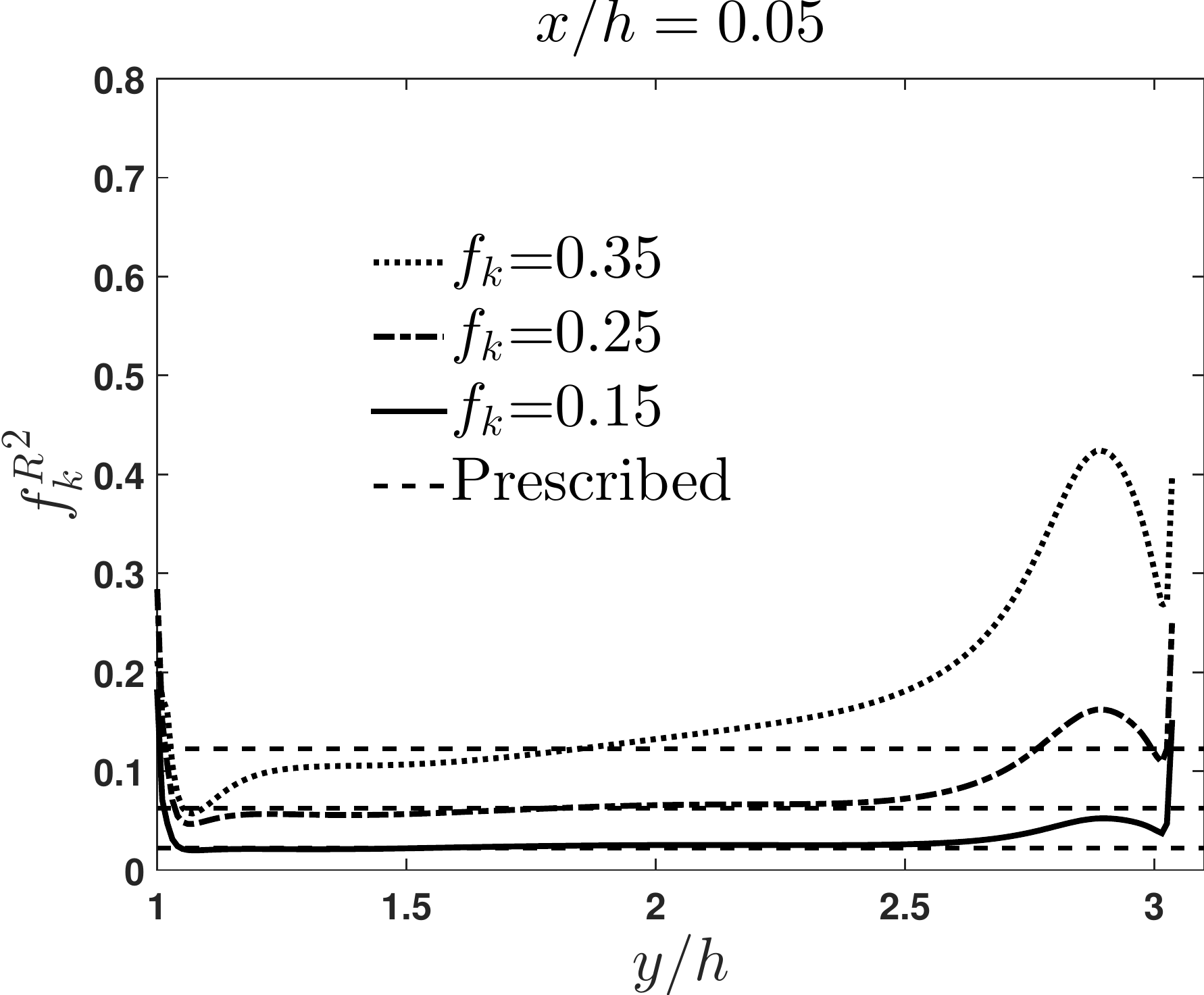}
  \vspace{-8pt}
  \caption{}
\end{subfigure}%
\begin{subfigure}{.5\textwidth}
  \centering
  \includegraphics[scale=0.28]{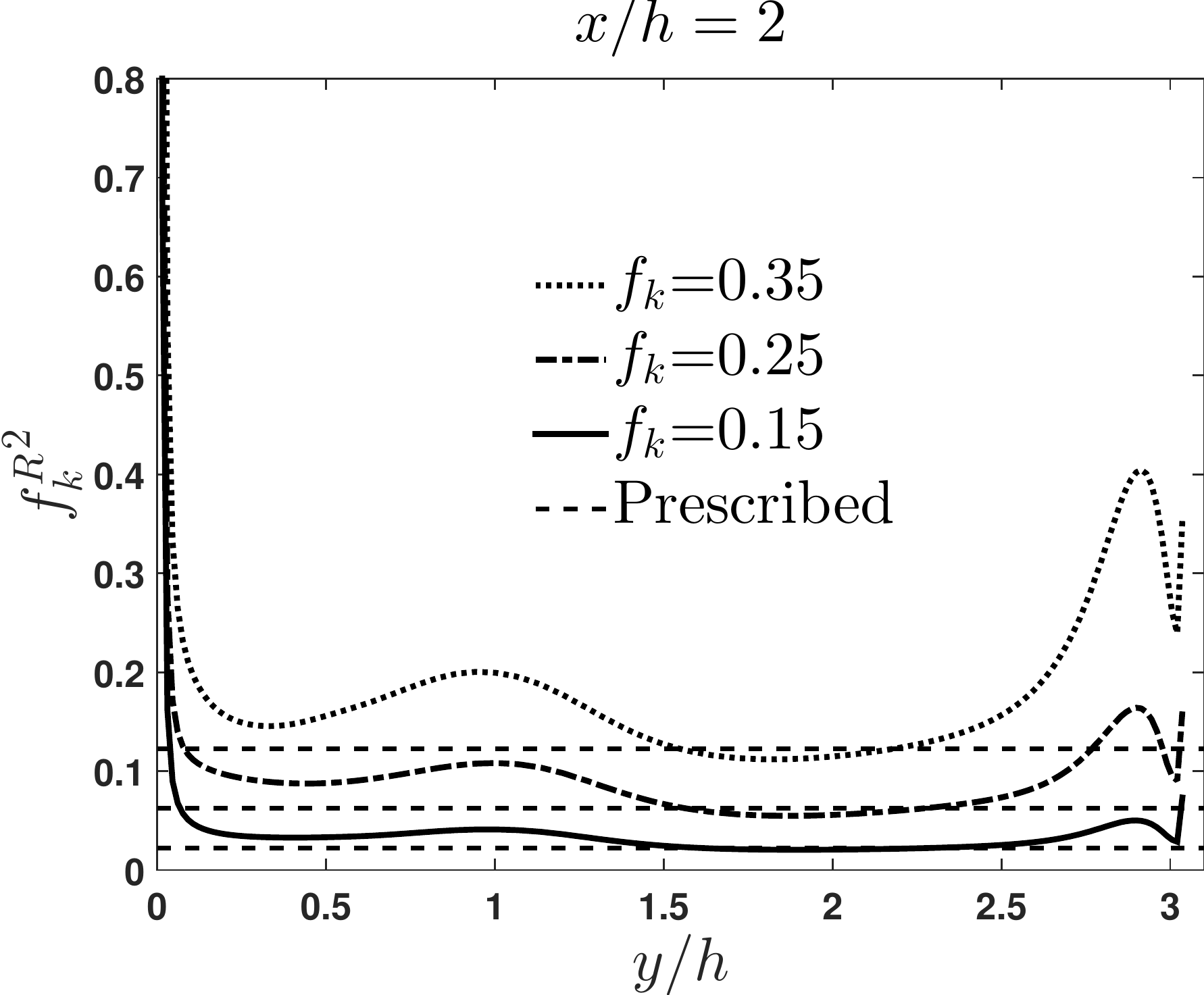}
  \vspace{-8pt}
  \caption{}
\end{subfigure}
\\
\begin{subfigure}{.5\textwidth}
   \centering
   \includegraphics[scale=0.28]{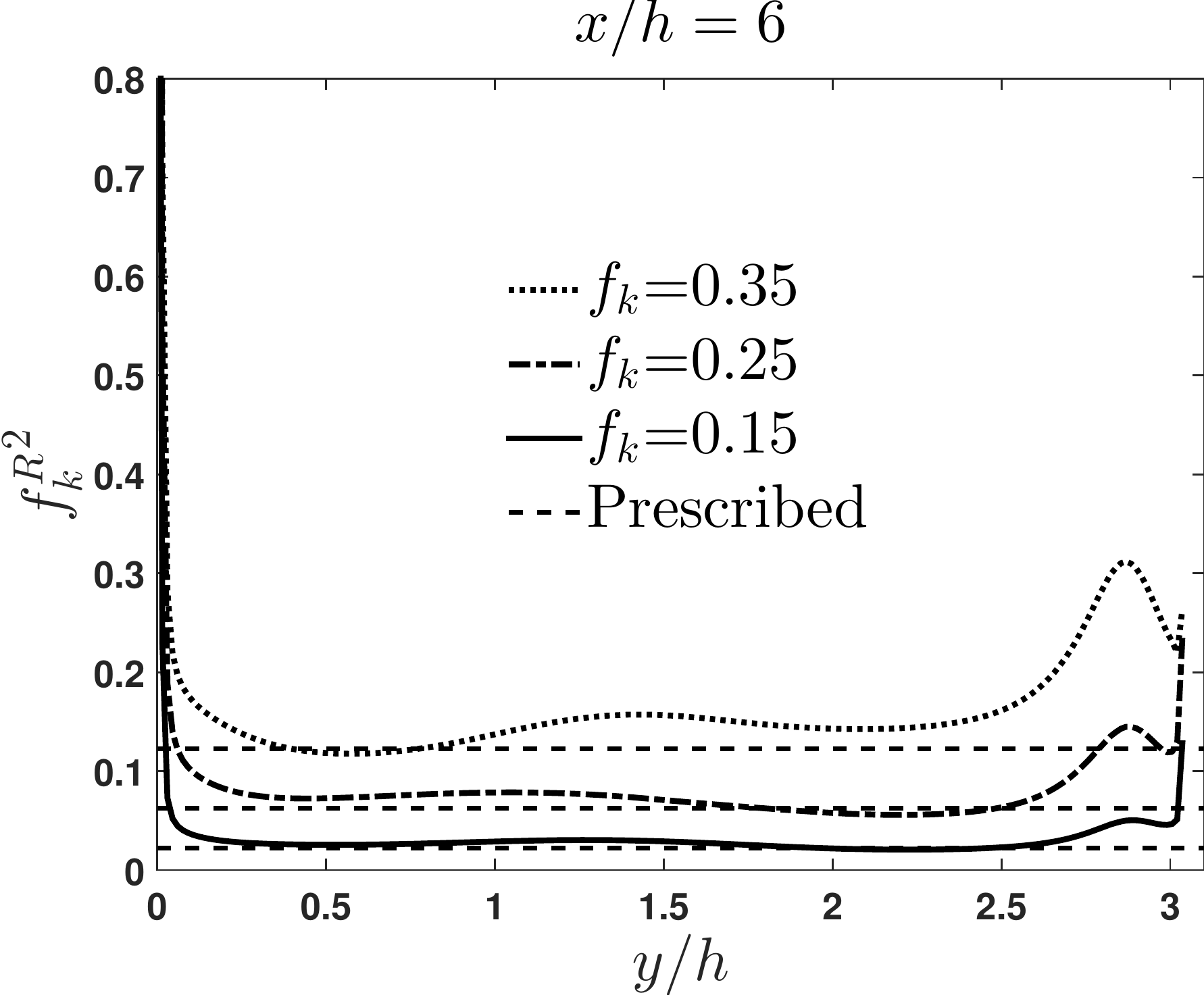}
   \vspace{-8pt}
   \caption{}
\end{subfigure}%
\begin{subfigure}{.5\textwidth}
   \centering
   \includegraphics[scale=0.28]{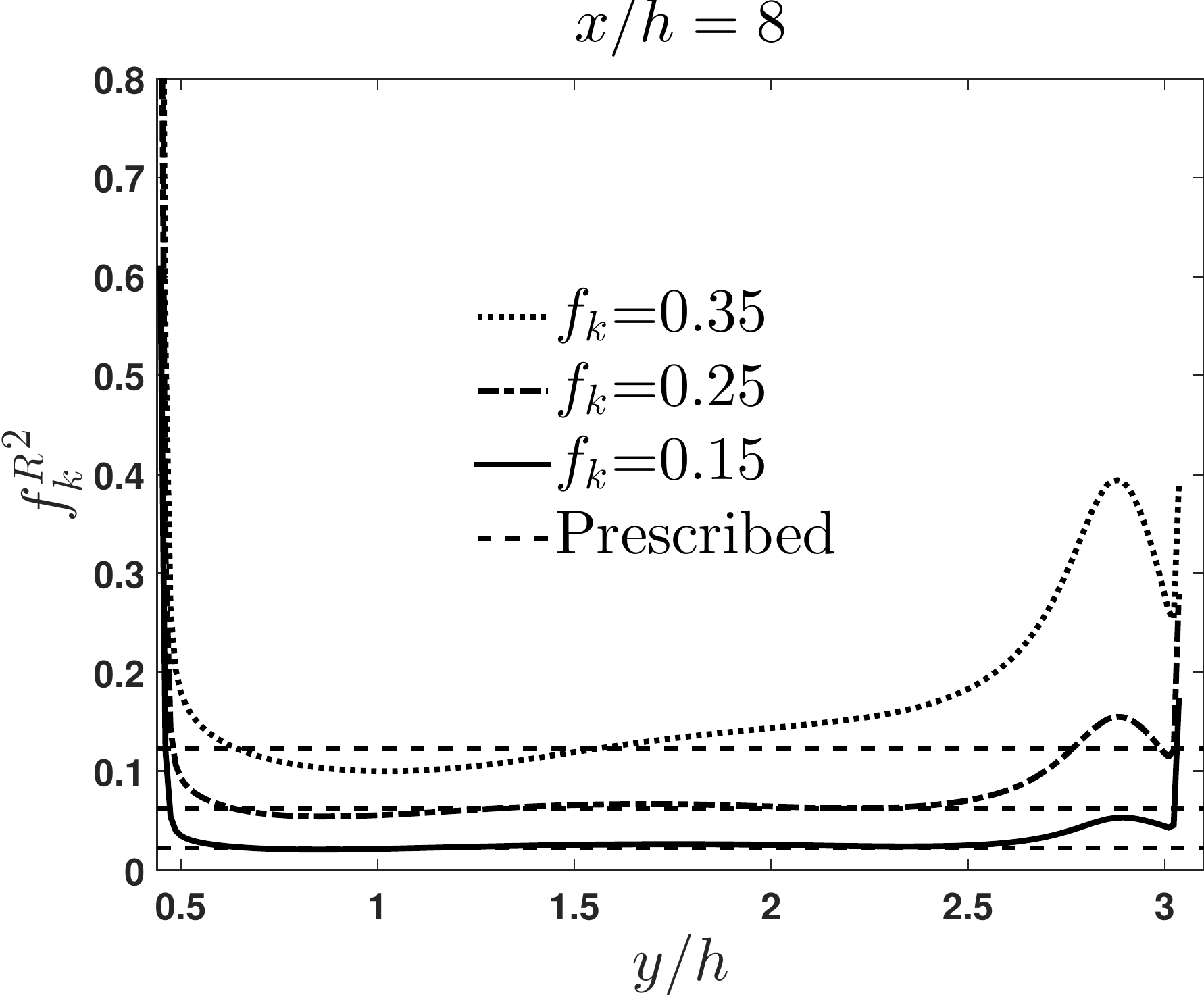}
   \vspace{-8pt}
   \caption{}
\end{subfigure}
\vspace{-8pt}
\caption{Recovery of the $f_k^2$ at different locations for Re=10590}
\label{fkr}
\end{figure}

Variation of $\nu_u/\nu$ along the normal direction at streamwise locations of $x/h$=0.05, 2, 6 and 8 are shown in Fig. \ref{fkr} for the PANS simulations with input $f_k$ values of 0.35, 0.25 and 0.15 at $Re$=10590. The prescribed $f_k^2$ is also shown in this plot for comparison. This figure indicates that the contribution of modelled stresses is reduced as $f_k$ decreases. The recovery of the filter parameter is well observed for almost the entire domain particularly in the middle region away from the walls for the PANS simulations. The poor recovery of the cut-off parameter close to the walls is because of the fact that the present PANS model is developed for the fully turbulent region and the low Reynolds number effects are not included in the equations. 

\subsubsection{Re=10590}

PANS simulations with $f_{k}$=0.35, 0.25 and 0.15 are performed on 0.9 million grid nodes for Re=10590 to obtain the flow statistics for mean velocity and turbulence quantities. The results are then compared against LES calculation on 4.7 million grid nodes \cite{Frohlich} and experimental data \cite{Rapp}. It is worth mentioning that that all the simulations are performed with second order accuracy in time and space. However, both RANS and PANS calculations are performed on a grid size which is around 5 times coarser than the LES grid.

\paragraph*{\textbf{Variation of $f_k$ study}}

\begin{figure}[H]
\centering
\begin{subfigure}{.5\textwidth}
  \centering
  \includegraphics[scale=0.3]{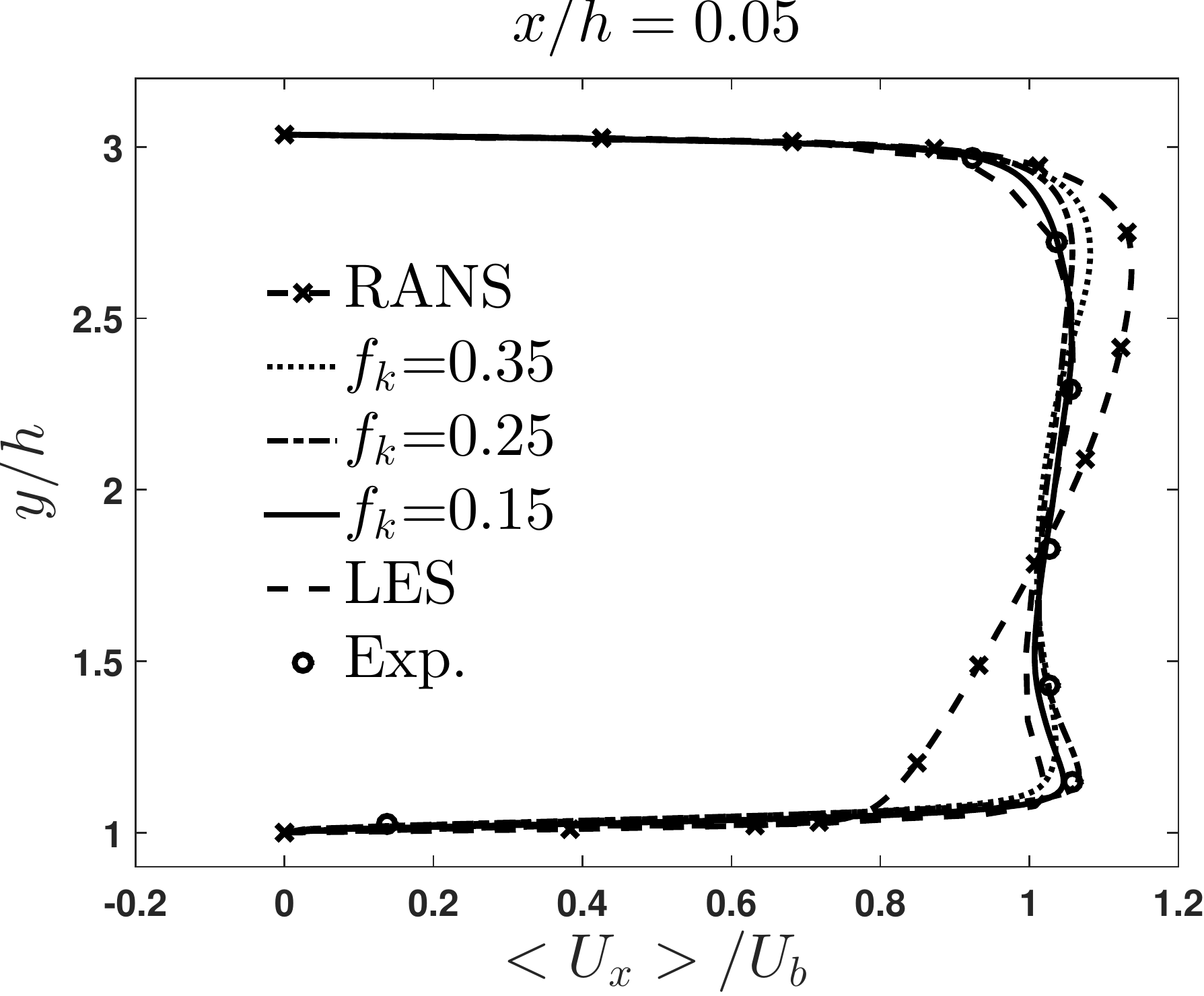}
  \vspace{-8pt}
  \caption{}
\end{subfigure}%
\begin{subfigure}{.5\textwidth}
  \centering
  \includegraphics[scale=0.3]{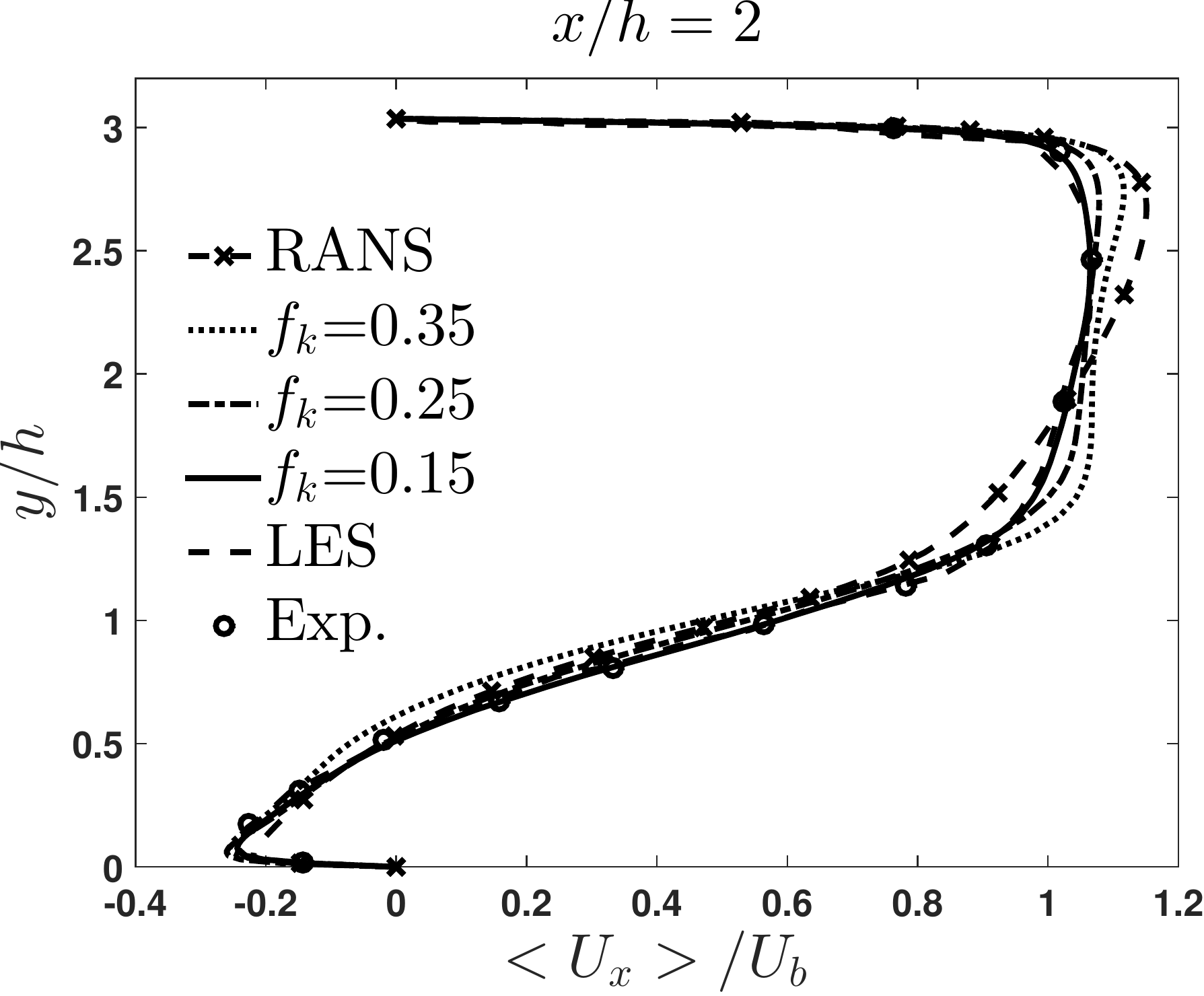}
  \vspace{-8pt}
  \caption{}
\end{subfigure}
\\
\begin{subfigure}{.5\textwidth}
   \centering
   \includegraphics[scale=0.3]{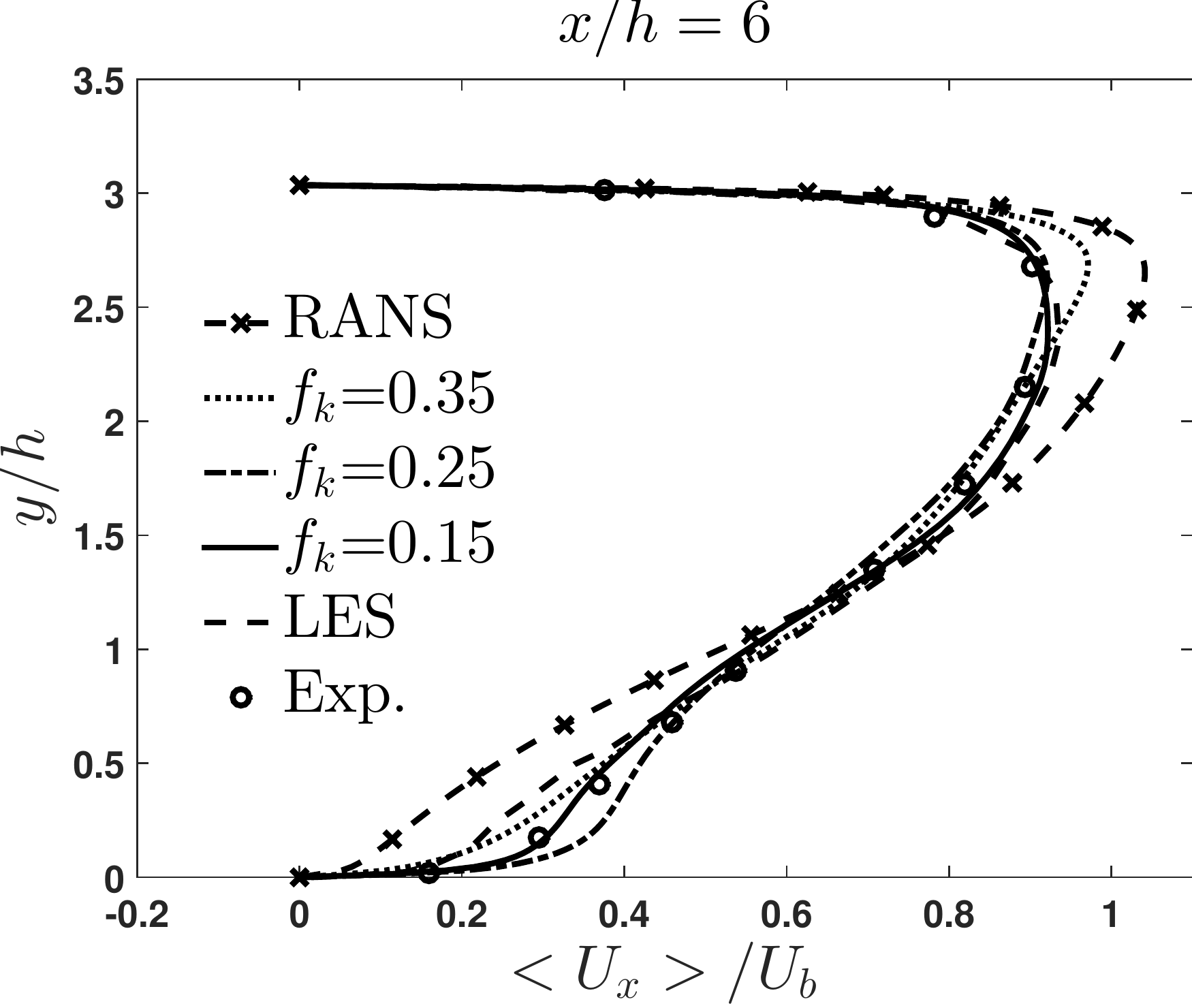}
   \vspace{-8pt}
   \caption{}
\end{subfigure}%
\begin{subfigure}{.5\textwidth}
   \centering
   \includegraphics[scale=0.3]{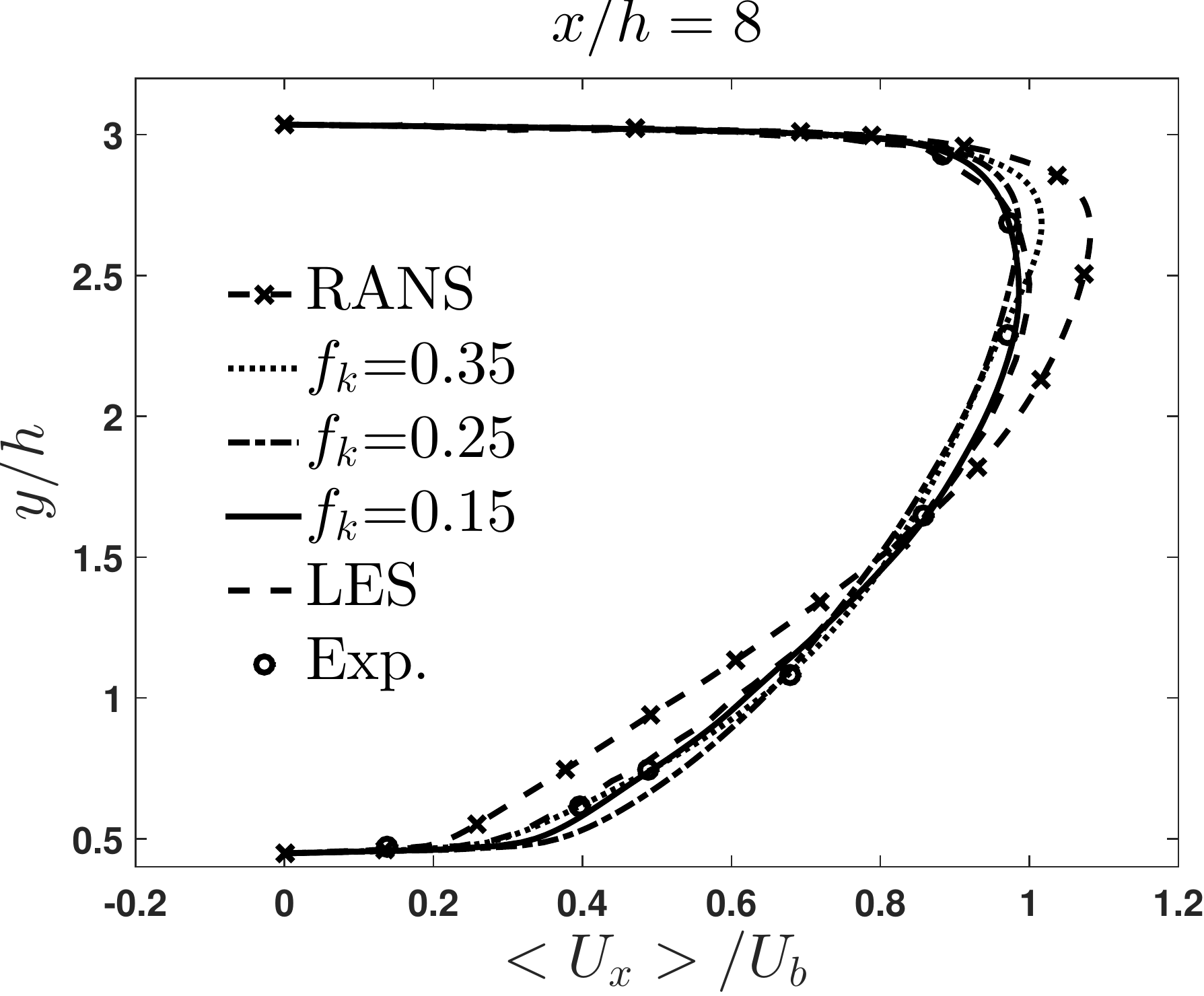}
   \vspace{-8pt}
   \caption{}
\end{subfigure}
\vspace{-8pt}
\caption{Streamwise velocity profiles at different locations for Re=10590}
   \label{ULES}
\end{figure}

Figure \ref{ULES} depicts the mean velocity profile at four streamwise locations of $x/h$=0.05, 2, 6 and 8. Based on the experimental study \cite{Rapp}, the selected positions are associated with the most important physics occurring in this flow configuration. At x/h=0.05, there is a peak in near wall streamwise velocity which is attributed to the flow acceleration towards the windward slope of the hill. As can be seen from Fig. \ref{ULES}a, RANS model completely fails to predict the flow acceleration near the wall, whereas PANS results at all of the selected $f_k$ values are in good agreement with experimental data. 

The next location, x/h=2, is in the middle of recirculation zone where the boundary layer is detached and there is an interaction between the free shear layer separating from the hill crest and the reverse flow below that. Although the near wall velocity is accurately recovered by all turbulence models, poor prediction of the RANS simulation for streamwise velocity near the top boundary is seen in Fig. \ref{ULES}b. 
The flow is in the post reattachment region at x/h=6 where flow recovery from the low energy separated region is very well represented by the PANS simulation with $f_k$=0.15. At x/h=8, the flow is accelerated on the windward slope of the hill. This feature is again well captured by PANS simulations. As can be seen from Fig. \ref{ULES}d, the near wall velocity for RANS simulation is noticeably lower than the corresponding PANS values. 

Comparison of the PANS results with LES data reveals that prediction of mean velocity profile for the PANS simulation with $f_k$=0.15 is superior than the LES study for the entire streamwise locations. However, deviation of the PANS results from experimental data with other $f_k$ values is visible in the vicinity of the upper wall, while LES calculation is close to the experimental values almost over the entire domain. Application of appropriate wall function close to the upper wall in LES calculation along with no wall treatments for the PANS simulation could be responsible for such a behaviour. 

\begin{figure}[H]
\centering
\begin{subfigure}{.5\textwidth}
  \centering
  \includegraphics[scale=0.3]{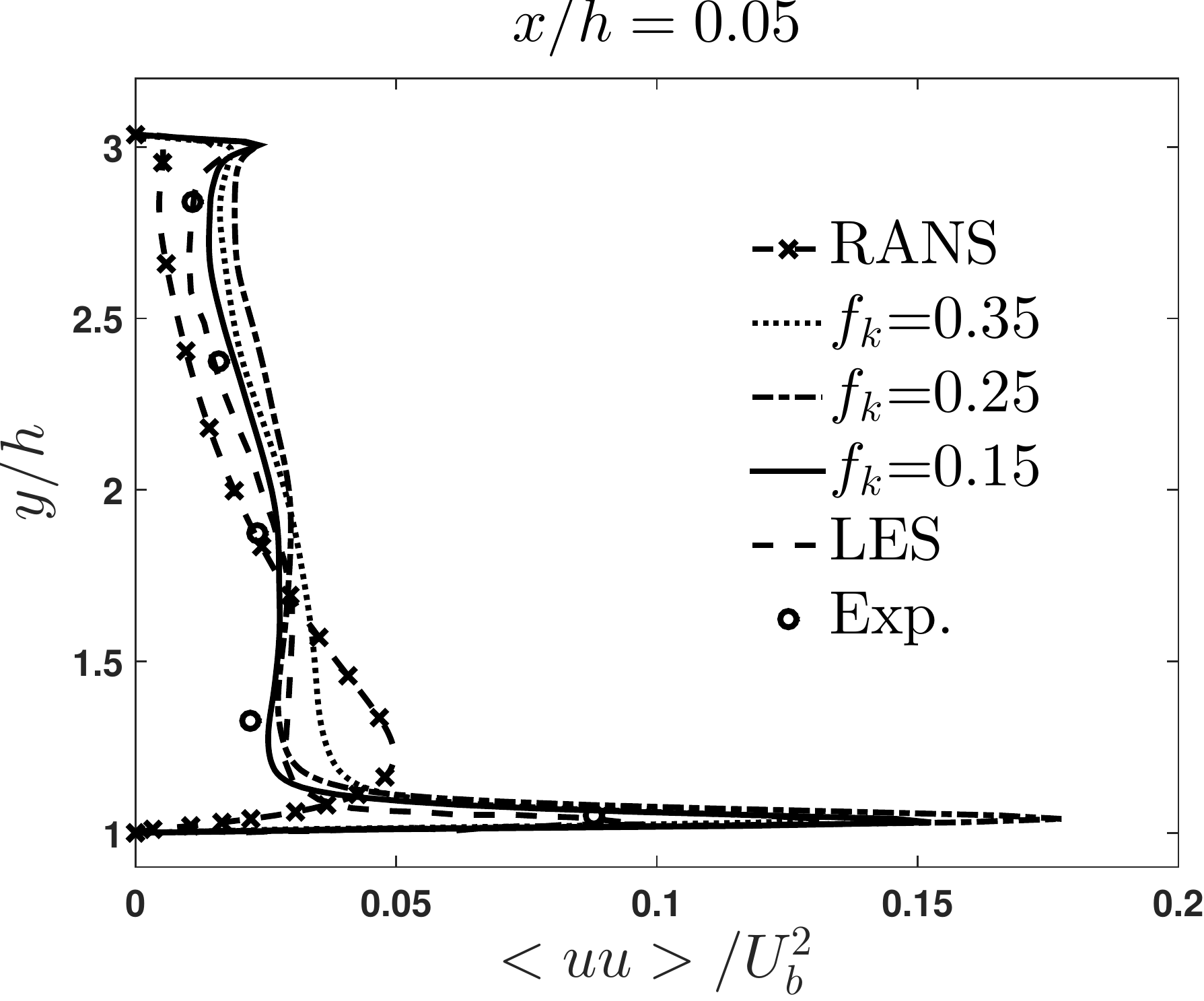}
  \vspace{-8pt}
  \caption{}
\end{subfigure}%
\begin{subfigure}{.5\textwidth}
  \centering
  \includegraphics[scale=0.3]{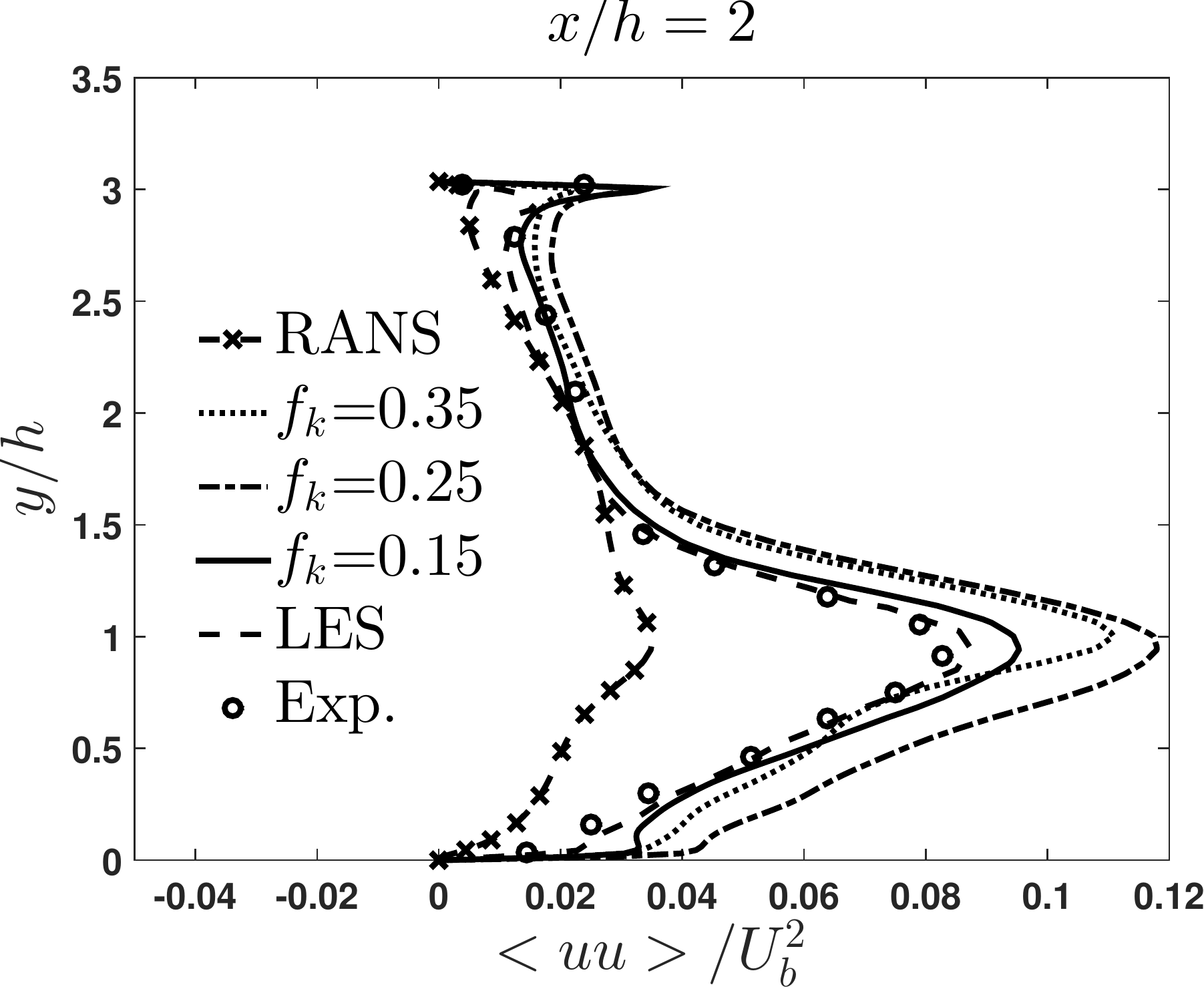}
  \vspace{-8pt}
  \caption{}
\end{subfigure}
\\
\begin{subfigure}{.5\textwidth}
   \centering
   \includegraphics[scale=0.3]{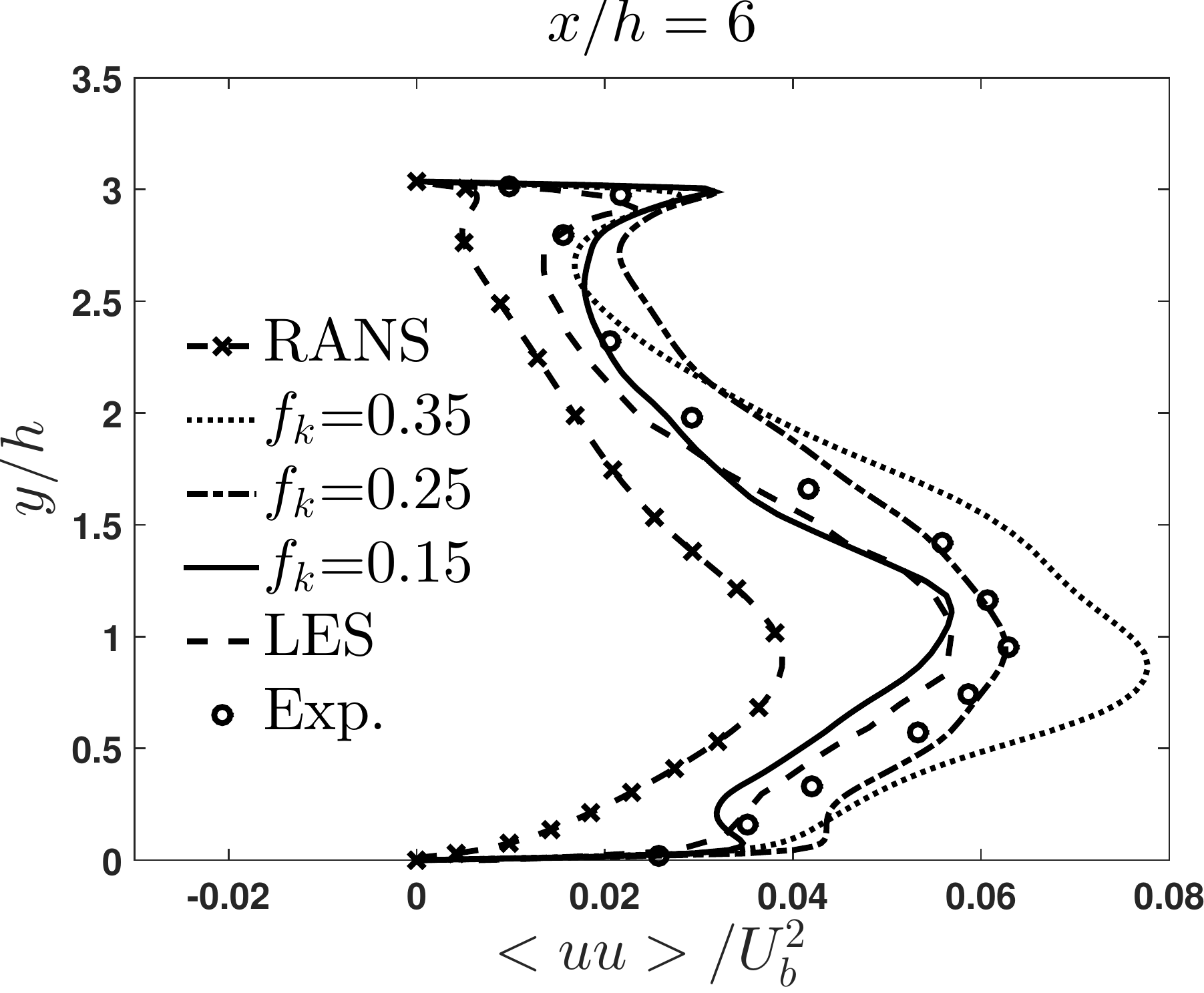}
   \vspace{-8pt}
   \caption{}
\end{subfigure}%
\begin{subfigure}{.5\textwidth}
   \centering
   \includegraphics[scale=0.3]{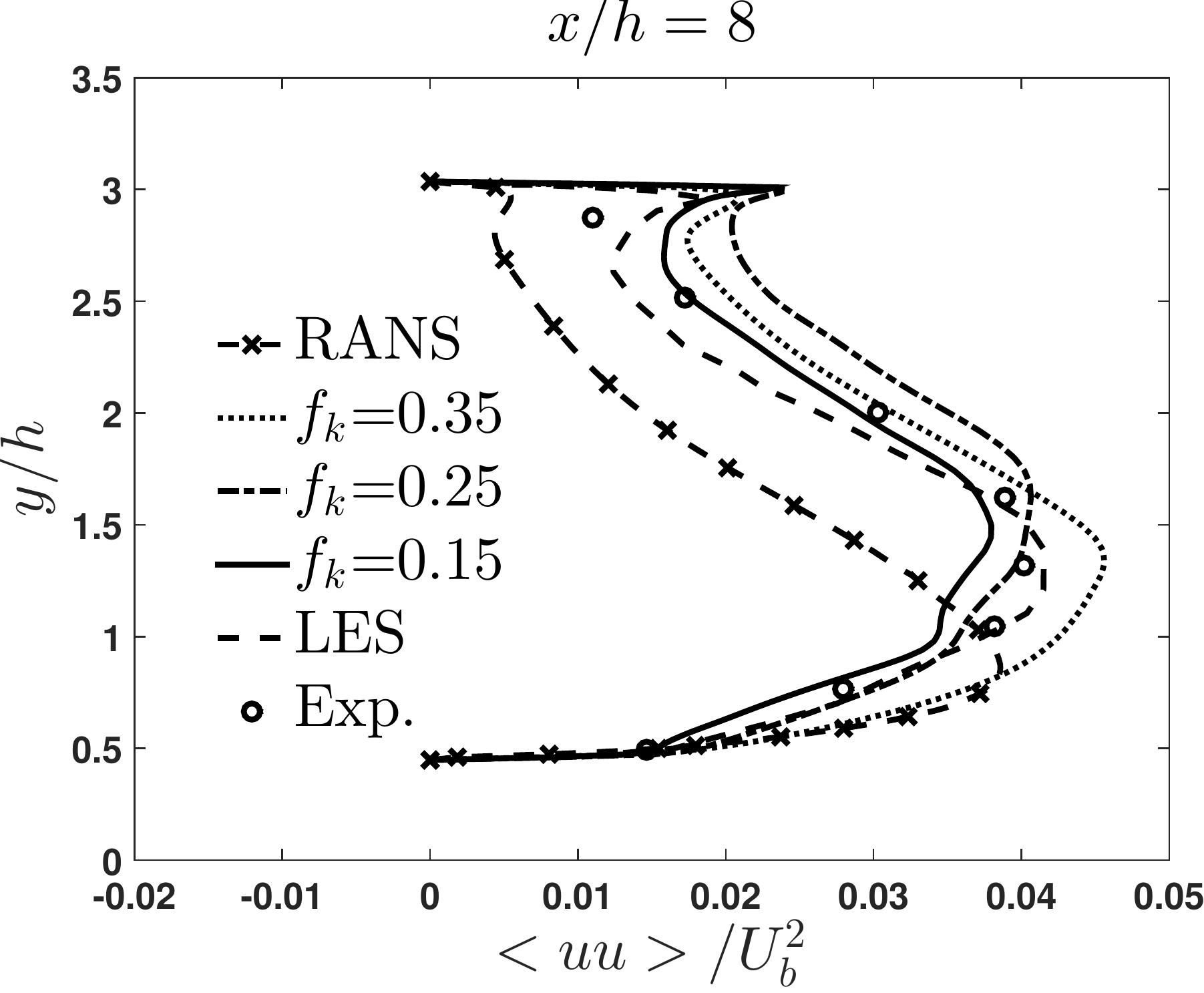}
   \vspace{-8pt}
   \caption{}
\end{subfigure}
\vspace{-8pt}
\caption{Streamwise stress profiles at different locations for Re=10590}
   \label{UULES}
\end{figure}

\begin{figure}[H]
\centering
\begin{subfigure}{.5\textwidth}
  \centering
  \includegraphics[scale=0.3]{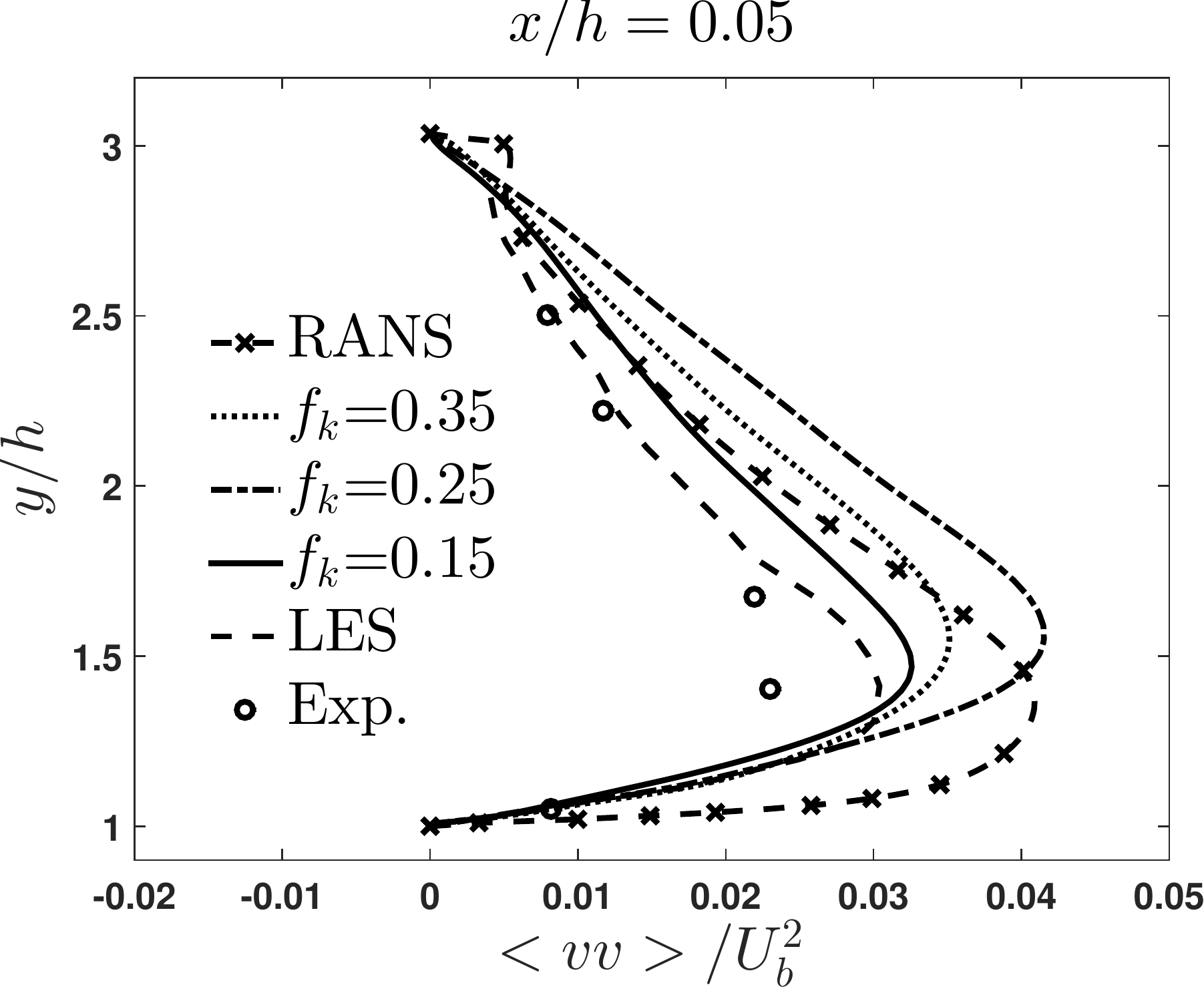}
  \vspace{-8pt}
  \caption{}
\end{subfigure}%
\begin{subfigure}{.5\textwidth}
  \centering
  \includegraphics[scale=0.3]{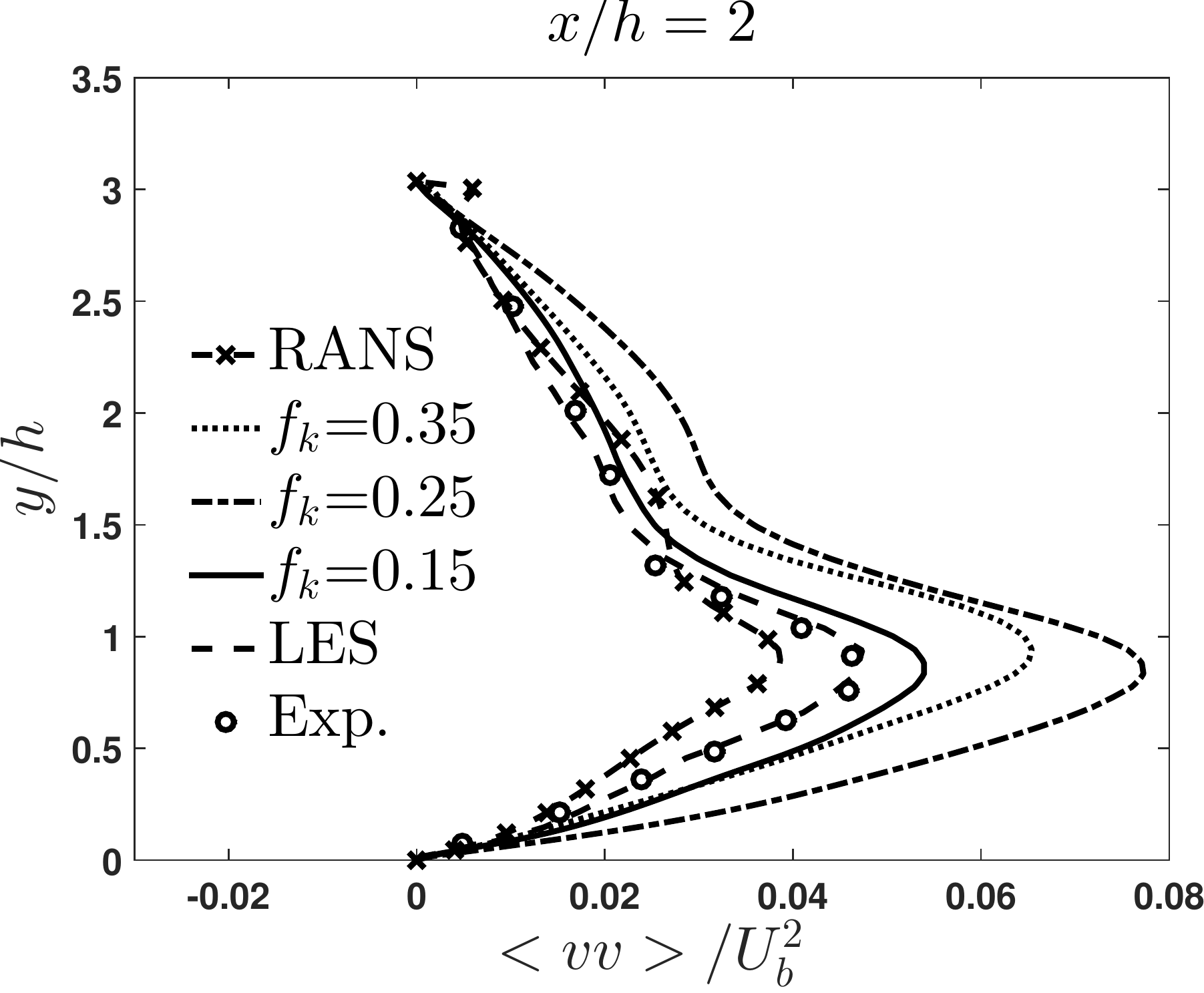}
  \vspace{-8pt}
  \caption{}
\end{subfigure}
\\
\begin{subfigure}{.5\textwidth}
   \centering
   \includegraphics[scale=0.3]{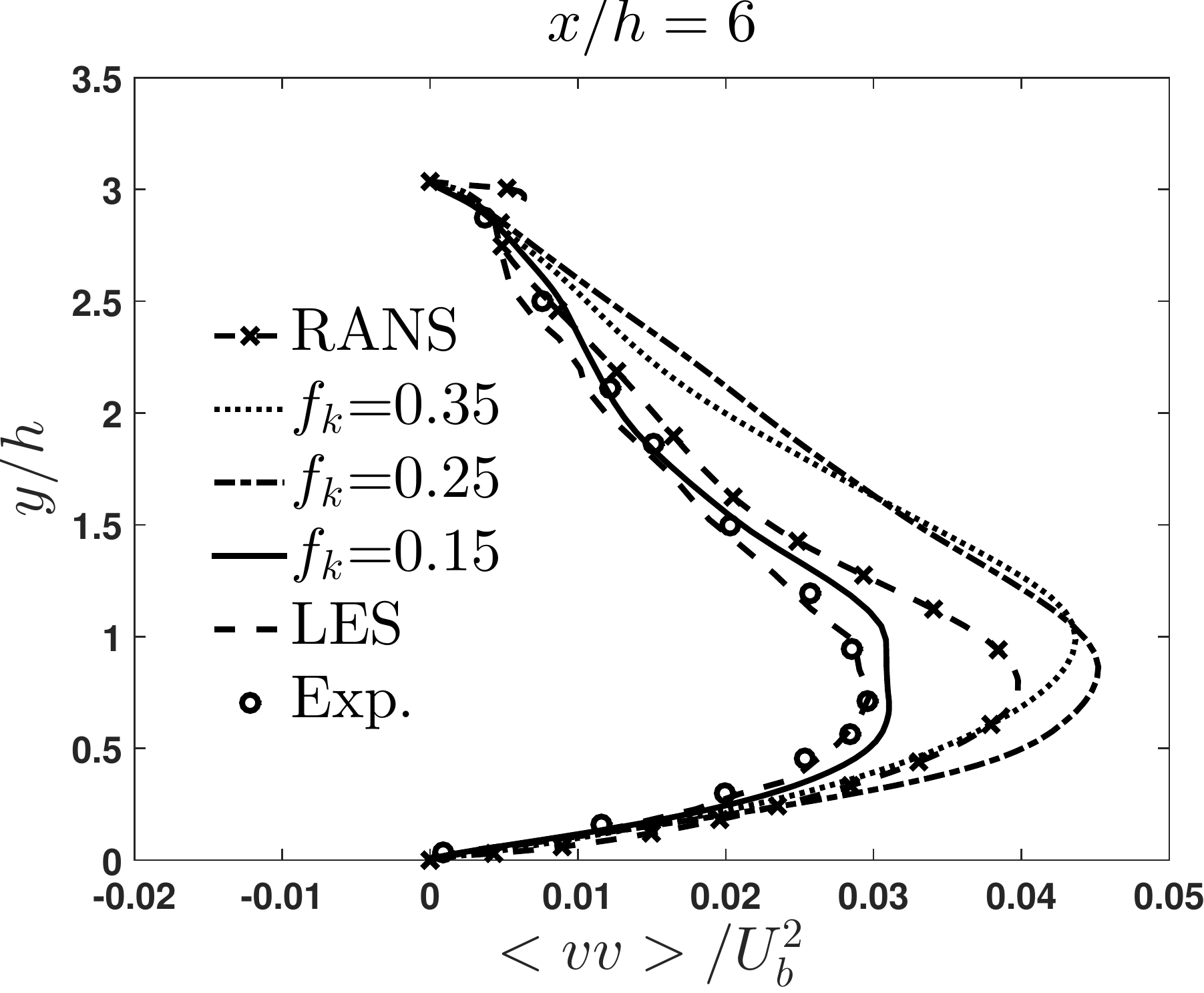}
   \vspace{-8pt}
   \caption{}
\end{subfigure}%
\begin{subfigure}{.5\textwidth}
   \centering
   \includegraphics[scale=0.3]{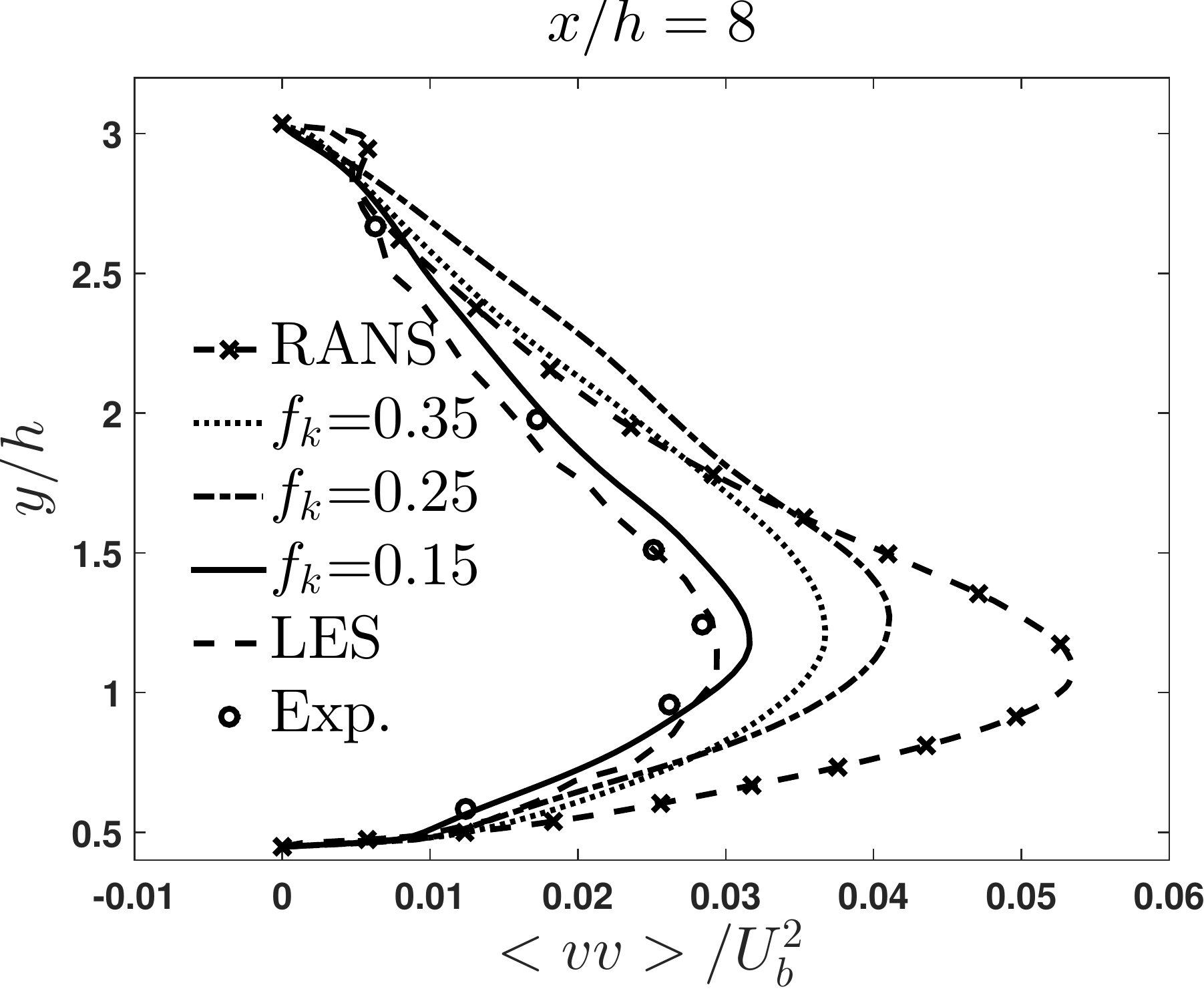}
   \vspace{-8pt}
   \caption{}
\end{subfigure}
\vspace{-8pt}
\caption{Normal stress profiles at different locations for Re=10590}
   \label{VVLES}
\end{figure}

\begin{figure}[H]
\centering
\begin{subfigure}{.5\textwidth}
  \centering
  \includegraphics[scale=0.3]{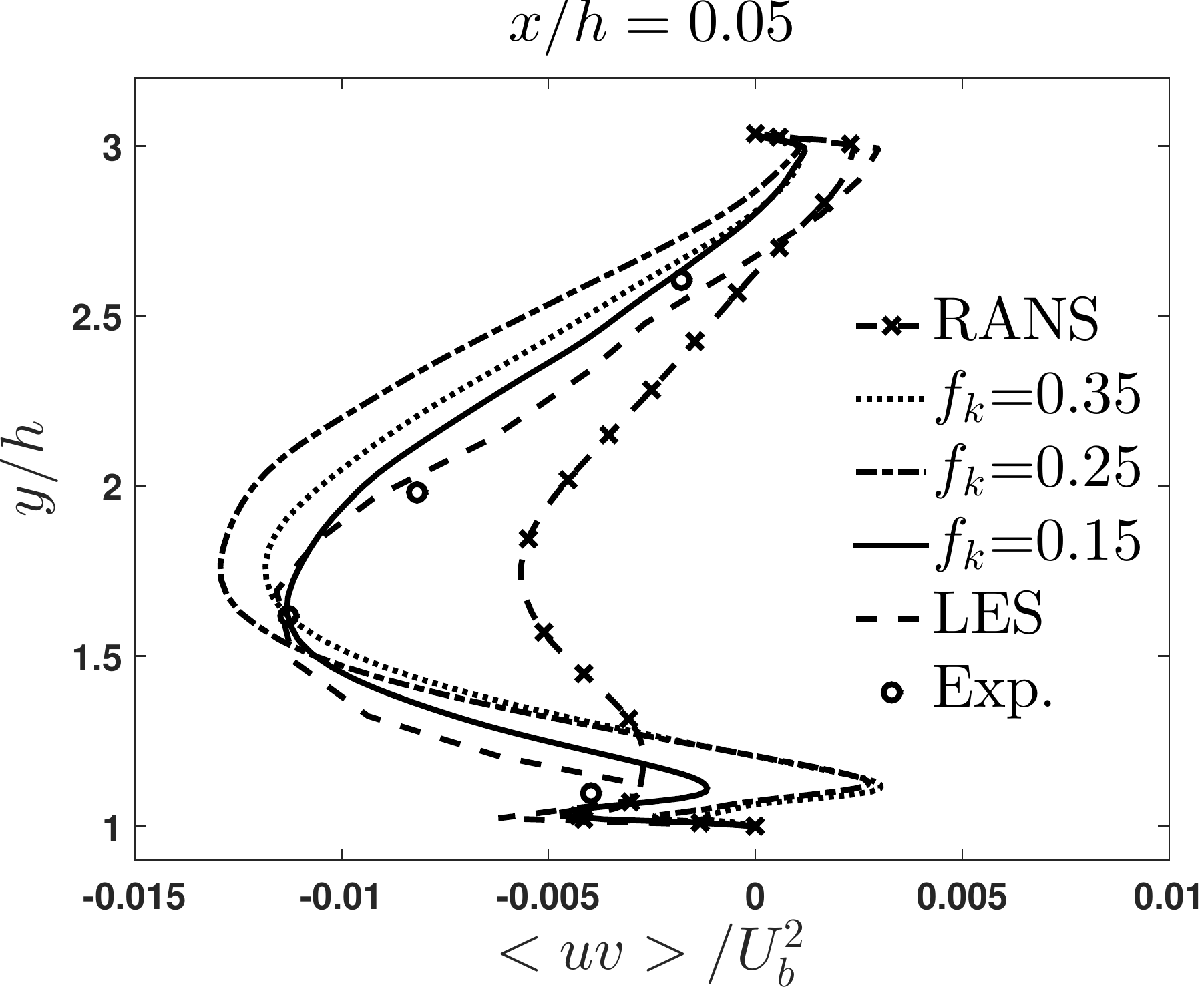}
  \vspace{-8pt}
  \caption{}
\end{subfigure}%
\begin{subfigure}{.5\textwidth}
  \centering
  \includegraphics[scale=0.3]{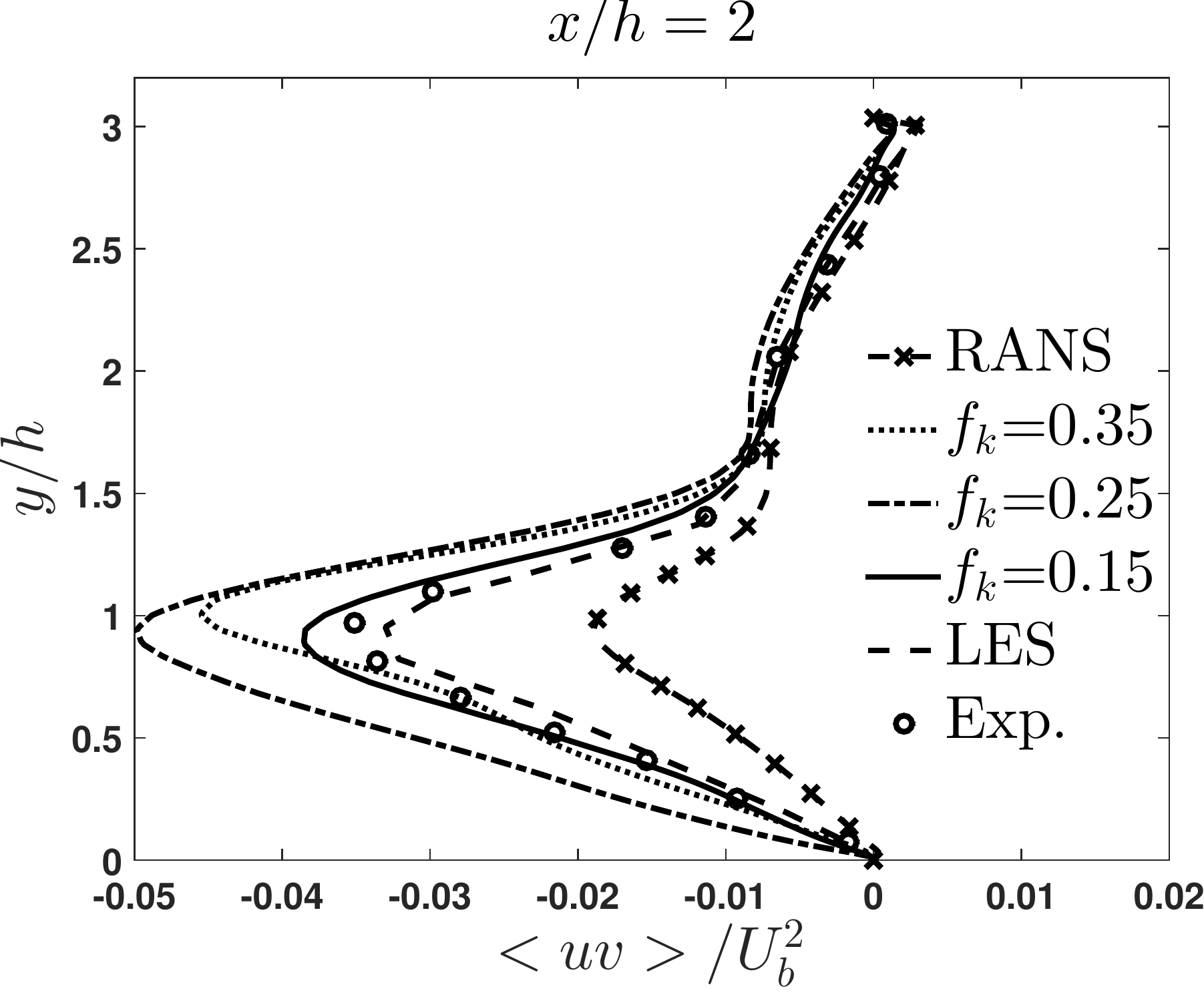}
  \vspace{-8pt}
  \caption{}
\end{subfigure}
\\
\begin{subfigure}{.5\textwidth}
   \centering
   \includegraphics[scale=0.3]{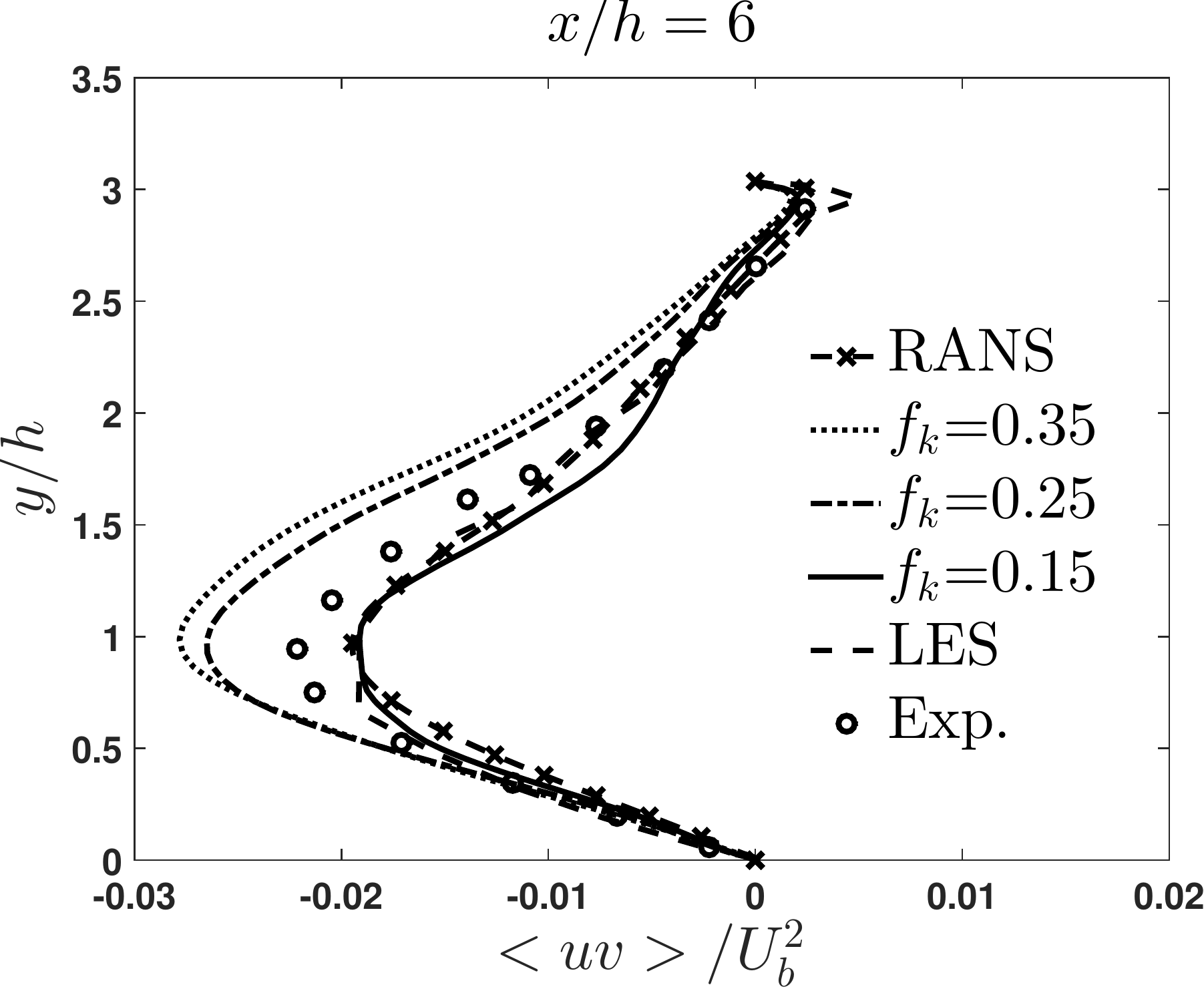}
   \vspace{-8pt}
   \caption{}
\end{subfigure}%
\begin{subfigure}{.5\textwidth}
   \centering
   \includegraphics[scale=0.3]{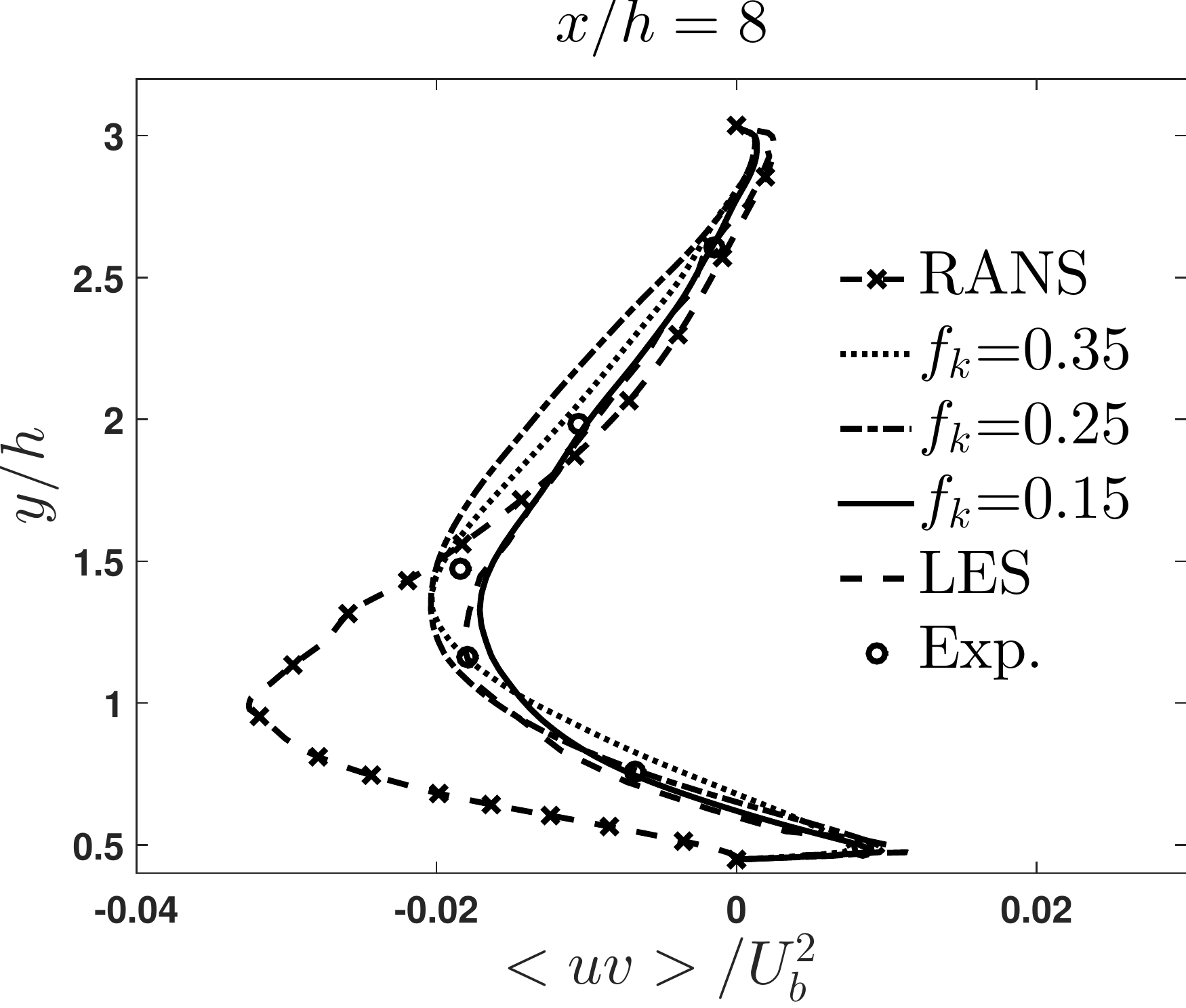}
   \vspace{-8pt}
   \caption{}
\end{subfigure}
\vspace{-8pt}
\caption{Shear stress profiles at different locations for Re=10590}
   \label{UVLES}
\end{figure}

Interpreting the behaviour of turbulence stresses provides a substantial aid to model validation and development. Figures \ref{UULES}-\ref{UVLES} show the streamwise, vertical normal stresses and shear stress profiles at the same successive locations, respectively. The total stress is computed as the sum of resolved and modeled stresses. It is apparent from Figs. \ref{UULES}-\ref{UVLES} that the statistics obtained by the RANS model although following the shape of the profiles, they highly deviates from the reference data at most streamwise locations of the channel. Overall, good quantitative agreement is observed for the stress profiles returned by the PANS simulation considering that the calculations are performed on a much coarser grid than LES. Specifically, the stress components predicted by the PANS simulation with $f_k$=0.15 is in the same order of accuracy compared to the LES data. Notable deviation of the PANS results with $f_k$=0.25 and 0.35 compared to LES and experiment is seen for wall-normal stress profiles. This is because the wall-normal statistics has lower values than the streamwise counterpart and therefore they are more sensitive to the numerical accuracy and grid resolution. Besides, it was seen from figure \ref{ULES} that the streamwise velocity profile is over-predicted by the PANS calculations with $f_k$=0.25 and 0.35 close to the upper wall. This can influence flow development in the channel which is controlled by the bulk velocity at the hill crest. Nevertheless, the outcome of the present PANS calculations for averaged streamwise velocity and stress profiles is fully satisfactory. 

\paragraph*{\textbf{Flow anisotropy}}

The value of this section lies in its usefulness for evaluating PANS model for resolving anisotropy of the flow.   
The Reynolds stress anisotropy tensor is defined by

\begin{equation}
b_{ij}=\frac{<u_iu_j>}{<u_ku_k>}-\frac{1}{3}\delta_{ij}
\end{equation}
\label{b}

Since the trace of $b_{ij}$ is zero, the anisotropy tensor has two independent invariants which is preferred to be identified by parameters, $\xi$ and $\eta$ given by

\begin{equation}
\xi=\left(\frac{b_{ij}b_{jk}b_{ki}}{6}\right)^\frac{1}{3} \quad \eta=\left(-\frac{b_{ij}b_{ij}}{3}\right)^\frac{1}{2}
\label{eta}
\end{equation}

For periodic hill flow, $\xi$ and $\eta$ can be obtained from the total Reynolds stress consisting of modelled and resolved stresses. The states of the Reynolds stress tensor correspond to the specific points in the $\xi$-$\eta$ plane. As discussed by Lumley \& Newman \cite{lumley1977} and Lumley \cite{lumley1978}, the realizable states of turbulence lie within a curvilinear triangle in the $\xi$-$\eta$ plane. The limits of the triangular domain corresponds to axisymmetric contraction, axisymmetric expansion and two-component turbulence. 

\begin{figure}[H]
\centering
\begin{subfigure}{.5\textwidth}
  \centering
  \includegraphics[scale=0.3]{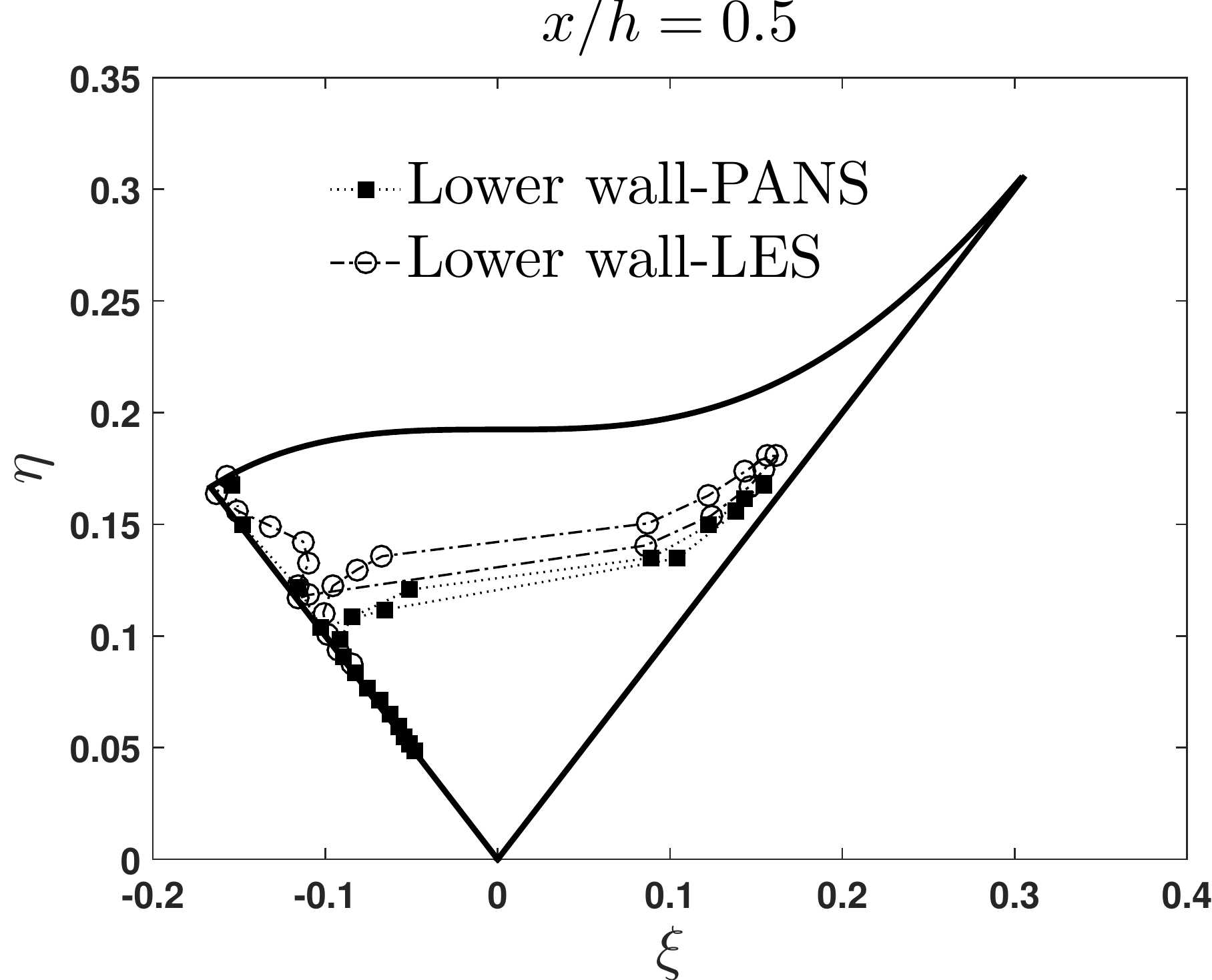}
  \vspace{-8pt}
  \caption{}
\end{subfigure}%
\begin{subfigure}{.5\textwidth}
  \centering
  \includegraphics[scale=0.3]{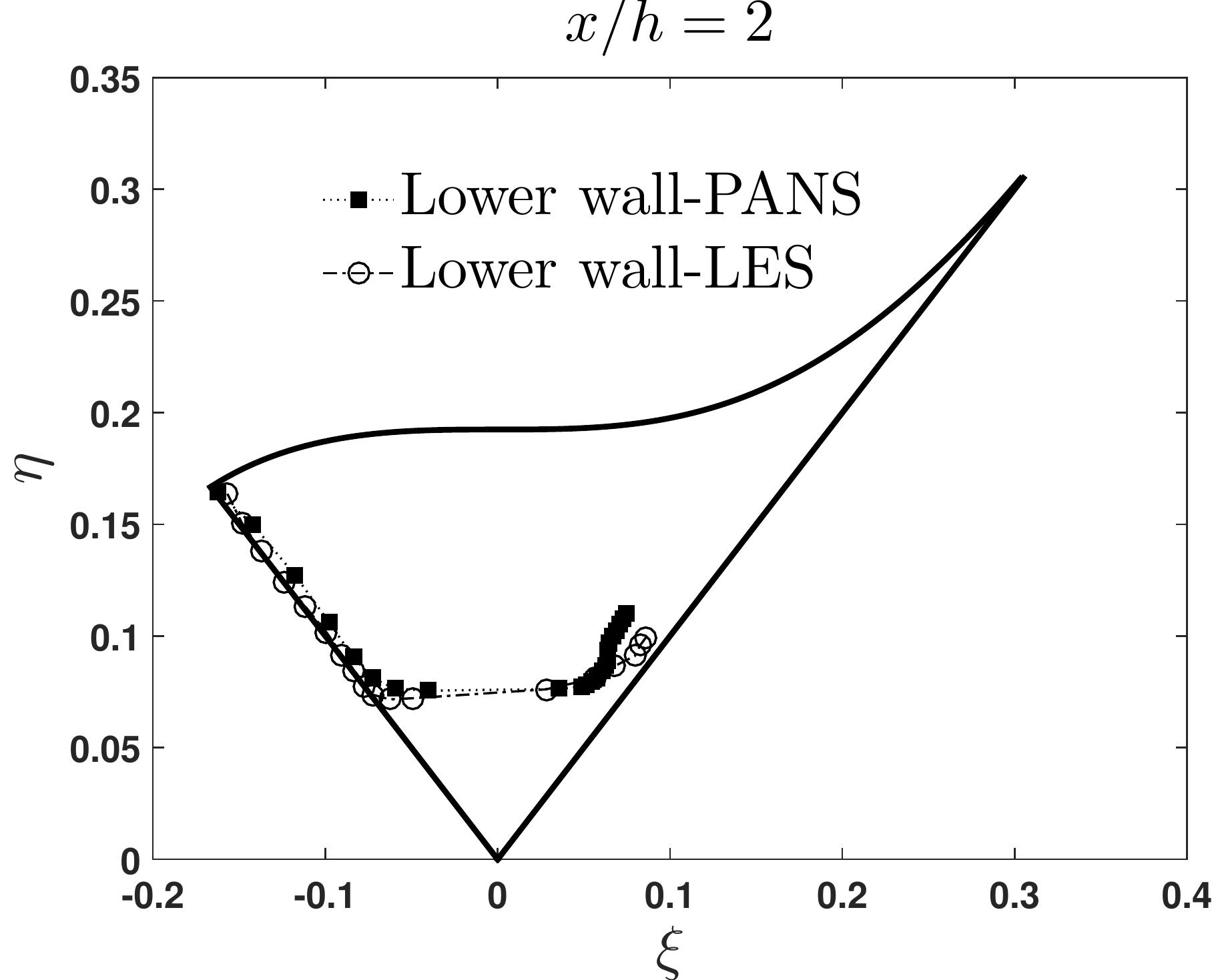}
  \vspace{-8pt}
  \caption{}
\end{subfigure}
\\
\begin{subfigure}{.5\textwidth}
   \centering
   \includegraphics[scale=0.3]{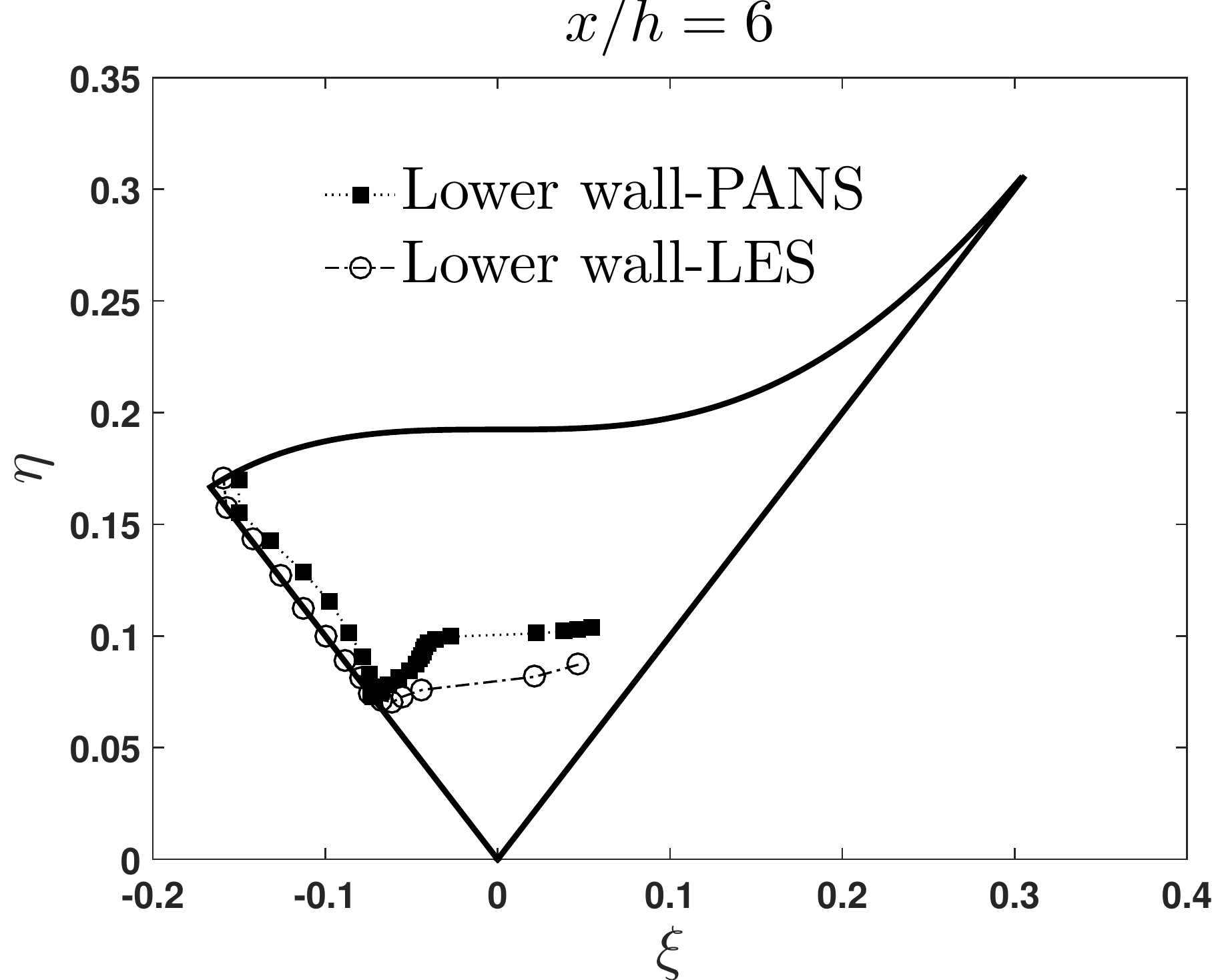}
   \vspace{-8pt}
   \caption{}
\end{subfigure}%
\begin{subfigure}{.5\textwidth}
   \centering
   \includegraphics[scale=0.3]{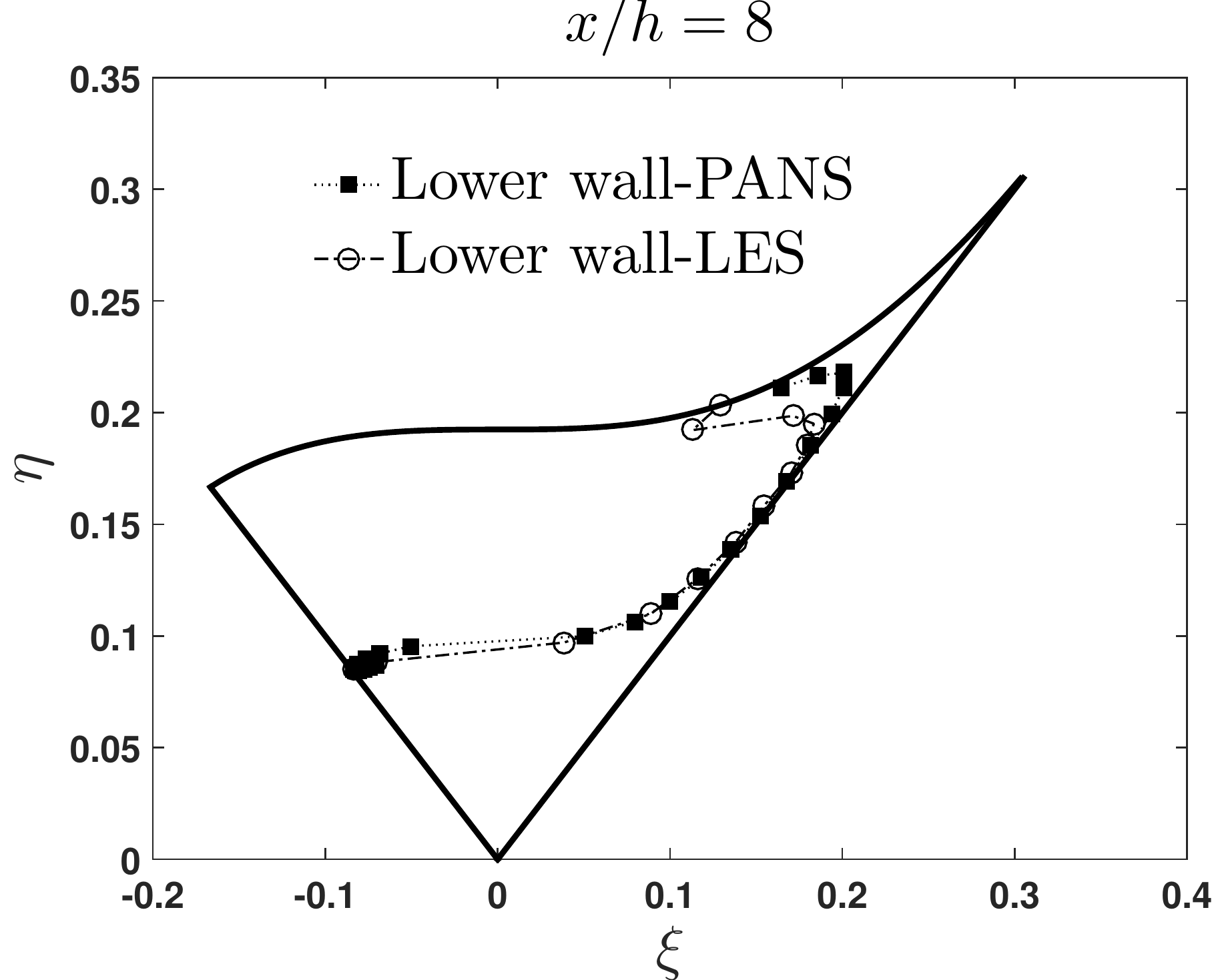}
   \vspace{-8pt}
   \caption{}
\end{subfigure}
\vspace{-8pt}
\caption{Invariant map along vertical direction at four streamwise locations for Re=10590}
   \label{Lumley}
\end{figure}

Figure \ref{Lumley} shows the flow anisotropy at several locations in the streamwise direction. By looking at this plot, important observations can be inferred. First, all the data for Reynolds stress tensor invariants are delimited by the Lumley triangle which proves that the realizability constraint is satisfied for PANS calculation at all the locations. Looking at the flow anisotropy at the region close to the bottom wall reveals that the flow becomes anisotropic as we move toward the wall. As seen from Fig. \ref{Lumley}, the near wall values for $\xi$ and $\eta$ are approaching the two-component turbulence state on the Lumley triangle. However, for the lower wall region, the way that two-component turbulence state is approached is different for separated region and far beyond the reattachment location. Figure \ref{Lumley} indicates that this way is passed through axisymmetric contraction for $x/h\leq$6, while at $x/h$=8, this approach has occurred along the axisymmetric expansion. This is consistent with LES findings \cite{Frohlich}.  

While the anisotropy level of the flow can be qualitatively observed by the Reynolds stress invariants through Lumley triangle, it can be quantitatively measured by calculating the flatness parameter. This parameter, proposed by Lumley combines the invariants of the Reynolds stress tensor reducing to the following equation for the flatness parameter

\begin{equation}
A=1+9\left(\frac{b_{ij}b_{jk}b_{ki}}{3}-\frac{b_{ij}b_{ij}}{2}\right) 
\label{A}
\end{equation}

Value of A goes to one for isotropic flows and it goes to zero at two-component turbulence state. Figure \ref{A} shows the flatness parameter for PANS and LES calculations at several streamwise locations. Fairly good agreement of the PANS data with LES simulation is observed at all the locations with the value of A around 0.8 in the central region. This plot further confirms that the flow is anisotropic near the wall, while in the core region becomes isotropic.  

This section demonstrates that PANS calculation satisfies the realizability condition and is able to anticipate anisotropy of the flow at different regions inside the channel.

\begin{figure}[H]
\centering
\begin{subfigure}{.5\textwidth}
  \centering
  \includegraphics[scale=0.3]{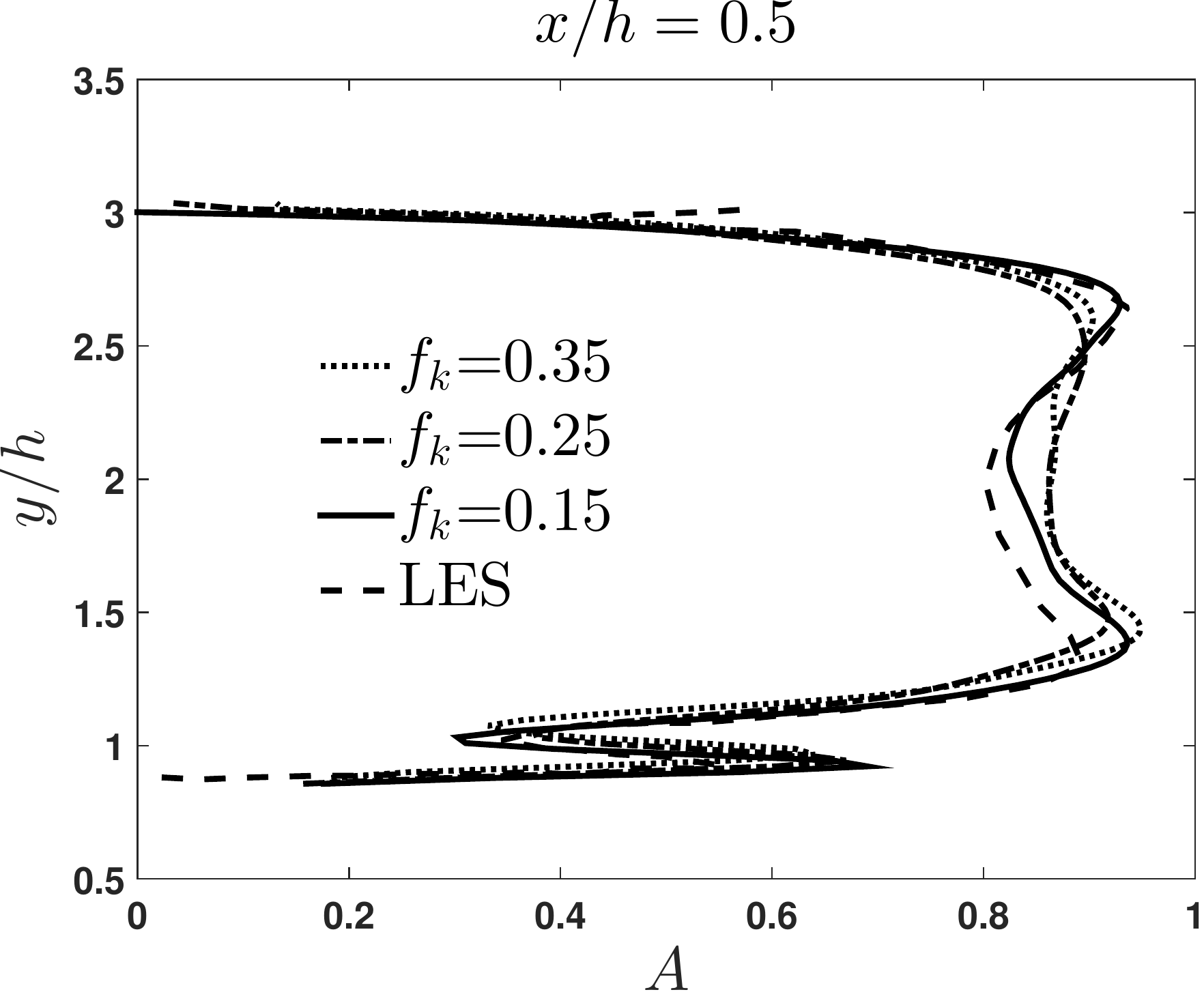}
  \vspace{-8pt}
  \caption{}
\end{subfigure}%
\begin{subfigure}{.5\textwidth}
  \centering
  \includegraphics[scale=0.3]{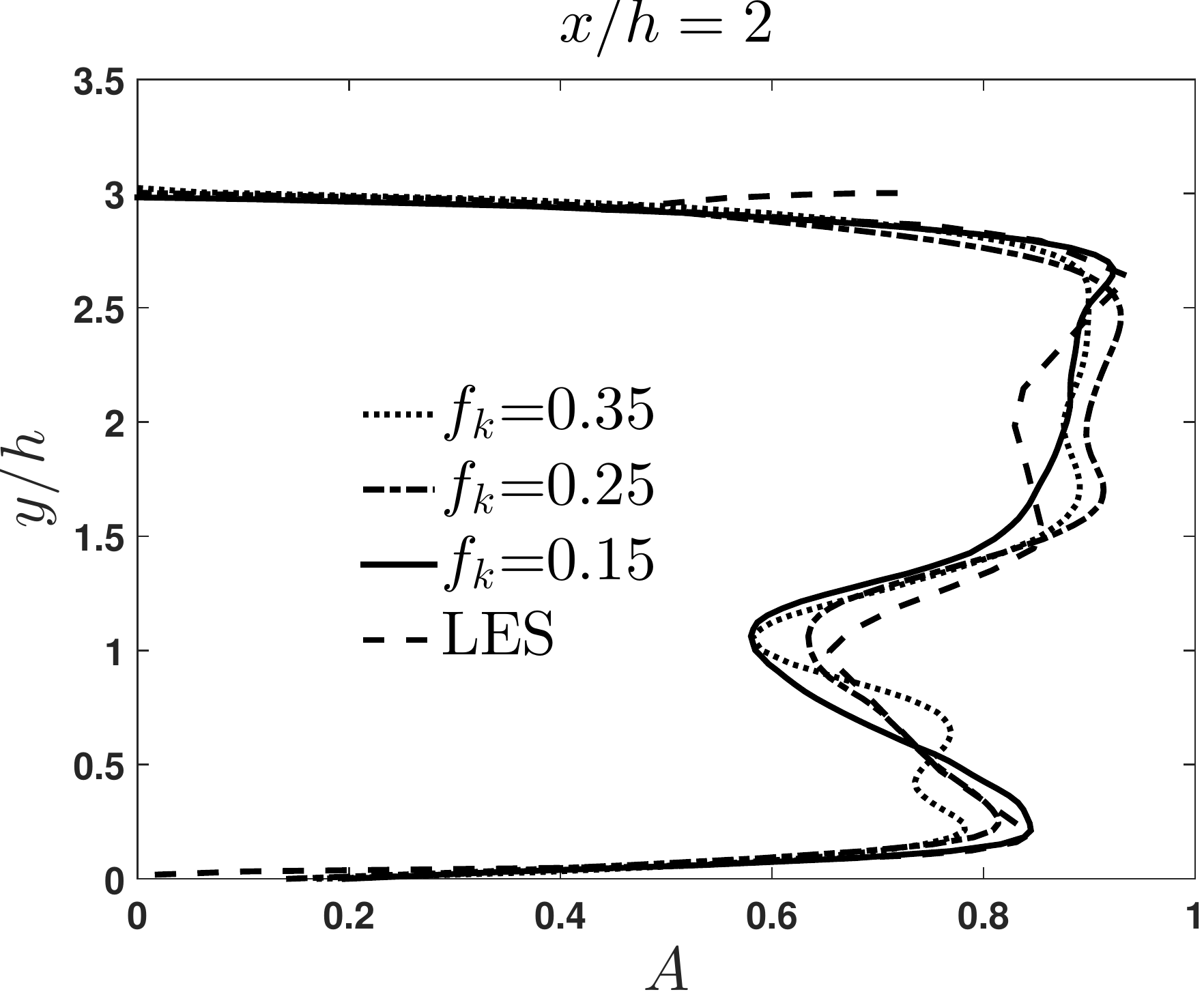}
  \vspace{-8pt}
  \caption{}
\end{subfigure}
\\
\begin{subfigure}{.5\textwidth}
   \centering
   \includegraphics[scale=0.3]{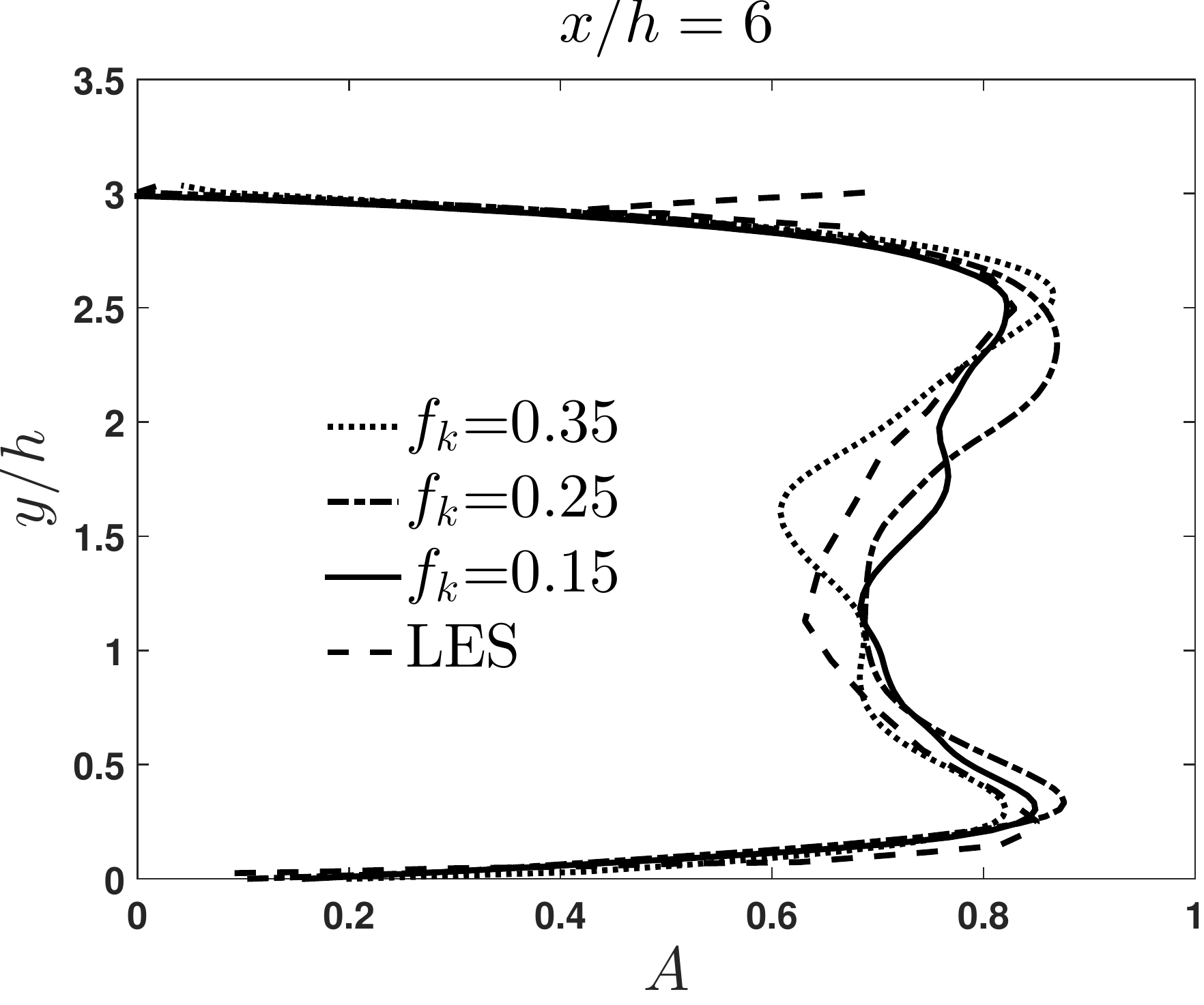}
   \vspace{-8pt}
   \caption{}
\end{subfigure}%
\begin{subfigure}{.5\textwidth}
   \centering
   \includegraphics[scale=0.3]{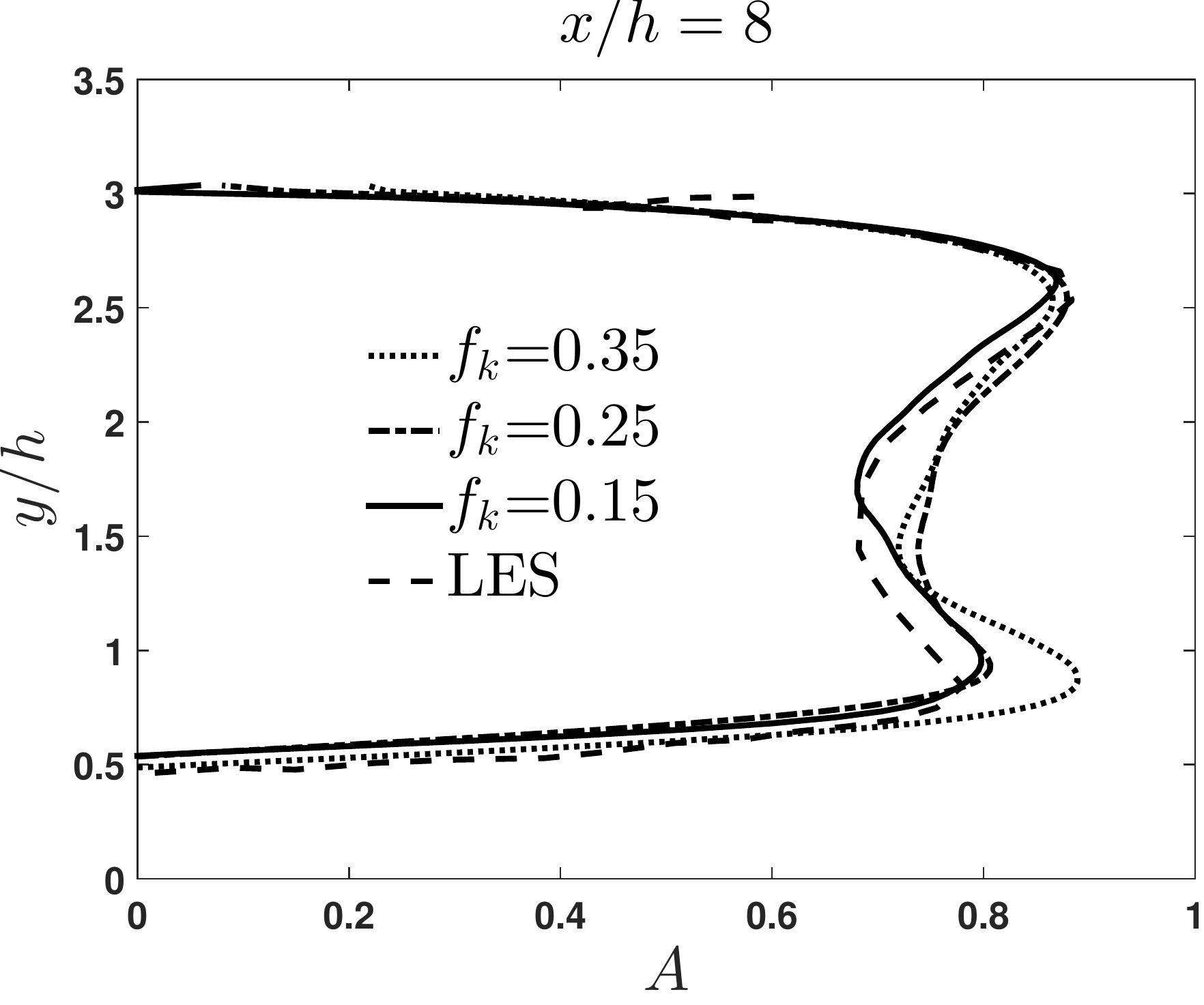}
   \vspace{-8pt}
   \caption{}
\end{subfigure}
\vspace{-8pt}
\caption{Distribution of the flatness parameter, A, at four streamwise locations for Re=10590}
   \label{A}
\end{figure}

\subsubsection{Re=37000}

Fairly good performance of the PANS model compared to LES \cite{Frohlich} at Re=10590 is further investigated at higher Reynolds number where calculations like LES/DNS are extremely costly and impractical to carry on. Additionally, due to the wider range of scales existing at higher Re, a range of cut-off parameters can be examined for resolving different levels of scales and their influence on flow statistics and structure.  
In this section, two categories of simulations are performed to investigate the effects of physical resolution as well as computational resolution on flow behaviour. At high enough computational resolution, it is very illustrative to compare simulations of different $f_k$ values to demonstrate the capabilities of PANS closure model. Further, it is important to establish that for high $f_k$ values, good results can be obtained on coarse grids. Thus, both $f_k$ and grid resolution studies yield important insight into the closure model performance.

\paragraph*{\textbf{Variation of $f_k$ study}}

\begin{figure}[H]
\centering
\begin{subfigure}{.5\textwidth}
  \centering
  \includegraphics[scale=0.3]{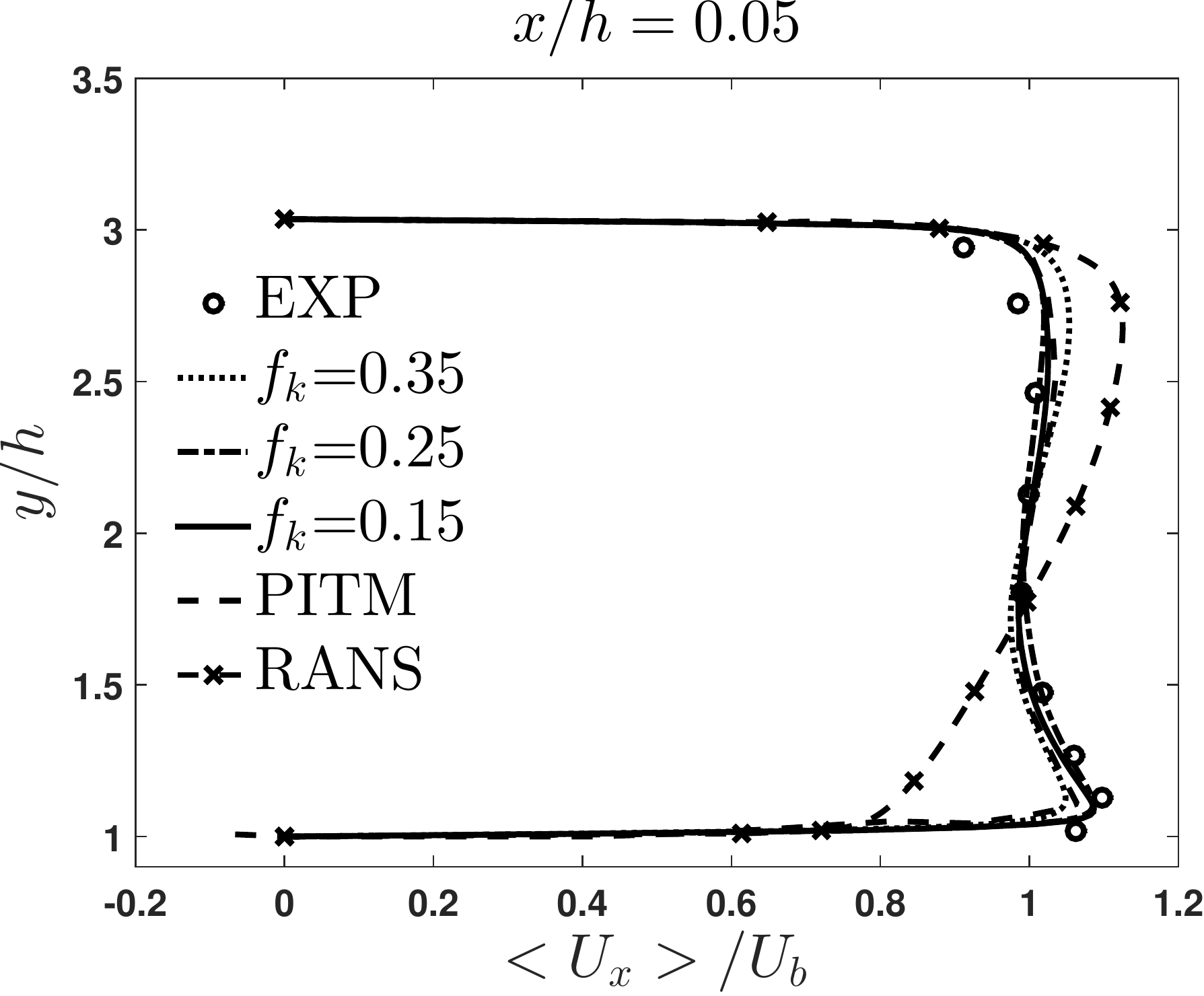}
  \vspace{-8pt}
  \caption{}
\end{subfigure}%
\begin{subfigure}{.5\textwidth}
  \centering
  \includegraphics[scale=0.3]{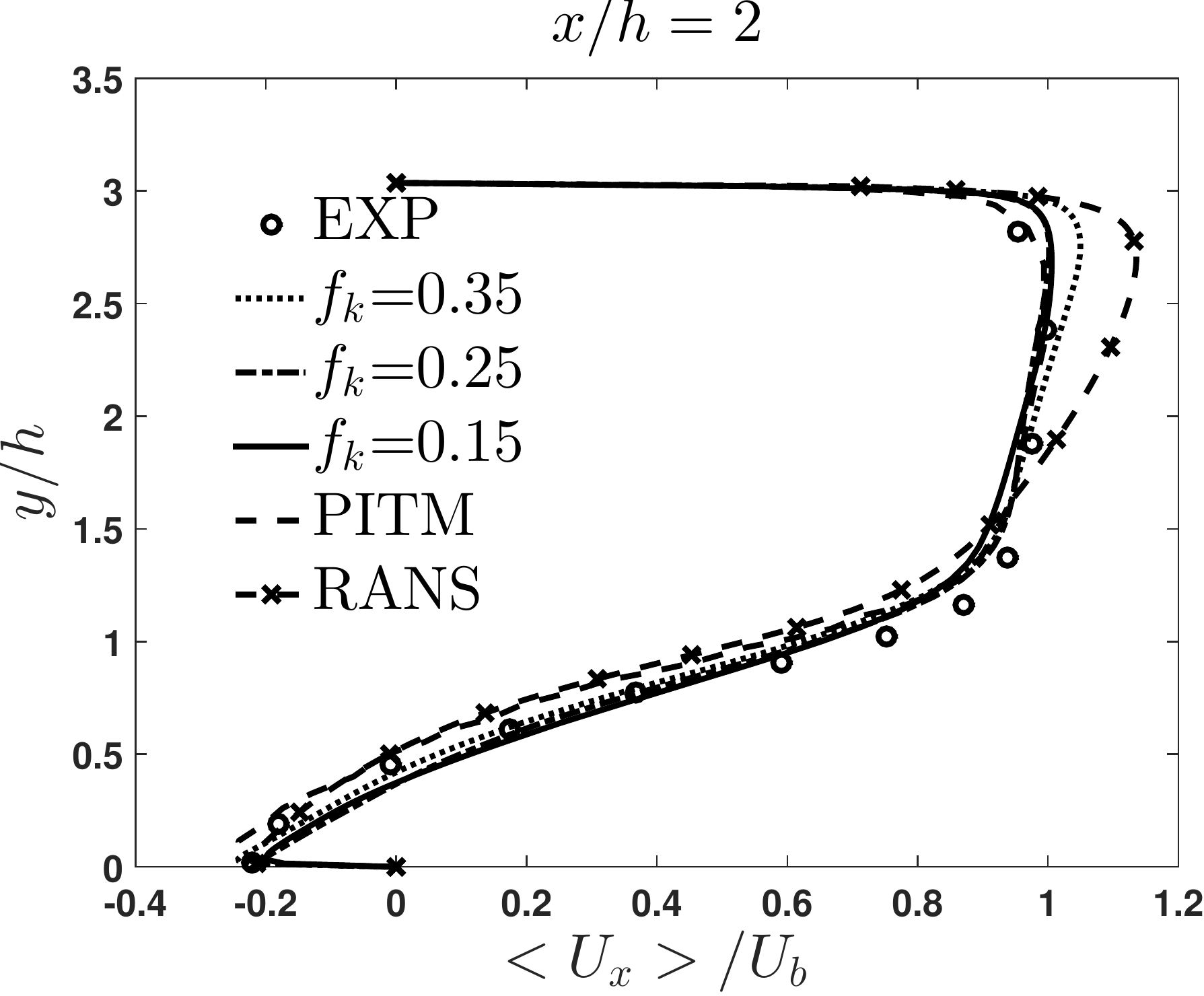}
  \vspace{-8pt}
  \caption{}
\end{subfigure}
\\
\begin{subfigure}{.5\textwidth}
   \centering
   \includegraphics[scale=0.3]{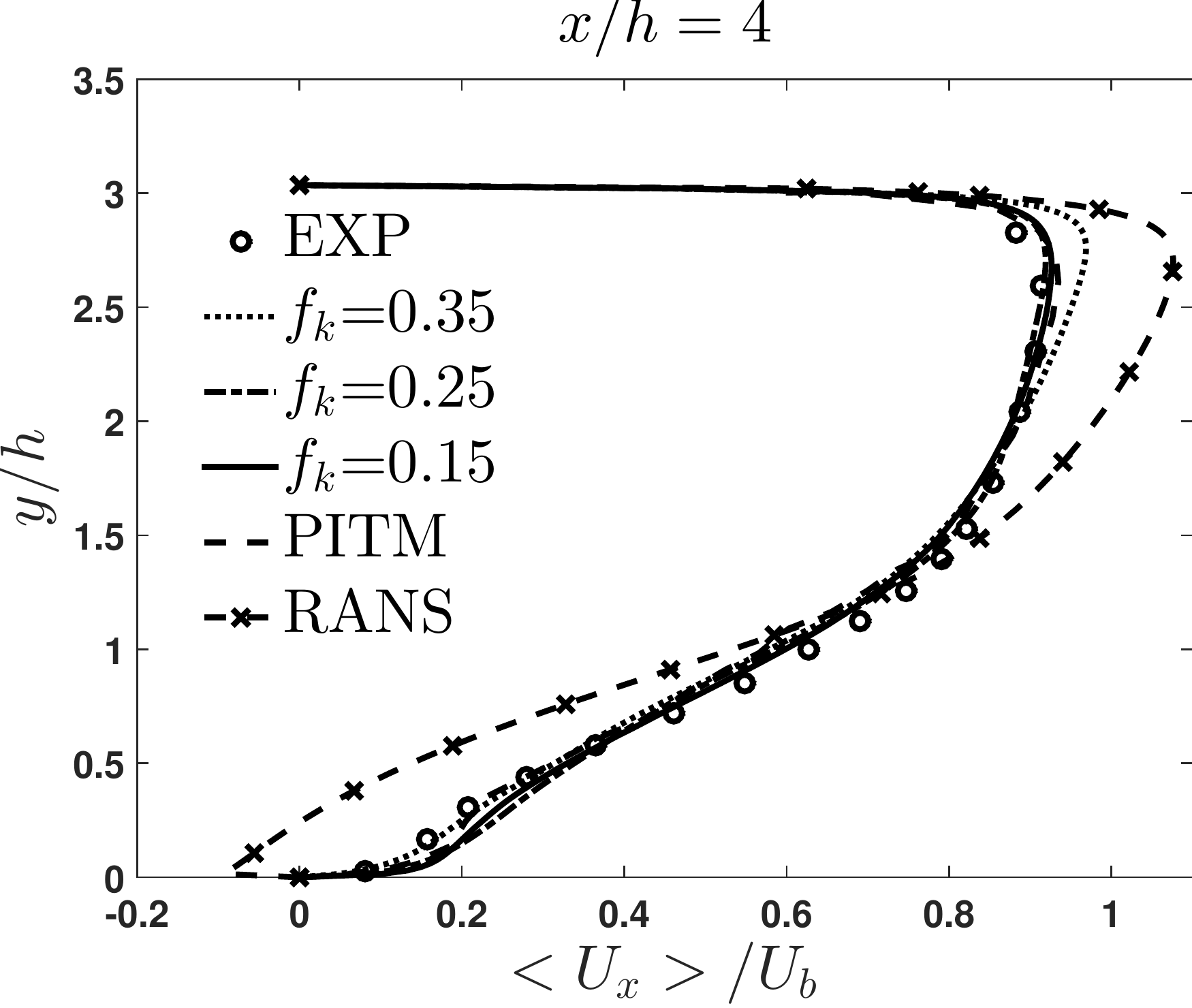}
   \vspace{-8pt}
   \caption{}
\end{subfigure}%
\begin{subfigure}{.5\textwidth}
   \centering
   \includegraphics[scale=0.3]{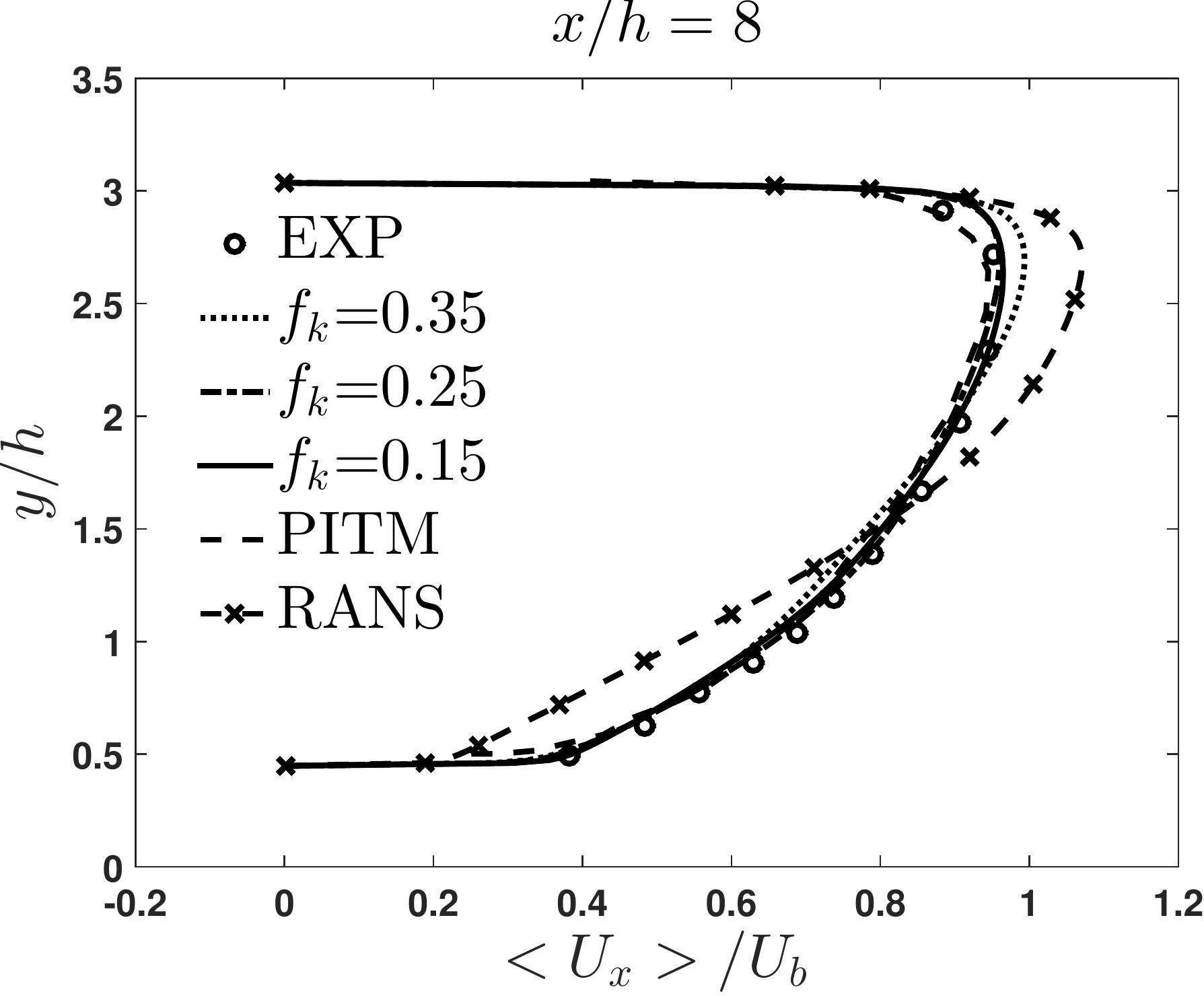}
   \vspace{-8pt}
   \caption{}
\end{subfigure}
\vspace{-8pt}
\caption{Streamwise velocity profiles at different locations for Re=37000}
\label{fku}
\end{figure}

We first investigate the effect of varying $f_k$ on flow statistics. In this section, all the studies are performed on the finest grid of about 0.9 million grid nodes. This grid permits computations of $f_k>0.15$.
Figure \ref{fku} shows the streamwise velocity profiles at four locations of x/h=0.05, 2, 4 and 8. The results are shown for resolutions: $f_k$=0.35, 0.25 and 0.15. The PANS results are compared against RANS $k-\omega$ model, PITM method \cite{PITM} and experimental data \cite{Rapp}. It can be observed from these plots that PANS results agree well with literature data particularly at the lowest cut-off parameter, $f_k$=0.15. The flow is expected to be reattached at x/h=4 which is well predicted by PANS and PITM method. However, a more elongated separation bubble is predicted by the RANS simulation.

An important observation from Fig. \ref{fku} is that the agreement on streamwise velocity profile with experimental data improves progressively with lower $f_k$ values. Among the RANS, PITM and PANS data, the best agreement is achieved for $f_k$=0.15 wherein a wider range of turbulence length scales are resolved in the highly unsteady regions.

\begin{figure}[H]
\centering
\begin{subfigure}{.5\textwidth}
  \centering
  \includegraphics[scale=0.3]{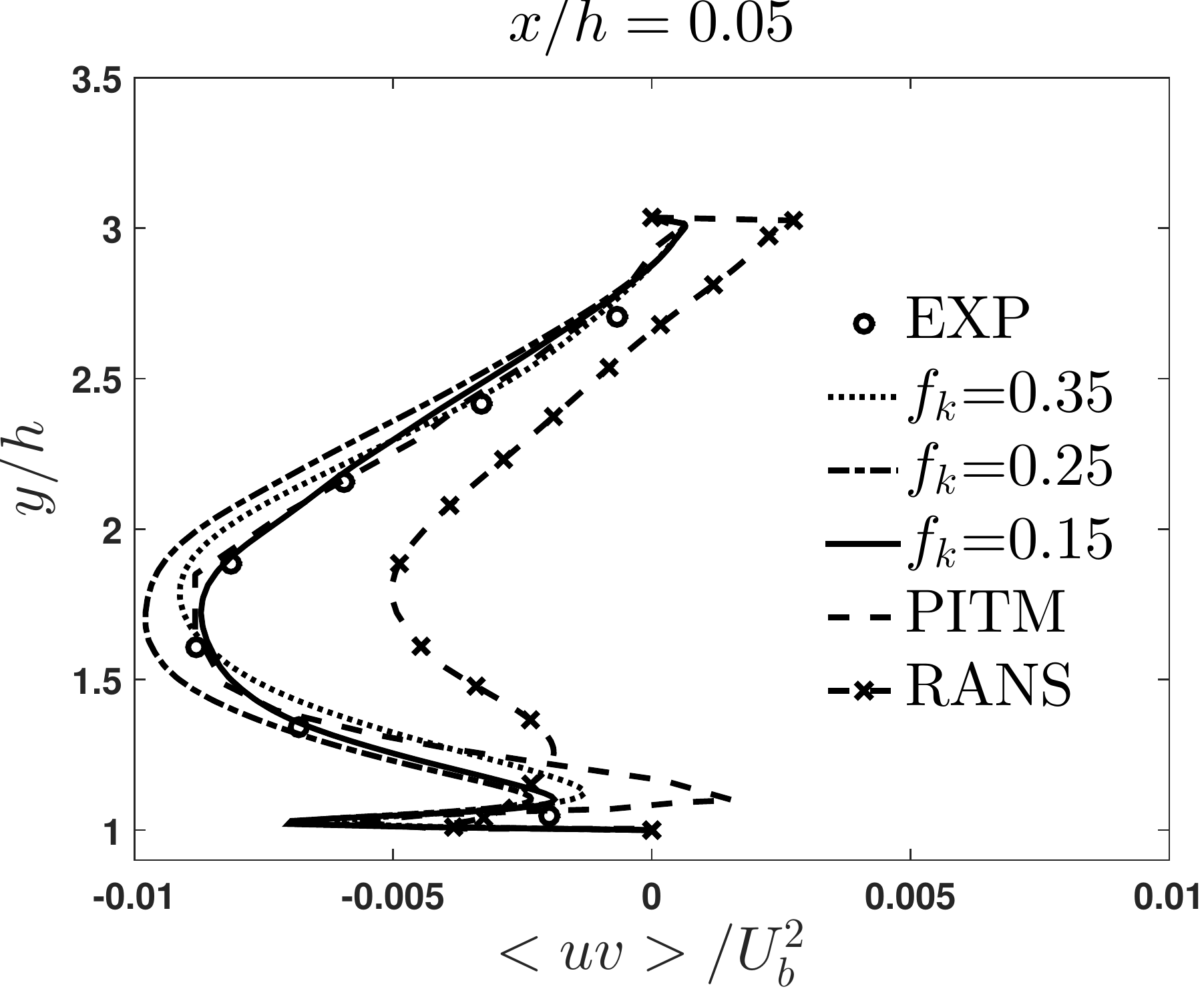}
  \vspace{-8pt}
  \caption{}
\end{subfigure}%
\begin{subfigure}{.5\textwidth}
  \centering
  \includegraphics[scale=0.3]{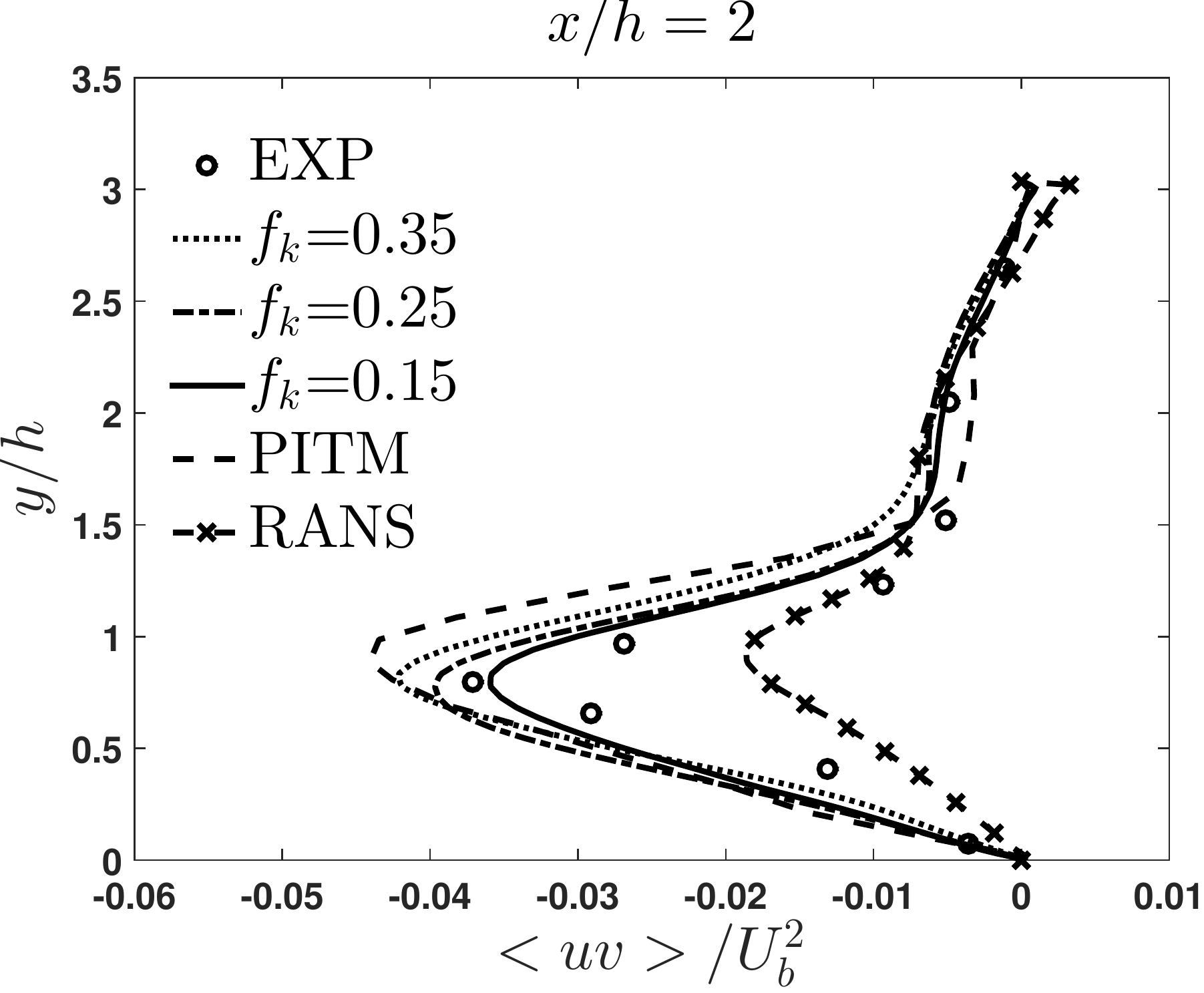}
  \vspace{-8pt}
  \caption{}
\end{subfigure}
\\
\begin{subfigure}{.5\textwidth}
   \centering
   \includegraphics[scale=0.3]{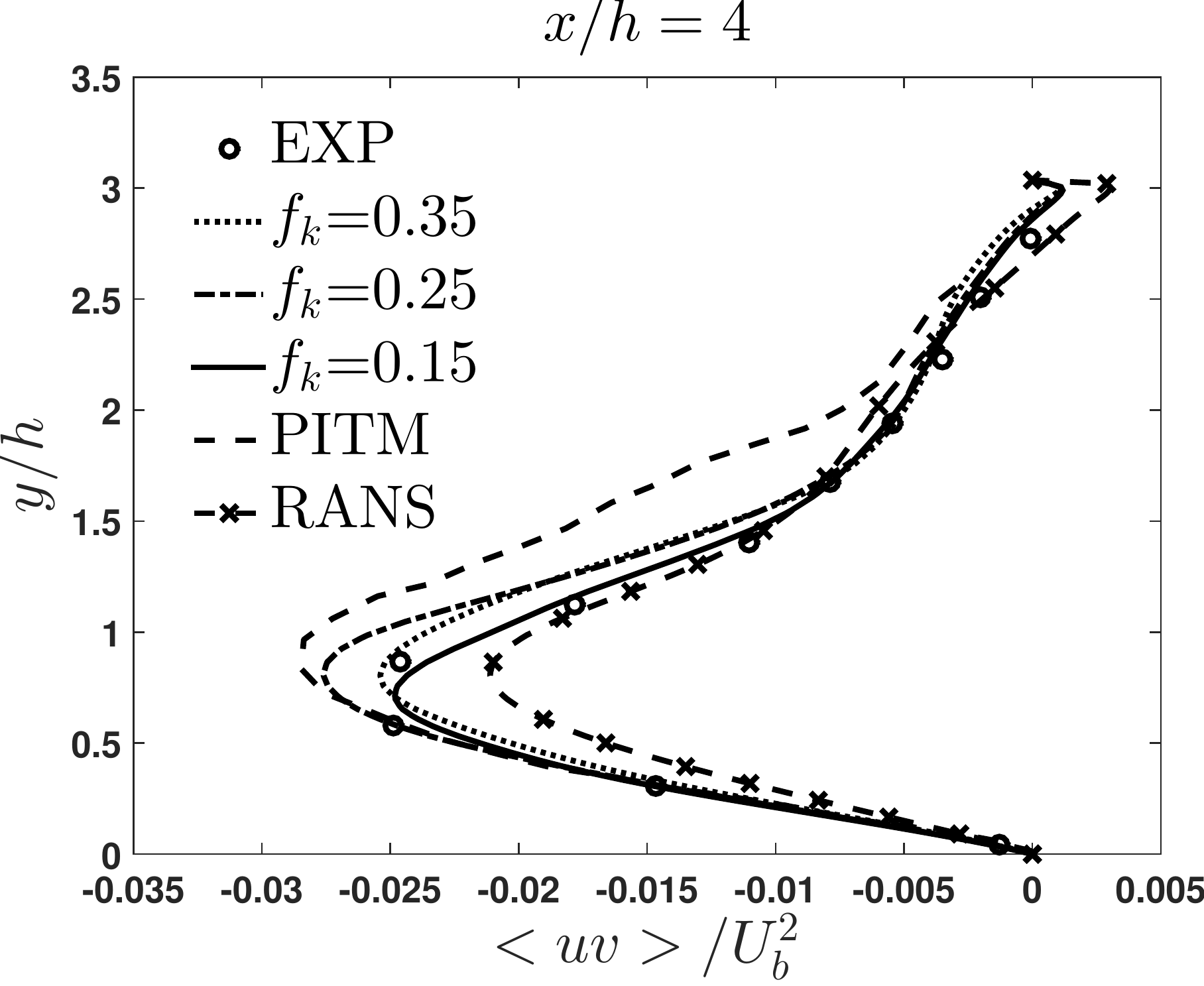}
   \vspace{-8pt}
   \caption{}
\end{subfigure}%
\begin{subfigure}{.5\textwidth}
   \centering
   \includegraphics[scale=0.3]{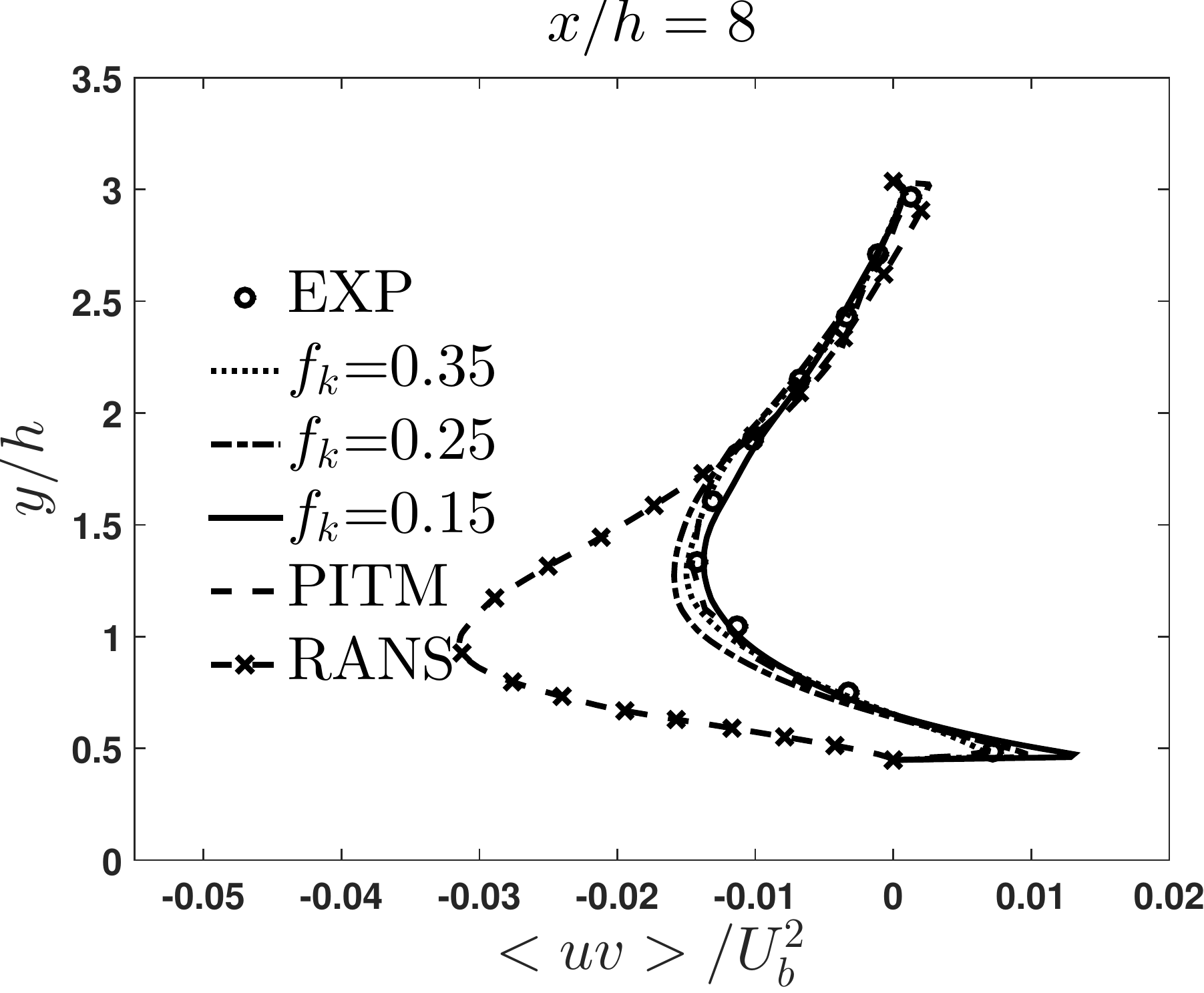}
   \vspace{-8pt}
   \caption{}
\end{subfigure}
\vspace{-8pt}
\caption{Shear stress profiles at different locations for Re=37000}
\label{fkuv}
\end{figure}

\begin{figure}[H]
\centering
\begin{subfigure}{.5\textwidth}
  \centering
  \includegraphics[scale=0.3]{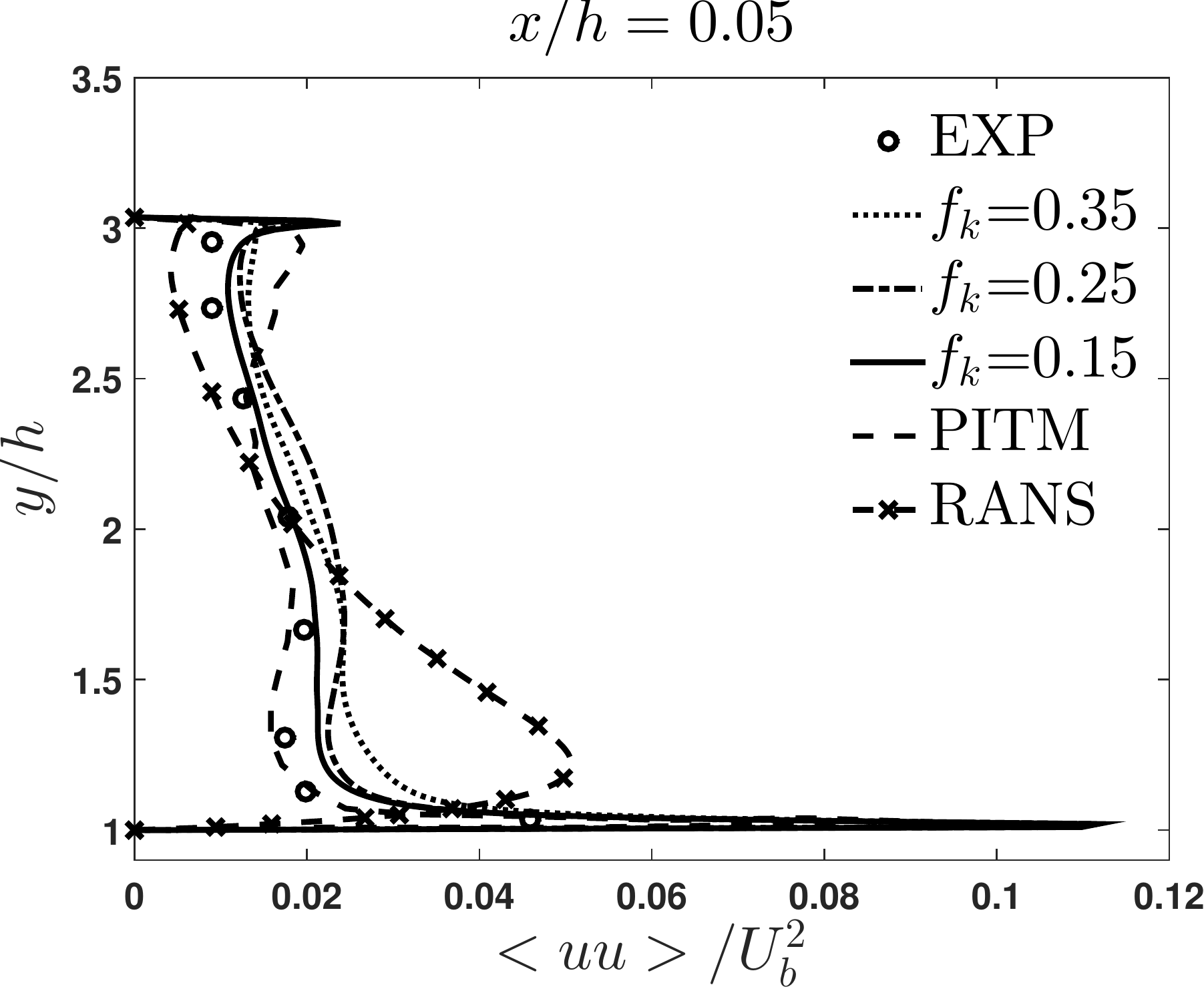}
  \vspace{-8pt}
  \caption{}
\end{subfigure}%
\begin{subfigure}{.5\textwidth}
  \centering
  \includegraphics[scale=0.3]{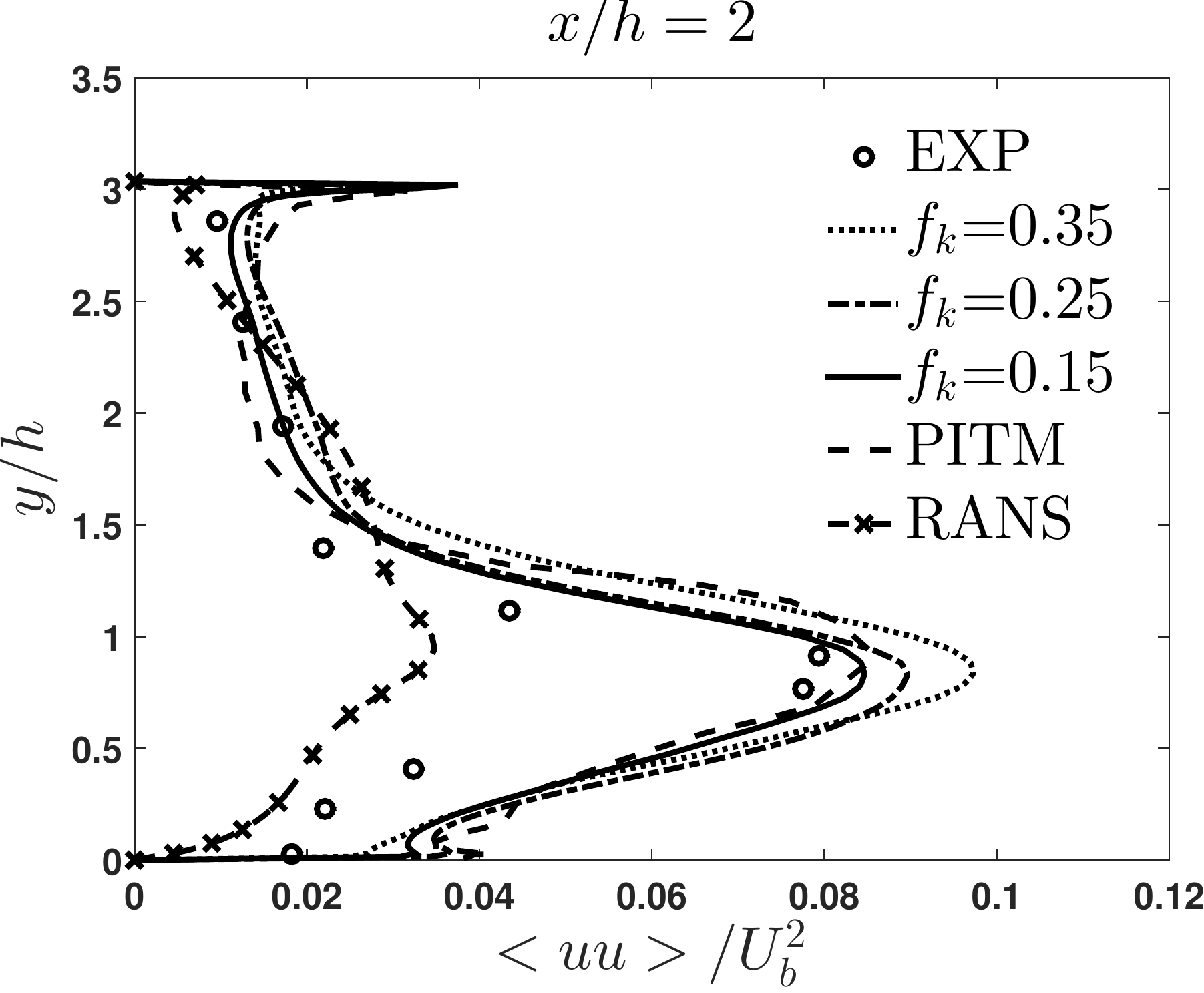}
  \vspace{-8pt}
  \caption{}
\end{subfigure}
\\
\begin{subfigure}{.5\textwidth}
   \centering
   \includegraphics[scale=0.3]{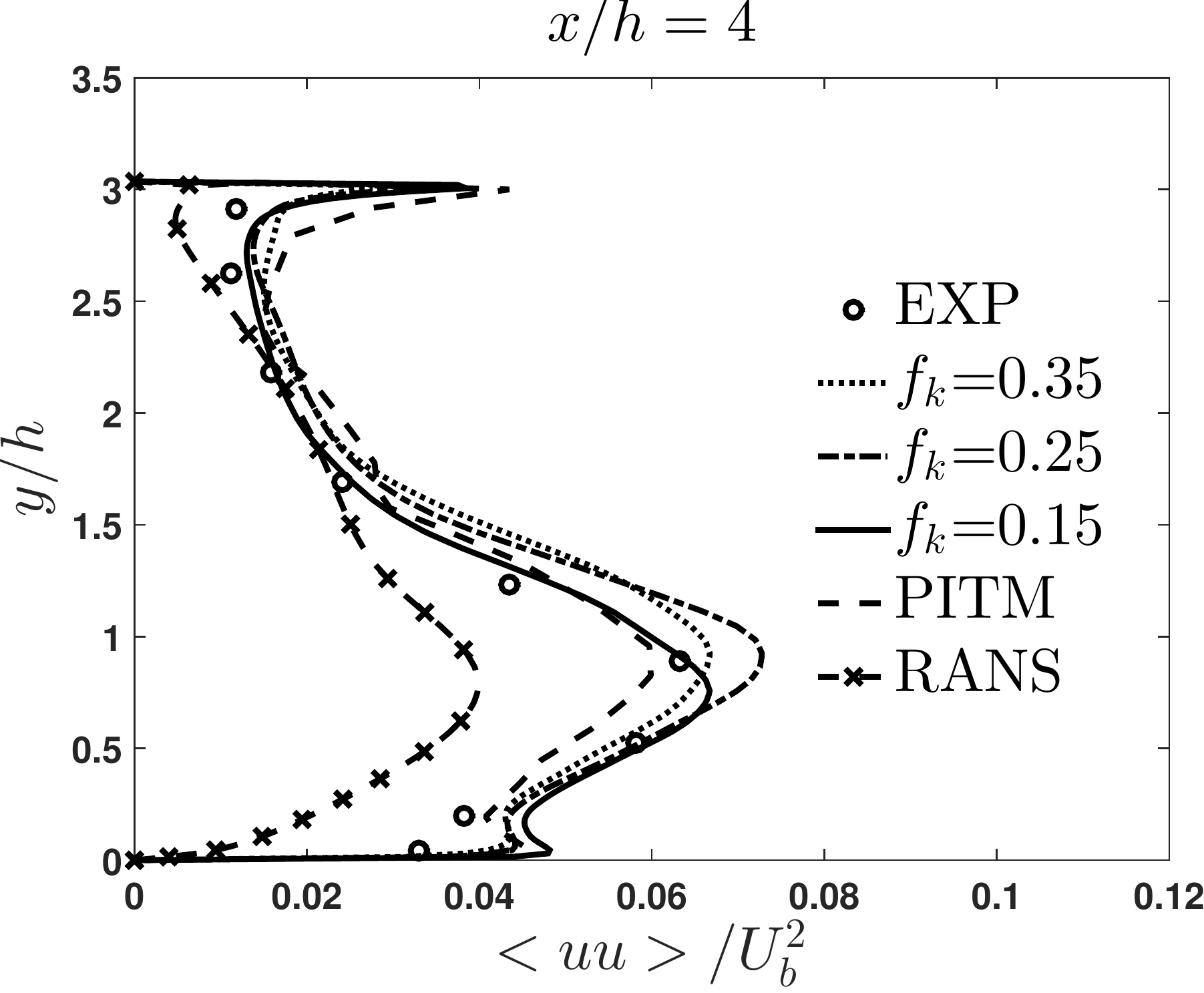}
   \vspace{-8pt}
   \caption{}
\end{subfigure}%
\begin{subfigure}{.5\textwidth}
   \centering
   \includegraphics[scale=0.29]{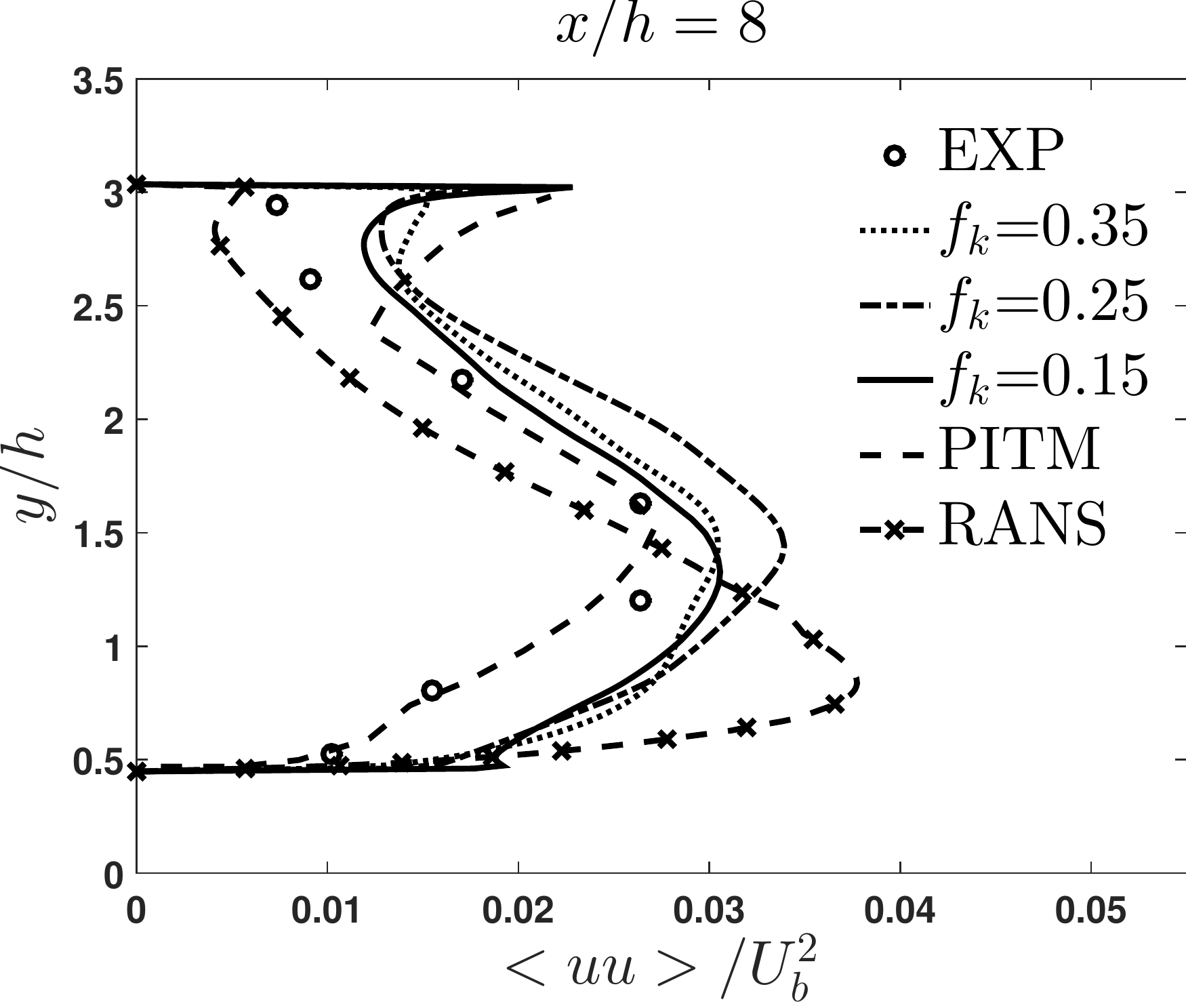}
   \vspace{-8pt}
   \caption{}
\end{subfigure}
\vspace{-8pt}
\caption{Streamwise stress profiles at different locations for Re=37000}
\label{fkuu}
\end{figure}

\begin{figure}[H]
\centering
\begin{subfigure}{.5\textwidth}
  \centering
  \includegraphics[scale=0.3]{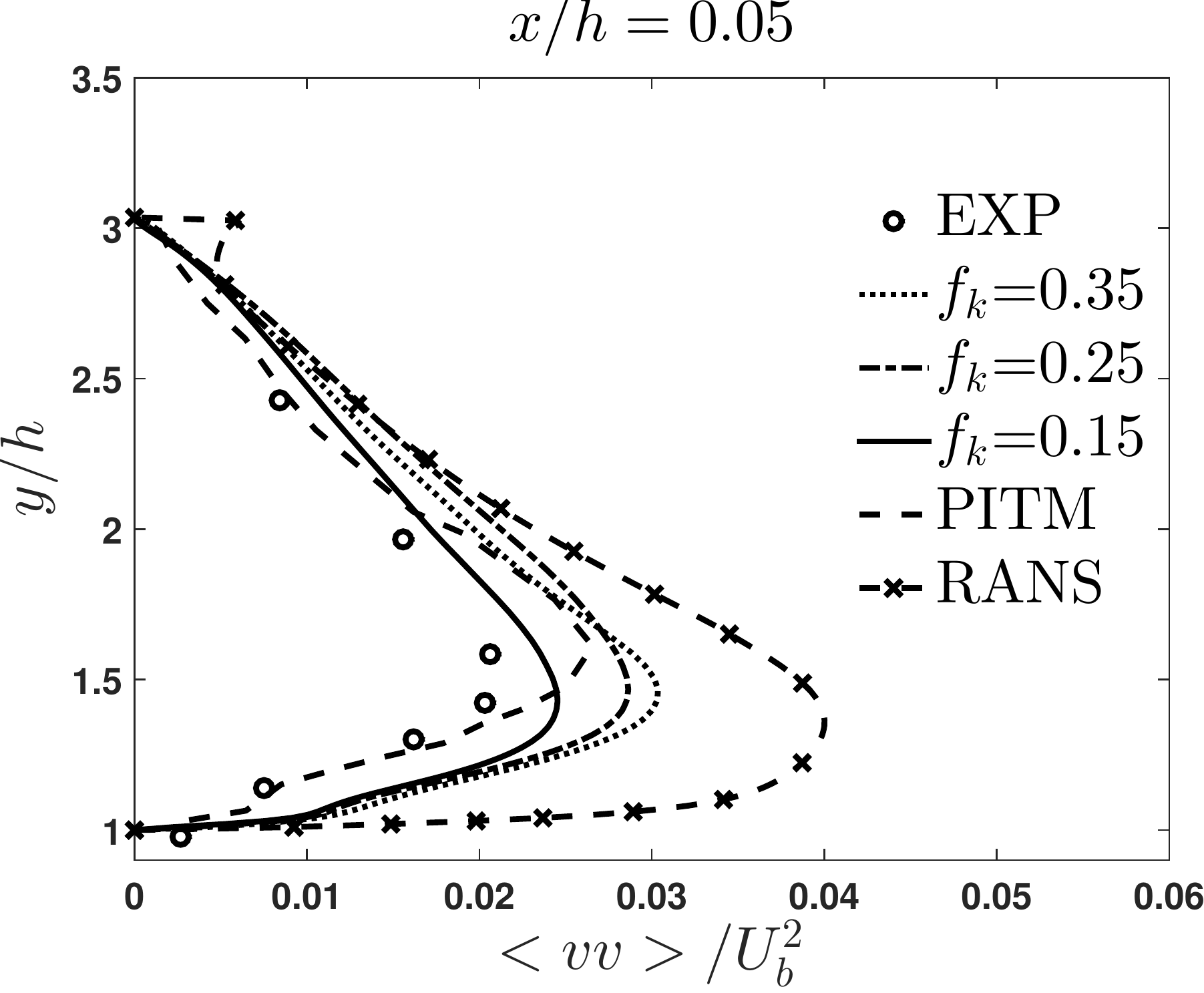}
  \vspace{-8pt}
  \caption{}
\end{subfigure}%
\begin{subfigure}{.5\textwidth}
  \centering
  \includegraphics[scale=0.3]{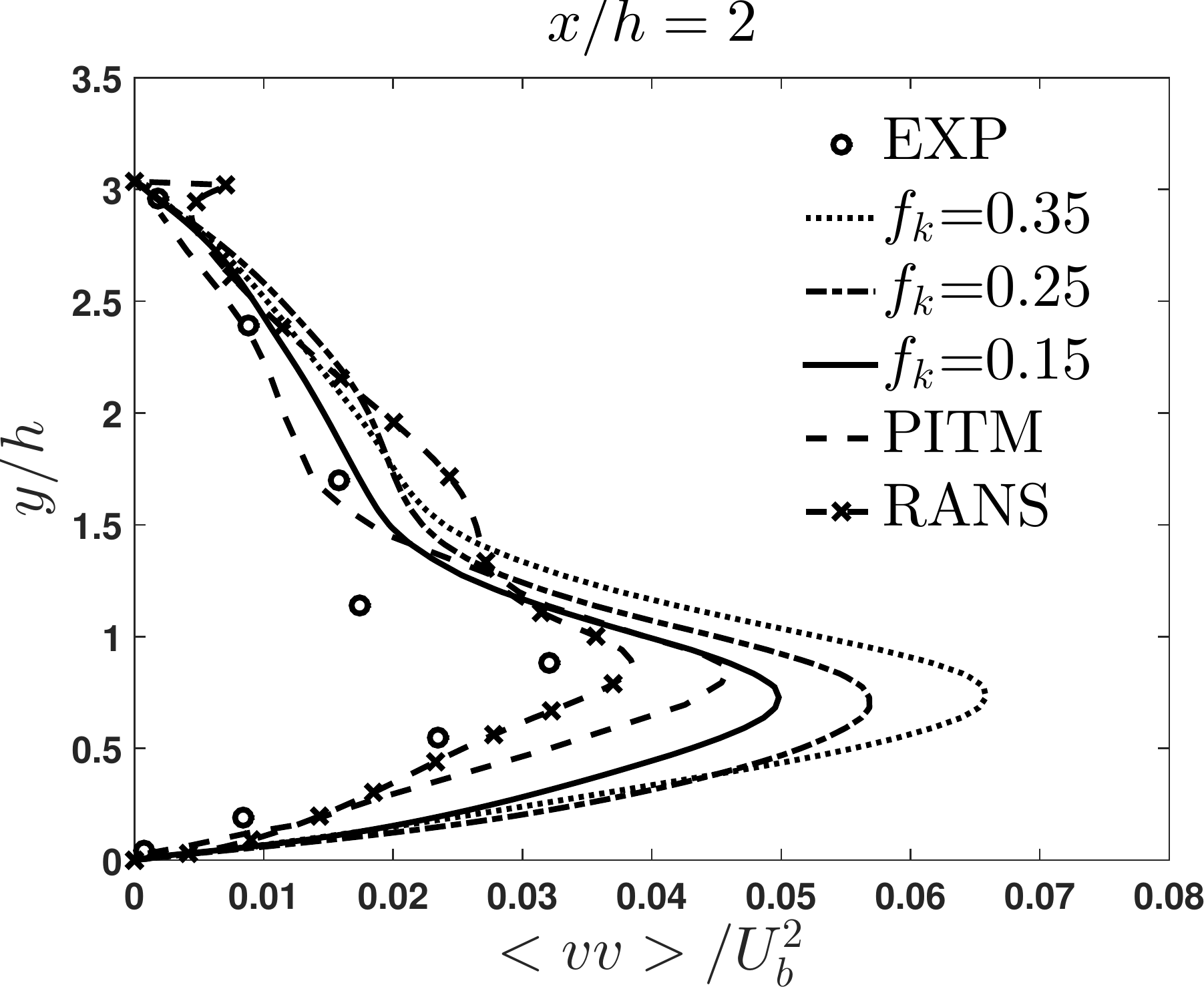}
  \vspace{-8pt}
  \caption{}
\end{subfigure}
\\
\begin{subfigure}{.5\textwidth}
   \centering
   \includegraphics[scale=0.3]{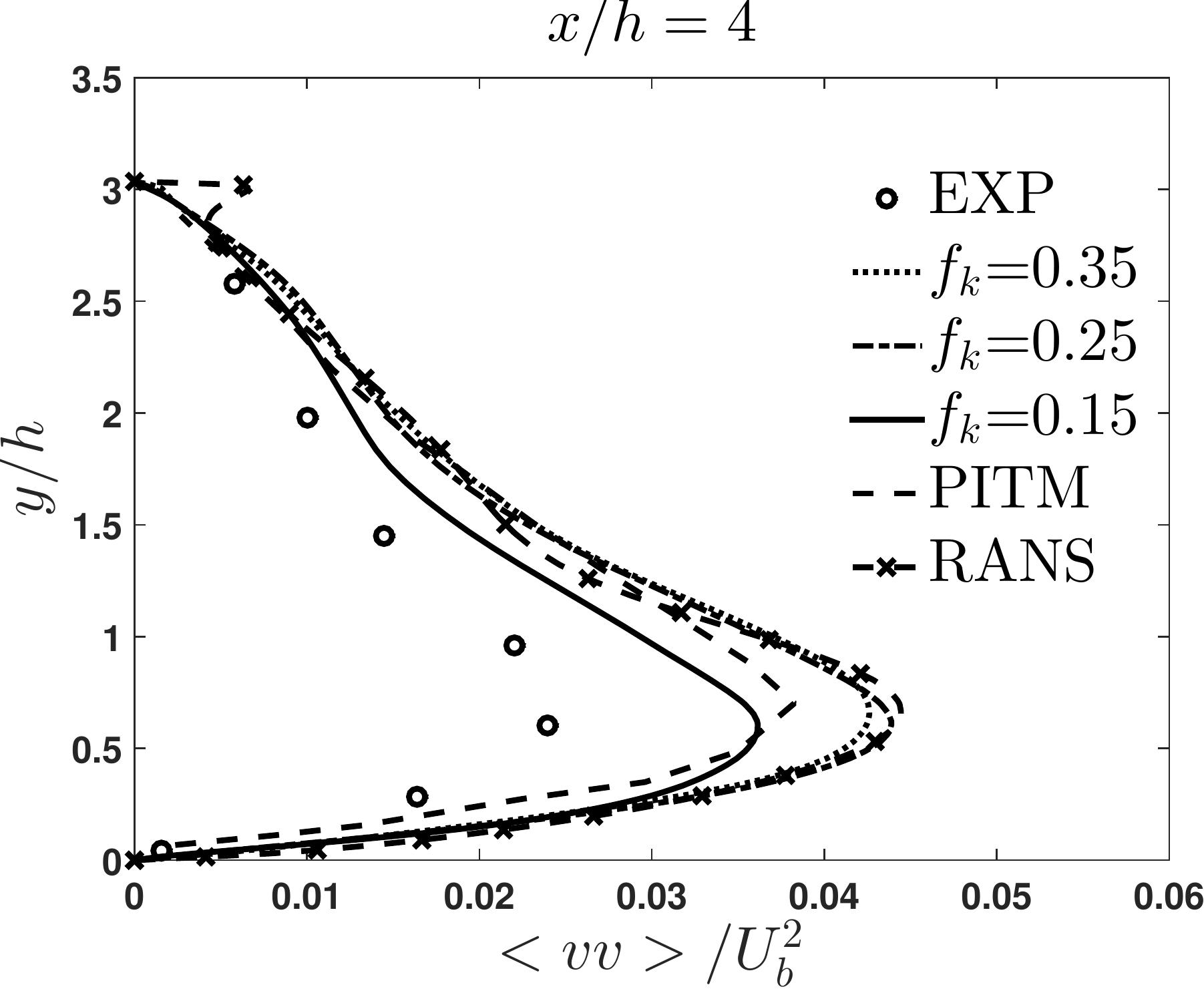}
   \vspace{-8pt}
   \caption{}
\end{subfigure}%
\begin{subfigure}{.5\textwidth}
   \centering
   \includegraphics[scale=0.3]{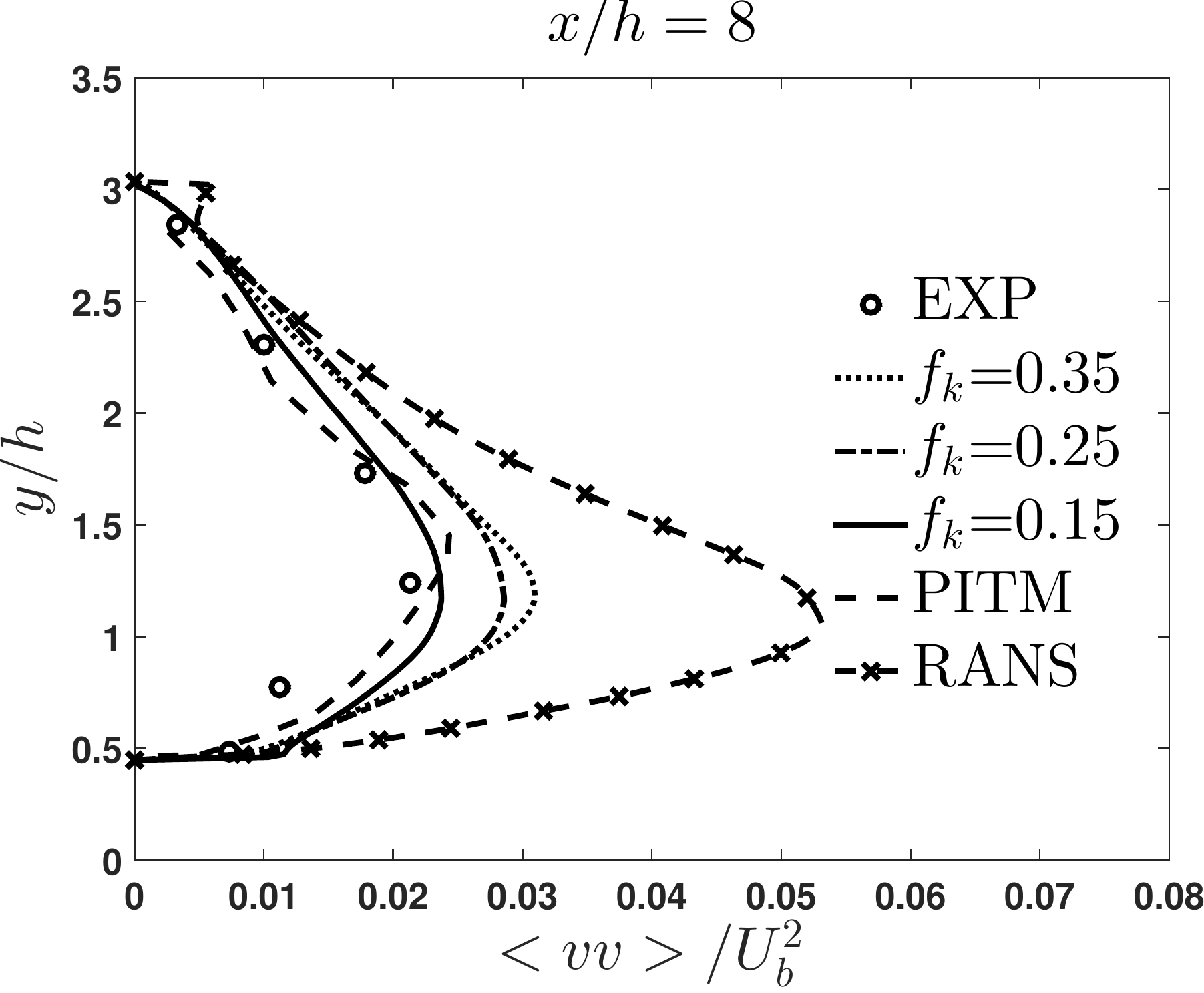}
   \vspace{-8pt}
   \caption{}
\end{subfigure}
\vspace{-8pt}
\caption{Vertical stress profiles at different locations for Re=37000}
\label{fkvv}
\end{figure}

Figures \ref{fkuv}, \ref{fkuu} and \ref{fkvv} show the components  of stress tensor at the four streamwise locations. For shear stress profiles, at x/h=0.05, the maximum value for $<u'v'>$ at $y/h\approx1.6$ is related to the local minimum of streamwise velocity. As can be seen, this peak value is well captured by PANS simulation for $f_k$=0.15. At the next two locations, x/h=2 and x/h=4, the peak value of shear stress is over predicted by the PITM simulation, whereas PANS results are in close agreement with experimental data. Location at x/h=8 is regarded as post reattachment region where the flow is recovering from upstream separation with flow components of different history. Poor prediction of the separated region by the RANS model results in over-prediction of shear stress at this location. Overall, Figs. \ref{fku} and \ref{fkuv} indicate that the mean flow quantities are well predicted by PANS simulation at all the cut-off ratios investigated here. However, the results for $f_k$=0.15 compares the best with the experimental data. 

For the streamwise Reynolds stress component, the experimental data and PANS results are in close agreement. Contradictory to the RANS calculation, the location of the peak values and distribution of the data are well estimated by the PANS simulations and have been successively improved by lowering the cut-off parameter to $f_{k}$=0.15. However, for the vertical Reynolds stress component, both the hybrid methods, PANS \& PITM, have notable deviations from experimental data at some streamwise locations. As explained in the previous section, misprediction of the streamwise velocity close to the upper wall affecting the mass flow rate imposed to the domain and second order accuracy of the schemes could be responsible for this error.

\subsubsection{Flow Structure}

\begin{figure}[H]
\centering
\begin{subfigure}{.5\textwidth}
  \centering
  \includegraphics[trim=1cm 0.8cm 9.5cm 14.0cm, clip=true, scale=0.4]{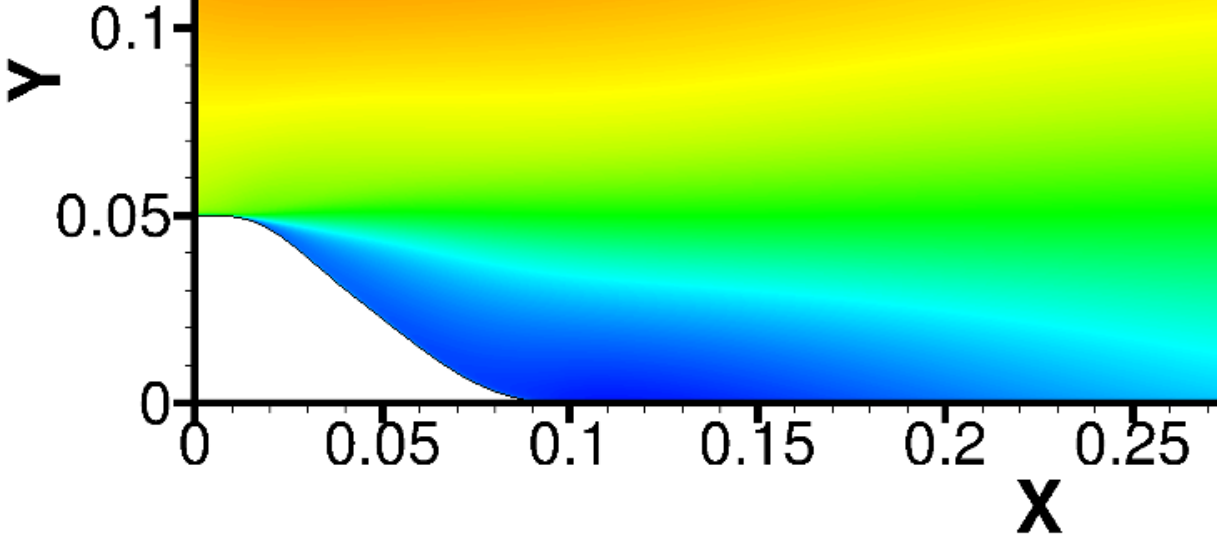}
  \vspace{-5pt}
  \caption{}
\end{subfigure}%
\begin{subfigure}{.5\textwidth}
  \centering
  \includegraphics[trim=1cm 0.8cm 9.5cm 14.0cm, clip=true, scale=0.4]{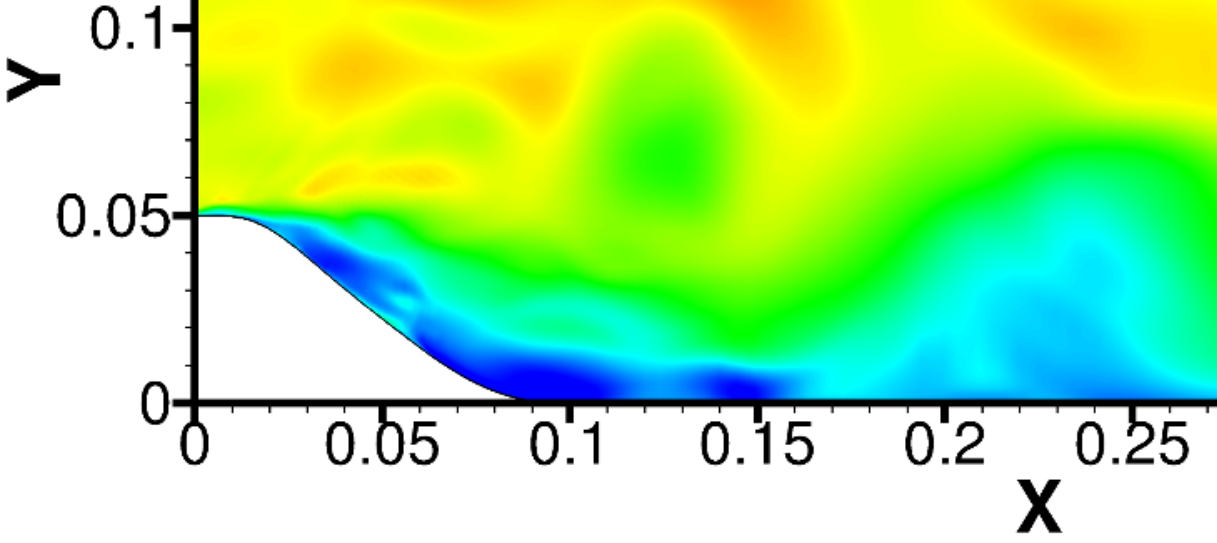}
  \vspace{-5pt}
  \caption{}
\end{subfigure}
\\
\begin{subfigure}{.5\textwidth}
   \centering
   \includegraphics[trim=1cm 0.8cm 9.5cm 14.0cm, clip=true, scale=0.4]{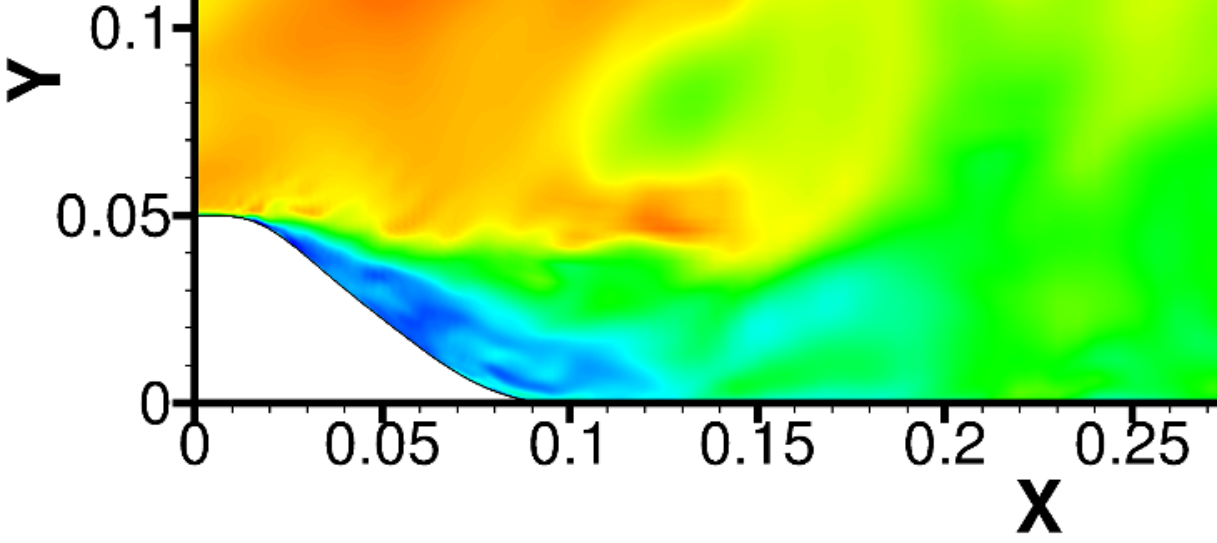}
   \vspace{-5pt}
   \caption{}
\end{subfigure}%
\begin{subfigure}{.5\textwidth}
   \centering
   \includegraphics[trim=1cm 0.8cm 9.5cm 14.0cm, clip=true, scale=0.4]{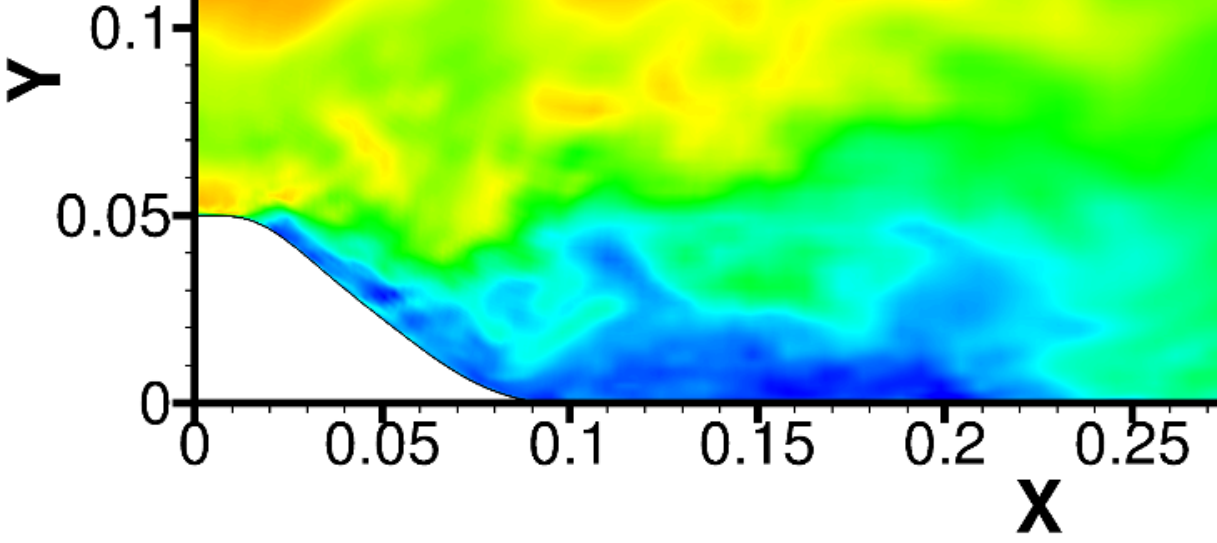}
   \vspace{-5pt}
   \caption{}
\end{subfigure}
\vspace{-5pt}
\caption{Instantaneous velocity profile, Re=37000: (a) RANS, (b) $f_k$=0.35, (c) $f_k$=0.25, (d) $f_k$=0.15 }
\label{SI}
\end{figure}

\begin{figure}[H]
\centering
\begin{subfigure}{.5\textwidth}
  \centering
  \includegraphics[trim=1cm 3cm 1cm 3cm, clip=true, scale=0.2]{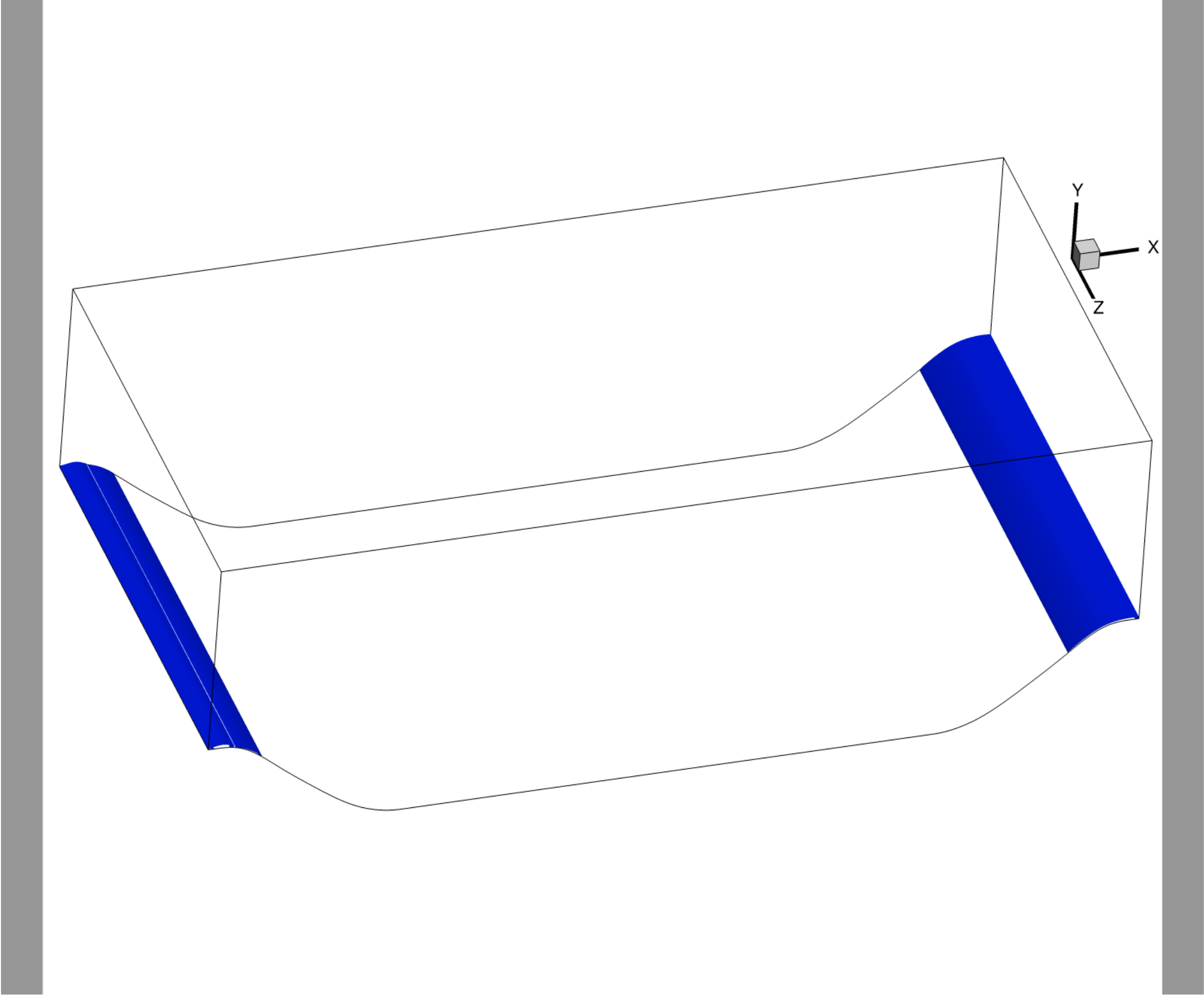}
  \vspace{-5pt}
  \caption{}
\end{subfigure}%
\begin{subfigure}{.5\textwidth}
  \centering
  \includegraphics[trim=6.5cm 3cm 6.5cm 3cm, clip=true, scale=0.2]{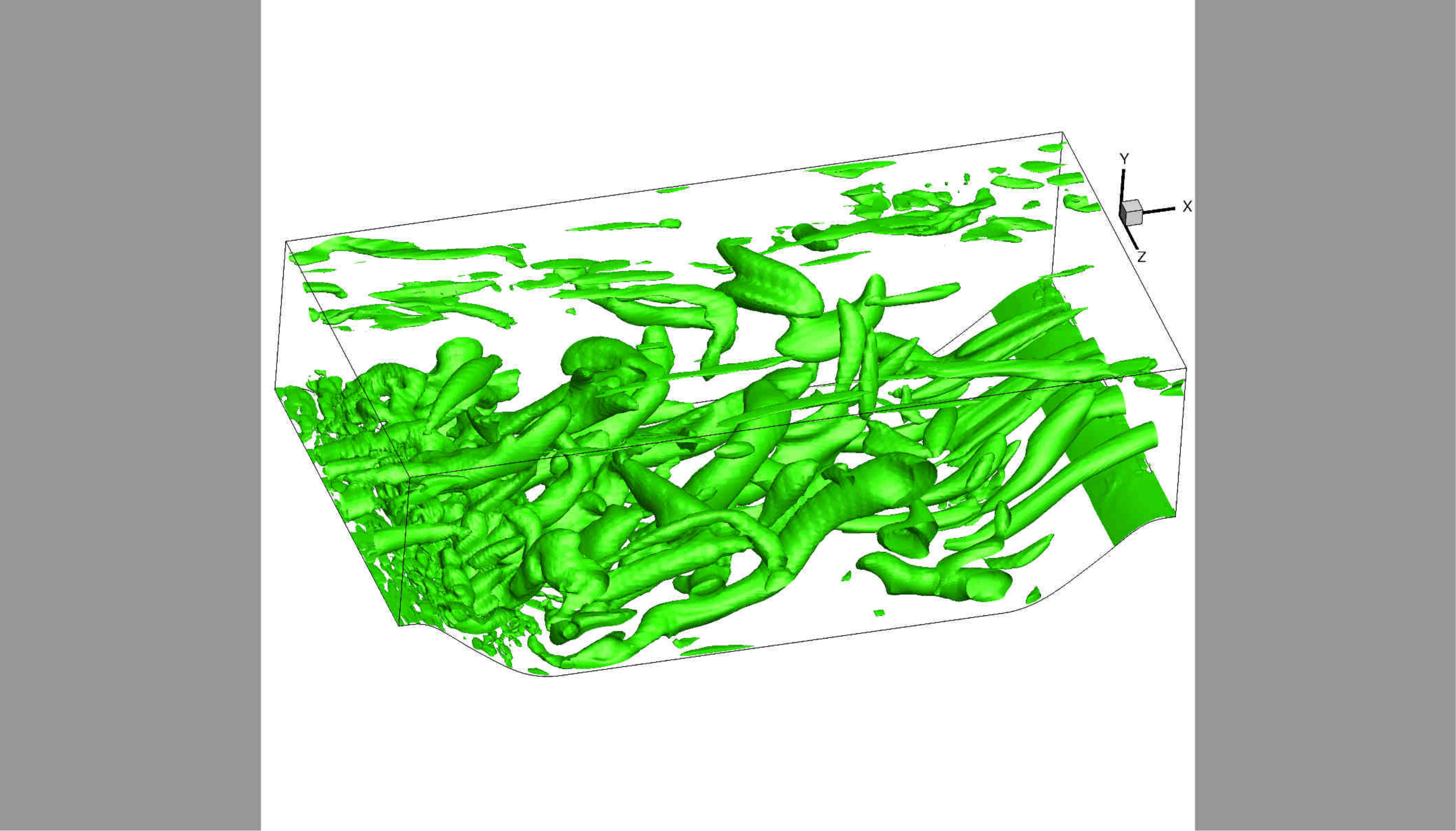}
  \vspace{-5pt}
  \caption{}
\end{subfigure}
\\
\begin{subfigure}{.5\textwidth}
   \centering
   \includegraphics[trim=6.5cm 3cm 6.5cm 3cm, clip=true, scale=0.2]{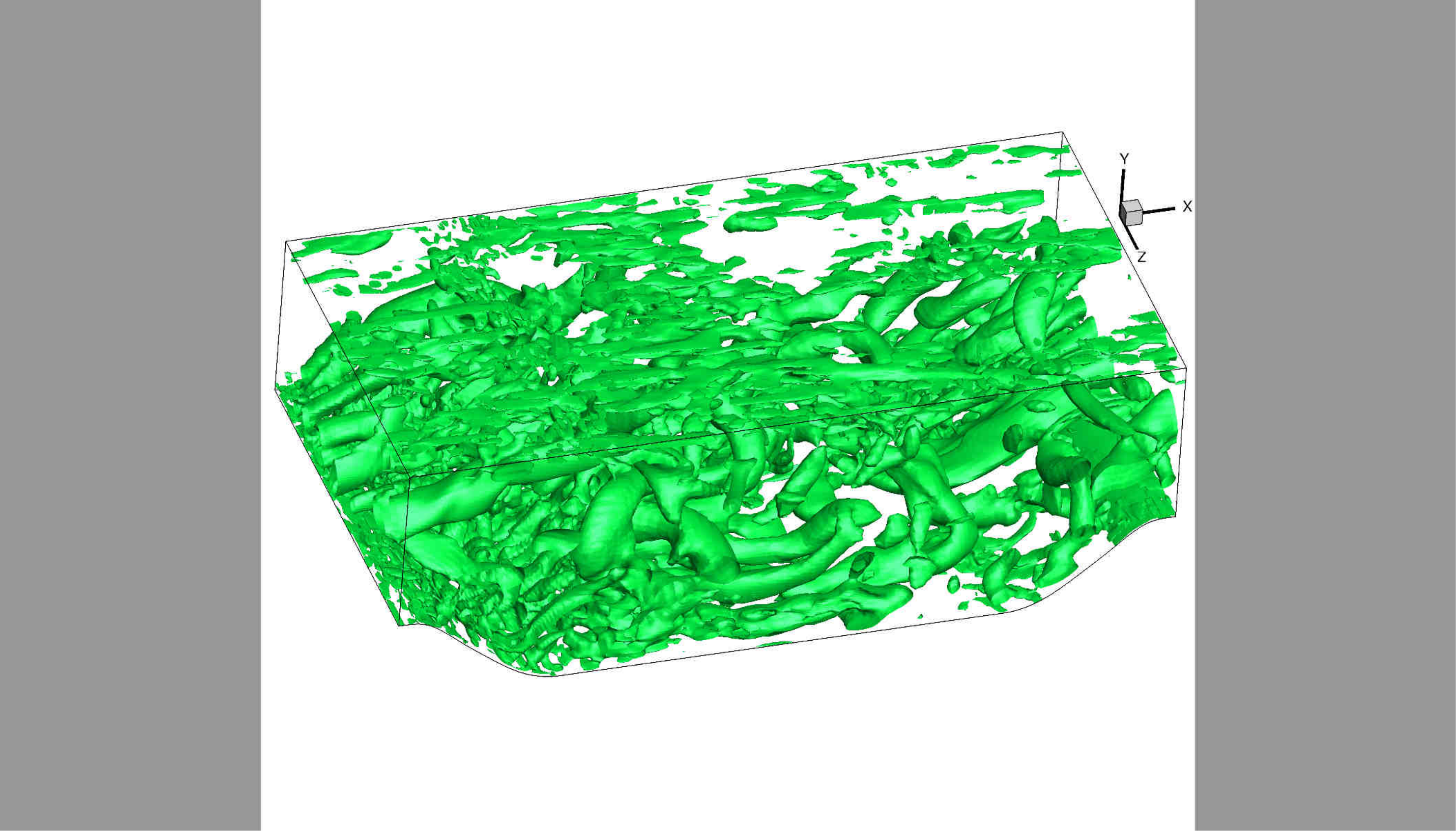}
   \vspace{-5pt}
   \caption{}
\end{subfigure}%
\begin{subfigure}{.5\textwidth}
   \centering
   \includegraphics[trim=6.5cm 3cm 6.5cm 3cm, clip=true, scale=0.2]{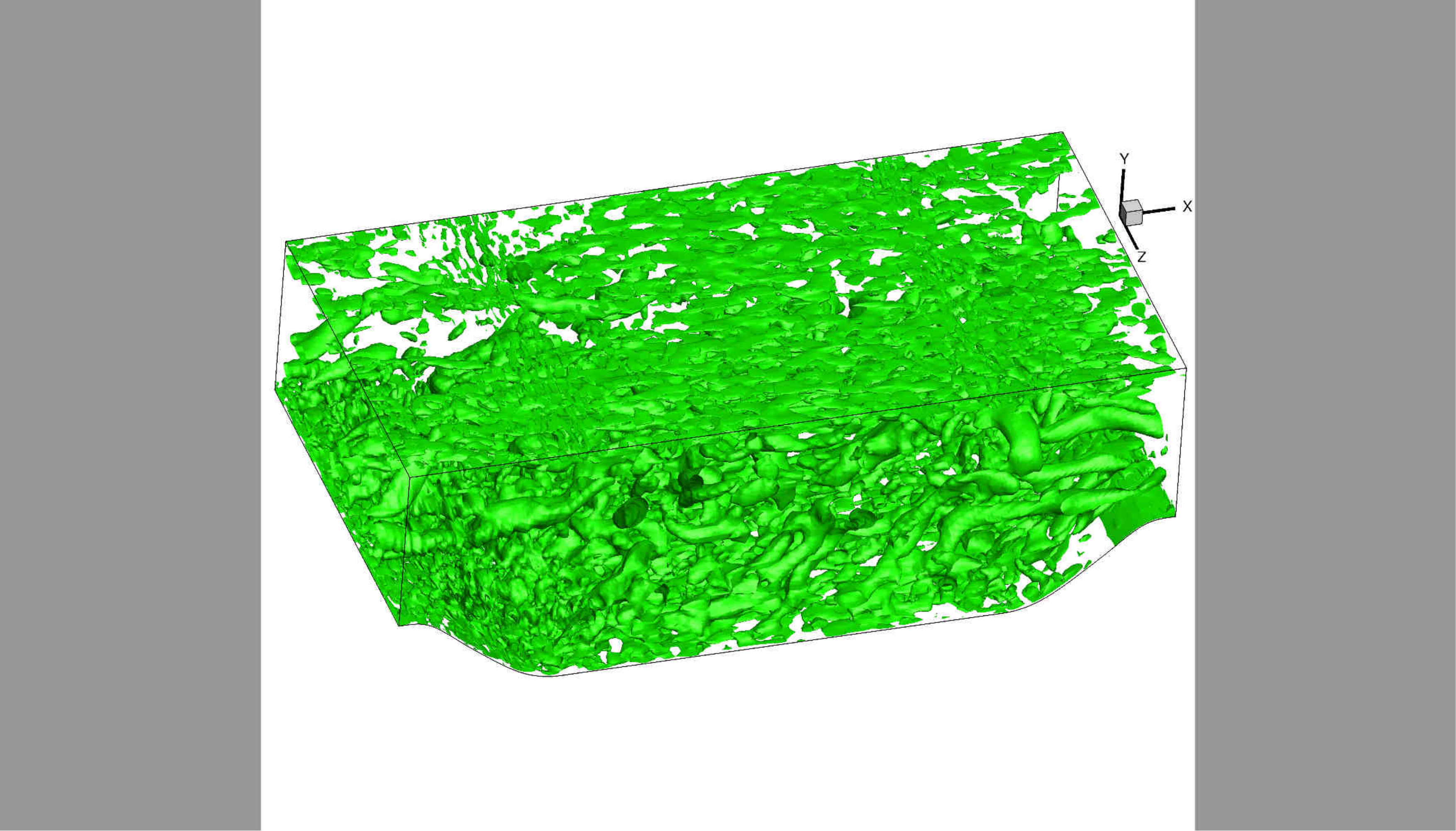}
   \vspace{-5pt}
   \caption{}
\end{subfigure}
\vspace{-5pt}
\caption{Q isosurfaces, Re=37000: (a) RANS, (b) $f_k$=0.35, (c) $f_k$=0.25, (d) $f_k$=0.15}
\label{SQ}
\end{figure}

This section is aimed to get insight into the flow structure of the current flow geometry. Contours of the instantaneous velocity field for the RANS and PANS calculations are depicted in Fig. \ref{SI}. The unsteadiness of the flow field is captured by the PANS calculations, whereas RANS solution looks steady. Failure of the RANS model to resolve  the wide range of scales inherent in the flow field and three dimensionality nature of the flow is sought via illustrating Q isosurfaces. The scalar Q defined by $-\frac{1}{2}\left(S_{ij}S_{ij}-\Omega_{ij}\Omega_{ij}\right)$ shows the balance between the rotation rate and strain rate and it provides a better visualization of the turbulent structures. Figure \ref{SQ} shows existence of eddies with wide range of scales for the PANS calculations, while for the RANS simulation, no flow structure is seen. Apparently, by reducing the filter parameter, $f_k$, more scales of motion are resolved and flow unsteadiness is better captured. The PANS simulations show that the whole recirculation region is influenced by large-scale energetic eddies with strong deformation and three dimensional interactions, which are ill-described by the RANS calculation.       

\begin{figure}[H]
\centering
\begin{subfigure}{.5\textwidth}
  \centering
  \includegraphics[trim=1cm 0.8cm 9.5cm 14.0cm, clip=true, scale=0.5]{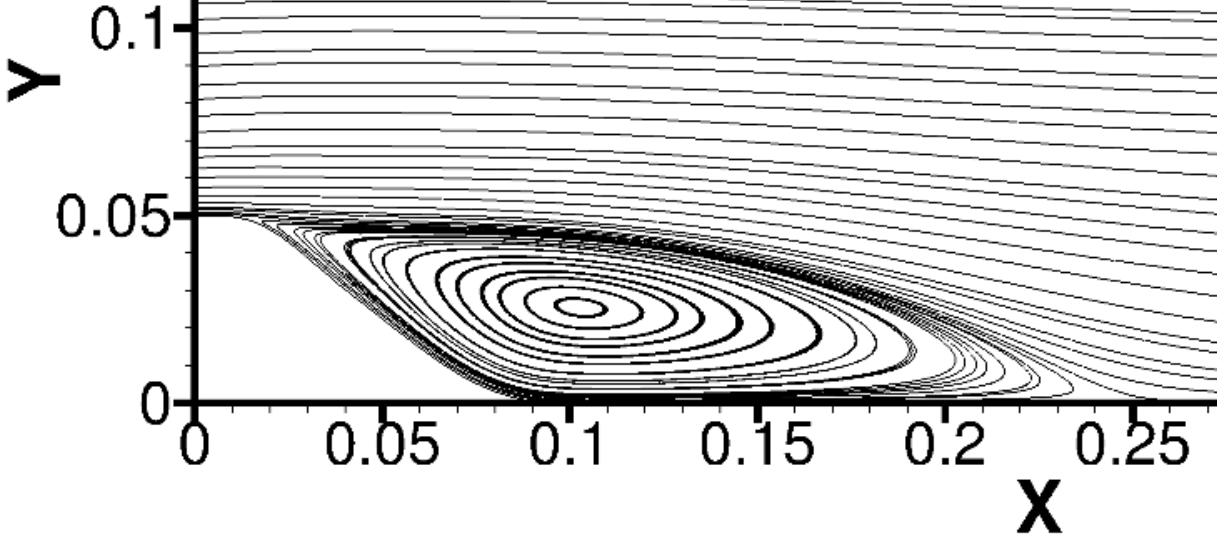}
  \vspace{-5pt}
  \caption{}
\end{subfigure}%
\begin{subfigure}{.5\textwidth}
  \centering
  \includegraphics[trim=1cm 0.8cm 9.5cm 14.0cm, clip=true, scale=0.5]{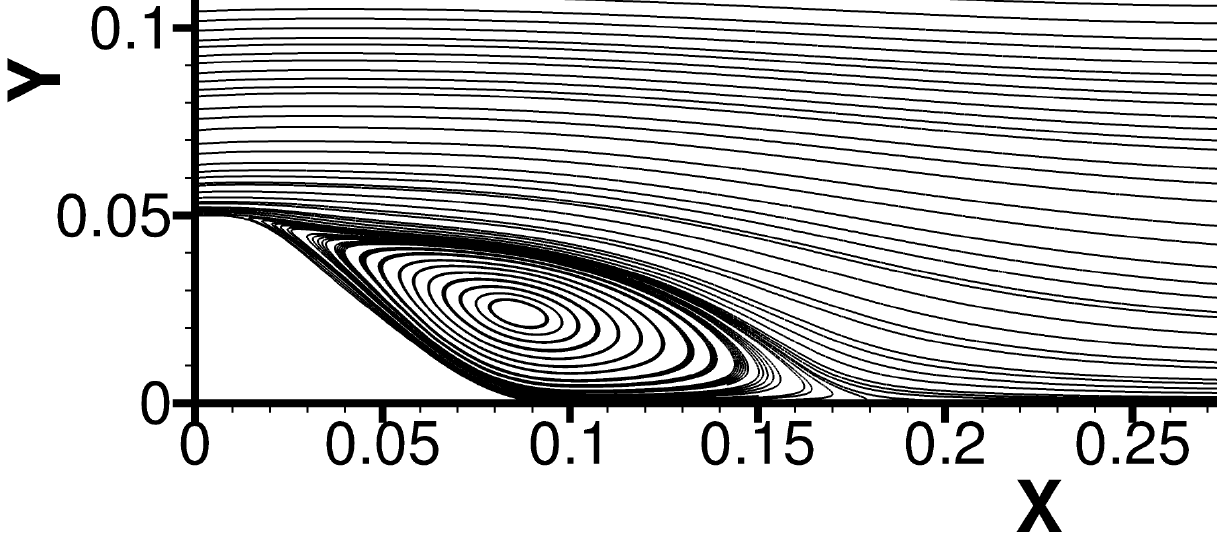}
  \vspace{-5pt}
  \caption{}
\end{subfigure}
\begin{subfigure}{.5\textwidth}
   \centering
   \includegraphics[trim=1cm 0.8cm 9.5cm 14.0cm, clip=true, scale=0.5]{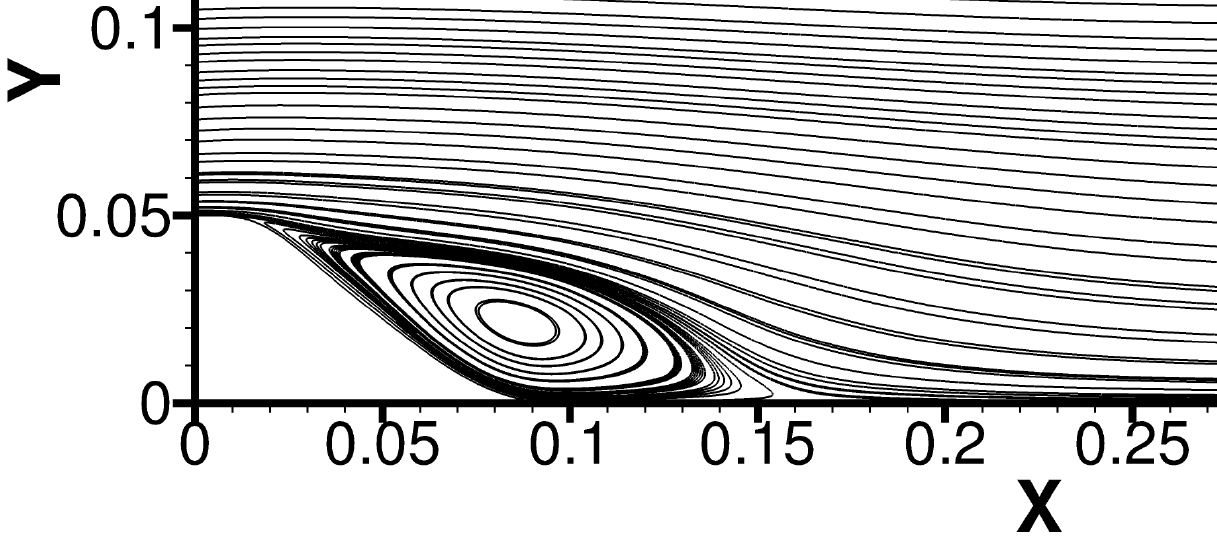}
   \vspace{-5pt}
   \caption{}
\end{subfigure}%
\begin{subfigure}{.5\textwidth}
   \centering
   \includegraphics[trim=1cm 0.8cm 9.5cm 14.0cm, clip=true, scale=0.5]{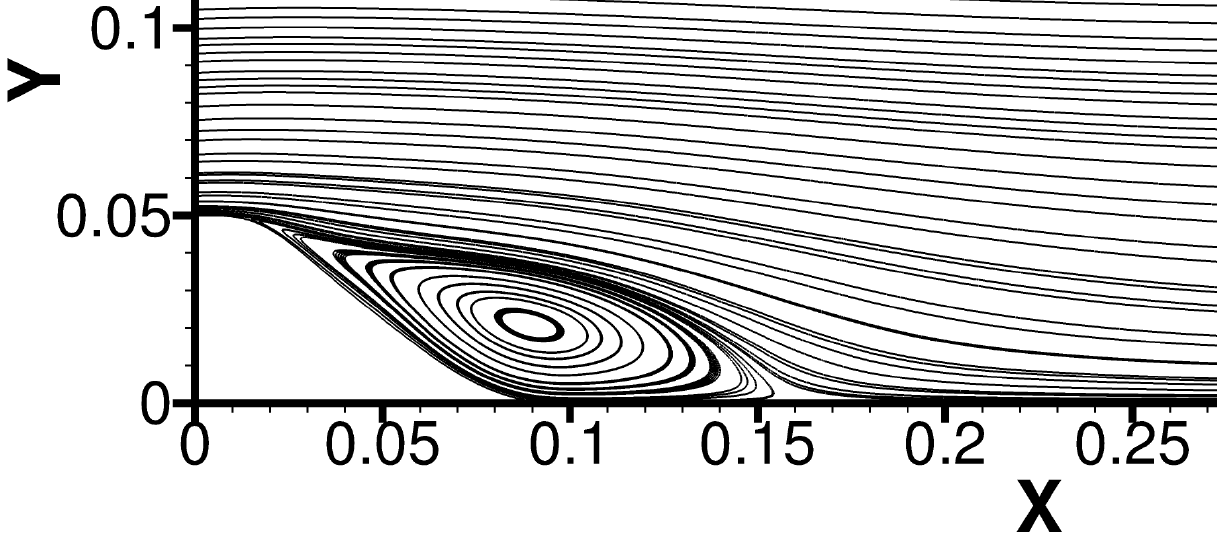}
   \vspace{-5pt}
   \caption{}
\end{subfigure}
\vspace{-5pt}
\caption{Time-averaged streamlines, Re=37000: (a) RANS, (b) $f_k$=0.35, (c) $f_k$=0.25, (d) $f_k$=0.15}
\label{SS}
\end{figure}

The physical point of view is addressed in the above discussion by looking at instantaneous flow field. However, it is of the practical and engineering aspect to investigate the size of separation bubble and point of separation for the mean flow. For this purpose, the time-averaged streamline contours are shown in figure \ref{SS} for the RANS and PANS calculations. Figure \ref{SS} indicates that the size of separation bubble is bigger in the RANS calculation which occupies more than 50 percent of the streamwise direction. By reducing $f_{k}$, the separated region is narrowed down in both x and y directions. 


\subsubsection{Separation and reattachment locations}
To qualitatively identify the location of separation and reattachments points, Fig. \ref{cf} is plotted for friction coefficient. These locations are obtained based on the fact that the friction coefficient vanishes at the point of separation and reattachment. Figure \ref{cf} shows that separation is delayed for the RANS simulation and is extended to farther downstream direction to reattach. However, both PITM and PANS calculations seem to follow the same pattern for flow separation and reattachment. Another interesting observation of this plot is the flow behaviour right after reattachment. After reattachment and partial recovery, the flow seems to be prone to separation at around $x/h$=7.2 where it is decelerated moving towards the downstream hill resulting in a local minimum in the friction coefficient plot. However, the flow accelerates on the windward slope of the hill and that is the reason for sharp rise of the friction coefficient just upstream of the hill crest. 

\begin{figure}
        \centering
 		\captionsetup{justification=centering}                                               
 		        \begin{subfigure}[b]{0.4\textwidth}
                \includegraphics[width=\textwidth]{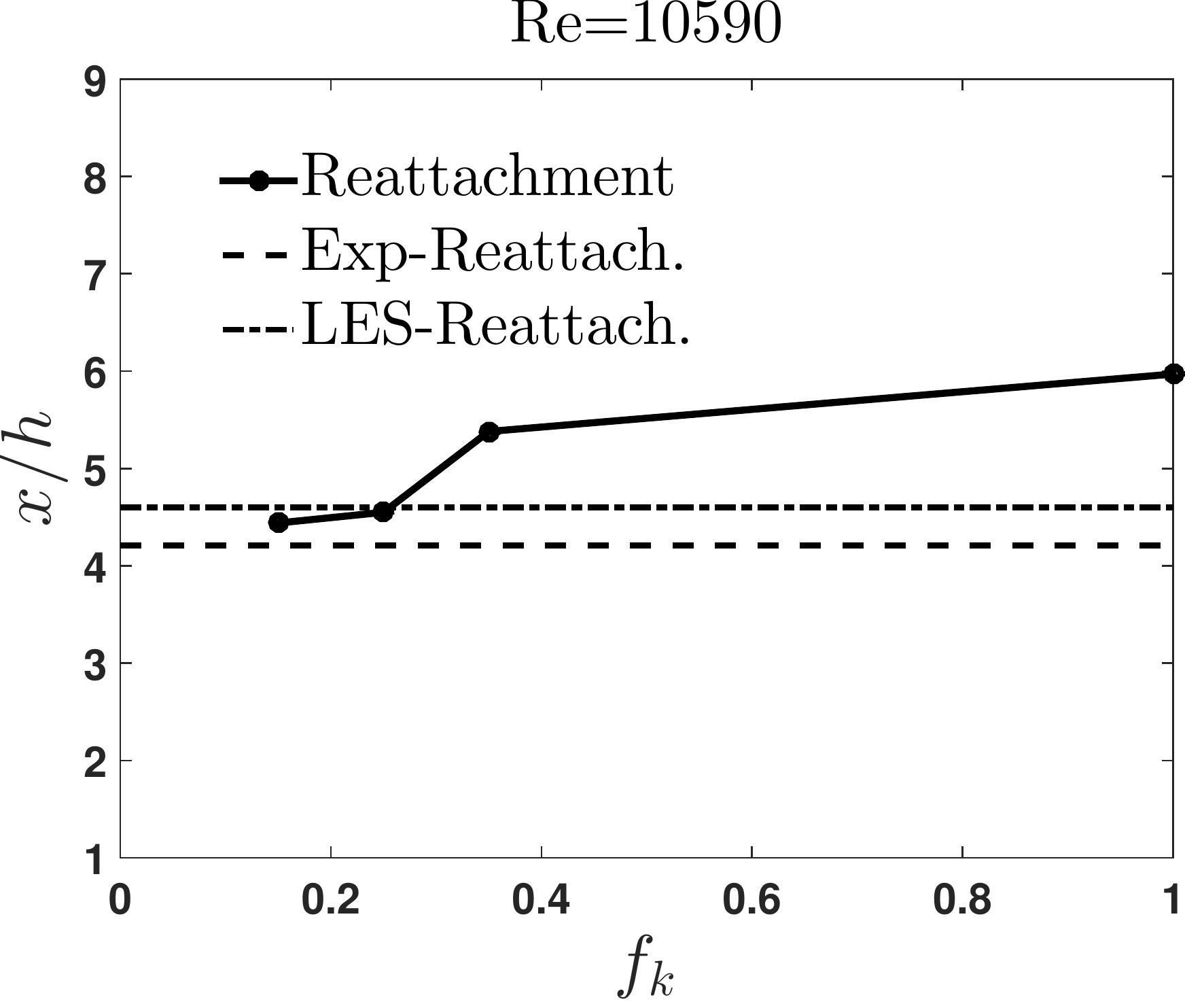}\begin{picture}(0,0)\put(-138,0){(a)}\end{picture}
        \end{subfigure}
        			\begin{subfigure}[b]{0.4\textwidth}
                \includegraphics[width=\textwidth]{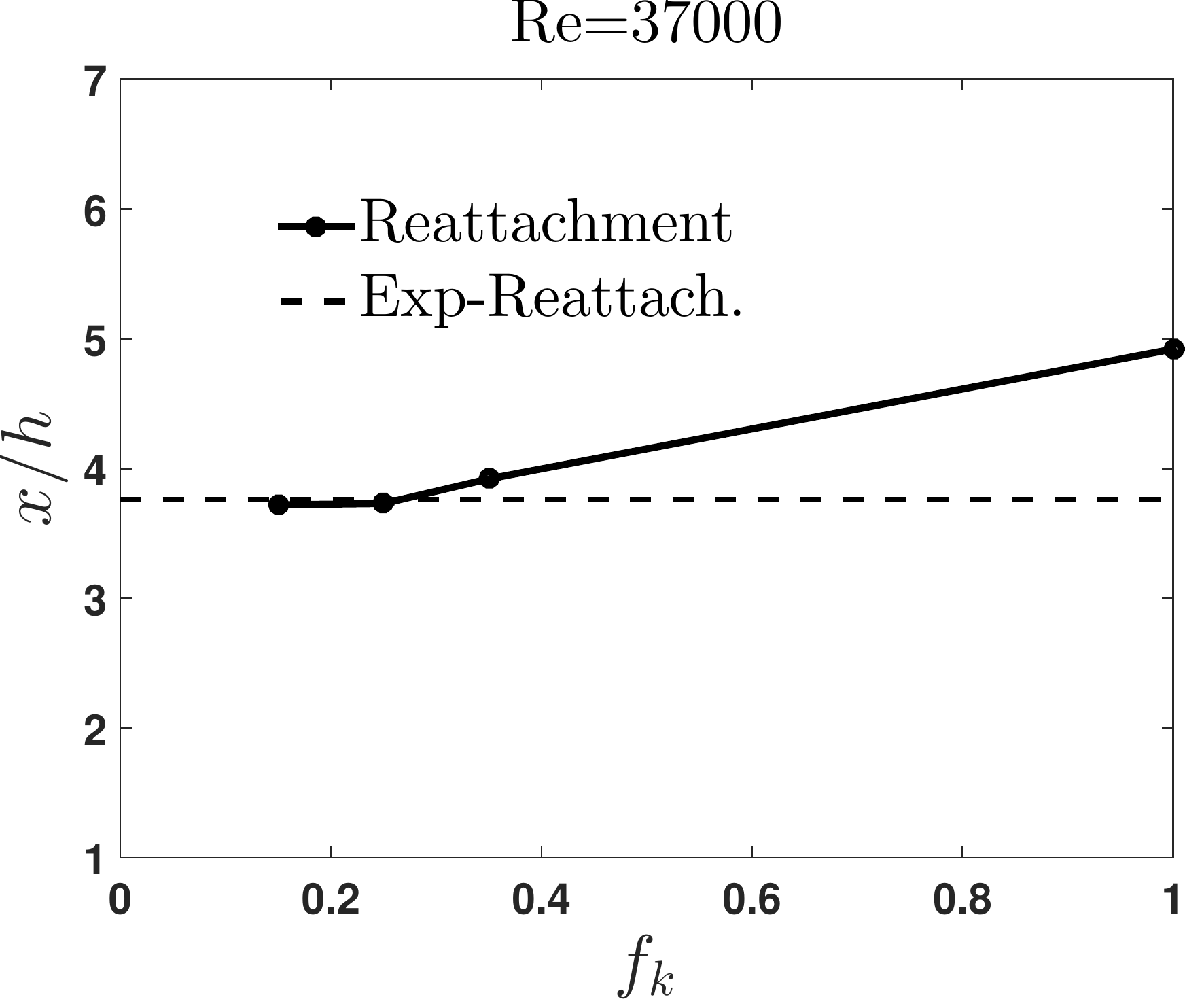}\begin{picture}(0,0)\put(-138,0){(b)}\end{picture}
        \end{subfigure}
  \caption{Reattachment location for different $f_k$ (a) $Re=10590$ (b) $Re=37000$}   
\label{reattach}                         
\end{figure}

Table \ref{sep.} summarizes the reattachment locations for all the simulations along with the experimental study which are also plotted in Fig. \ref{reattach}. It is seen from Fig. \ref{reattach} (a) that the prediction of the reattachment location improves dramatically with decreasing $f_k$. RANS is in error by about 40$\%$ whereas $f_k$ = 0.15 case is within 5$\%$ of the experimental data. High quality predictions of the reattachment location by reducing $f_k$ is also observed in Fig. \ref{reattach} (b) for the higher Reynolds number case.

\begin{figure}
\centering
  \includegraphics[scale=0.5]{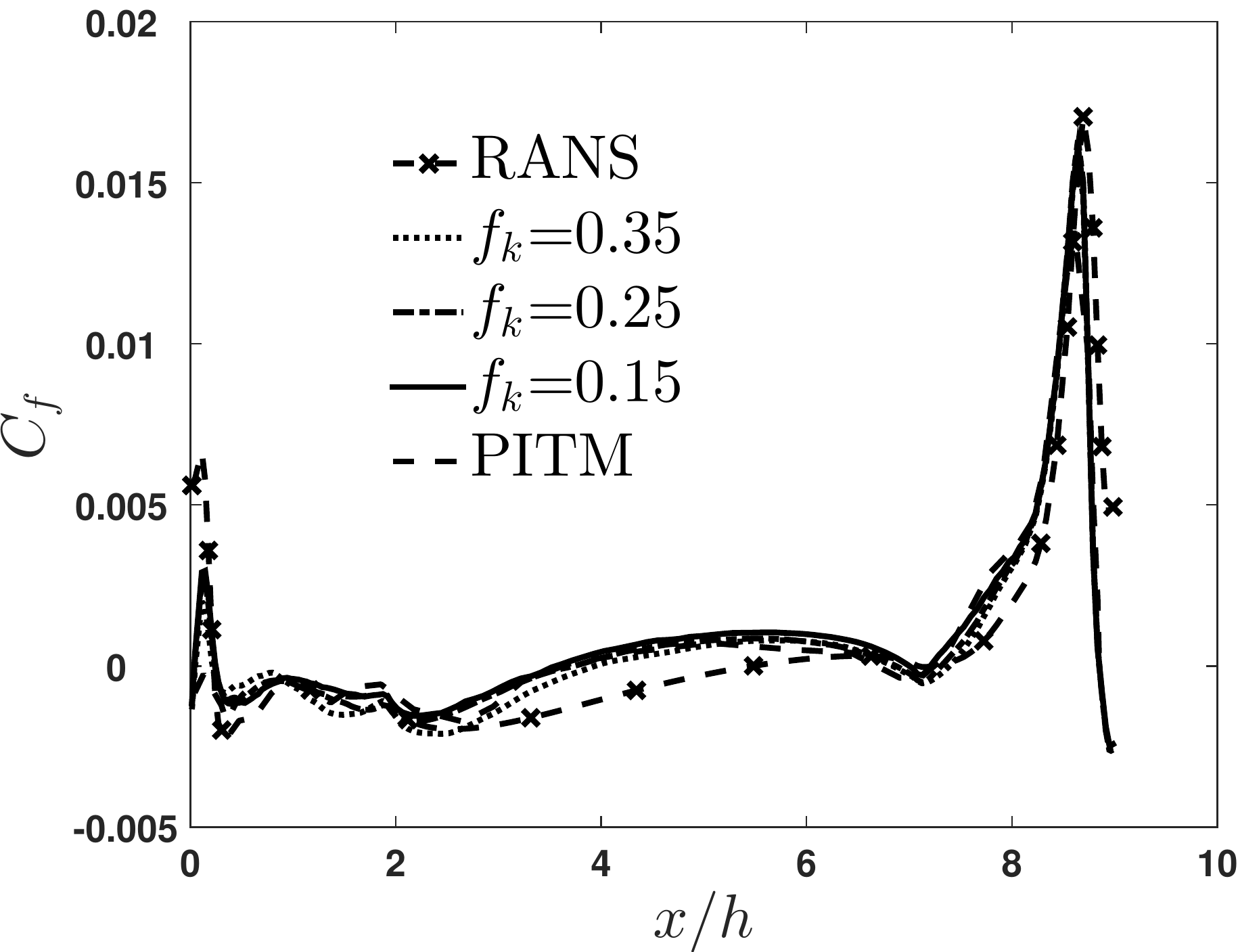}
\caption{Friction Coefficient}
\label{cf}
\end{figure}

\begin{table}
\centering
\caption{Reattachment locations for different simulations}
\begin{tabular}{ l l l l l}
\hline\noalign{\smallskip}
\textbf{Study} & \textbf{$f_k$} & \textbf{$f_\epsilon$} &  \textbf{$\left(x/h\right)_{Reattach.}$}\\ \hline\noalign{\smallskip}

\textbf{Re=37000} \\ \hline\noalign{\smallskip}
RANS & 1 & 1 & 4.92 \\ \hline\noalign{\smallskip}
PANS & 0.35 & 1 & 3.92 \\ \hline\noalign{\smallskip}
PANS & 0.25 & 1 & 3.73 \\ \hline\noalign{\smallskip}
PANS & 0.15 & 1 & 3.72 \\ \hline\noalign{\smallskip}
PITM \cite{PITM}& - & - & 3.63 \\ \hline\noalign{\smallskip}
Exp. \cite{Rapp}& - & - & 3.76 \\ \hline\noalign{\smallskip}
\textbf{Re=10590} \\ \hline\noalign{\smallskip}
RANS & 1 & 1 & 5.97 \\ \hline\noalign{\smallskip}
PANS & 0.35 & 1 & 5.38 \\ \hline\noalign{\smallskip}
PANS & 0.25 & 1 & 4.55 \\ \hline\noalign{\smallskip}
PANS & 0.15 & 1 & 4.44 \\ \hline\noalign{\smallskip}
LES \cite{Frohlich}& - & - & 4.6 \\ \hline\noalign{\smallskip}
Exp. \cite{Rapp}& - & - & 4.21 \\ \hline\noalign{\smallskip}

\end{tabular}
\label{sep.}
\end{table}

\subsubsection{Resolution study}

In this section, the grid sensitivity of the PANS simulations with cut-off ratio of $f_k$=0.15 is investigated for two grid resolutions. The detail on the grid sizes is summarized in Table \ref{case}. As seen in Table \ref{case}, the coarse grid is generated by reducing the grid nodes in the streamwise and spanwise directions while keeping the resolution fixed in the normal direction. The results from the PANS simulations are compared with PITM results at almost the same grid sizes which are specified in Table \ref{case}. 

\begin{figure}
\centering
\begin{subfigure}{.5\textwidth}
  \centering
  \includegraphics[scale=0.3]{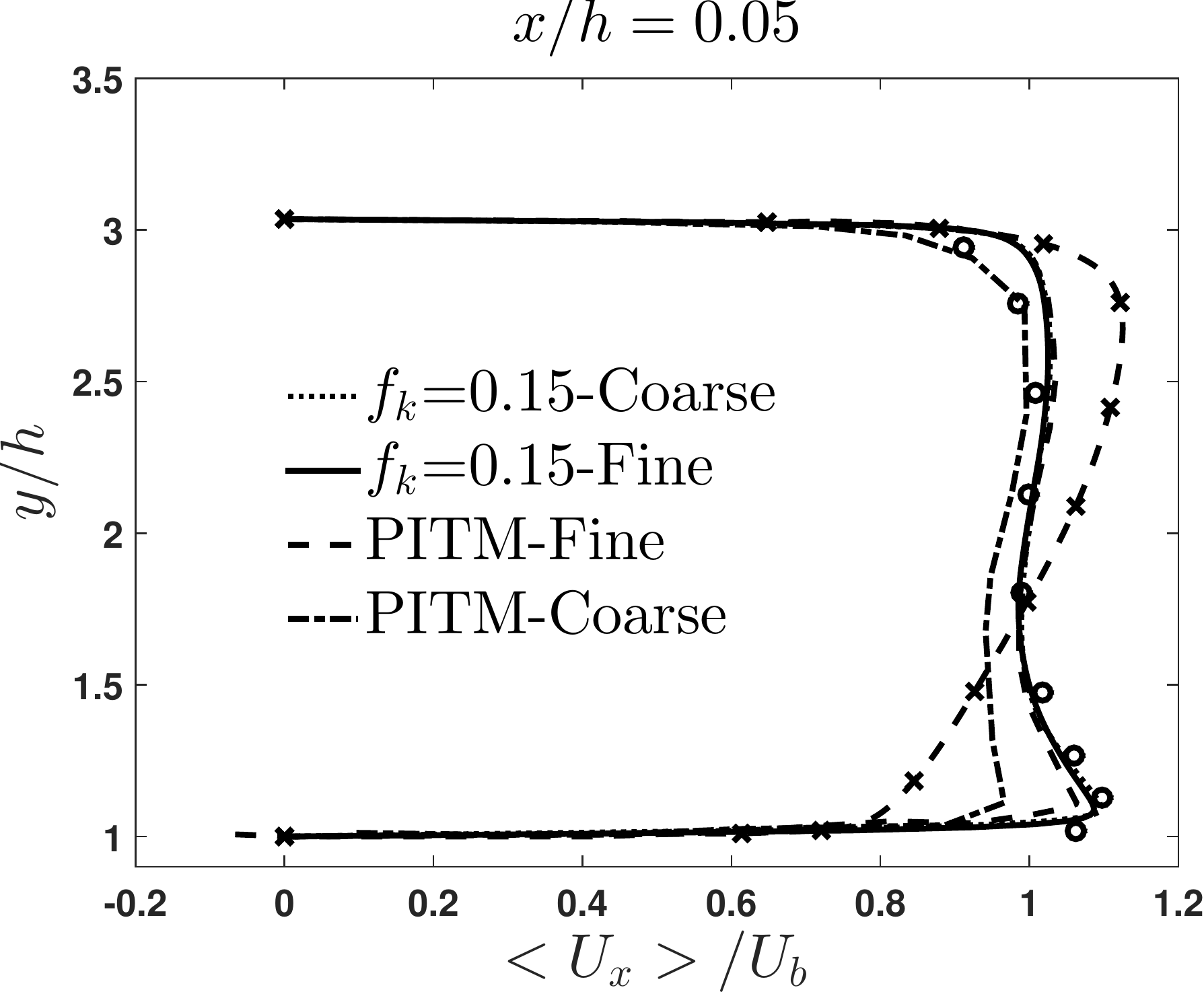}
  \vspace{-8pt}
  \caption{}
\end{subfigure}%
\begin{subfigure}{.5\textwidth}
  \centering
  \includegraphics[scale=0.3]{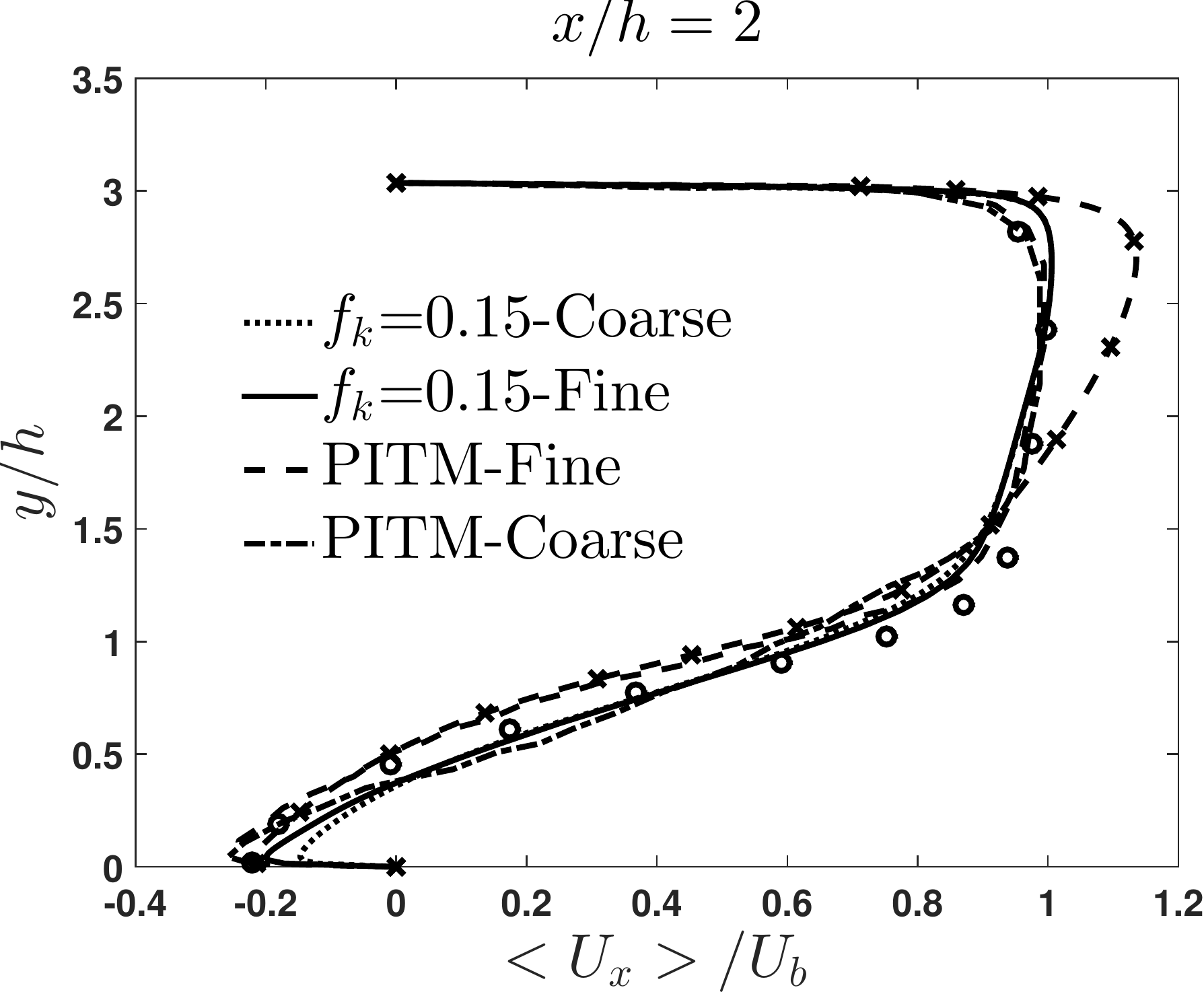}
  \vspace{-8pt}
  \caption{}
\end{subfigure}
\\
\begin{subfigure}{.5\textwidth}
   \centering
   \includegraphics[scale=0.3]{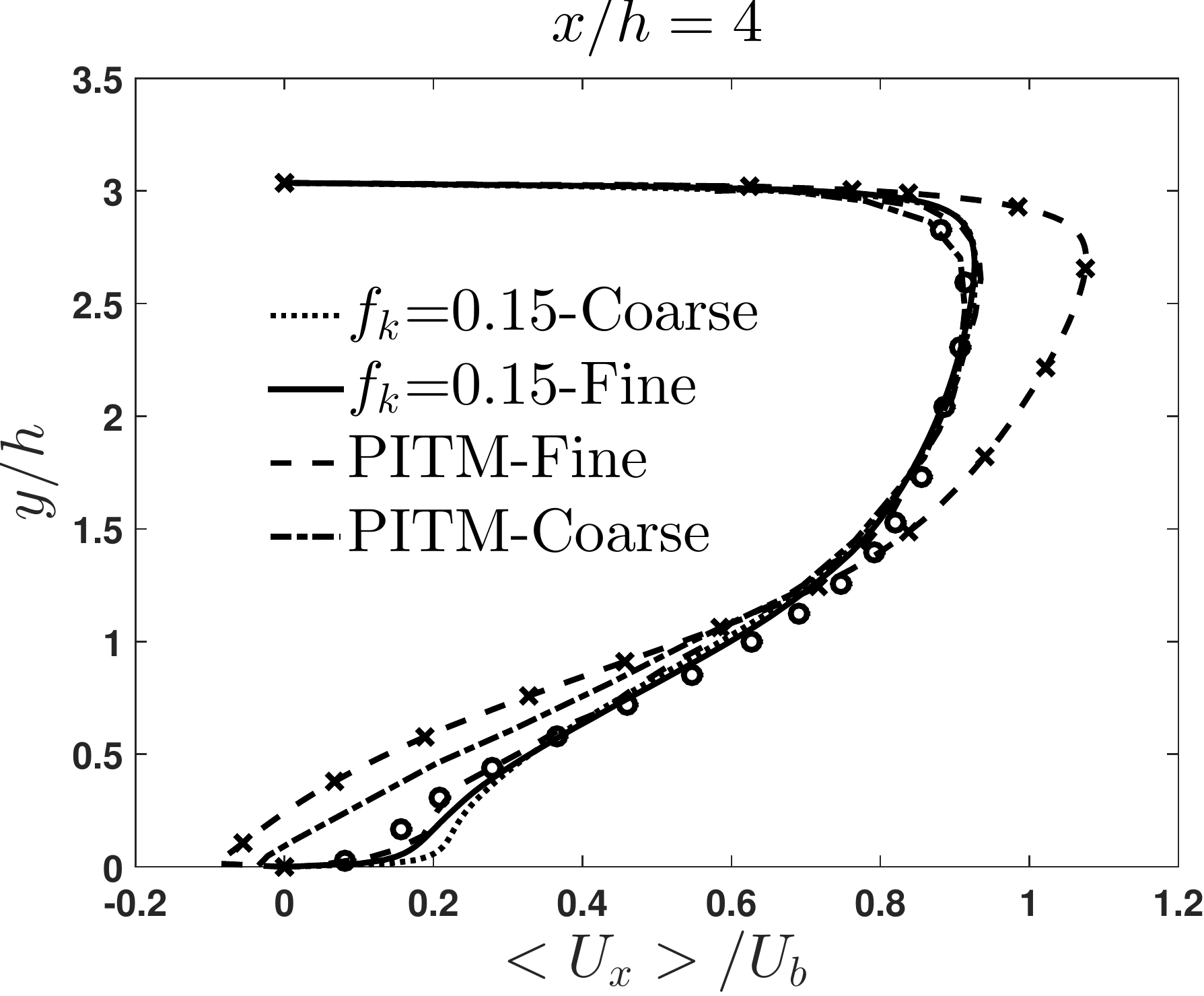}
   \vspace{-8pt}
   \caption{}
\end{subfigure}%
\begin{subfigure}{.5\textwidth}
   \centering
   \includegraphics[scale=0.3]{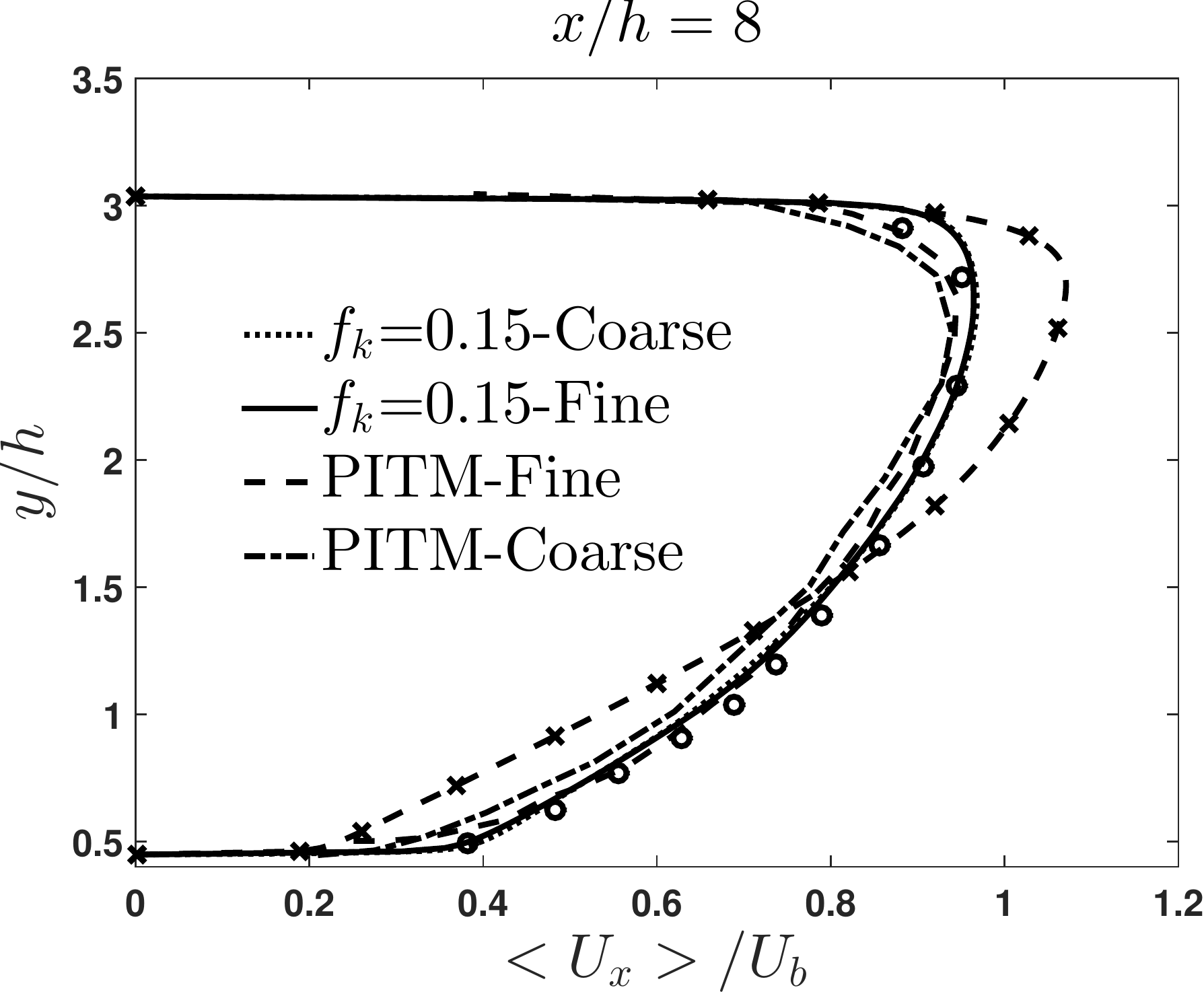}
   \vspace{-8pt}
   \caption{}
\end{subfigure}
\vspace{-8pt}
\caption{Streamwise velocity (symbols for Exp. and RANS are consistent with Fig.  \ref{fku})}
\label{gridu}
\end{figure}

\begin{figure}
\centering
\begin{subfigure}{.5\textwidth}
  \centering
  \includegraphics[scale=0.3]{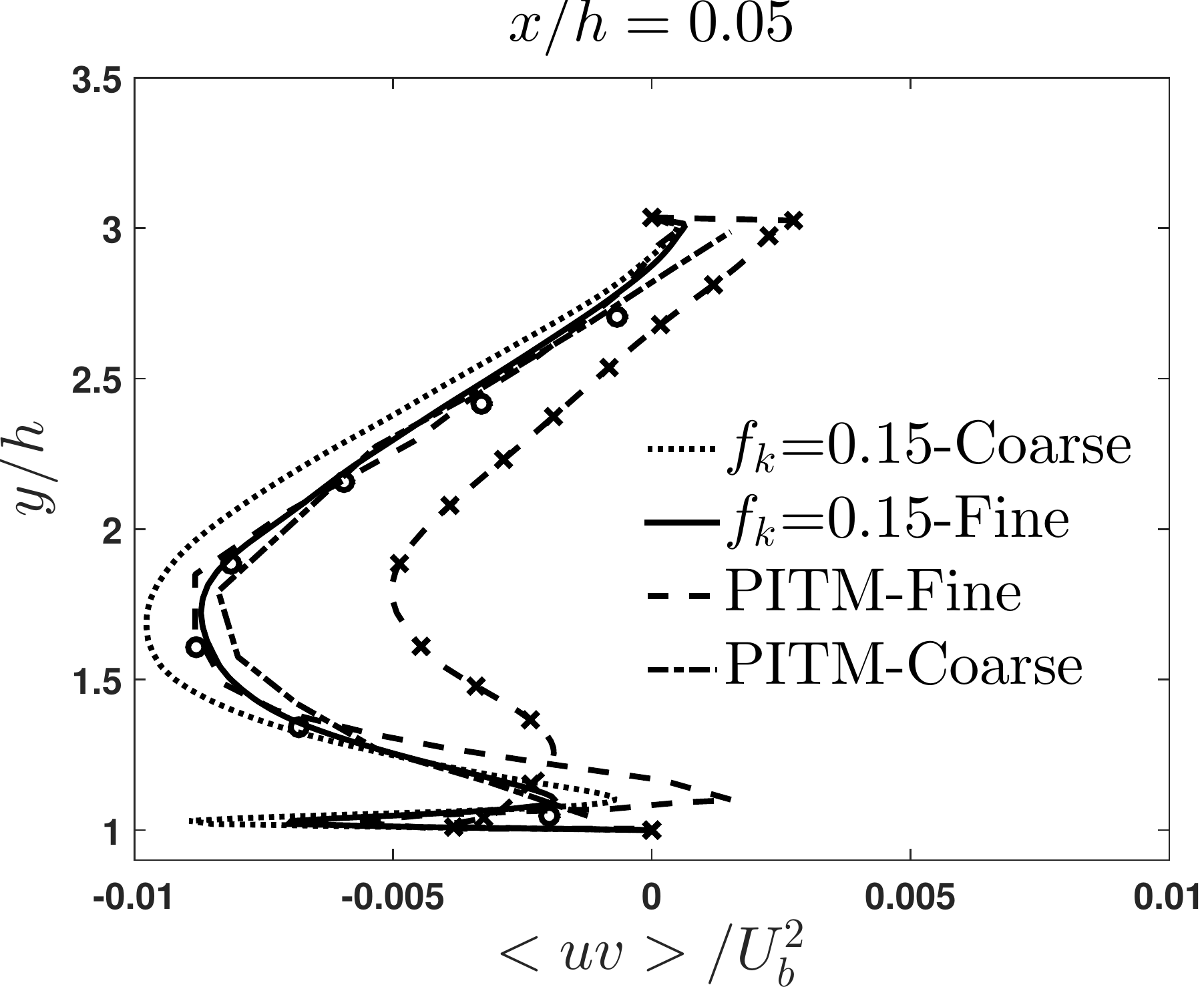}
  \vspace{-8pt}
  \caption{}
\end{subfigure}%
\begin{subfigure}{.5\textwidth}
  \centering
  \includegraphics[scale=0.3]{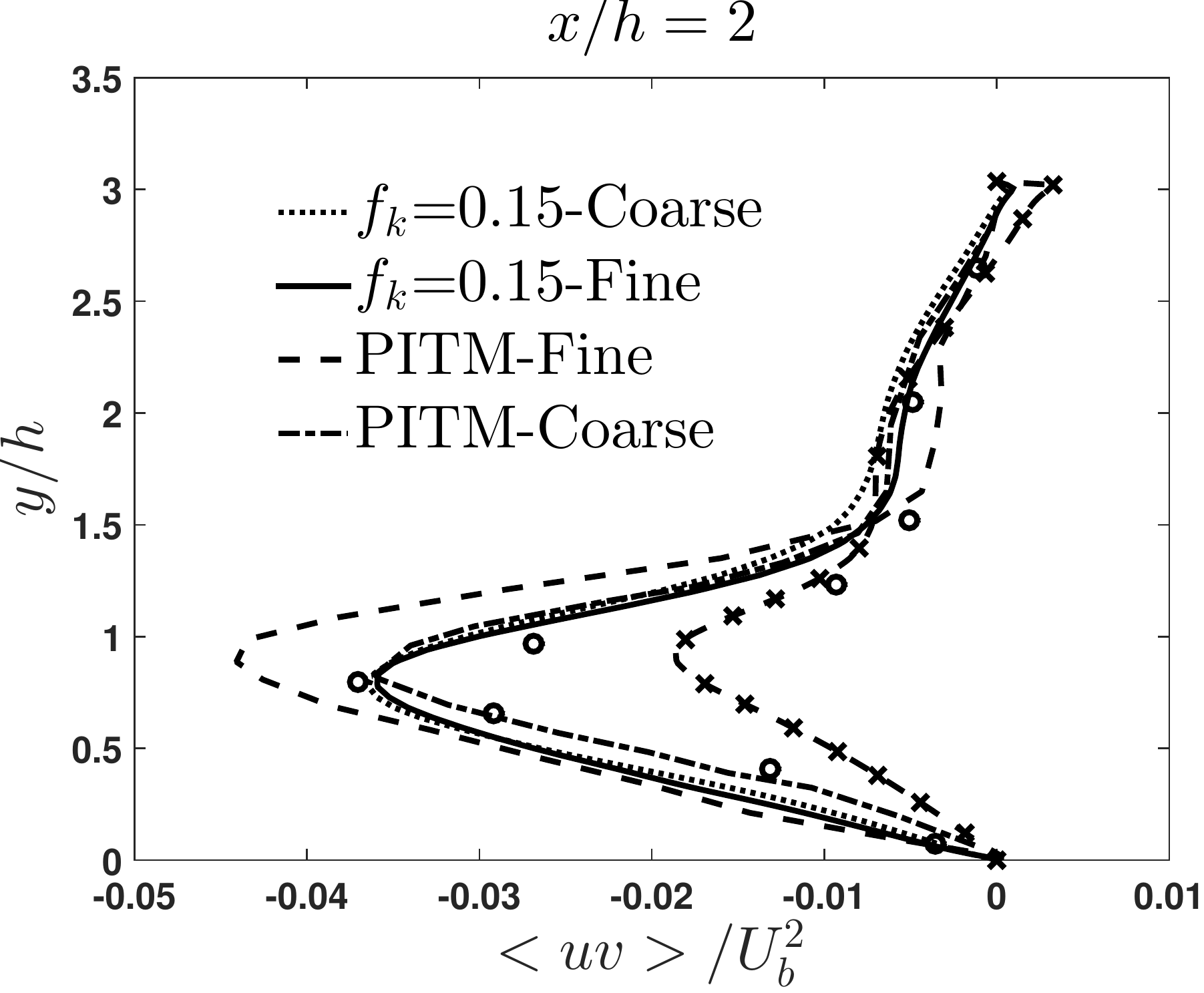}
  \vspace{-8pt}
  \caption{}
\end{subfigure}
\\
\begin{subfigure}{.5\textwidth}
   \centering
   \includegraphics[scale=0.3]{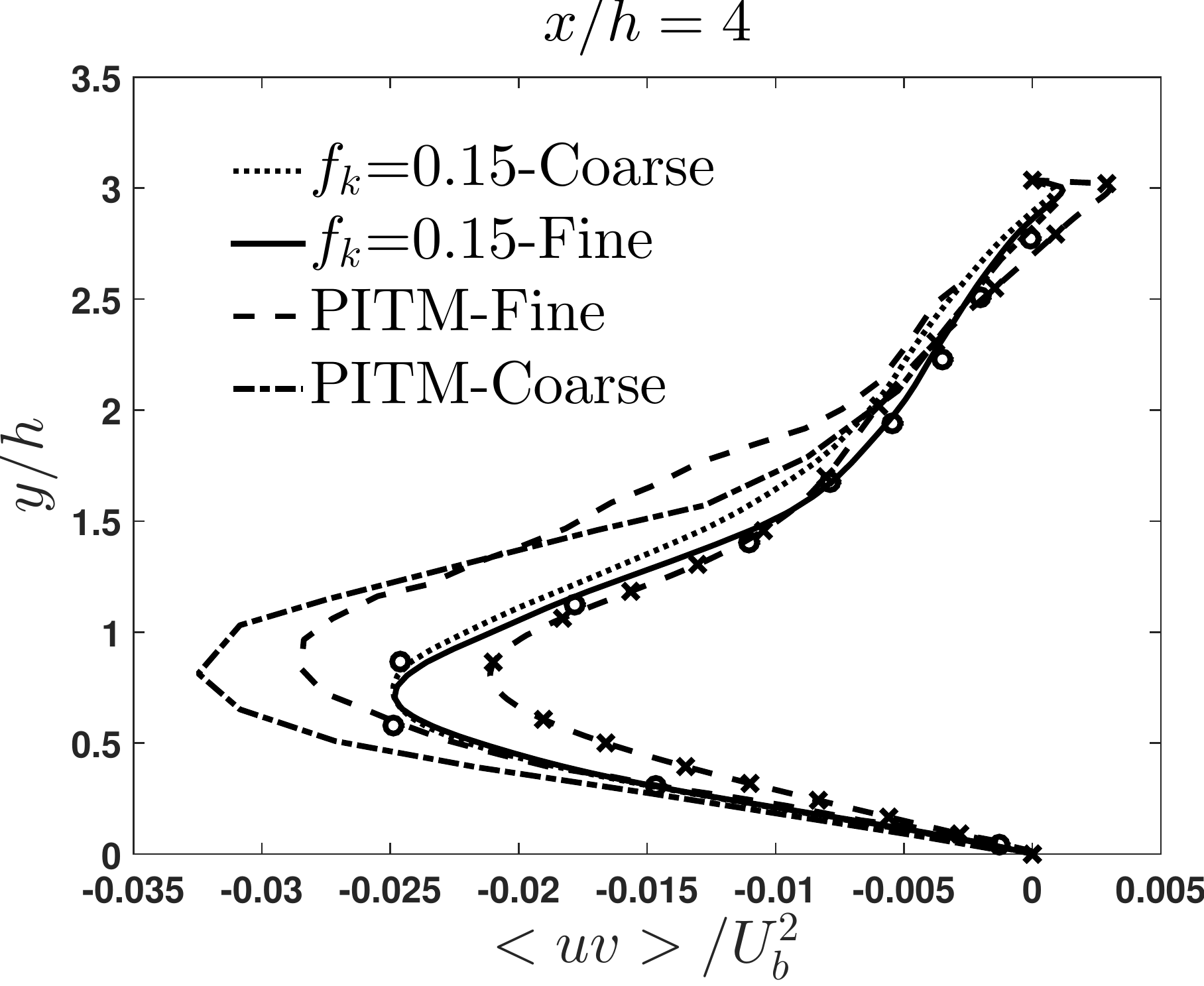}
   \vspace{-8pt}
   \caption{}
\end{subfigure}%
\begin{subfigure}{.5\textwidth}
   \centering
   \includegraphics[scale=0.3]{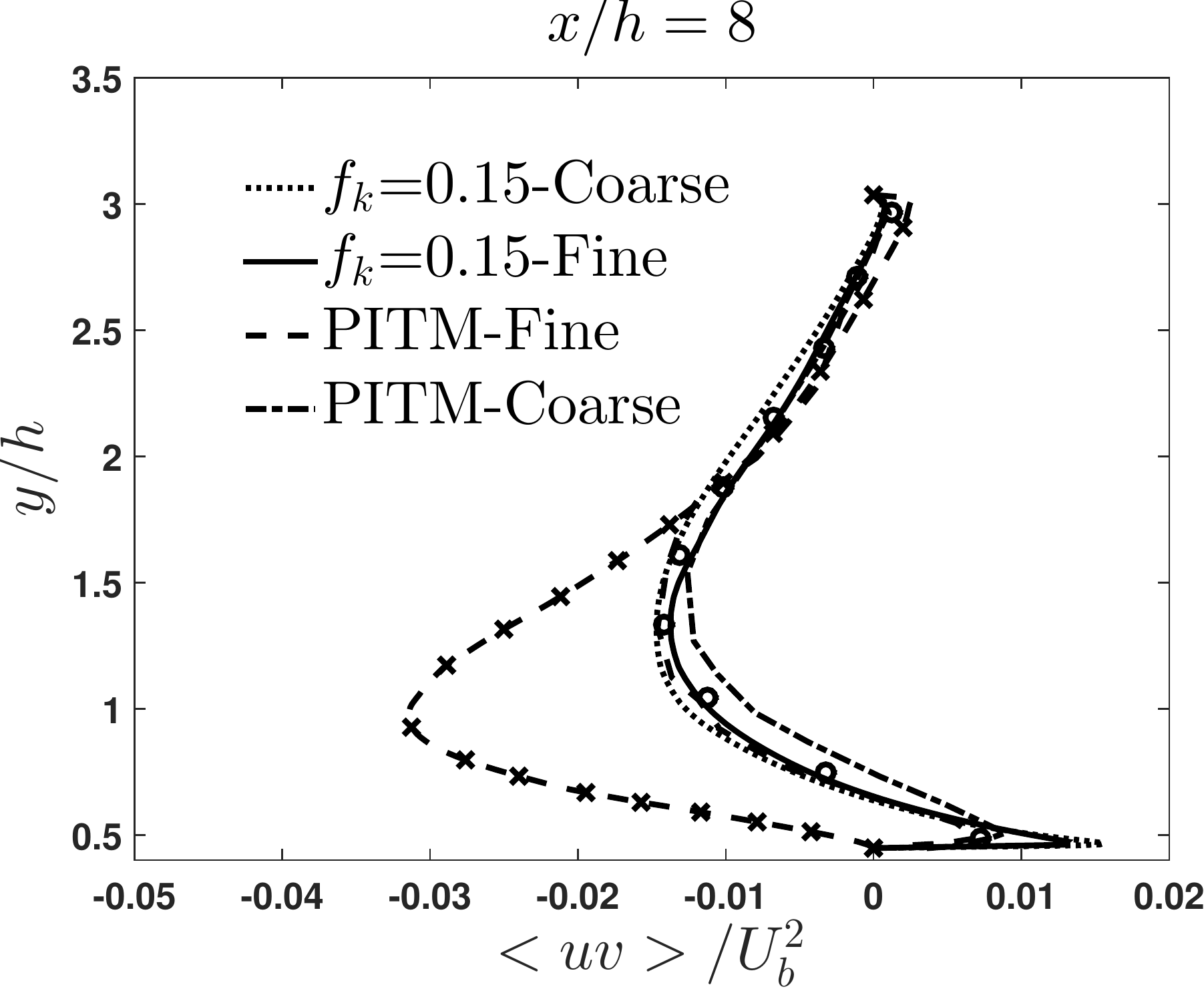}
   \vspace{-8pt}
   \caption{}
\end{subfigure}
\vspace{-8pt}
\caption{Shear stress (symbols for Exp. and RANS are consistent with Fig. \ref{fkuv})}
\label{griduv}
\end{figure}

\begin{figure}
\centering
\begin{subfigure}{.5\textwidth}
  \centering
  \includegraphics[scale=0.3]{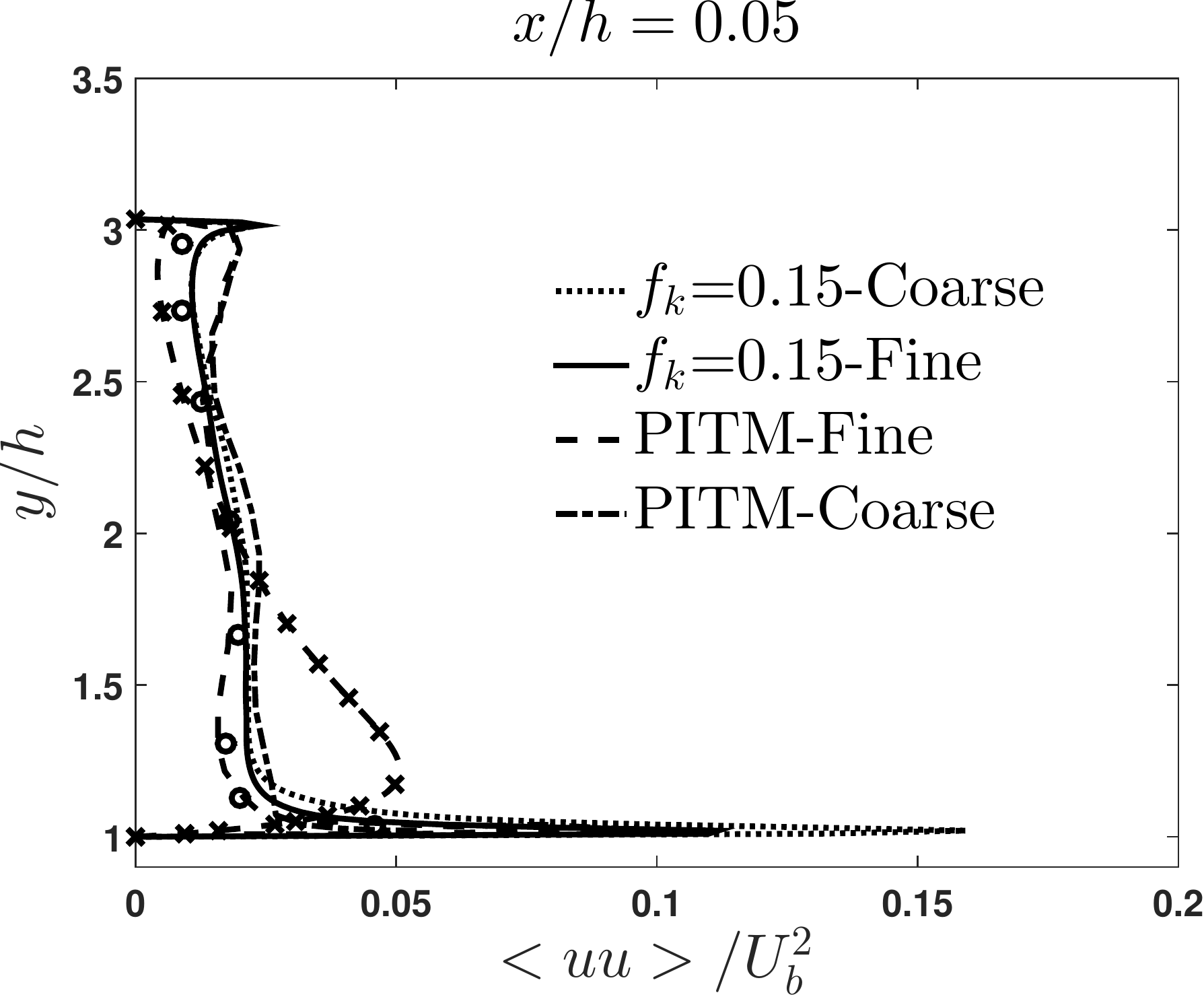}
  \vspace{-8pt}
  \caption{}
\end{subfigure}%
\begin{subfigure}{.5\textwidth}
  \centering
  \includegraphics[scale=0.3]{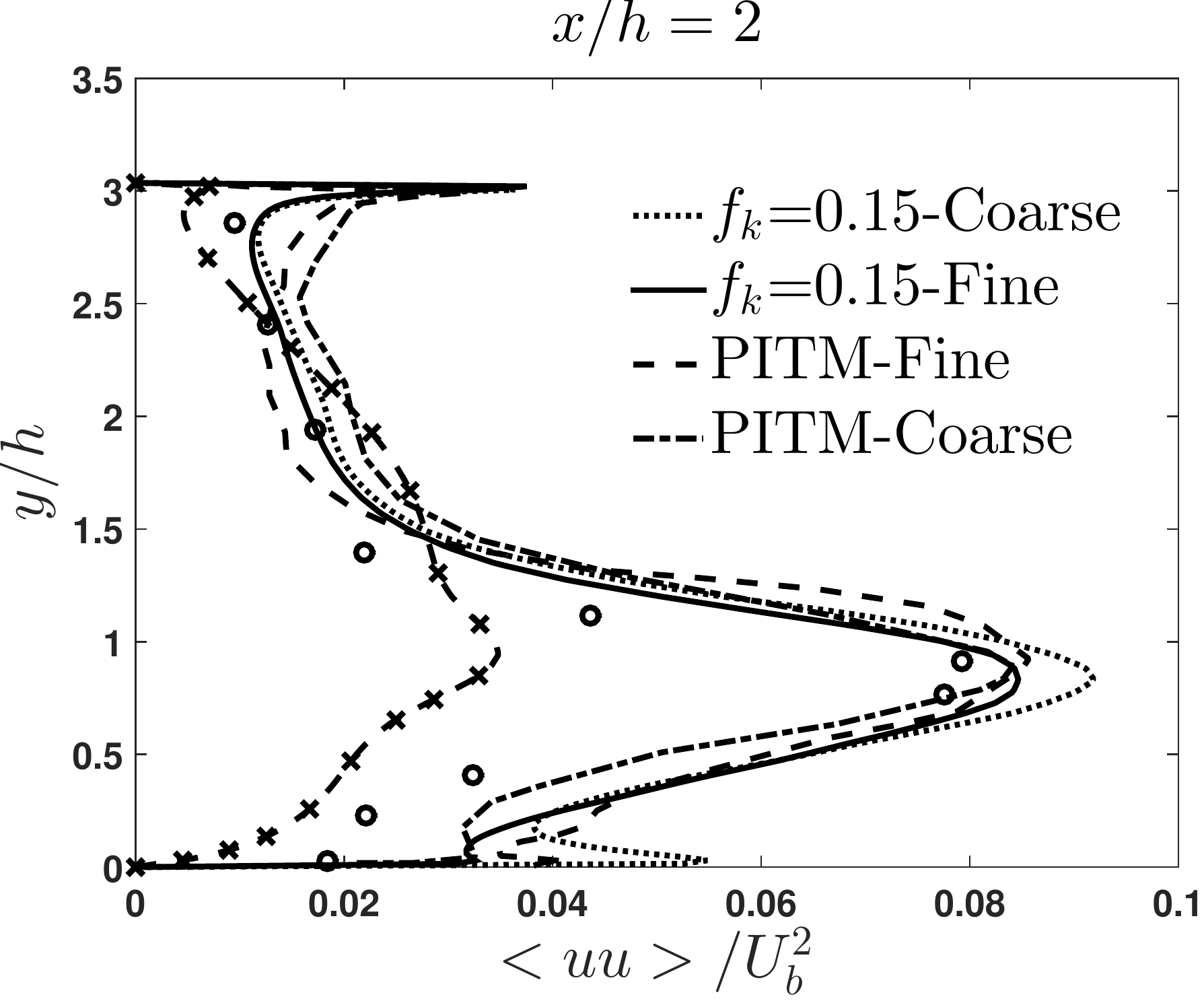}
  \vspace{-8pt}
  \caption{}
\end{subfigure}
\\
\begin{subfigure}{.5\textwidth}
   \centering
   \includegraphics[scale=0.3]{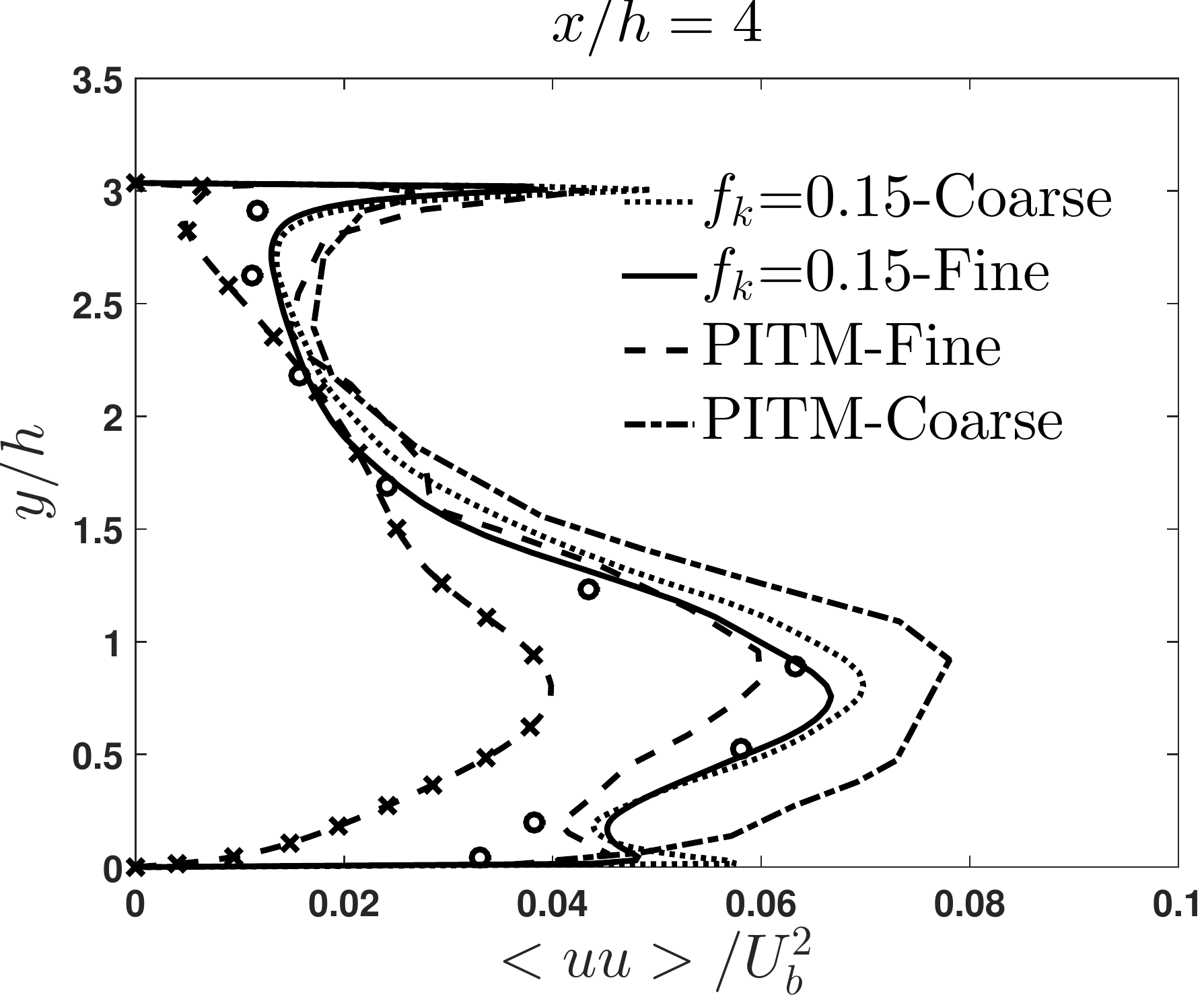}
   \vspace{-8pt}
   \caption{}
\end{subfigure}%
\begin{subfigure}{.5\textwidth}
   \centering
   \includegraphics[scale=0.3]{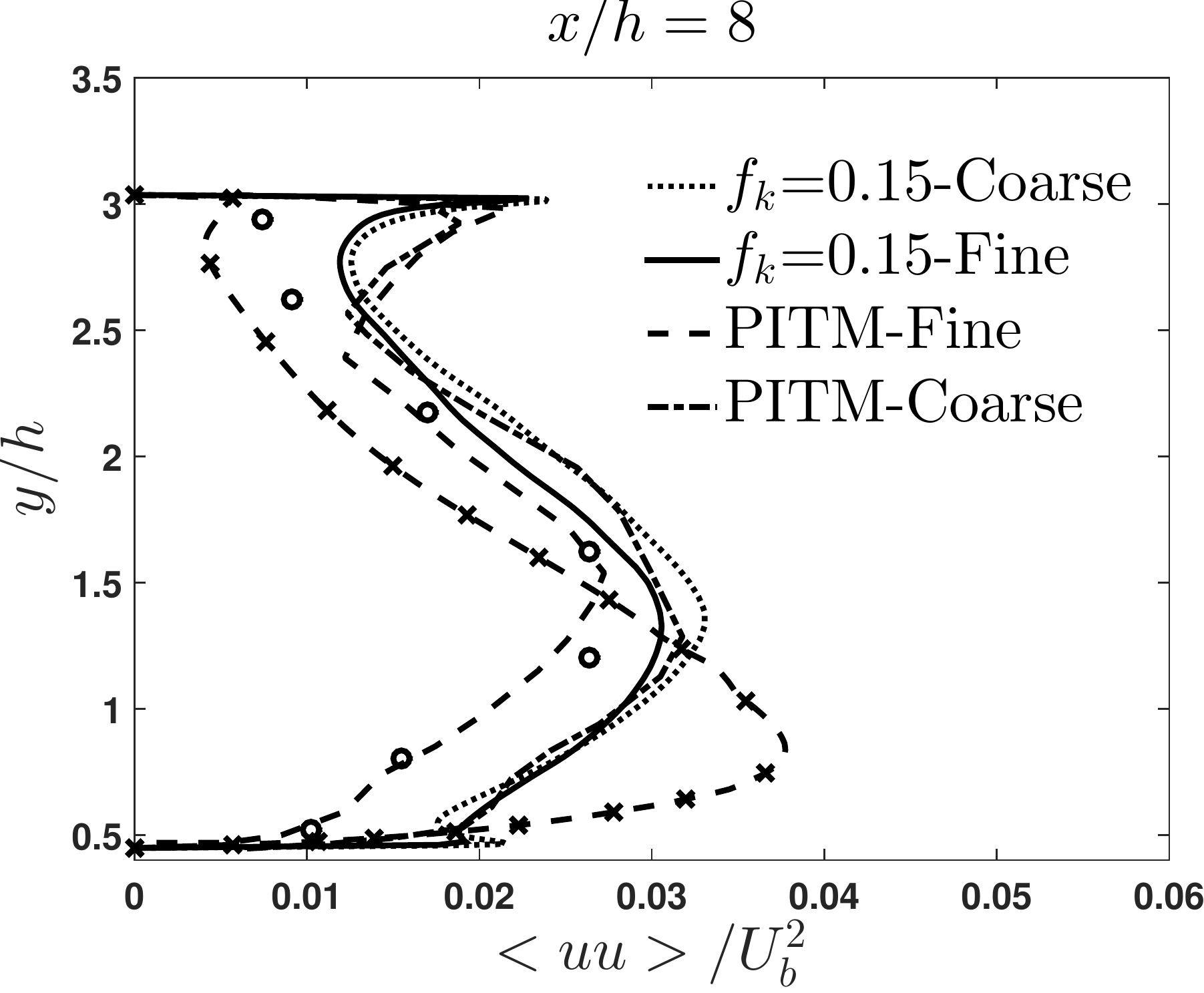}
   \vspace{-8pt}
   \caption{}
\end{subfigure}
\vspace{-8pt}
\caption{Streamwise stress component (symbols for Exp. and RANS are consistent with Fig. \ref{fkuv})}
\label{griduu}
\end{figure}

\begin{figure}
\centering
\begin{subfigure}{.5\textwidth}
  \centering
  \includegraphics[scale=0.3]{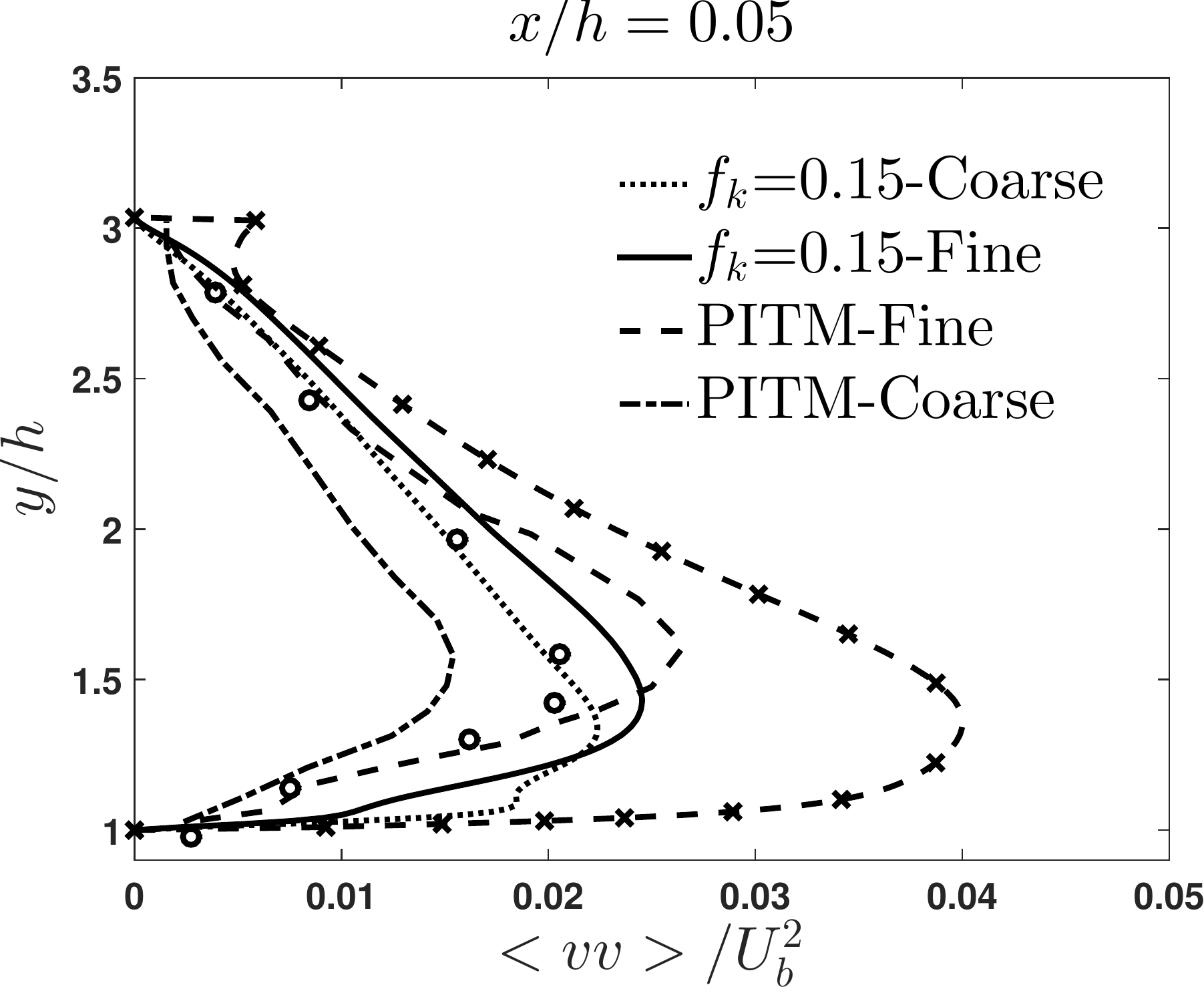}
  \vspace{-8pt}
  \caption{}
\end{subfigure}%
\begin{subfigure}{.5\textwidth}
  \centering
  \includegraphics[scale=0.3]{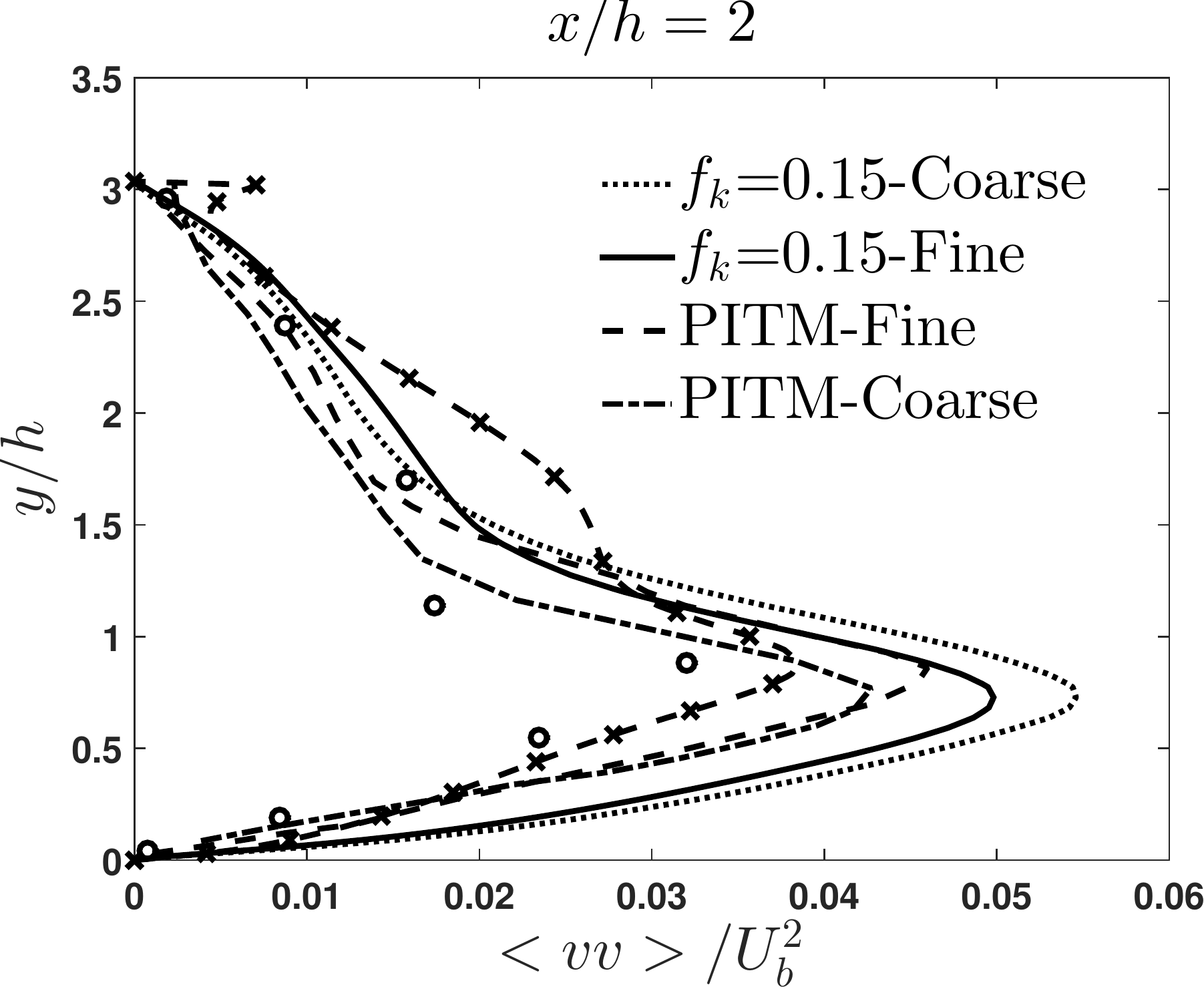}
  \vspace{-8pt}
  \caption{}
\end{subfigure}
\\
\begin{subfigure}{.5\textwidth}
   \centering
   \includegraphics[scale=0.3]{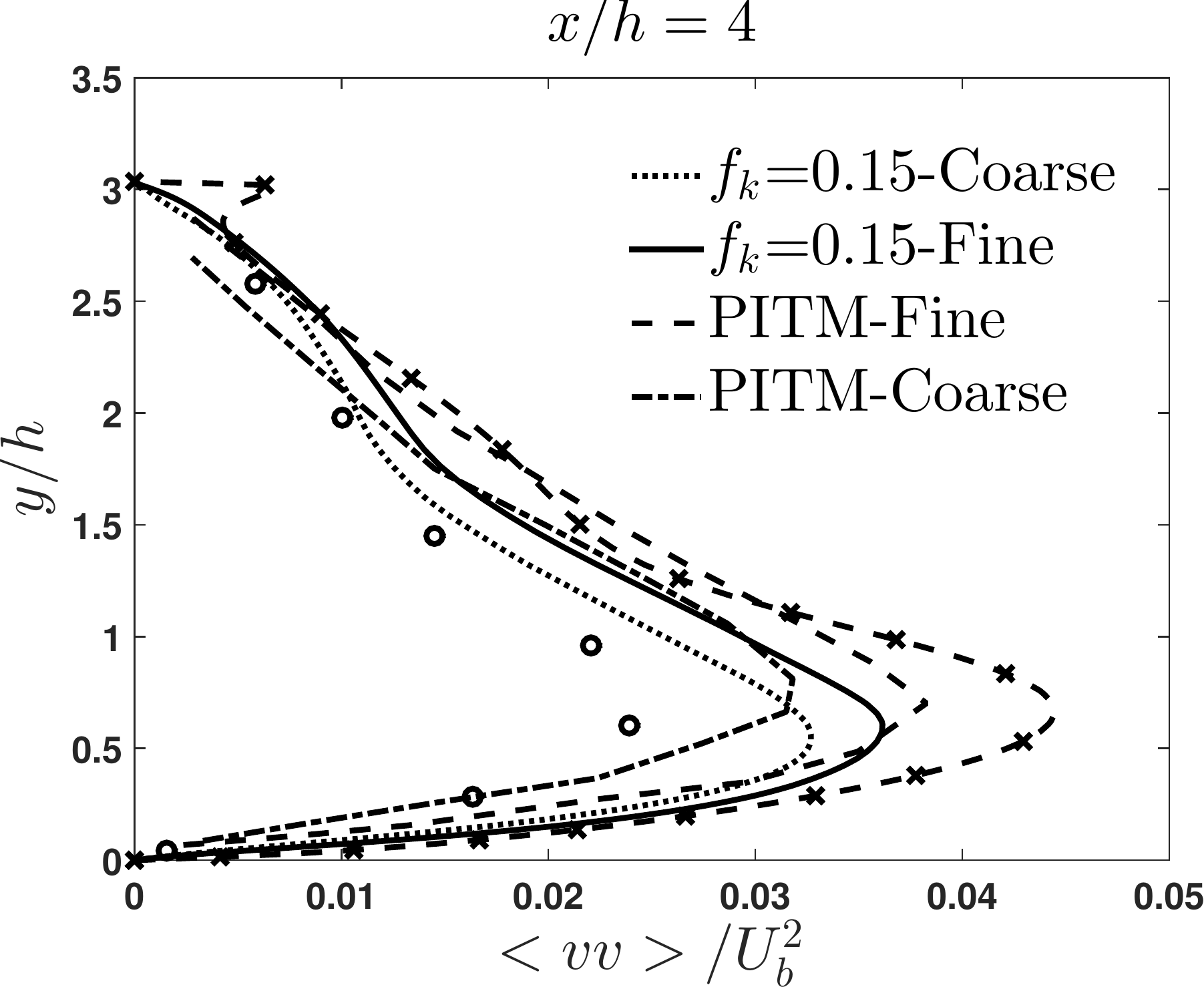}
   \vspace{-8pt}
   \caption{}
\end{subfigure}%
\begin{subfigure}{.5\textwidth}
   \centering
   \includegraphics[scale=0.3]{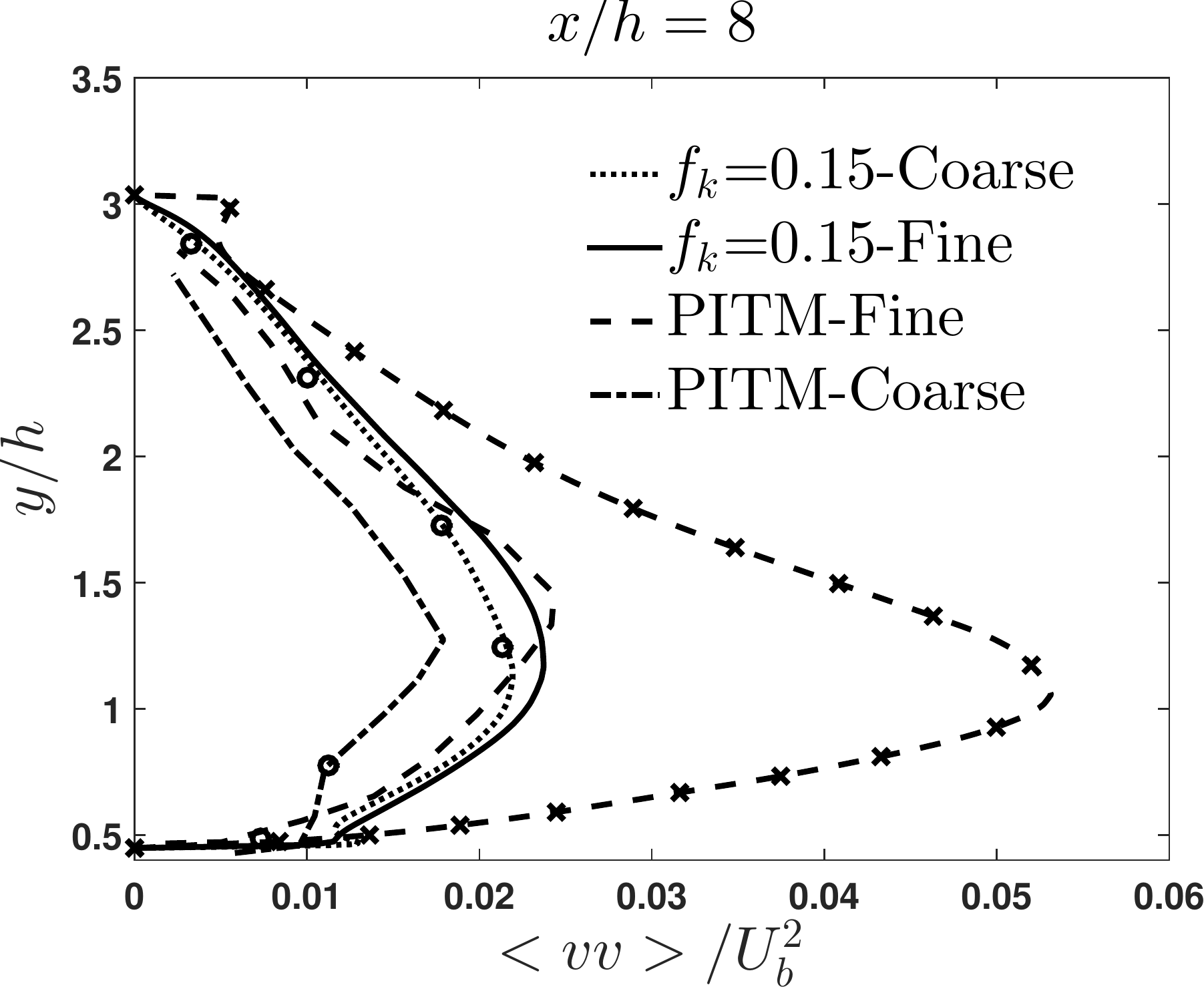}
   \vspace{-8pt}
   \caption{}
\end{subfigure}
\vspace{-8pt}
\caption{Vertical stress component (symbols for Exp. and RANS are consistent with Fig. \ref{fkuv})}
\label{gridvv}
\end{figure}

Figures \ref{gridu}-\ref{gridvv} show the mean streamwise velocity and stress profiles for both PANS and PITM calculations on the coarse and fine grid sizes. As seen in these plots, at almost all the locations, the mean quantities are hardly distinguishable for the PANS simulations on the both grid sizes. Regarding the PANS calculations, the only noticeable difference between the results of the two grid resolutions is observed at x/h=2 and x/h=4 for $f_k$=0.15. This is expected since as $f_k$ is reduced, the grid should be fine enough to capture the small turbulence length scales. Another important finding is that PITM results are damaged when the grid resolution is reduced. For instance, for the coarse grid, PITM is not able to recover the peak velocity near the wall at $x/h$=0.05. The separation bubble is also extended farther than $x/h$=4 for the PITM calculation. The same conclusion can be made about the stress profiles for the PANS and PITM data. the over-prediction of PITM data for the coarse grid is notable at $x/h$=4. This strong sensitivity of PITM results to grid resolution is related to the filter parameter in this method which is dependent on the grid size.

\section{Flow over wall-mounted hump}
\label{hump-study1}
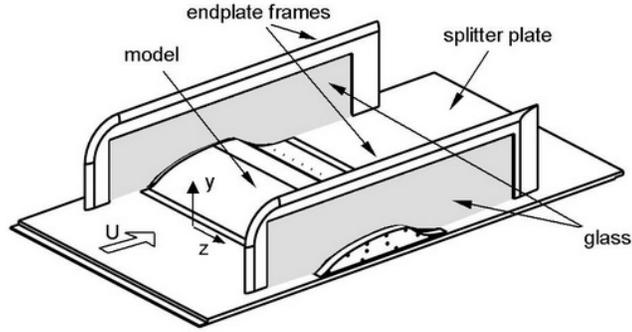
\begin{figure}
        \centering
 		\captionsetup{justification=centering}                                               
                \includegraphics[width=0.55\textwidth]{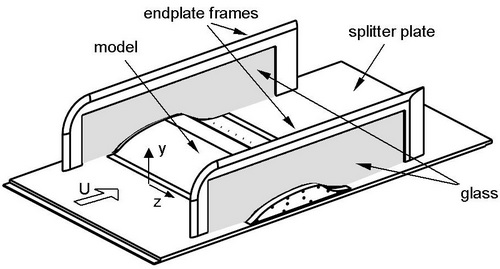}\begin{picture}(0,0)\put(-138,0){}\end{picture}
\caption{Experimental set-up for hump flow configuration}   
\label{exp-setup}                         
\end{figure}

The experiment for this flow geometry was originally designed and investigated by Seifert and Pack \cite{seifert2002active}. In this experiment, flow passes over a curved surface, and it separates as a result of strong adverse pressure gradient. As shown in figure \ref{exp-setup}, the experimental configuration consists of a Glauert-Goldschmied type body mounted over a splitter plate with two side end-plates. This flow is nominally two-dimensional although there are side-wall effects near the end-plates (3D features). For the experimental set-up, the hump has a chord length of 0.42 m. The free stream Mach number is 0.1, low enough to consider the flow incompressible. The flow Reynolds number is approximately $2.23\times10^6$ per meter, or $9.36\times10^5$ based on the hump chord length. The experimental database is used for the development and validation of new and existing turbulence models.

A validation workshop on synthetic jets and turbulent separation control was held in Williamsburg, Virginia in 2004 \cite{rumsey2004summary}. Flow over mounted hump simulation was selected as one of the three test cases where CFD results from several investigators were presented. This problem was chosen to compare results with experimental data \cite{seifert2002active,greenblatt2006} as existing RANS/LES/hybrid methods continue to have problems in accurately predicting the separation bubble characteristics and recovery of the flow. Three configurations of the wall-mounted hump case were considered in the workshop: (1) flow separation over the hump with no flow-control (base line case), (2) the effects of suction though a slot, and (3) an oscillatory zero-net mass-flux jet through the slot. The outcome of this workshop was that the RANS methods are wrongly predicting the flow features as they do not account for the spanwise variations and structures. Therefore, the urge of high fidelity modeling approach was recognized by the researchers. In this study, the no flow-control configuration or baseline case is investigated. 

Various RANS modeling approaches such as the Spalart-Allmaras(SA) \cite{SA2004}, $k-\omega$ \cite{ke2008}, shear stress transport model (SST)\cite{SST2006}, and several versions of $k-\epsilon$ models \cite{ke2008,ke2007} are used in the past to model flow separation for the current flow geometry. Noticeable mismatch for the mean velocity, stress components and surface friction coefficient compared to the experimental data was seen for these RANS family models. Several other investigators found a better prediction of the reattachment location using the DES, DDES, and zonal hybrid RANS/LES models \cite{DES2006}. However, still there were discrepancies with experimental data for the second order statistics.

Some researches have evaluated LES and ILES models for the current flow simulation. It is important to mention that the LES simulations were performed on remarkably much more expensive computational grids that then corresponding RANS and hybrid RANS/LES approaches. You et al. \cite{LES_hump1} simulated this problem with LES model and obtained good prediction of reattachment length,  mean velocity and stresses for this flow configuration. Avdis et al. \cite{avdis} although predicted the mean velocity and reattachment size closely to experiment, the turbulent stresses were significantly over predicted by their ILES model. In LES study of Saric et al. \cite{saric2006_hump}, the parameters agreed well with experiment only within the separation bubble and they deviated from experiment in the post-reattachment region. 

The unsteadiness generated for the flow separation studied in Sec. \ref{G1-hill} is induced by the flow geometry and maintained by defining the periodic inflow/outflow boundary conditions. The purpose of this section is to further evaluate G1-PANS method for computing a turbulent, separated flow where the inflow/outflow boundary conditions are not periodic and the proper simulation of the incoming flow upstream of the curved surface is critical.  

\subsection{simulation procedure}

\label{simulation detail}
\begin{figure}
\centering
  \includegraphics[trim=1cm 1cm 0cm 14cm, clip=true, scale=0.55]{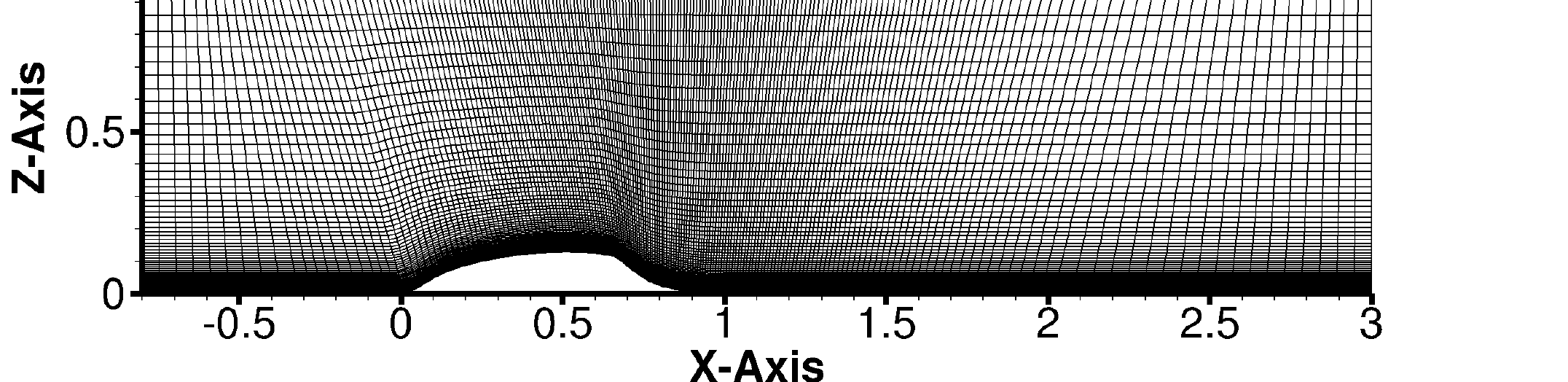}
\vspace{-6pt}
\caption{$201\times96\times30$ grid resolution of the hump flow simulation}
\label{grid-hump}
\end{figure}

The Reynolds number based on the free-stream velocity $U_{\infty}$ and the hump chord $C$ is 935,892. A schematic of the non-orthogonal, body-fitted mesh is shown in Fig. \ref{grid-hump}. Table \ref{case} lists the computational studies directed towards the configuration shown in Fig. \ref{grid-hump}. For each study, the numbers of cells in the streamwise, wall-normal and spanwise directions are given, along with the numerical strategy (LES, PANS, RANS). The dimensionless wall distance of the closest computational nodes, $y^+ \approx 1$ is considered for the lower wall.  

\begin{table}[H]
\centering
\caption{Details of the test cases simulated}
\begin{tabular}{ l l l l l}
\hline\noalign{\smallskip}
\textbf{Study} & \textbf{$f_k$} & \textbf{$f_\epsilon$} & \textbf{Grid} & \textbf{Averaging Period}\\ \hline\noalign{\smallskip}
\textbf{Re=935,892} \\ \hline\noalign{\smallskip}
RANS & 1 & 1 & $201\times60\times20$ & 10T-15T \\ \hline\noalign{\smallskip}
PANS & 0.2 & 1 & $201\times96\times30$ & 10T-15T \\
\hline\noalign{\smallskip}
LES & - & - & $768\times96\times128$ & - \\ 
\end{tabular}
\label{case}
\end{table}

\begin{figure} 
        \centering
 		\captionsetup{justification=centering}                                               
                \includegraphics[width=0.75\textwidth]{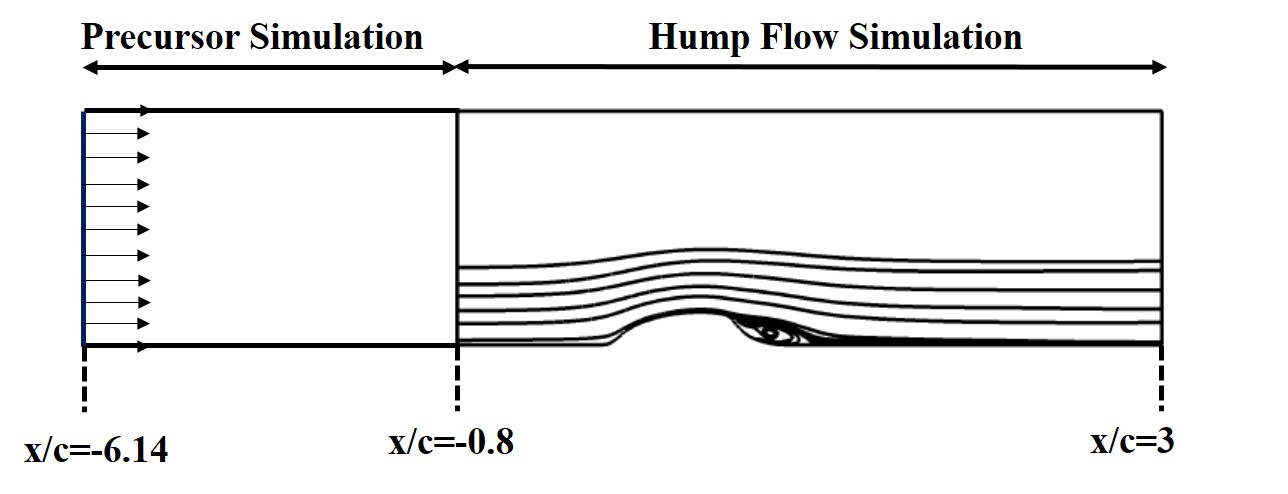}\begin{picture}(0,0)\put(-138,0){}\end{picture}
\caption{Schematic of the hump configuration, including the domains used in the main simulation and the simulation generating inflow conditions}   
\label{simulation}                         
\end{figure}
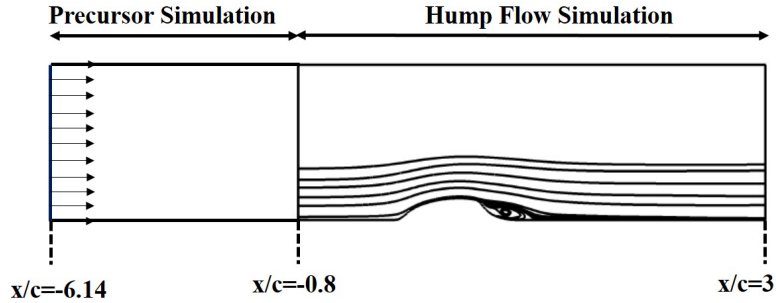

Same as the procedure taken in LES simulation \cite{avdis}, in order to reduce the size of computational domain, a precursor channel flow simulation is performed to allow development of a turbulent boundary layer upstream of the hump. Then, the outflow profiles of the channel flow simulation are fed to the hump flow simulation as the inflow condition. Therefore, for all the RANS and PANS computations, steady RANS profiles were imposed at the inlet plane placed at $x/c=-0.8C$ upstream of the hump leading edge. This procedure is shown in Fig. \ref{simulation}. 

The flow is considered to be periodic in spanwise direction. The convective outlet is applied at the outflow, while the inflow boundary condition is set from the precursor simulation. The bottom boundary is wall and either wall/symmetry plane boundary conditions are applied to the top boundary. It has been shown that the two choices of boundary type for the upper wall is not influencing the flow separation process for the hump flow \cite{saric2006_hump}. The computational domain extends from $x/C$= -0.8 to $x/C$ = 3 in the streamwise direction. In the cross-stream direction, $y/C = 0$ corresponds to the surface containing the hump, and the domain extends to $y/C = 0.909$, corresponding to the wall of the wind tunnel in the experiment. The flow is developed for 10 flow-through times, and the instantaneous fields are then averaged for a period of 5 flow-through times. The details of the boundary conditions and solvers for the hump flow simulation are given in table \ref{setup}.

For the precursor channel flow simulation, velocity at the inlet is set to $34.6 m/s$, and the pressure gradient is zero. At the outlet, gradients of all flow variables except pressure is zero. The pressure is set arbitrarily to zero at the outflow because the algorithm takes into account variations in pressure, and not the absolute value. The lower wall is a viscous no-slip wall. At the wall, $k$ is set to zero, and $\omega$ is set to the value suggested in \cite{wilcox1998}. 

\begin{table}[H]
\centering
\caption{Simulation set-up in OpenFoam for the hump simulation}
\begin{tabular}{ l l }
\hline\noalign{\smallskip}
\textbf{Settings} & \textbf{Choice} \\
\hline\noalign{\smallskip}
Simulation type & 3D Unsteady  \\ 
Solver & Transient incompressible (pisoFoam)\\
Temporal discretization & Backward (second order accurate) \\
Spatial discretization & Gauss Linear (second order accurate) \\
Pressure-velocity coupling & PISO \\
Turbulence model & G2-PANS $k-\omega$ \\ \hline\noalign{\smallskip}
\textbf{Boundary} & \textbf{Type} \\ \hline\noalign{\smallskip}
Inflow & RANS inflow  \\
Outflow & Zero Gradient/Convective outlet  \\
Bottom patch & Inviscid wall  \\
Top patch & Inviscid wall/Symmetry  \\
Lateral & Periodic  \\ 
\end{tabular}
\label{setup}
\end{table}

\subsection{Results}

\begin{figure}
\centering
\begin{subfigure}{.5\textwidth}
  \centering
  \includegraphics[scale=0.3]{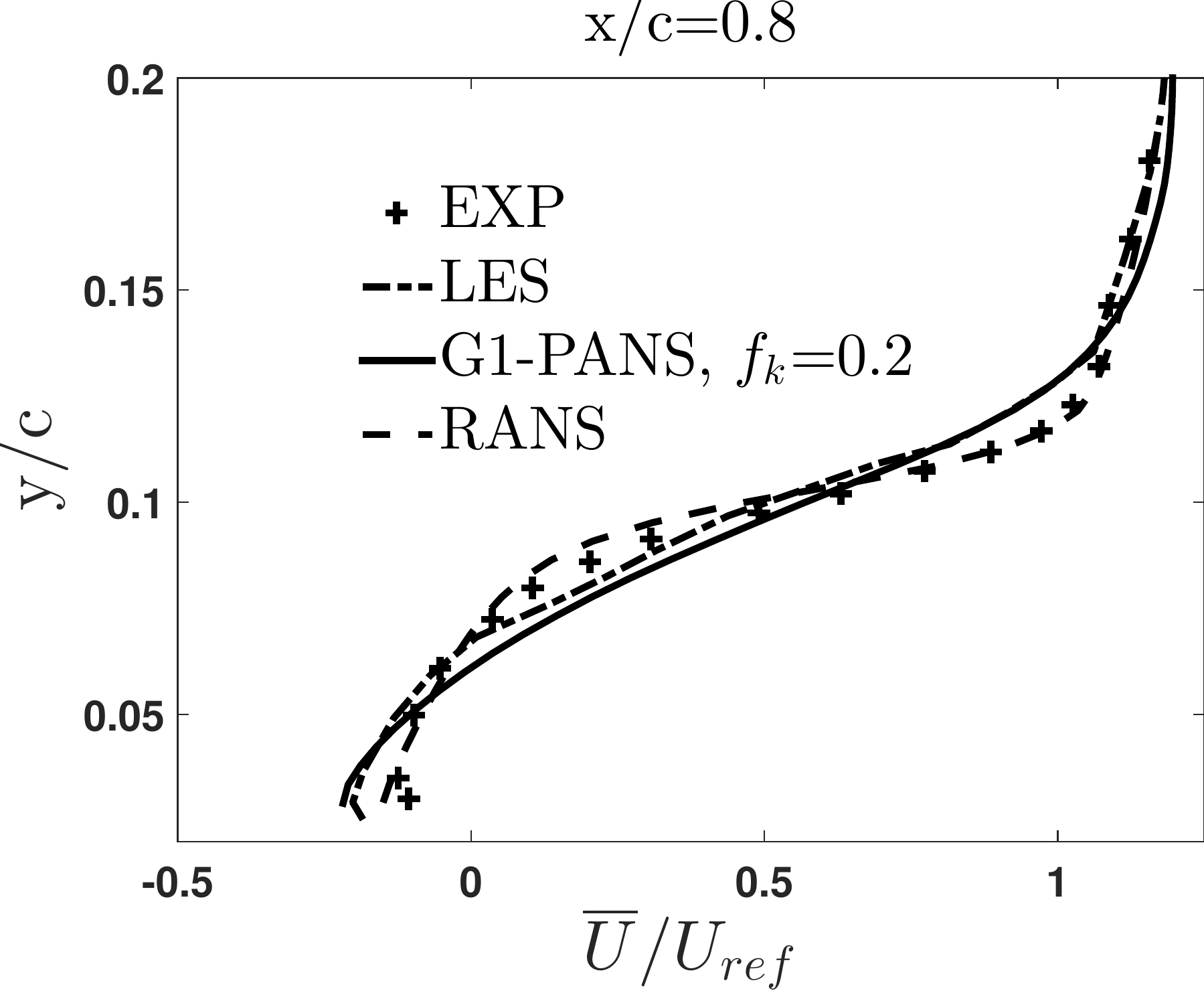}
  \vspace{-8pt}
  \caption{}
\end{subfigure}%
\begin{subfigure}{.5\textwidth}
  \centering
  \includegraphics[scale=0.3]{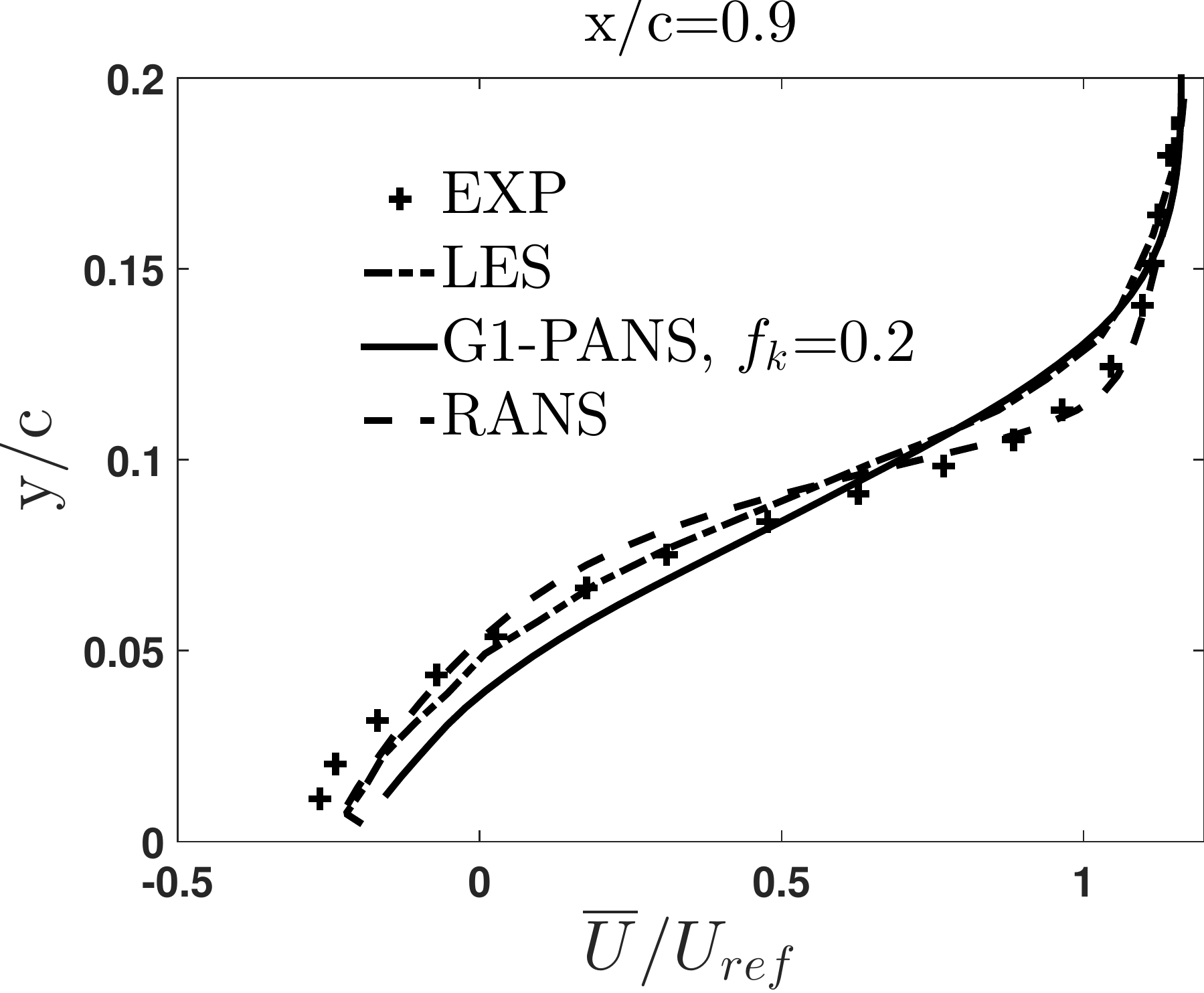}
  \vspace{-8pt}
  \caption{}
\end{subfigure}
\\
\begin{subfigure}{.5\textwidth}
   \centering
   \includegraphics[scale=0.3]{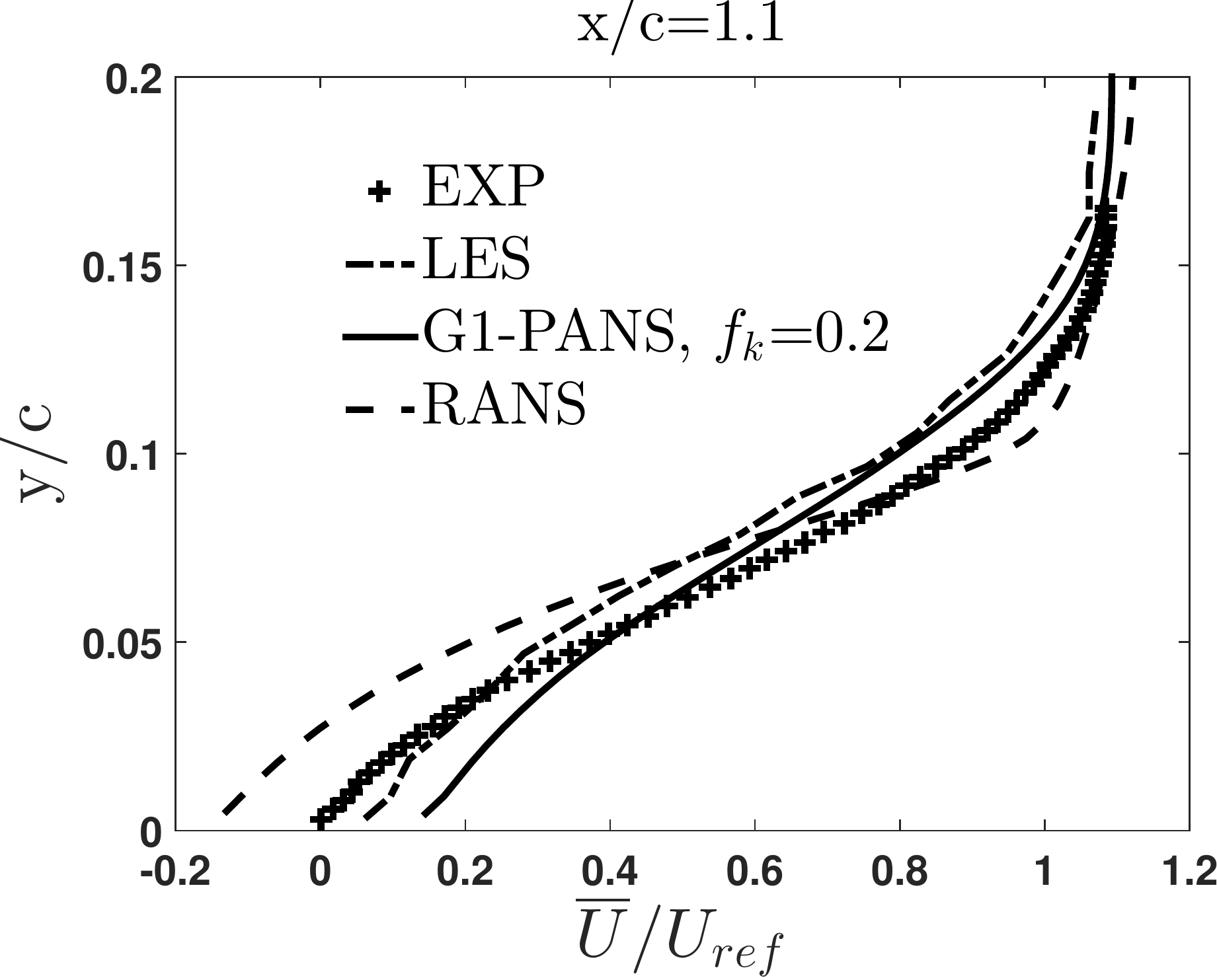}
   \vspace{-8pt}
   \caption{}
\end{subfigure}%
\begin{subfigure}{.5\textwidth}
   \centering
   \includegraphics[scale=0.3]{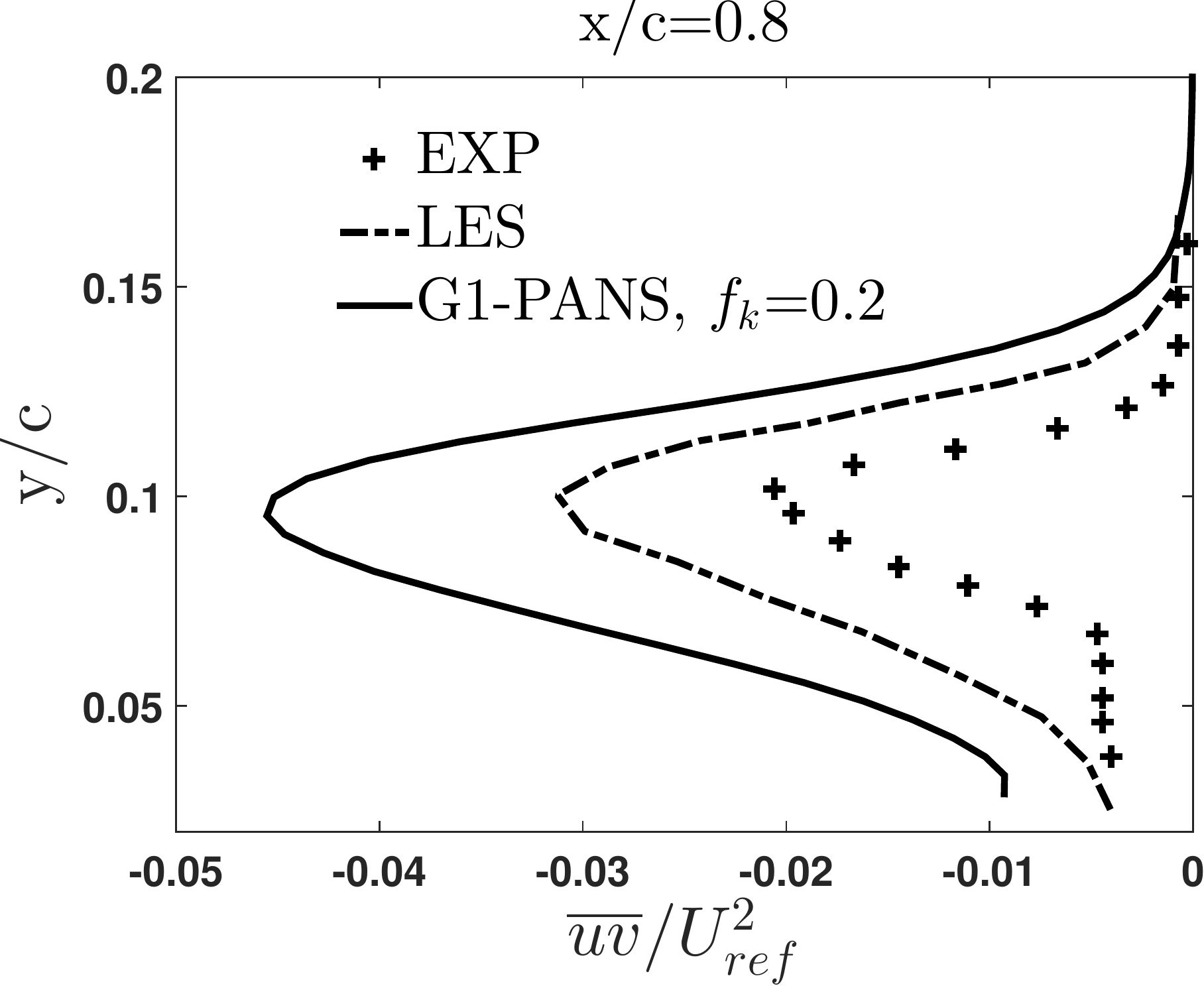}
   \vspace{-8pt}
   \caption{}
\end{subfigure}
\\
\begin{subfigure}{.5\textwidth}
   \centering
   \includegraphics[scale=0.3]{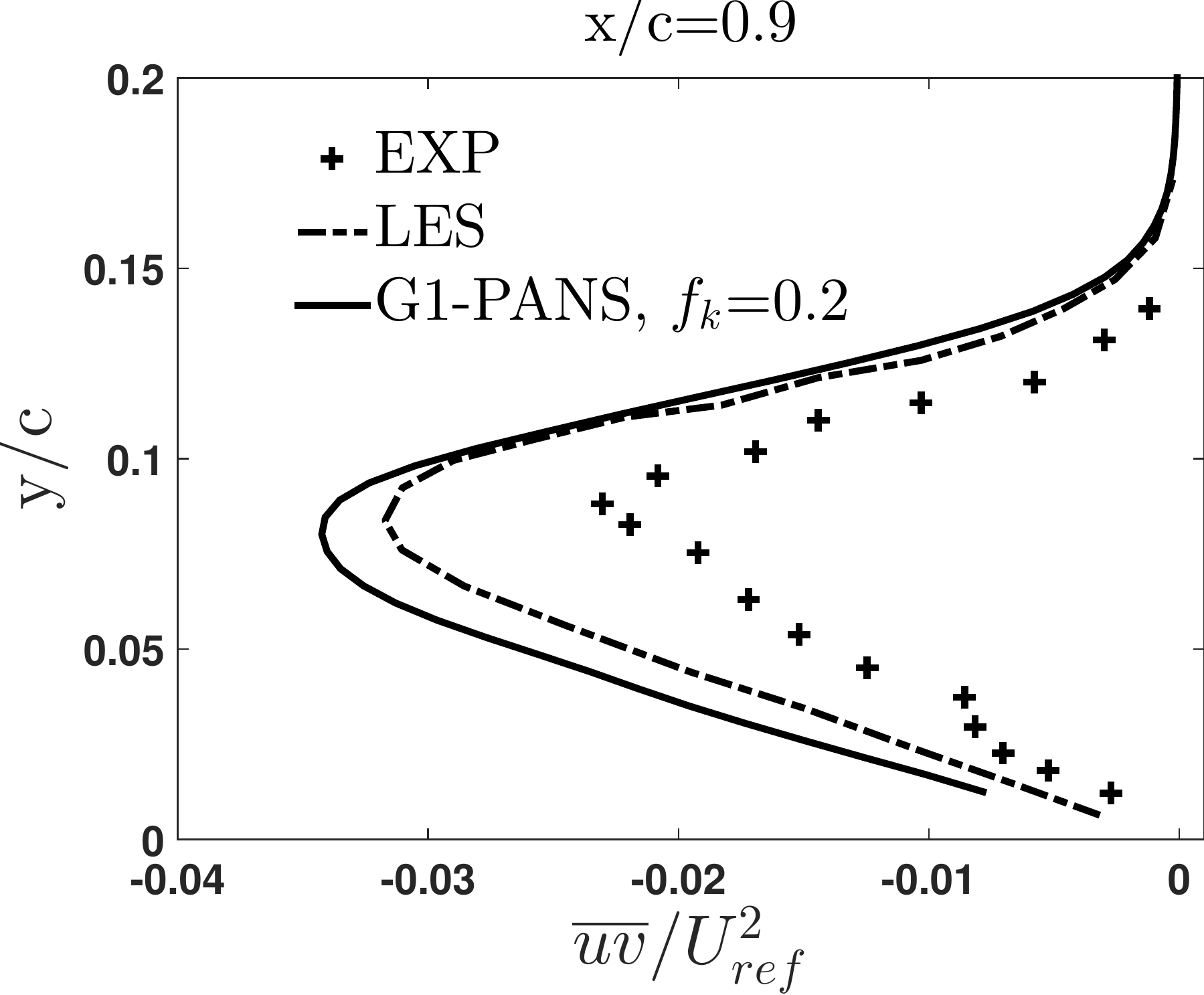}
   \vspace{-8pt}
   \caption{}
\end{subfigure}%
\begin{subfigure}{.5\textwidth}
   \centering
   \includegraphics[scale=0.3]{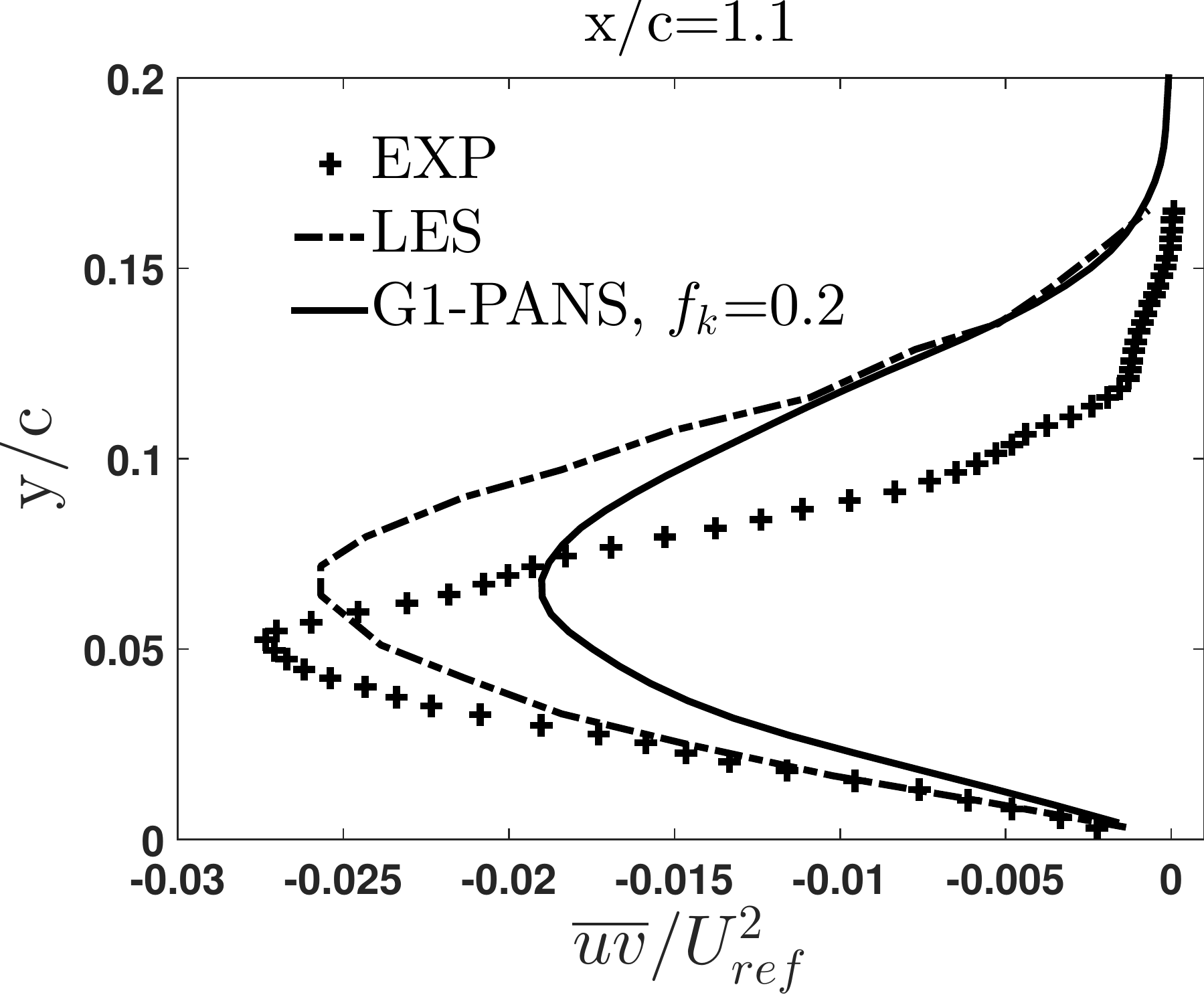}
   \vspace{-8pt}
   \caption{}
\end{subfigure}
\vspace{-8pt}
\caption{G1-PANS simulation of flow over mounted hump}
\label{G1-hump}
\end{figure}

Figure \ref{G1-hump} shows the mean flow statistics for the G1-PANS calculation of the hump flow as well as the corresponding results for RANS and LES \cite{avdis}. The statistics are shown at several streamwise locations in the separated region and reattachment location. As shown in this figure, for the early separated region, the mean velocity is predicted with a good accuracy by all models. However, the difference between different simulation approaches appears in the vicinity of the reattachment location, $x/c=1.1$ for the mean velocity profile. As seen in Fig. \ref{G1-hump} (c), early and delayed reattachment is predicted by the G1-PANS and RANS models, respectively. Looking at the stress profiles depicted in Figs. \ref{G1-hump} (d)-(f) reveals remarkable over-prediction of pick value of shear stress in the separation region and under-estimation of the peak value in the reattachment region for both G1-PANS and LES simulations. 

These results indicate that the key instability mechanisms inherent in this flow are not resolved accurately due to the lack of grid resolution for G1-PANS and LES calculations. In other words, the G1-PANS simulation of the developing boundary layer upstream of the hump and separated region downstream of the hump is highly sensitive to the grid resolution. Several strategies are addressed in \cite{higuera2015} to improve the G1-PANS predictions which mainly include adaptation of higher grid resolution and unsteady inflow generation using the Lund's recycling-rescaling method. Since using a high grid resolution is not practical for engineering flows, the near wall modeling of the PANS method is discussed in the next chapter.

%
%
%


\chapter{\uppercase {Near Wall Modeling of the PANS Method}}
\label{wall-model}

\section{Introduction}

An strategy to drastically reduce computational cost in high fidelity simulations for a wide range of industrial applications is to couple PANS with the RANS Model. The idea of developing a unified hybrid RANS/PANS model is inspired by the fact that the PANS model posses the same structure as RANS regarding the transport of unresolved kinetic energy, dissipation and definition of the unresolved eddy viscosity. In this approach, the near wall scales are modeled by the RANS model, and therefore the near wall grid size and time-step constraint are considerably relaxed. With this type of modelling, steady RANS calculation is used for near wall calculation, and unsteady PANS simulation is utilized where it is needed in the domain. Therefore, PANS closure model is employed throughout the whole domain and transition from RANS to PANS is controlled by smooth variation in the filter parameter, $f_k$.

As mentioned earlier, the first generation of the PANS model is obtained for a constant filter parameter. PANS model derivation for a variable resolution calculation is addressed first by Girimaji and Wallin \cite{girimaji2013} for a temporally varying resolution. They developed the model based on energy conservation rules and validated that for decaying isotropic turbulence. They discussed that the commutation errors arise in the governing equations for turbulence quantities and momentum equation as a result of resolution variation which can not be neglected. These errors are addressed by including extra terms in the governing equations which can be significant if the prescribed filter parameter variation is considerable. 

While the PANS closure model development for the temporal variation of filter parameter is addressed in \cite{girimaji2013}, the scope of this section is to develop appropriate turbulence closure model for bridging between different resolutions in space, and particularly near-wall region. Seamless transition from region of low-resolution near the wall to high-resolution away from the wall is controlled using the PANS filter parameter. Commutation effects as a result of filter variation are modelled using additional term in the turbulent kinetic energy equation. In addition, to conserve the total turbulent energy due to the interaction of unresolved and resolved flow fields, innovative strategies are proposed. This study identifies some important challenges regarding the numerical stability and appropriate implementation of the energy conservation principles.

Model development and analysis have been performed for turbulent channel flow simulation at low and high Reynolds numbers. The objective of the current study is twofold. First, it is aimed to demonstrate that the G1-PANS model developed for constant filter parameter is able to accurately capture the flow physics given the required amount of grid resolution for a specified Reynolds number. Second objective is to develop the second generation of the PANS model for variable resolution calculations near the wall. The ability of the developed model is then evaluated to simulate turbulent channel flow at high Reynolds numbers where the constant resolution approach  could be extremely costly and not viable to perform. The results are compared with DNS data of turbulent channel flow \cite{hoyas2008reynolds}.  

\section{Derivation of the PANS closure model-generation 2}

Assuming constant filter parameters, $f_{k}$ and $f_{\epsilon}$, the G1-PANS model given in Eqn. \ref{eq:NSfiltered} and Eqn. \ref{eq:PANS1} was obtained for the resolved and unresolved fields. Closure modeling in region of resolution variation is discussed next.  

\textbf{ Spatio-temporal $f_k$ variation}: The resolution ($f_k$) variation introduces commutation effects in the momentum, turbulent kinetic energy and dissipation equations. These effects are modeled as follows
\begin{eqnarray}
\label{eq:NS2}
\frac{\partial U_i}{\partial t} &+&
U_j \frac{\partial U_i}{\partial x_j} = -\frac{\partial \tau (V_i,V_j)}{\partial x_j} - \frac{\partial p}{\partial x_i} + \nu \frac{\partial^2 U_i}{\partial x_j \partial x_j}-F_i; \\
\label{eq:k2}
	\frac{Dk_u}{Dt} &=& P_u+D_{Tr}+P_{Tr} - \beta^* k_u \omega_u + \frac{\partial}{\partial x_j} \left[ \left(\nu + \nu_{u} / \sigma_{ku} \right) \frac{\partial k_u}{\partial x_j} \right]
	\\
\label{eq:w2}
\frac{D\omega_u}{Dt} &=& -\frac{\omega_u}{k_u}\left(P_{Tr}+D_{Tr}\right)+\alpha \frac{\omega_u}{k_u} P_u -\beta' \omega_u^2 + \frac{\partial}{\partial x_j} \left[ \left(\nu + \nu_{u} / \sigma_{\omega_u} \right) \frac{\partial \omega_u}{\partial x_j} \right]\\ \nonumber
\end{eqnarray}
The extra energy transfer terms, $D_{Tr}$ and $P_{Tr}$ in Eqns. \ref{eq:NS2}-\ref{eq:w2} are derived by taking variation of $f_k$ into account. These terms can be obtained by inspecting the evolution of the unresolved kinetic energy in the case of resolution variation. By definition, the advective term of the unresolved kinetic energy in the PANS method is evolved as

\begin{equation}
\label{eq:conv}
f_k(\frac{\partial{k}}{\partial{t}}+\overline{U_j}\frac{\partial{k}}{\partial{x_j}})=\frac{\partial{k_u}}{\partial{t}}+\overline{U_j}\frac{\partial{k_u}}{\partial{x_j}}-P_{Tr}
\end{equation} 

Where, $k$ is the total kinetic energy, $k_u$ is the modelled or unresolved kinetic energy and $P_{Tr}$ is 

\begin{equation}
\label{eq:ptr}
P_{Tr}=\frac{k_u}{F_k}\frac{Df_k}{Dt}
\end{equation} 

And the diffusion term is obtained as
 
\begin{equation}
\label{eq:diff}
f_k\frac{\partial{}}{\partial{x_k}}[(\nu+\frac{\nu_t}{\sigma_k})\frac{\partial{k}}{\partial{x_k}}]=\frac{\partial{}}{\partial{x_k}}[(\nu+\frac{\nu_u}{\sigma_{ku}})\frac{\partial{k_u}}{\partial{x_k}}]+D_{Tr}
\end{equation}

Where, $D_{Tr}$ is

\begin{equation}
\label{eq:dtr}
D_{Tr}=-\frac{k_u}{f_k}\frac{\partial{}}{\partial{x_k}}(\nu_u^*\frac{\partial{f_k}}{\partial{x_k}})-\frac{2\nu_u^*}{f_k}(\frac{\partial{k_u}}{\partial{x_k}}-\frac{k_u}{f_k}\frac{\partial{f_k}}{\partial{x_k}})\frac{\partial{f_k}}{\partial{x_k}};~~\nu_u^*=\nu+\frac{\nu_u}{\sigma_{ku}}
\end{equation} 

Here, the subscript $u$ denotes the unresolved parameters. If the cut-off is outside of the dissipative range, changing $f_k$ will not affect the dissipation and merely adds an additional term in the $\omega$ equation as a result of the transformation from the $k-\epsilon$ equations as seen in Eqn. \ref{eq:w2}. 

The model coefficients are consistent with Eqn. \ref{eq:PANS1}.  The $P_{Tr}$ term defined in Eqn. \ref{eq:ptr} is associated with the transfer of energy between resolved and unresolved scales when there is a change of resolution in time and/or streamwise direction, whereas the $D_{Tr}$ term is accounting for this energy exchange in the case of wall-normal change in the resolution. 
$F_i$ in Eqn. \ref{eq:NS2} is the commutation term in the momentum equation responsible for interscale energy transfer which can be modeled by invoking energy conservation rules. By multiplying $U_i$ to both sides of Eqn. \ref{eq:NS2}, the equation for the resolved kinetic energy will be obtained

\begin{equation}
\label{eq:mom_res}
\frac{DE}{Dt}=-U_i\frac{1}{\rho}\frac{\partial{P}}{\partial{x_i}}+U_i\frac{\partial{}}{\partial{x_k}}[(\nu+\nu_t)\frac{\partial{U_i}}{\partial{x_k}}]-U_iF_i
\end{equation} 

 Energy conservation principles dictate that the additional terms present in the equations for the unresolved kinetic energy \ref{eq:k2} and resolved kinetic energy  \ref{eq:mom_res} should be in balance, or in the mathematical form, the following equality must be ensured:

\begin{equation}
\label{eq:balance}
U_iF_i=P_{Tr}+D_{Tr}
\end{equation}  

The above condition suggest the following equation for  $F_i$
\begin{equation}
\label{eq:F1}
F_i=\frac{U_i}{U_iU_i}(P_{Tr}+D_{Tr})
\end{equation} 

For ease of implementation of $F_i$ in the momentum equation, Eqn. \ref{eq:F1} can also be interpreted as 

\begin{equation}
\label{eq:F2}
F_i=-\frac{\partial{}}{\partial{x_k}}(\nu_{Tr}\frac{\partial{U_i}}{\partial{x_k}})
\end{equation} 

Where $\nu_{Tr}$ is called the commutation viscosity and is added to the molecular and eddy viscosity to construct the viscous term in the momentum equation and is given by

\begin{equation}
\label{eq:nutr}
\nu_{Tr}=\frac{P_{Tr}+D_{Tr}}{2S_{ij}S_{ij}}, S_{ij}=\frac{1}{2}(\frac{\partial{U_i}}{\partial{x_j}}+\frac{\partial{U_i}}{\partial{x_i}})
\end{equation} 

Energy transfer to the resolved scales can be obtained by a negative $\nu_{Tr}$, while with positive value, energy is taken from resolved scales. Using the above derivation for the PANS model, we will perform the analysis of changing resolution near the wall to investigate wall-bounded turbulent flows at high Reynolds numbers. The PANS filter parameter, $f_k$ is changing from 1 near the wall to arbitrary value away from the wall. Since the small turbulence length scales associated with the dissipative range are not resolved in the present study, $f_{\epsilon}$ is set to one for all calculations. Fully developed channel flow is computed for Reynolds numbers of $Re_{\tau}$=$u_{\tau}h/\nu$=180-8000.

\section{Simulation procedure of turbulent channel flow}

Turbulent channel flow simulations are performed using an incompressible finite volume solver in OpenFoam. The numerical schemes for discritizing the equations are second order accurate in time and space. The domain extends $4h$, $2h$ and $2h$ in the streamwise, spanwise and wall-normal directions, respectively where $h$ is the channel half width. Periodic boundary condition is applied in the streamwise as well as spanwise directions. Reynolds number, $Re_{\tau}$ is defined as $u_{\tau}h/\nu$ where $u_{\tau}$ is the friction velocity, and $\nu$ is the kinematic viscosity of the fluid. Because of the periodicity of domain, flow is driven by a constant pressure gradient which is added as a source term to the momentum equation. 

In order to study flow statistics and structure, several G1-PANS and G2-PANS simulations are performed for a range of Reynold numbers between 180 and 8000. The simulation results are compared with DNS data \cite{hoyas2008reynolds} for Reynolds numbers up to 4200. based on the best knowledge of the author, for the Reynolds numbers above 4200, no DNS data is available as the grid requirement becomes a critical issue. Therefore, to investigate the ability of G2-PANS model to recover the mean velocity profile at affordable computational cost, the higher Reynold number of 8000 is also included in this study. Table \ref{case1} summarizes different test cases with their specific grid resolution as well as those of DNS data.

\begin{table}[H]
\centering
\caption{Grid resolutions for turbulent channel flow simulations}
\begin{tabular}{|c|c|c|}
\hline
\multicolumn{3}{|l|}{\textbf{\hspace{3.5cm}Turbulent Channel Flow Simulation}}\\
\hline 
{\textbf{Reynolds number}}&\textbf{Model}&\textbf{Grid}\\ \hline 
\multirow{4}{*}{$Re_{\tau}=180$}
&RANS&$64^3$\\ \cline{2-3}
&G1-PANS ($f_{k}=0.2$)&$64\times101\times64$\\ \cline{2-3}
&DNS&$N_y=97$, $\Delta_x^+=9$, $\Delta_x^+=6.7$\\ \hline
\multirow{4}{*}{$Re_{\tau}=550$}
&RANS&$64^3$\\ \cline{2-3}
&G1-PANS ($f_{k}=0.2$)&$64\times101\times64$\\ \cline{2-3}
&DNS&$N_y=257$, $\Delta_x^+=13$, $\Delta_x^+=6.7$\\ \hline
\multirow{4}{*}{$Re_{\tau}=950$}
&RANS&$64^3$\\ \cline{2-3}
&G2-PANS ($f_{k}=0.2 \& 0.3$)&$64\times101\times64$\\ \cline{2-3}
&DNS&$N_y=385$, $\Delta_x^+=11$, $\Delta_x^+=5.7$\\ \hline
\multirow{4}{*}{$Re_{\tau}=2000$}
&RANS&$64\times80\times30$\\ \cline{2-3}
&G2-PANS ($f_{k}=0.2 \& 0.3$)&$64\times101\times64$\\ \cline{2-3}
&DNS&$N_y=633$, $\Delta_x^+=12$, $\Delta_x^+=6.1$\\ \hline
\multirow{3}{*}{$Re_{\tau}=4200$}
&RANS&$64^3$\\ \cline{2-3}
&G2-PANS ($f_{k}=0.2 \& 0.3$)&$64\times181\times64$\\ \cline{2-3}
&G1-PANS ($f_{k}=0.2$)&$64\times181\times64$\\ \cline{2-3}
&DNS&$N_y=1081$, $\Delta_x^+=12.8$, $\Delta_x^+=6.4$\\ \hline
\multirow{2}{*}{$Re_{\tau}=8000$}
&RANS&$64^3$\\ \cline{2-3}
&G2-PANS ($f_{k}=0.2$)&$64\times181\times64$\\ \hline
\end{tabular}
\label{case1}
\end{table}

\section{Results}

This section is divided into three separate important studies. First, the two approaches in obtaining the transport coefficients of the G1-PANS closure model are compared for the flow statistics and resolved flow scales at $Re_\tau$=180. Besides, the G1-PANS calculations at low and high Reynolds numbers are presented in the first part. 
In the second part, the G1-PANS equations are solved for a variable $f_k$ simulation. This approach is referred as G1.5-PANS model. Finally, in order to explore the effect of energy-scale transfer terms for a variable resolution simulation, the G2-PANS model results are discussed.  

\subsection{G1-PANS simulations of turbulent channel flow}
As discusses in Sec. \ref{hill}, two limiting cases known as ZTM and MTM are proposed to model the transport of the unresolved kinetic energy and dissipation. Besides, scale-dependent boundary-layer analysis for a partially-resolved boundary layer demonstrated that the ZTM was the appropriate model for this region. In order to provide a proof of concept for this analysis, the results for turbulent channel flow calculations are investigated for both G1-PANS ZTM and G1-PANS MTM models in the subsequent section.

\subsubsection{G1-PANS ZTM vs. G1-PANS MTM}

The two turbulence transport models (i.e. ZTM and MTM) in the context of the PANS method are evaluated for the turbulent boundary layer with $Re_\tau$=180.

\begin{figure}[H]
        \centering
 		\captionsetup{justification=centering}                                       
                \includegraphics[width=0.65\textwidth]{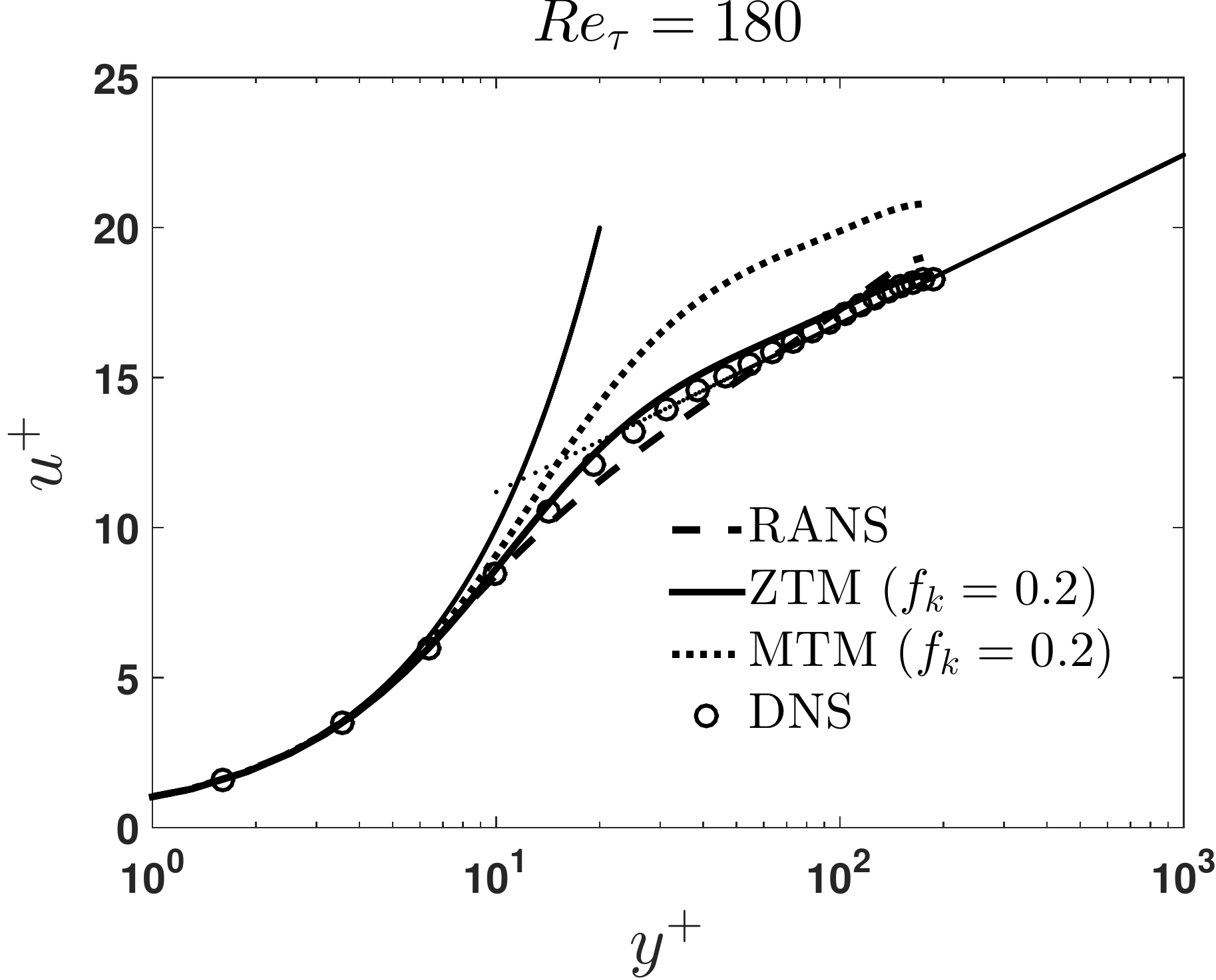}\begin{picture}(0,0)\put(-138,0){}\end{picture}
\caption{Velocity profile at $Re_\tau=180$}   
\label{MTM180-U}     
\end{figure}
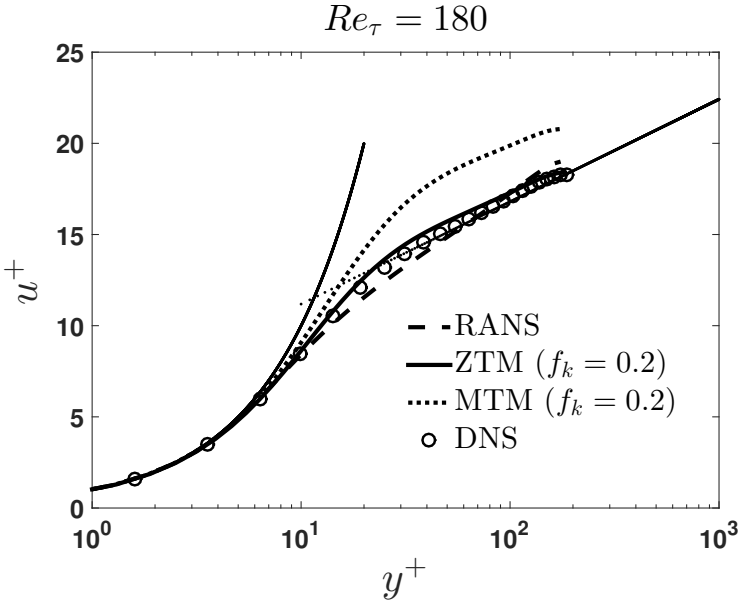

\begin{figure}[H]
        \centering
 		\captionsetup{justification=centering}                                       
 		        \begin{subfigure}[b]{0.45\textwidth}
                \includegraphics[width=\textwidth]{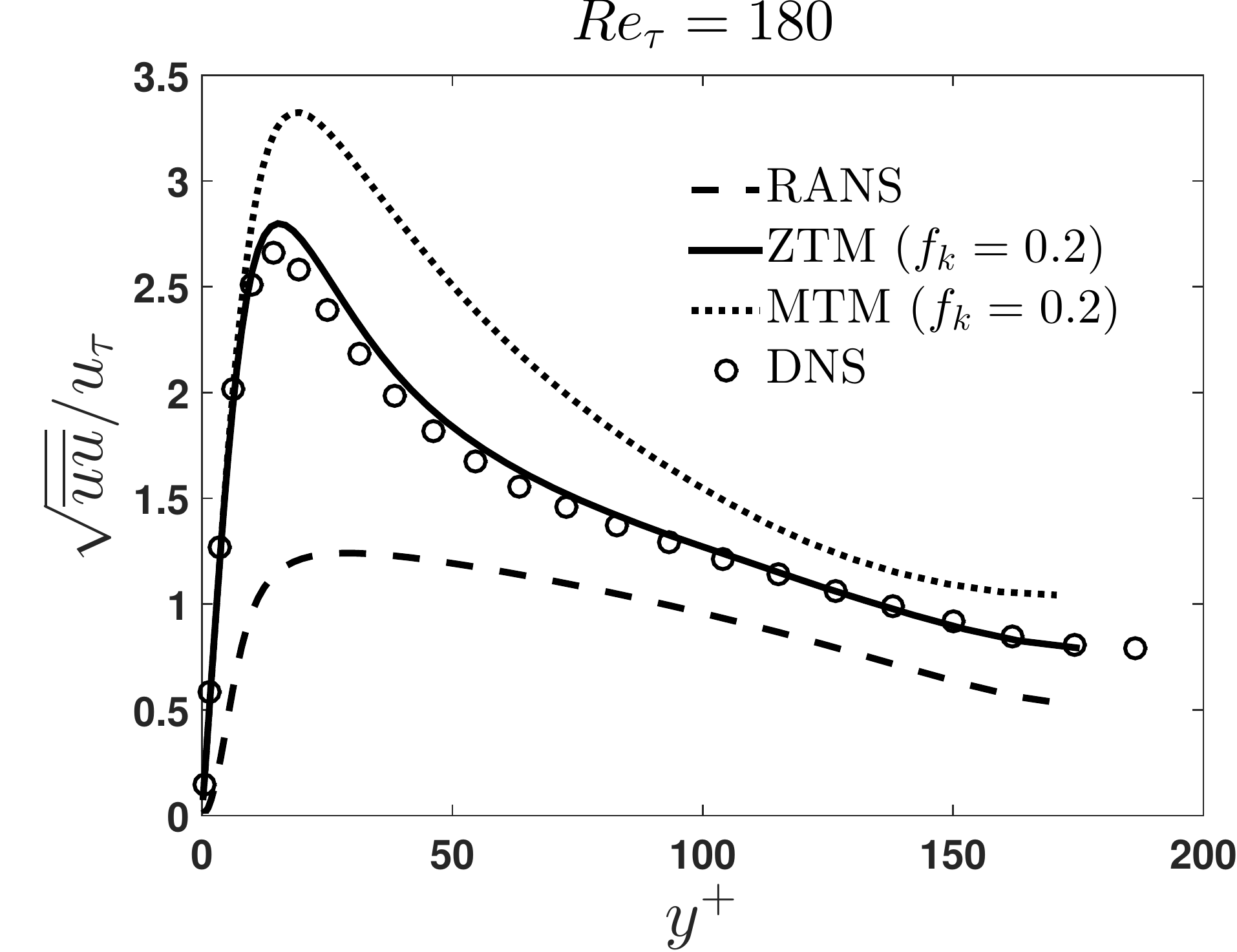}\begin{picture}(0,0)\put(-138,0){(a)}\end{picture}
        \end{subfigure}
        			\begin{subfigure}[b]{0.45\textwidth}
                \includegraphics[width=\textwidth]{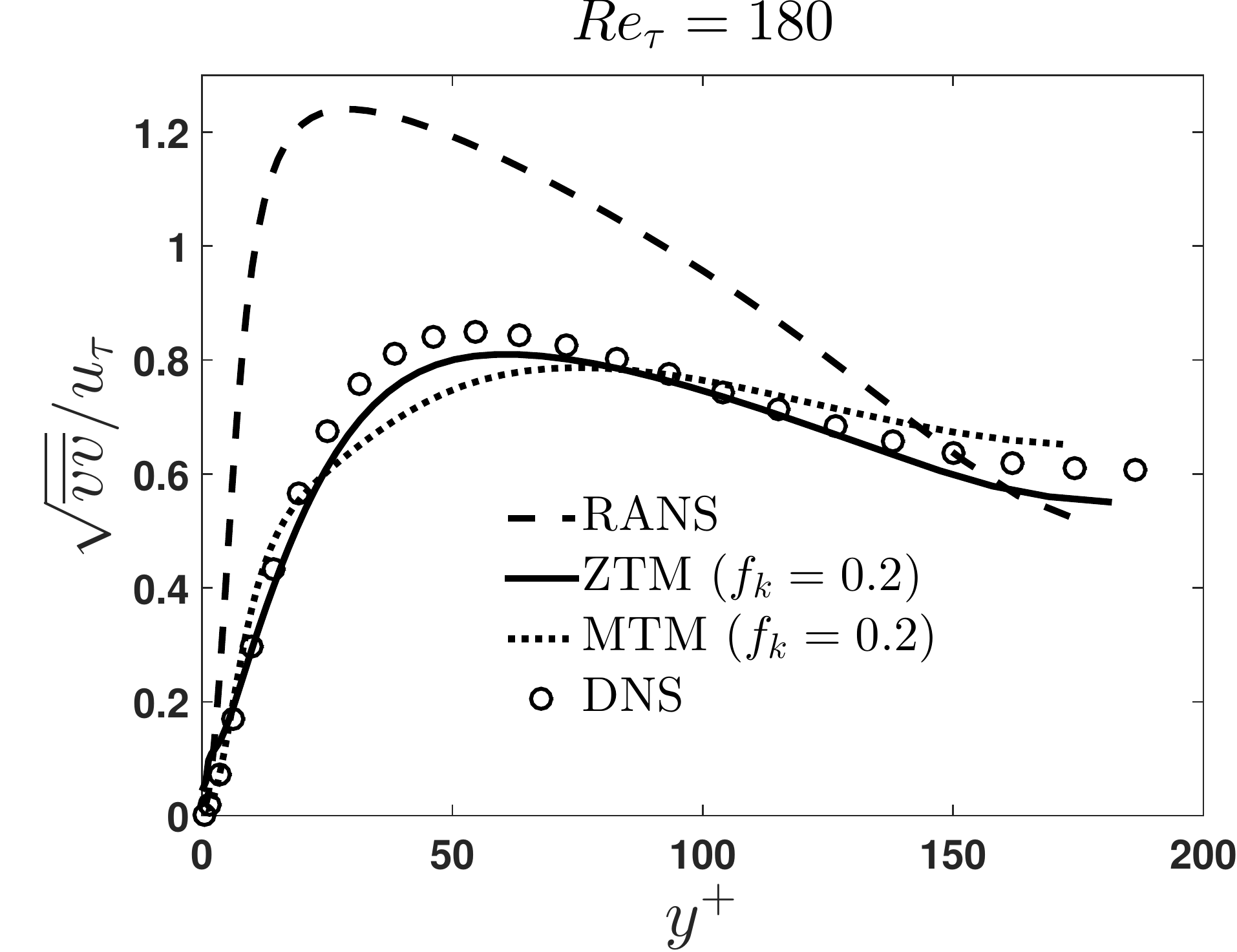}\begin{picture}(0,0)\put(-138,0){(b)}\end{picture}
        \end{subfigure}
                \begin{subfigure}[b]{0.45\textwidth}
                \includegraphics[width=\textwidth]{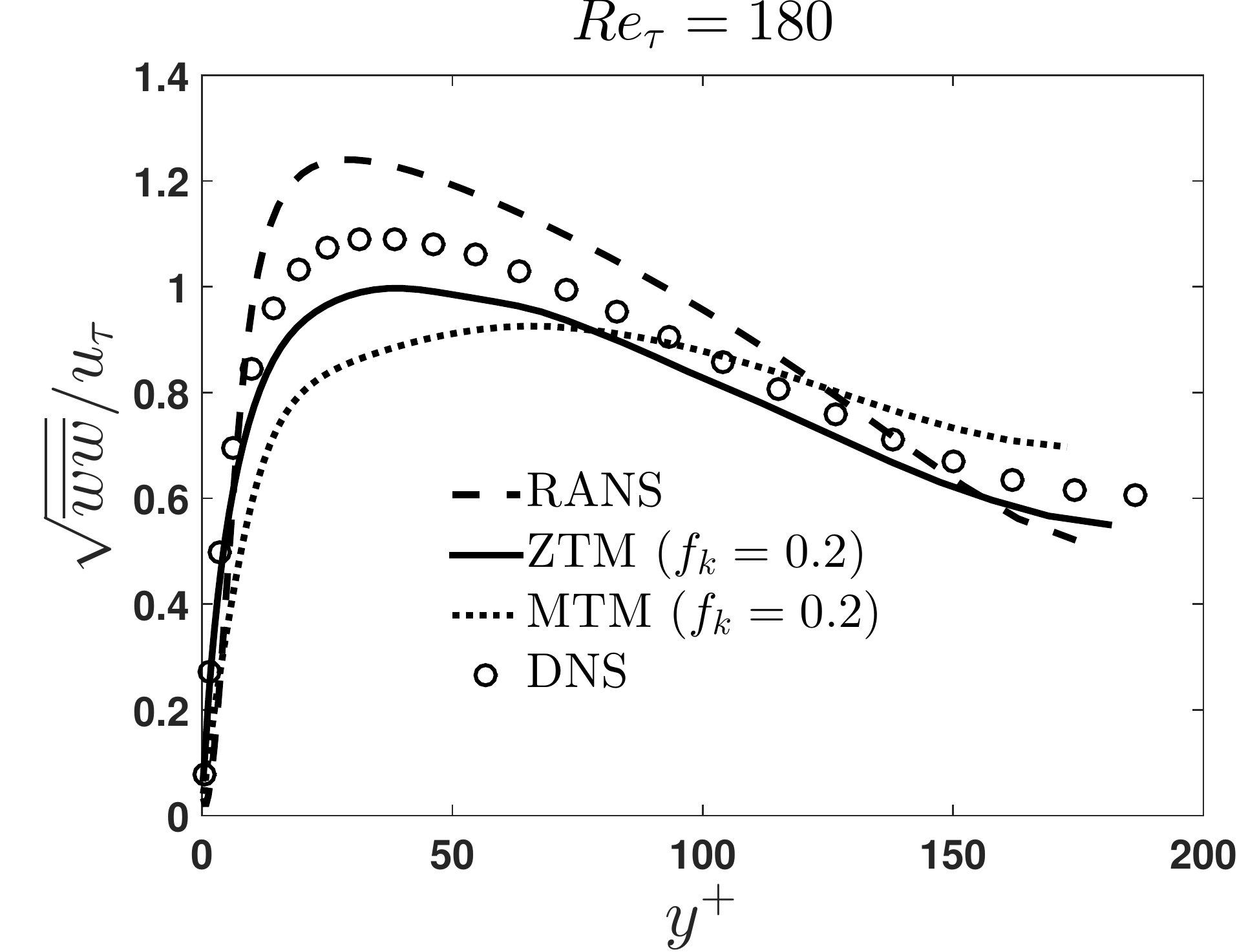}\begin{picture}(0,0)\put(-138,0){(c)}\end{picture}
        \end{subfigure}       
 		        \begin{subfigure}[b]{0.45\textwidth}
                \includegraphics[width=\textwidth]{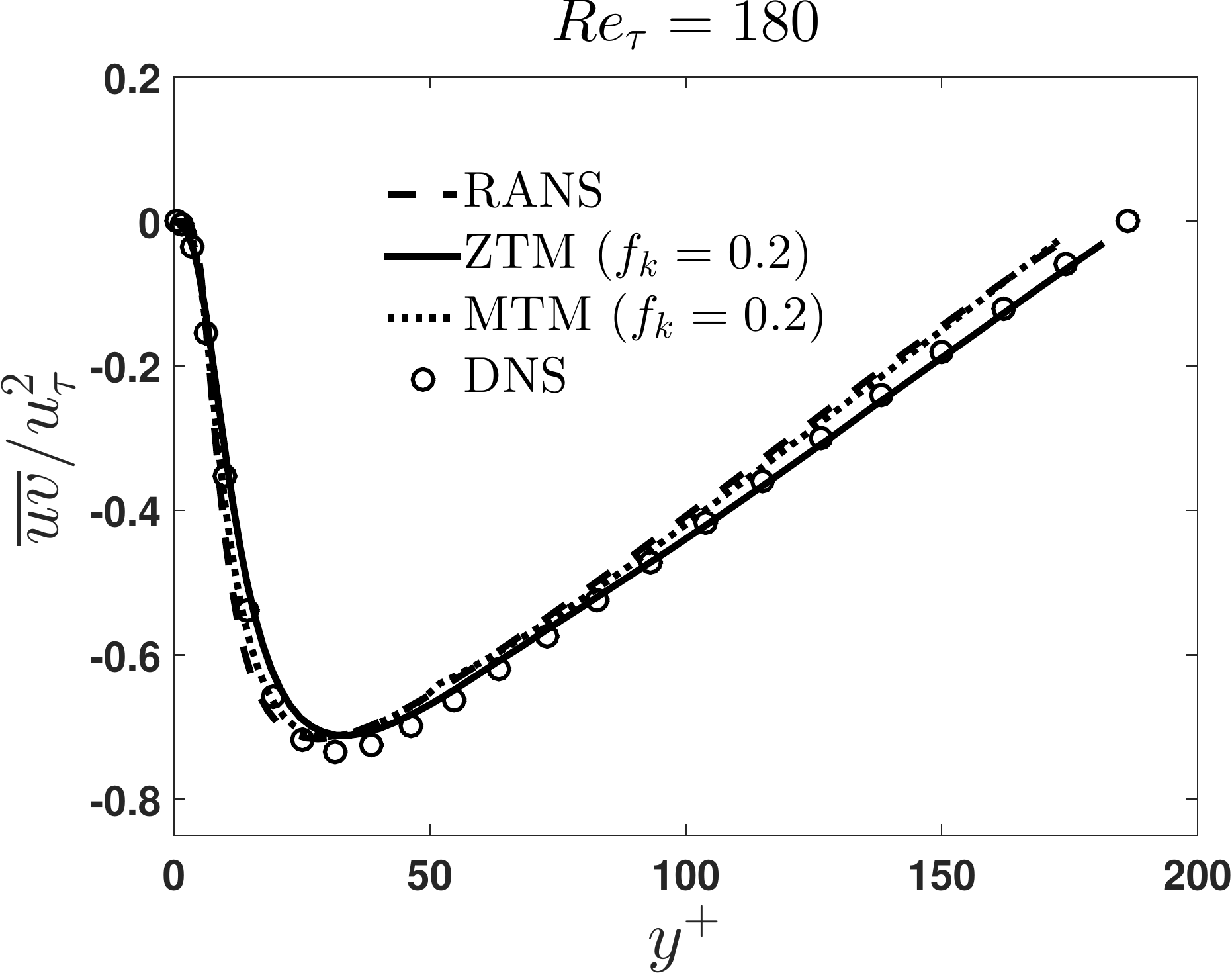}\begin{picture}(0,0)\put(-138,0){(d)}\end{picture}
        \end{subfigure}
\caption{G1-PANS simulation of turbulent channel flow for $Re_{\tau}=180$(a) Streamwise stress (b) Normal stress (c) Spanwise stress (d) Shear stress}   
\label{MTM180}     
\end{figure}

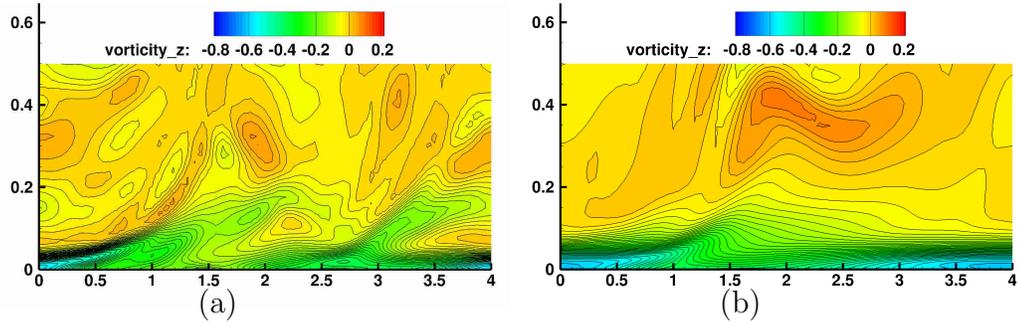
\begin{figure}[H]
        \centering
 		\captionsetup{justification=centering}                                       
 		        \begin{subfigure}[b]{0.45\textwidth}
                \includegraphics[trim=1.0cm 0.8cm 0cm 8cm, clip=true, scale=0.35]{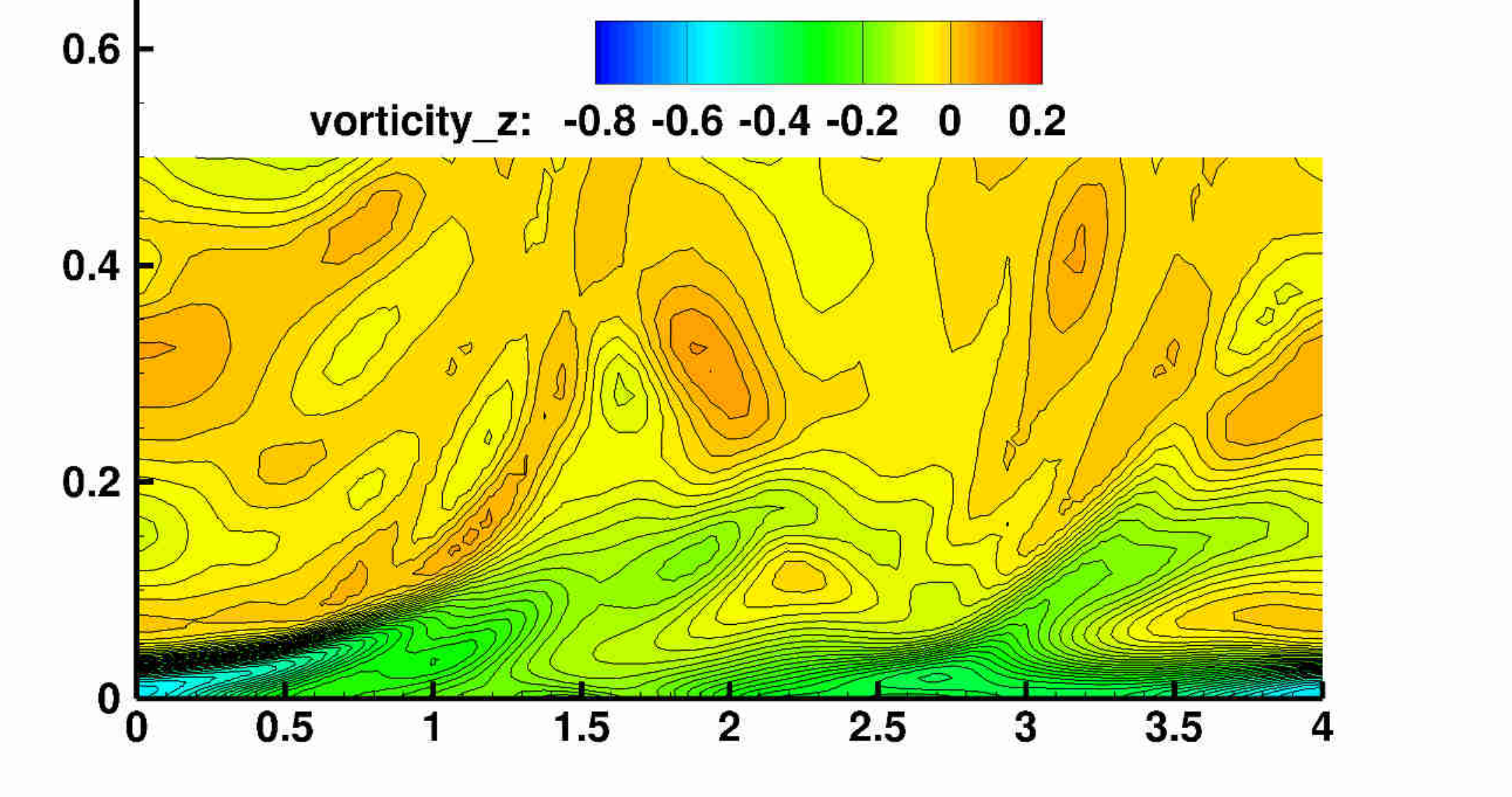}\begin{picture}(0,0)\put(-138,-2){(a)}\end{picture}
        \end{subfigure}
        			\begin{subfigure}[b]{0.45\textwidth}
                \includegraphics[trim=1cm 0.8cm 0cm 8cm, clip=true, scale=0.35]{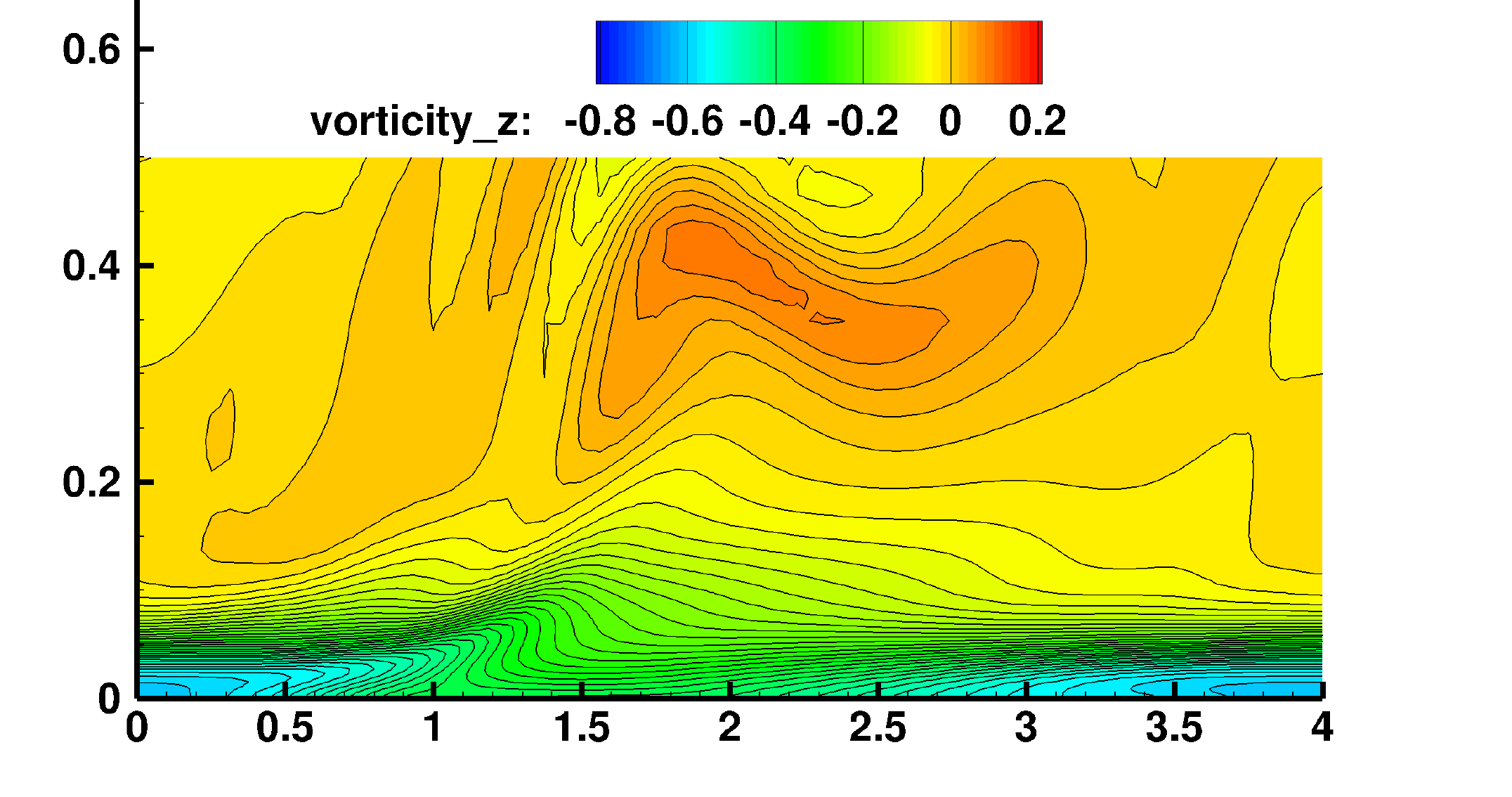}\begin{picture}(0,0)\put(-138,-2){(b)}\end{picture}
        \end{subfigure}
\caption{Vorticity contours for G1-PANS simulation of turbulent channel flow at $Re_{\tau}=180$(a) ZTM (b) MTM}   
\label{MTM180-S}     
\end{figure}

Figures \ref{MTM180-U} and \ref{MTM180} show the mean flow statistics for the two transport models alongside with the DNS data. It is observed from Fig. \ref{MTM180-U} that the slope of the log layer is accurately predicted by both models, but the MTM model fails to obtain the right velocity profile. Overshoot in the streamwise stress and under-estimation of wall-normal and spanwise stresses are observed in Fig. \ref{MTM180} for the MTM model. On the other hand, the ZTM model approximates the velocity profile and normal stresses very close to the DNS data. Furthermore, Fig. \ref{MTM180-S} compares the two approaches by depicting the Z-vorticity contours. This figure displays the fact that considerably more scales of flow specially close to the walls are resolved by the ZTM when compared to the MTM assumption. This study demonstrates that the ZTM approach is indeed the correct transport closure for the boundary layer analysis. 

\subsubsection{Motivations to develop the second generation of the PANS model}

\begin{figure}[H]
        \centering
 		\captionsetup{justification=centering}                                       
                \includegraphics[width=0.55\textwidth]{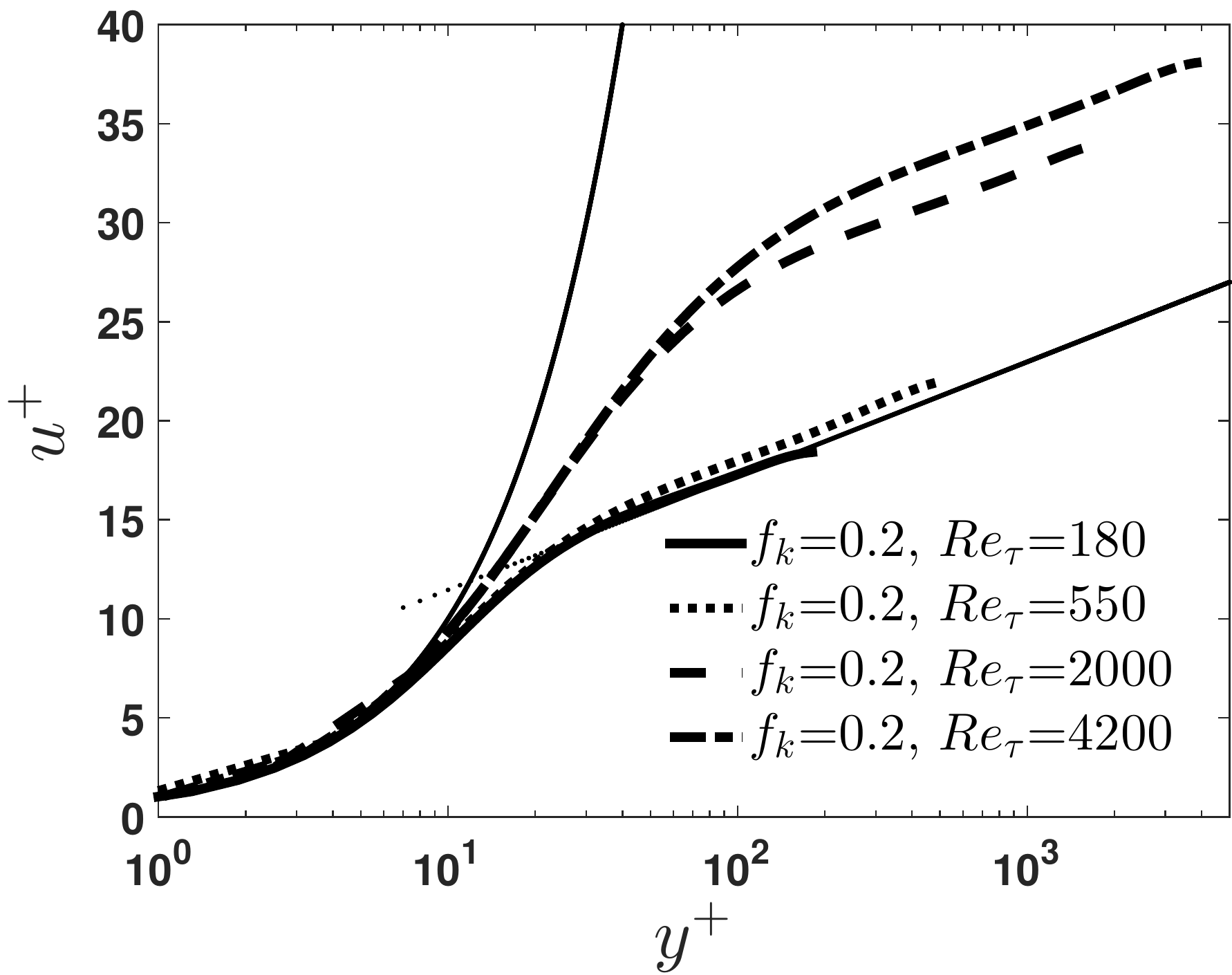}\begin{picture}(0,0)\put(-138,0){}\end{picture}
\caption{G1-PANS simulation of turbulent channel flow at different $Re_\tau$}   
\label{cons_fk_U}     
\end{figure}
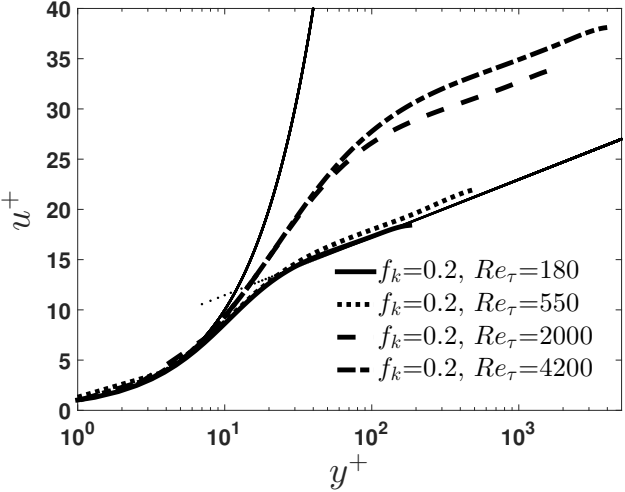

The mean velocity profile obtained from the G1-PANS calculations and DNS data for $Re_\tau$=180-4200 are shown in figure \ref{cons_fk_U}. Grid resolution of $64\times101\times64$ is utilized for all the Reynolds numbers. Accuracy of the G1-PANS model given the specified grid resolution is investigated for low and high Reynolds number flows. Figure \ref{cons_fk_U} reveals that the G1-PANS model with $f_k$=0.2 is able to recover the log layer very well for the two low Reynolds number cases. However, for the higher Reynolds numbers of 2000 and 4200, the log layer mismatch is remarkable. In fact, for these Reynolds numbers, the velocity profile is in good agreement with DNS data only in the laminar sublayer region ($y^+<5$), and mismatch occurs within the buffer layer region and thereafter the log layer region. 

\begin{figure}[H]
        \centering
 		\captionsetup{justification=centering}                                       
 		        \begin{subfigure}[b]{0.45\textwidth}
                \includegraphics[width=\textwidth]{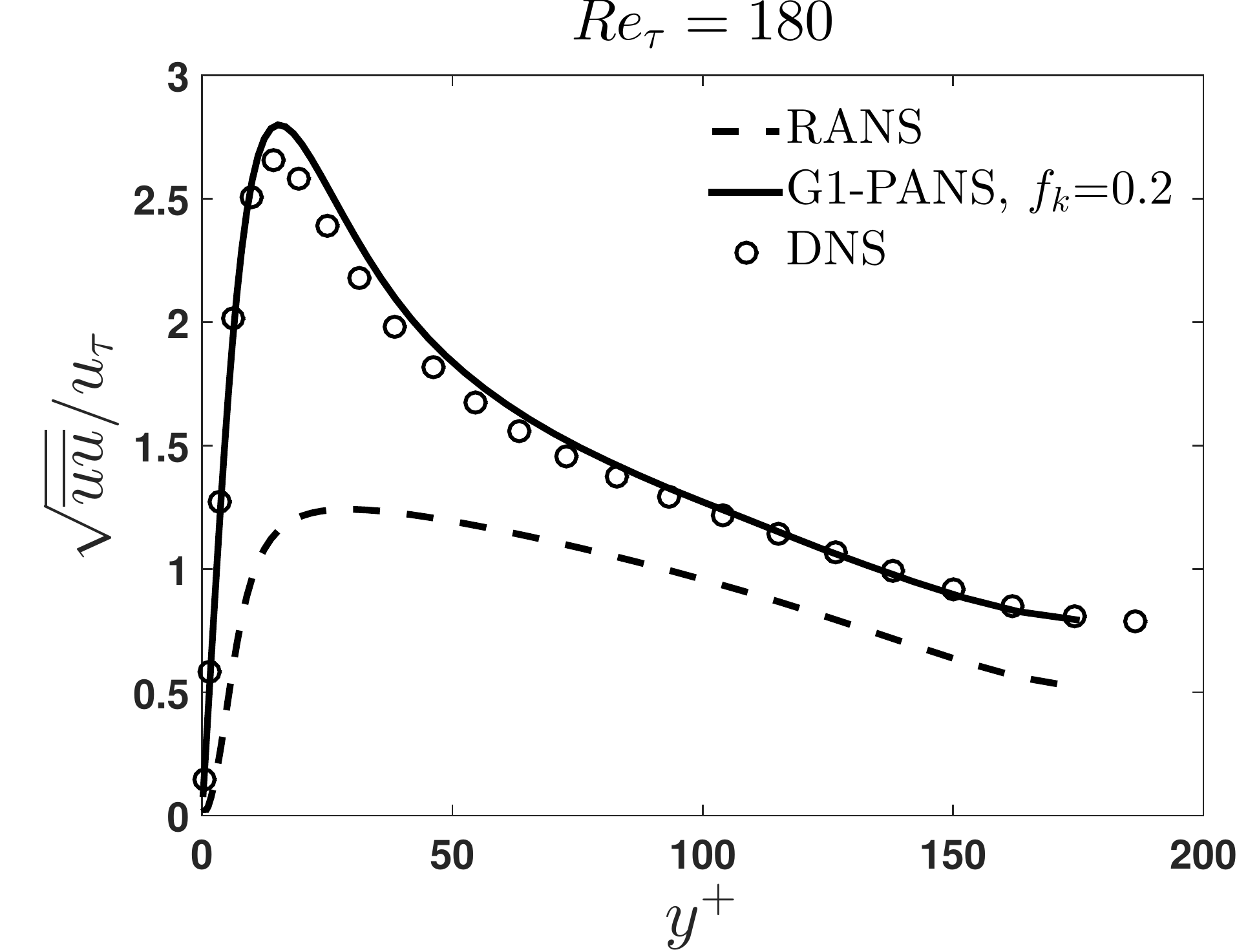}\begin{picture}(0,0)\put(-138,0){(a)}\end{picture}
        \end{subfigure}
        			\begin{subfigure}[b]{0.45\textwidth}
                \includegraphics[width=\textwidth]{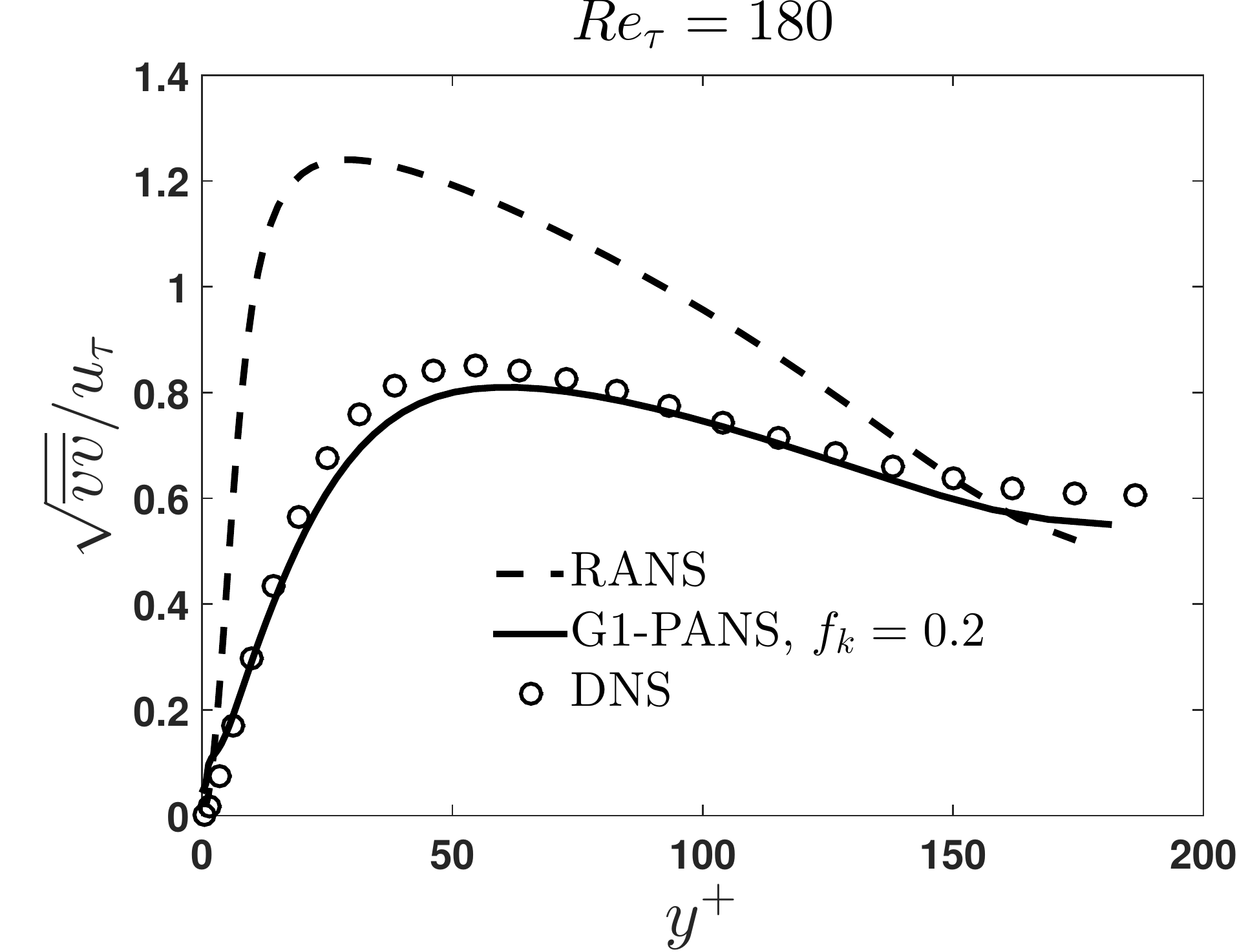}\begin{picture}(0,0)\put(-138,0){(b)}\end{picture}
        \end{subfigure}
                \begin{subfigure}[b]{0.45\textwidth}
                \includegraphics[width=\textwidth]{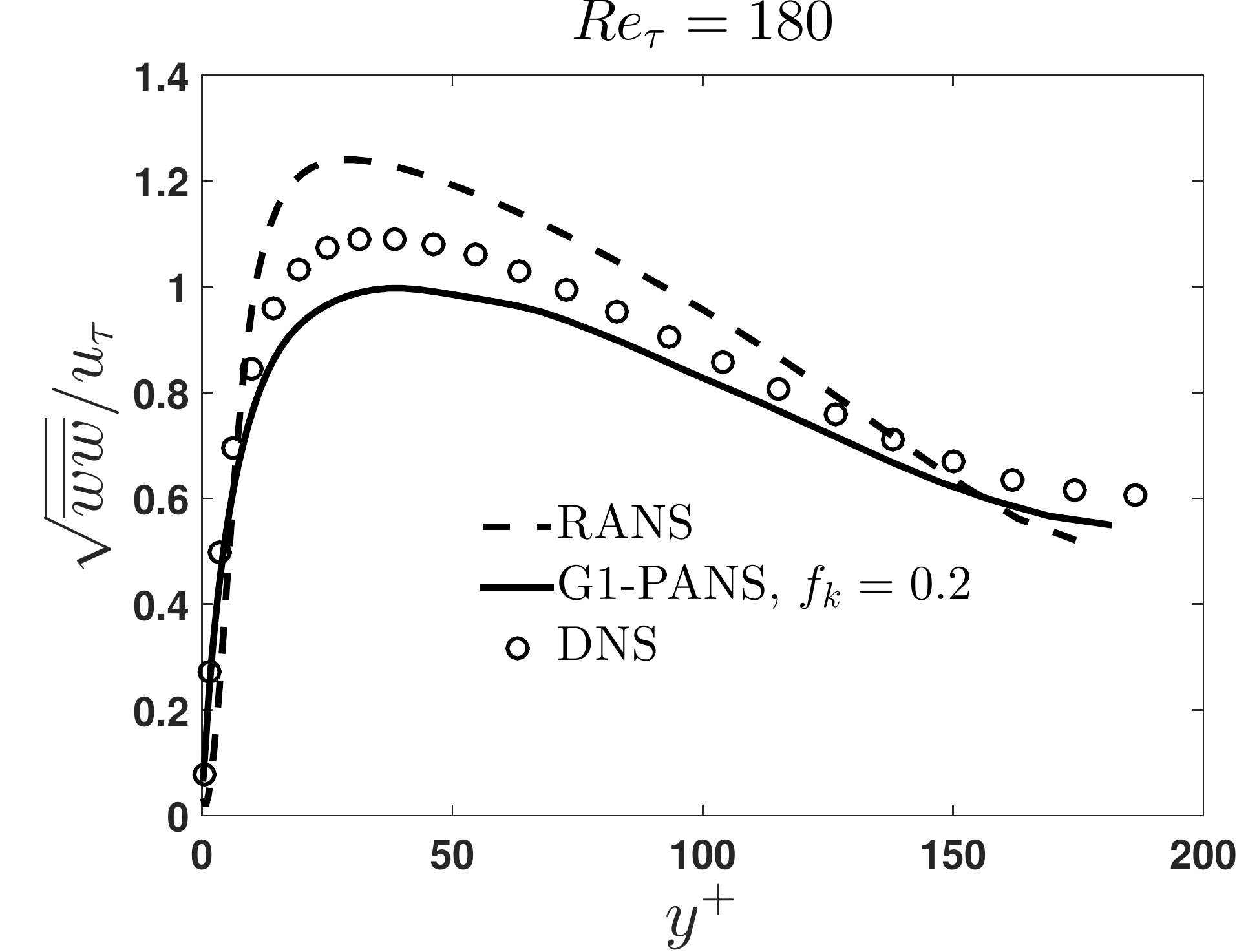}\begin{picture}(0,0)\put(-138,0){(c)}\end{picture}
        \end{subfigure}       
 		        \begin{subfigure}[b]{0.45\textwidth}
                \includegraphics[width=\textwidth]{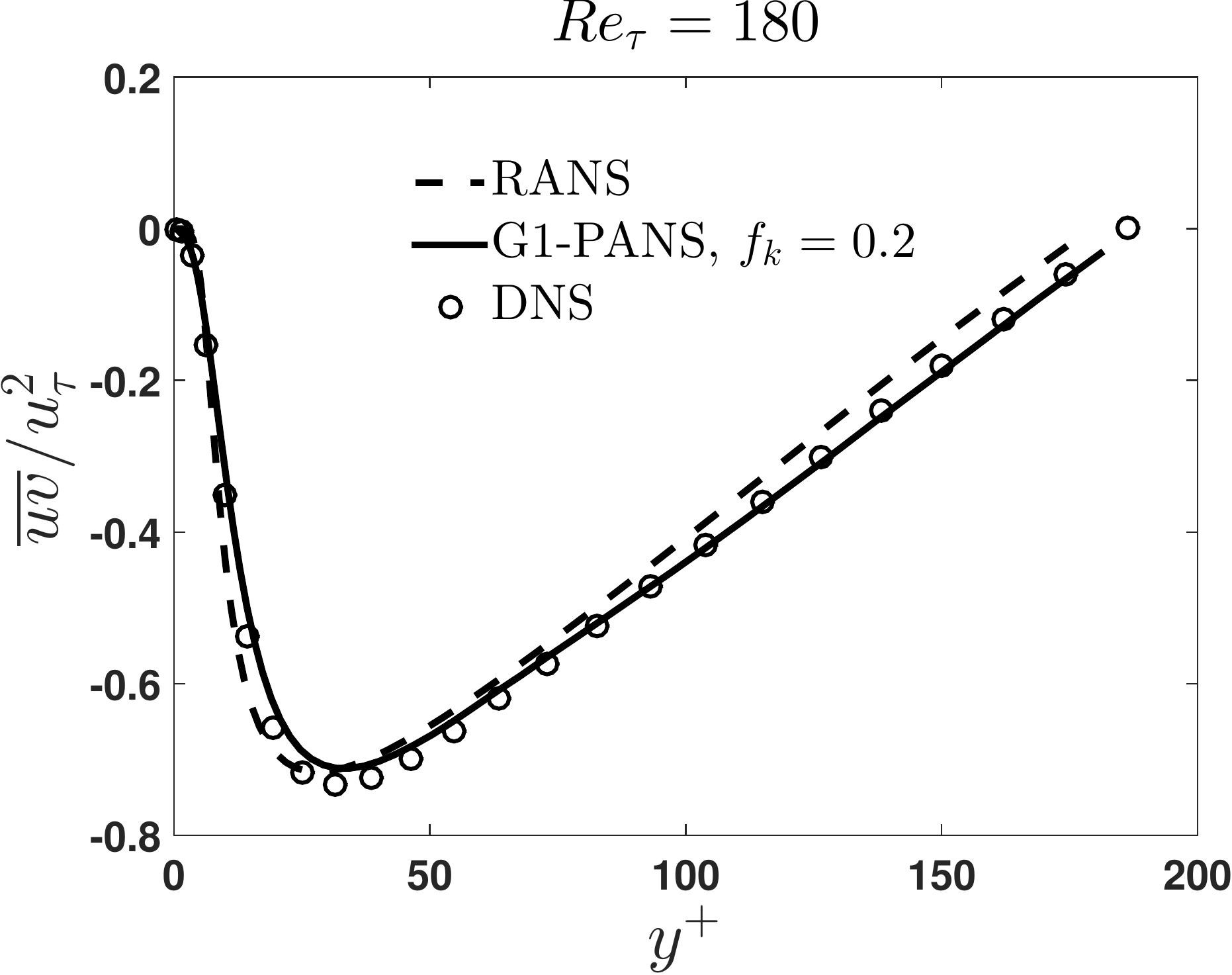}\begin{picture}(0,0)\put(-138,0){(d)}\end{picture}
        \end{subfigure}
\caption{G1-PANS simulation of turbulent channel flow for $Re_{\tau}=180$ (a) Streamwise stress (b) Normal stress (c) Spanwise stress (d) Shear stress}   
\label{uu180}     
\end{figure}

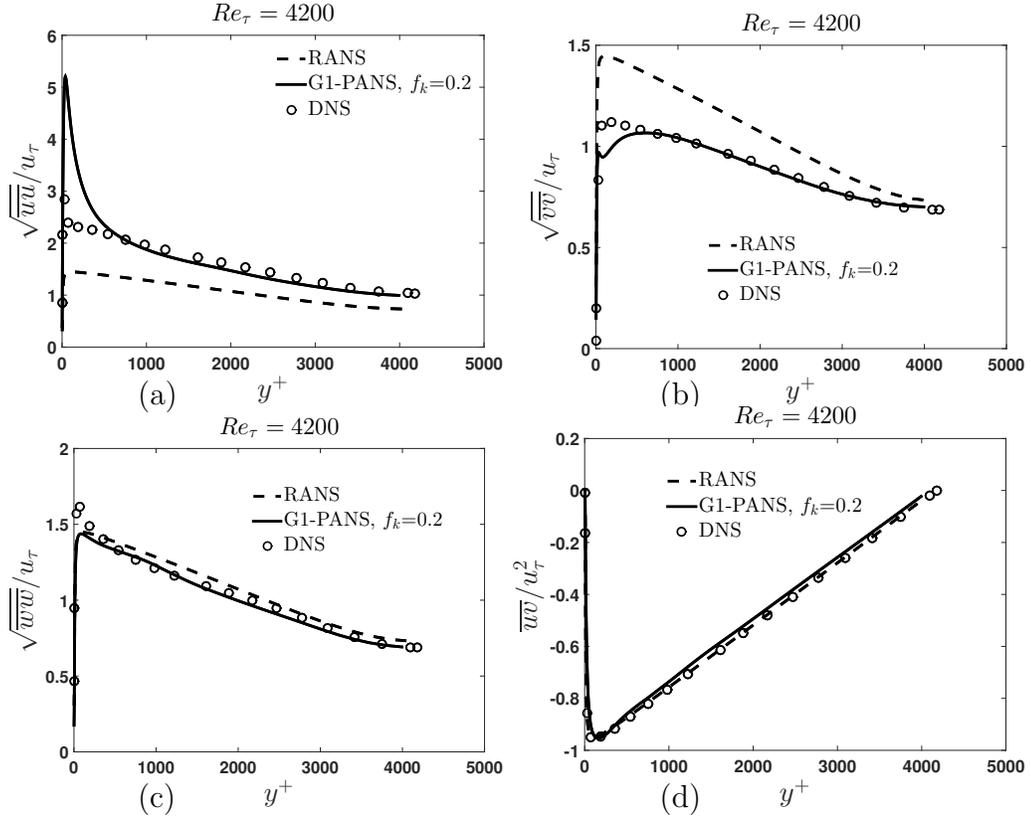
\begin{figure}[H]
        \centering
 		\captionsetup{justification=centering}                                       
 		        \begin{subfigure}[b]{0.45\textwidth}
                \includegraphics[width=\textwidth]{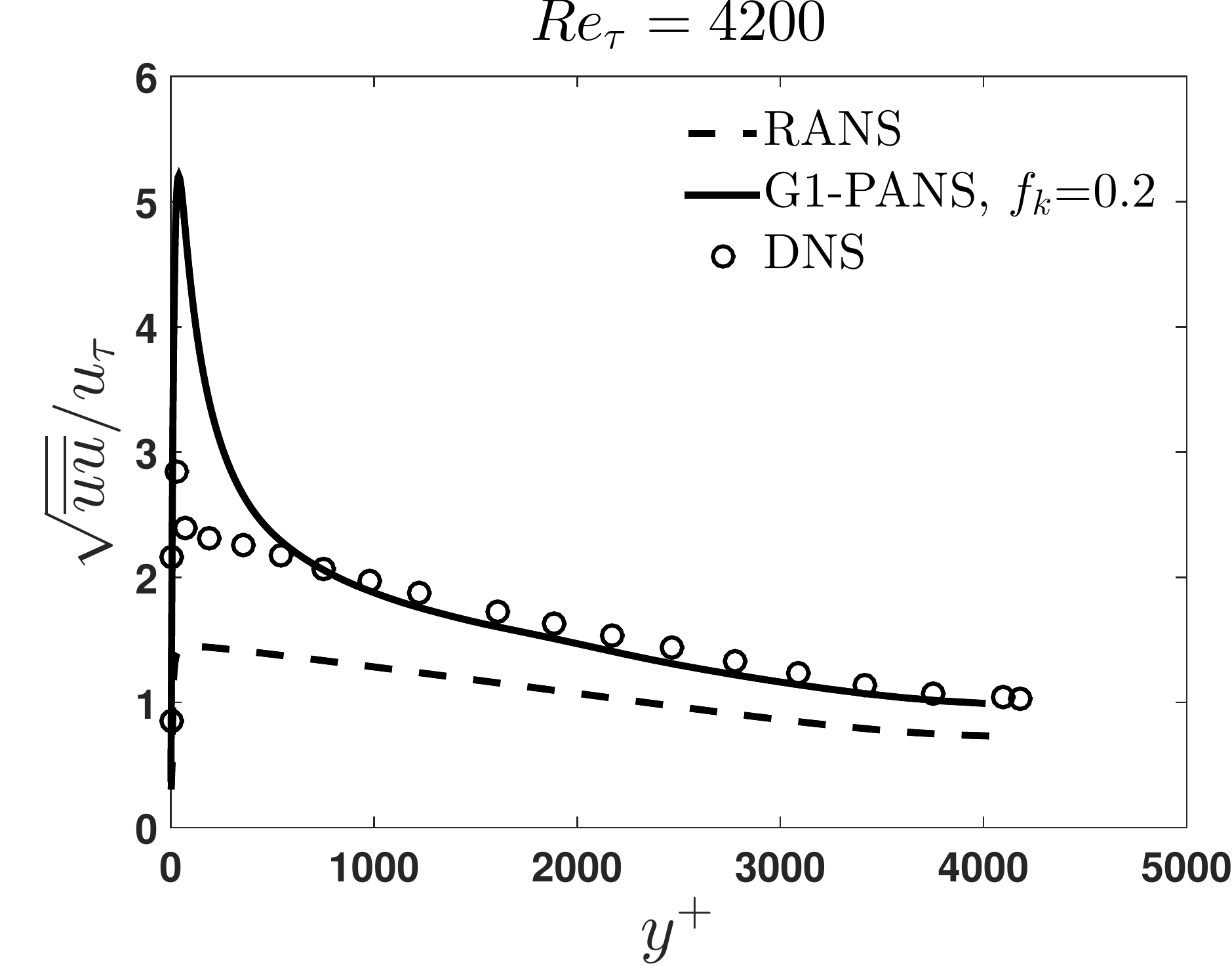}\begin{picture}(0,0)\put(-138,0){(a)}\end{picture}
        \end{subfigure}
        			\begin{subfigure}[b]{0.45\textwidth}
                \includegraphics[width=\textwidth]{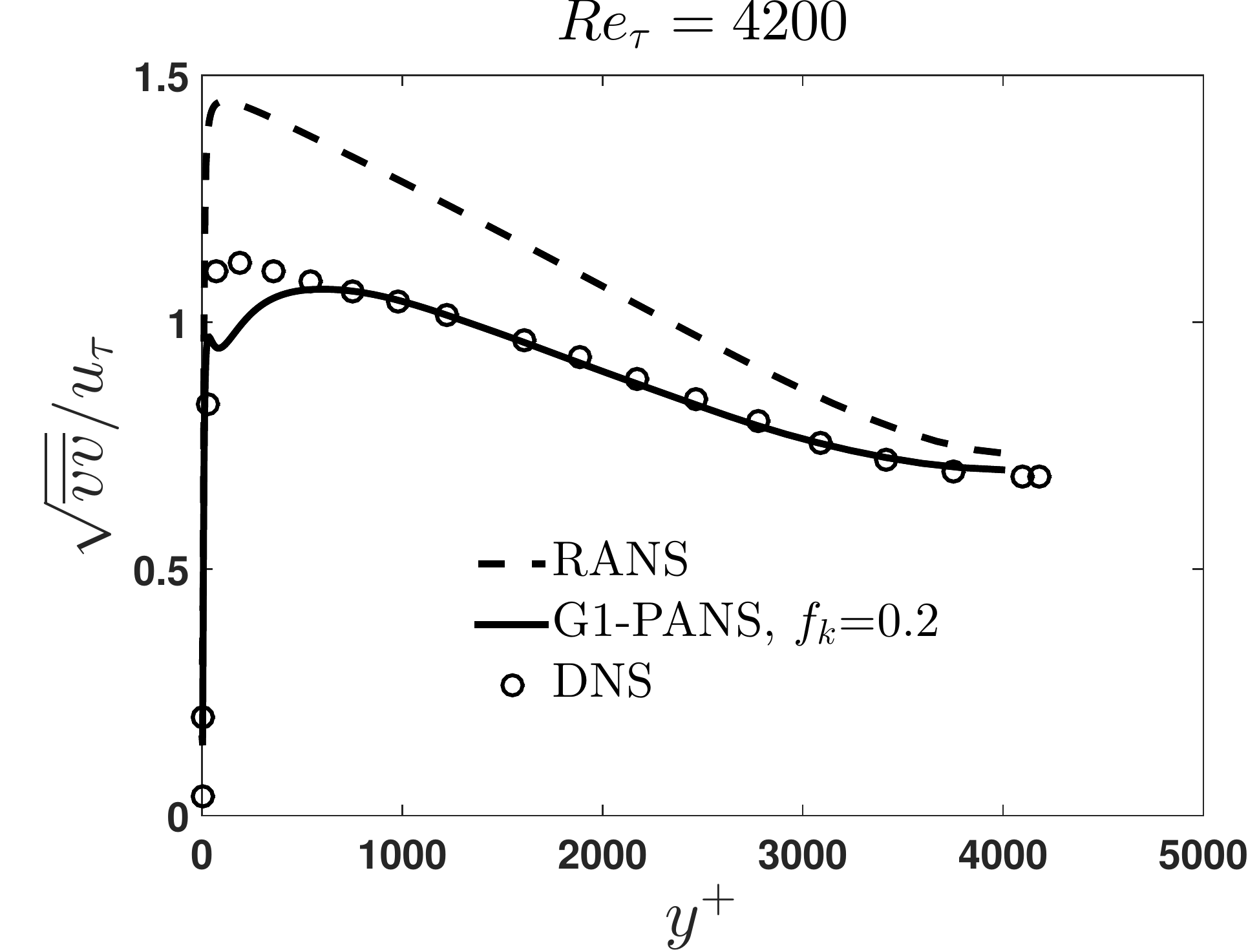}\begin{picture}(0,0)\put(-138,0){(b)}\end{picture}
        \end{subfigure}
                \begin{subfigure}[b]{0.45\textwidth}
                \includegraphics[width=\textwidth]{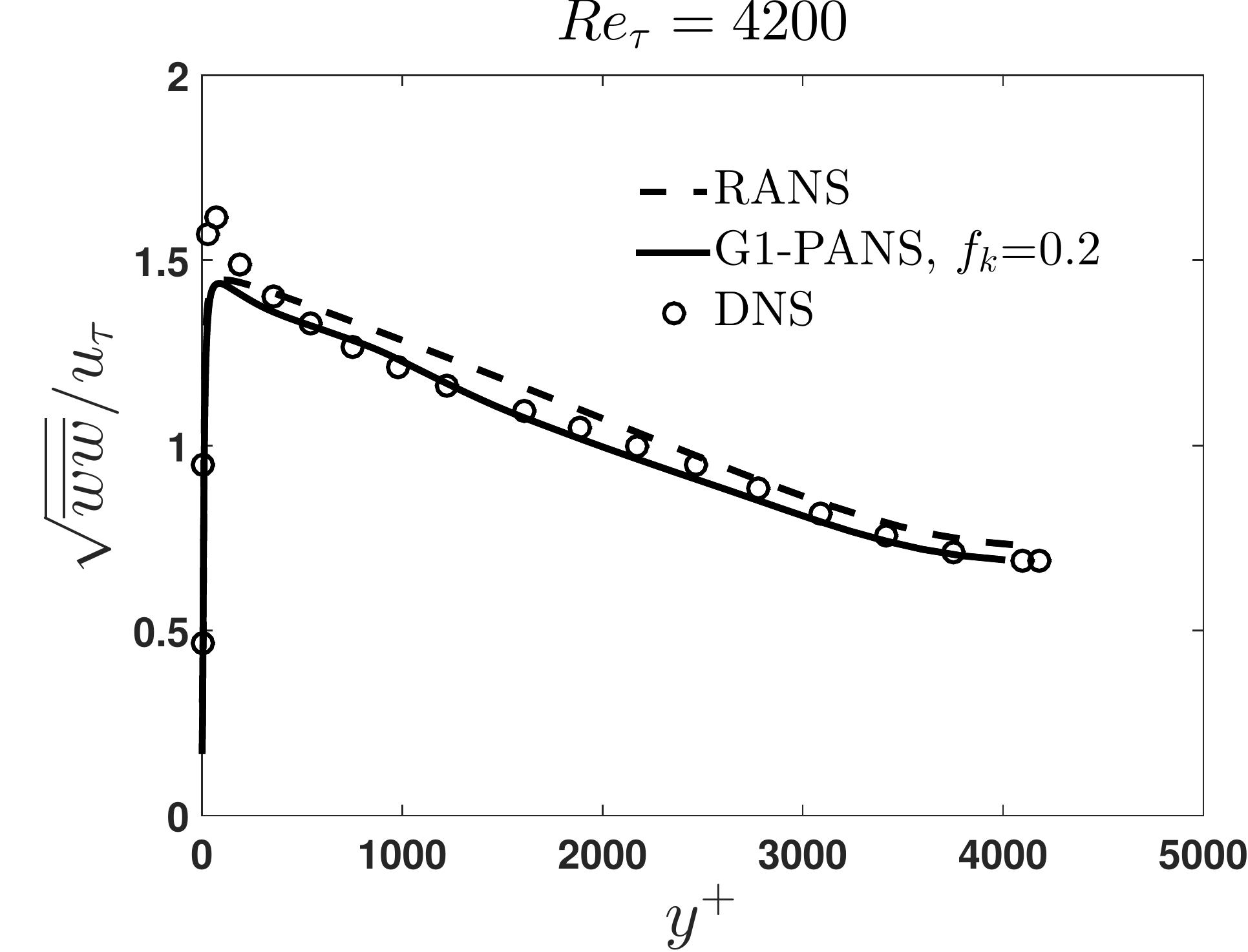}\begin{picture}(0,0)\put(-138,0){(c)}\end{picture}
        \end{subfigure}       
 		        \begin{subfigure}[b]{0.45\textwidth}
                \includegraphics[width=\textwidth]{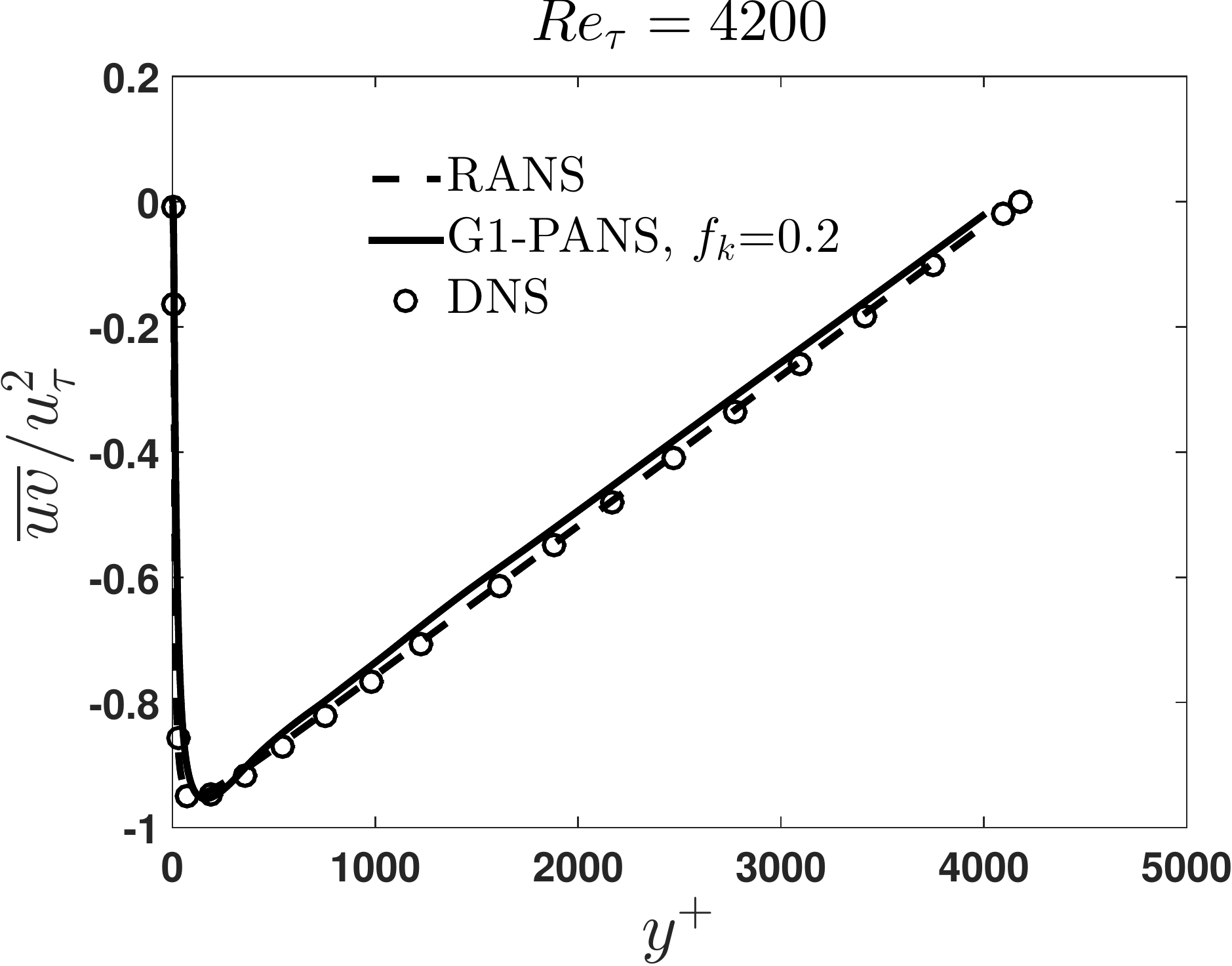}\begin{picture}(0,0)\put(-138,0){(d)}\end{picture}
        \end{subfigure}
\caption{G1-PANS simulation of turbulent channel flow for $Re_{\tau}=4200$ (a) Streamwise stress (b) Normal stress (c) Spanwise stress (d) Shear stress}   
\label{uu4200}     
\end{figure}

Figures \ref{uu180} and \ref{uu4200} show the second order statistics for $Re_\tau$=180 and 4200, respectively. It can be inferred from figure \ref{uu180} that the stress components are in very good agreement with DNS data for the G1-PANS calculation and the anisotropy of the flow is well predicted by the model at $Re_\tau$=180. Besides, Fig. \ref{uu4200} indicates significant over prediction and under prediction of the streamwise and normal stresses for the G1-PANS simulation particularly in near wall region at $Re_\tau$=4200. 

The velocity overshoot and wrong normal stress profiles obtained for higher Reynolds number of 4200 for G1-PANS calculations are mainly due to the insufficient grid resolution. While the grid resolution for specified filter parameter of $f_k$=0.2 seems to be adequate for the low Reynolds number cases, it is substantially low for the higher Reynolds number flows. The extent of grid refinement specifically in the normal and spanwise directions for the DNS studies given in table \ref{case1} further confirms that a huge computational domain is required in order to capture small scale structures near the wall at $Re_\tau$=2000 and 4200. Therefore, with the current grid, G1-PANS model is only able to capture the right flow physics at low Reynolds number cases and for high Reynolds number flows, excessive grid refinement particularly in the near wall region is required. As discussed earlier, in order to obviate the computational expense, variable resolution approach is followed next.  

\subsection{G1.5-PANS}

In previous section, it was observed that the near wall length scales are not resolved by increasing the resolution of the model, while not providing adequate number of grid nodes in that region. Therefore, in accordance with the scope of the hybrid models, an alternative approach is to model all the near wall scales with fully averaged RANS model, and resolve the necessary length scales away from the wall by lowering $f_k$. This method requires transition from a completely steady solution adjacent to the wall to an unsteady calculation away from the wall. 

%

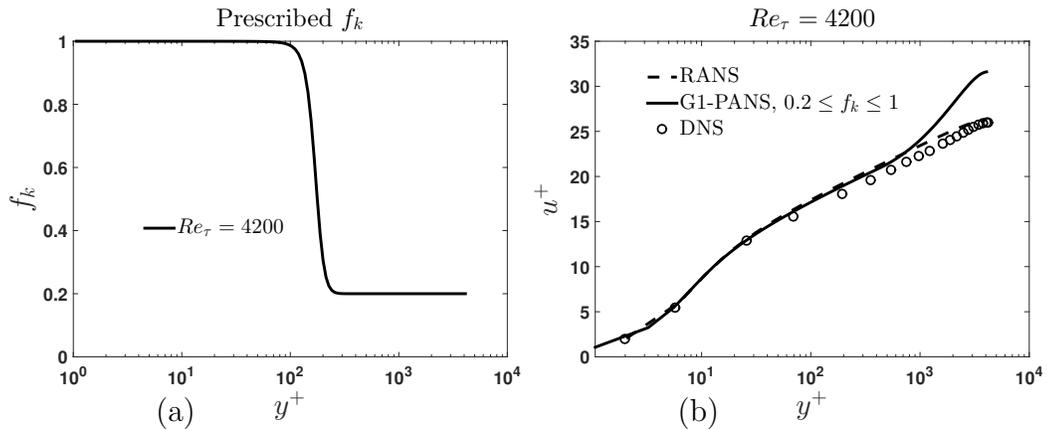
\begin{figure}[H]
        \centering
 		\captionsetup{justification=centering}                                       
		        \begin{subfigure}[b]{0.45\textwidth}
                \includegraphics[width=\textwidth]{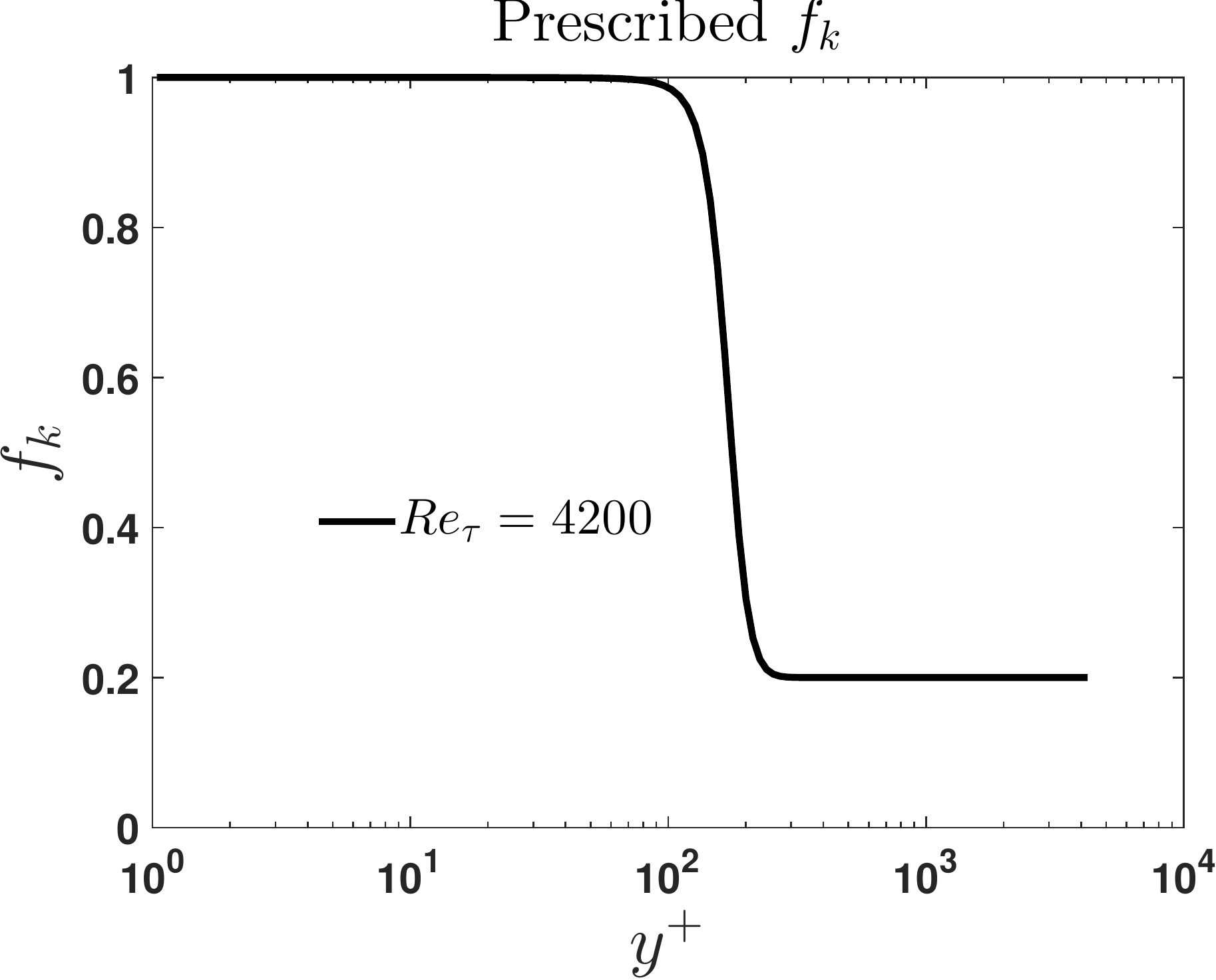}\begin{picture}(0,0)\put(-138,0){(a)}\end{picture}
        \end{subfigure}
        			\begin{subfigure}[b]{0.45\textwidth}
                \includegraphics[width=\textwidth]{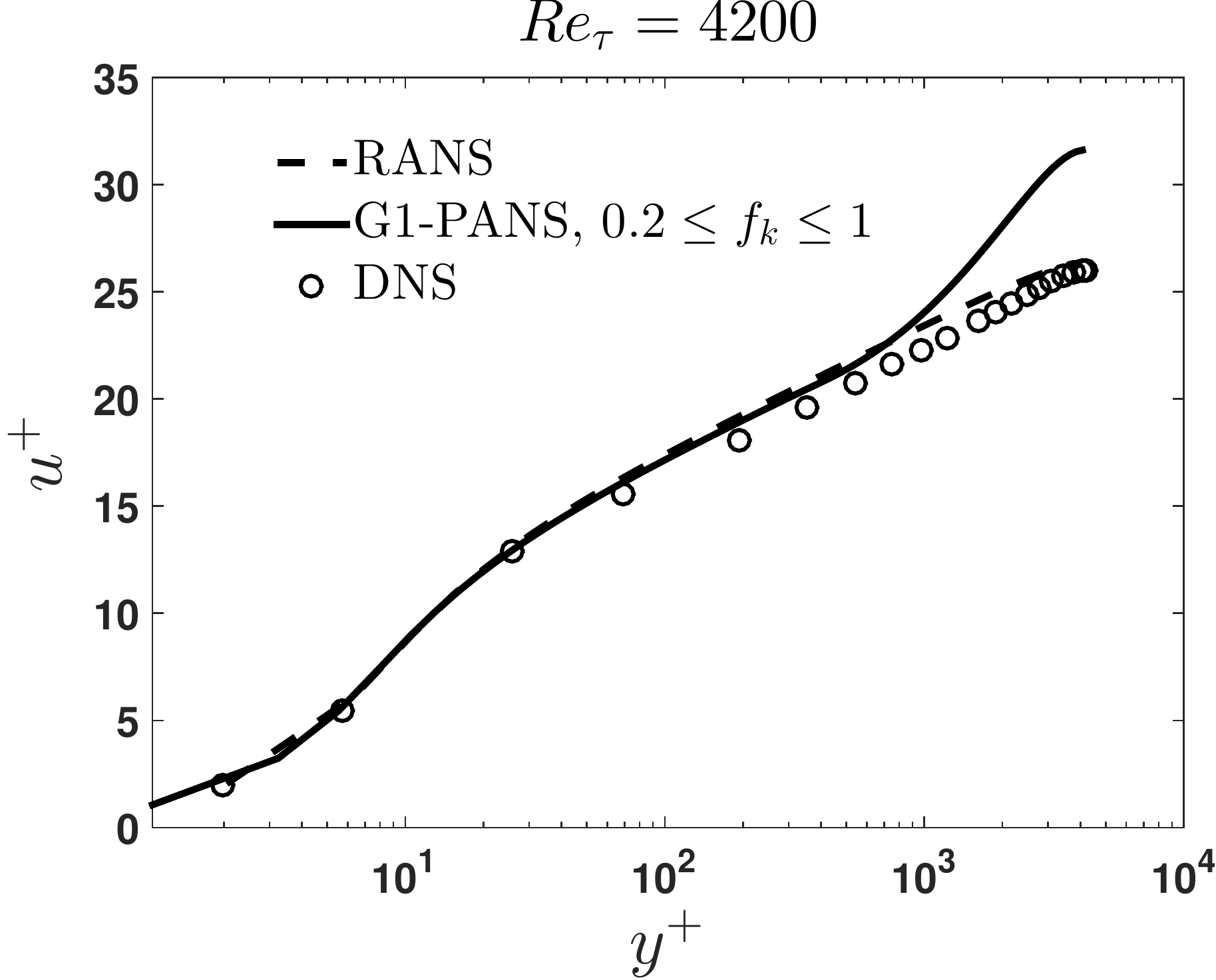}\begin{picture}(0,0)\put(-138,0){(b)}\end{picture}
        \end{subfigure}
\caption{G1.5-PANS simulation of turbulent channel flow for $Re_\tau=4200$(a) $f_k$ variation, (b) Mean velocity}   
\label{varfk_u}     
\end{figure}

The model equations of G1-PANS is used for the calculations, but with variable rather than constant $f_k$. This approach is given the name G1.5-PANS method where the commutation terms are not included. Turbulent channel flow simulation using the G1.5-PANS method is performed for $Re_\tau$=4200 where the near wall region is fully modelled. Figure \ref{varfk_u} (a) shows the prescribed variation of filter parameter in near wall region for $Re_\tau$=4200. As shown in this figure, the transition from the steady to unsteady regions happens at $100 < y^+ < 250$. 

The corresponding mean velocity profile is illustrated in figure \ref{varfk_u} (b). The mean velocity profile clearly shows that changing the resolution near the wall and using the G1-PANS formulation leads to a wrong slope of the profile in the log layer region and velocity overshoot. Also, the variation of stress components shows deviation of data from DNS and proximity to the RANS solution. The tendency of the current hybrid modeling approach to reach steady RANS solution indicates that a proper energy scale transfer has not be been achieved by only changing the resolution near the wall. Therefore, in the next section, the second generation of the PANS model which accounts for the resolution change accompanied by including the additional energy scale transfer terms is investigated. 

\begin{figure}[H]
        \centering
 		\captionsetup{justification=centering}                                       
 		        \begin{subfigure}[b]{0.45\textwidth}
                \includegraphics[width=\textwidth]{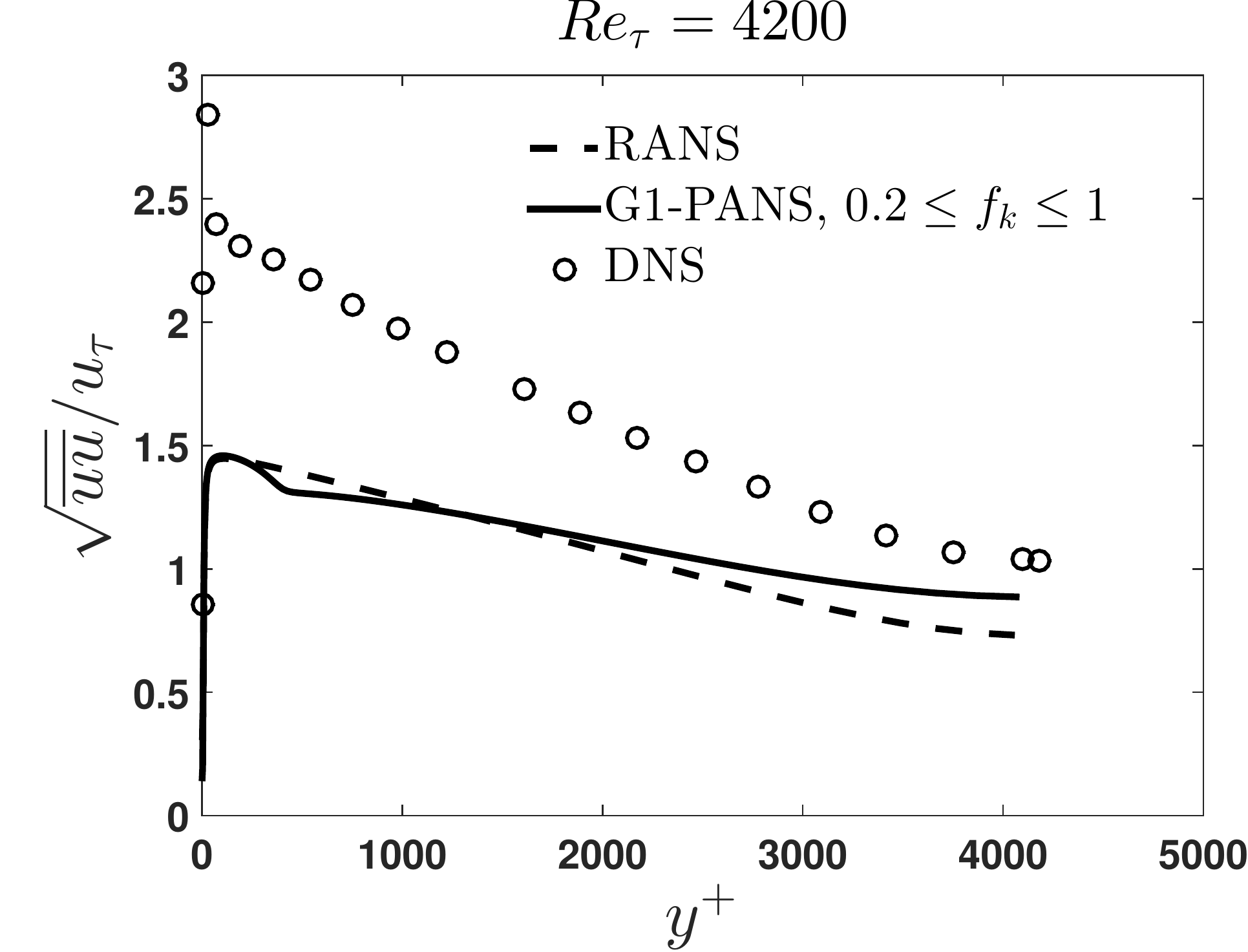}\begin{picture}(0,0)\put(-138,0){(a)}\end{picture}
        \end{subfigure}
        			\begin{subfigure}[b]{0.45\textwidth}
                \includegraphics[width=\textwidth]{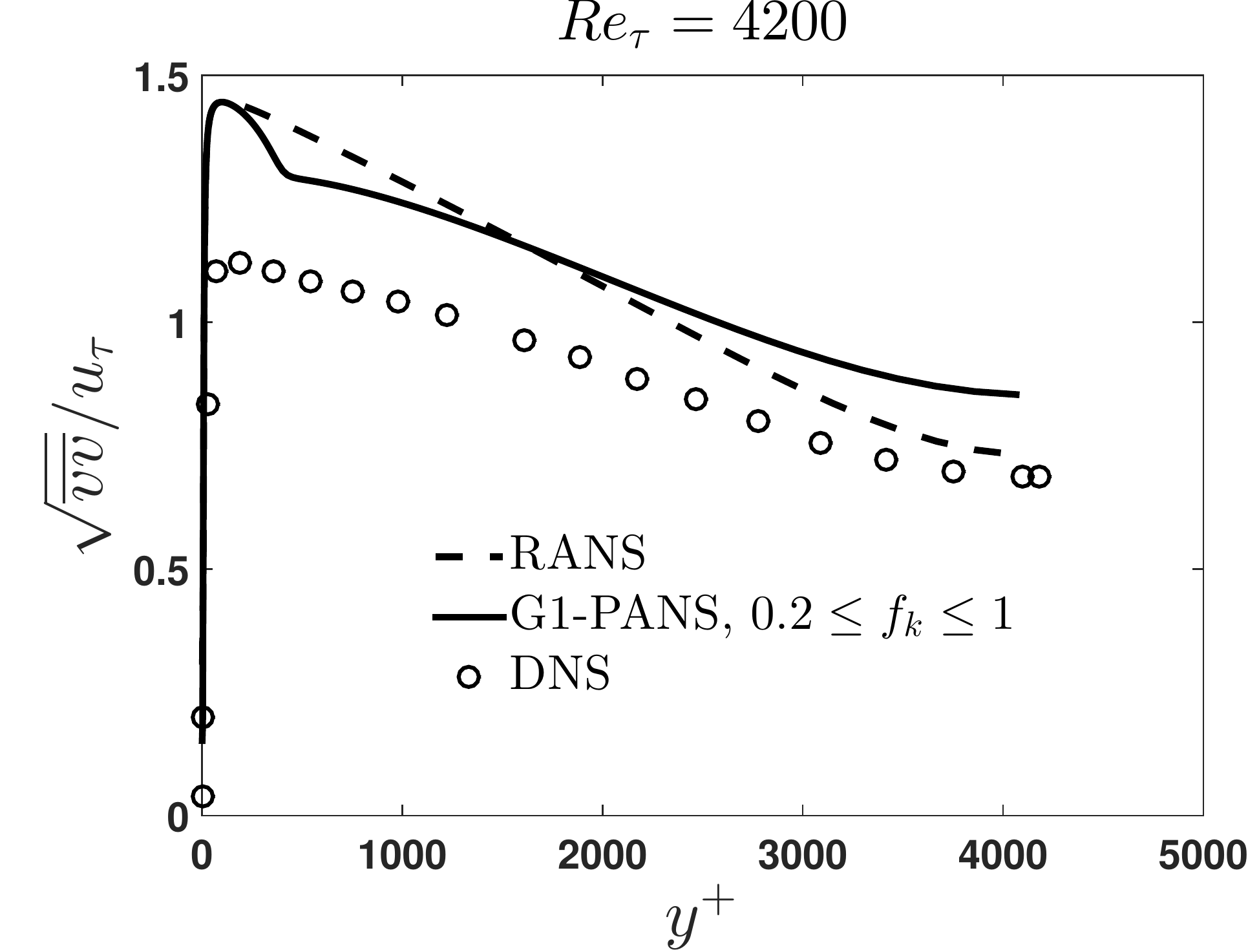}\begin{picture}(0,0)\put(-138,0){(b)}\end{picture}
        \end{subfigure}
                \begin{subfigure}[b]{0.45\textwidth}
                \includegraphics[width=\textwidth]{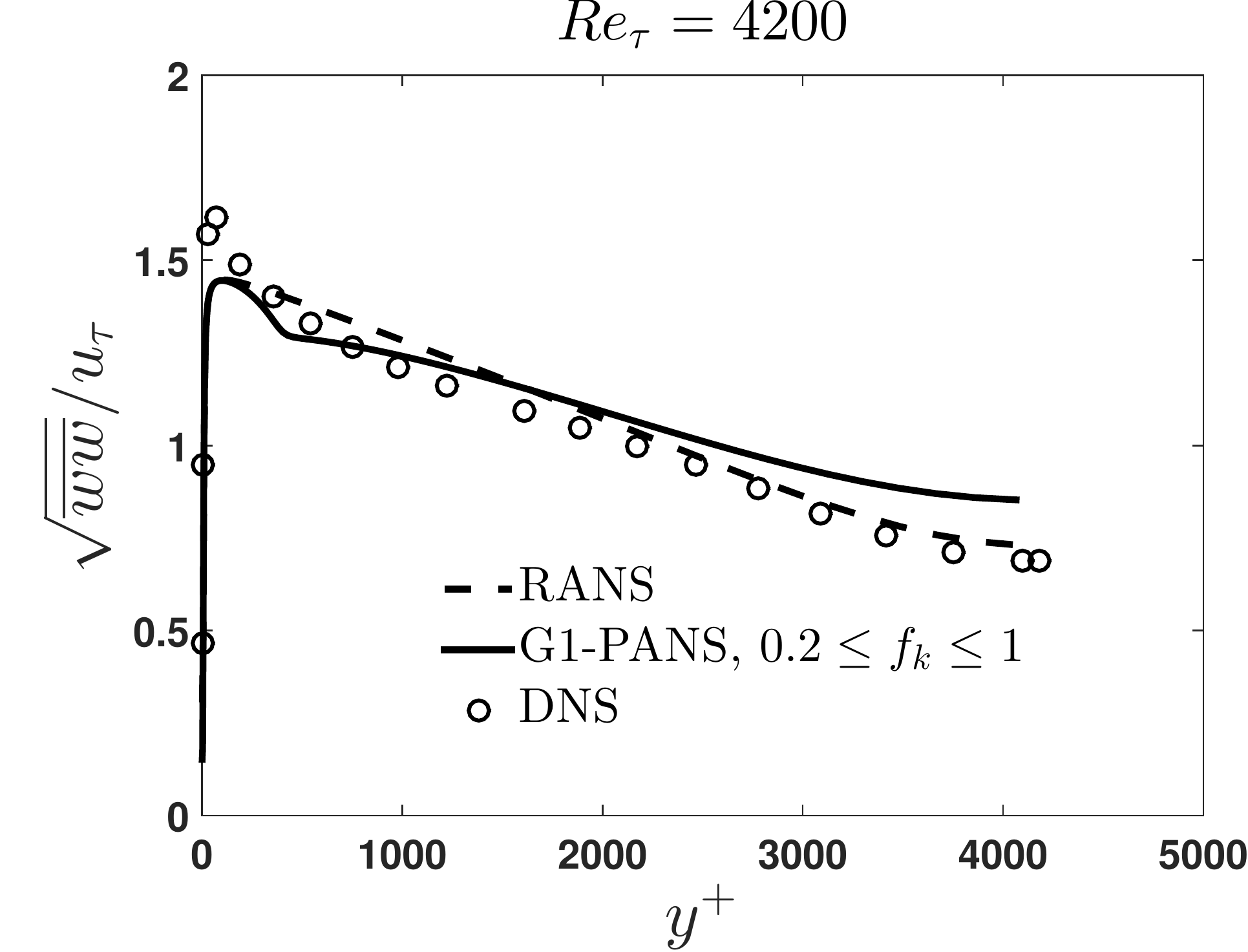}\begin{picture}(0,0)\put(-138,0){(c)}\end{picture}
        \end{subfigure}       
 		        \begin{subfigure}[b]{0.45\textwidth}
                \includegraphics[width=\textwidth]{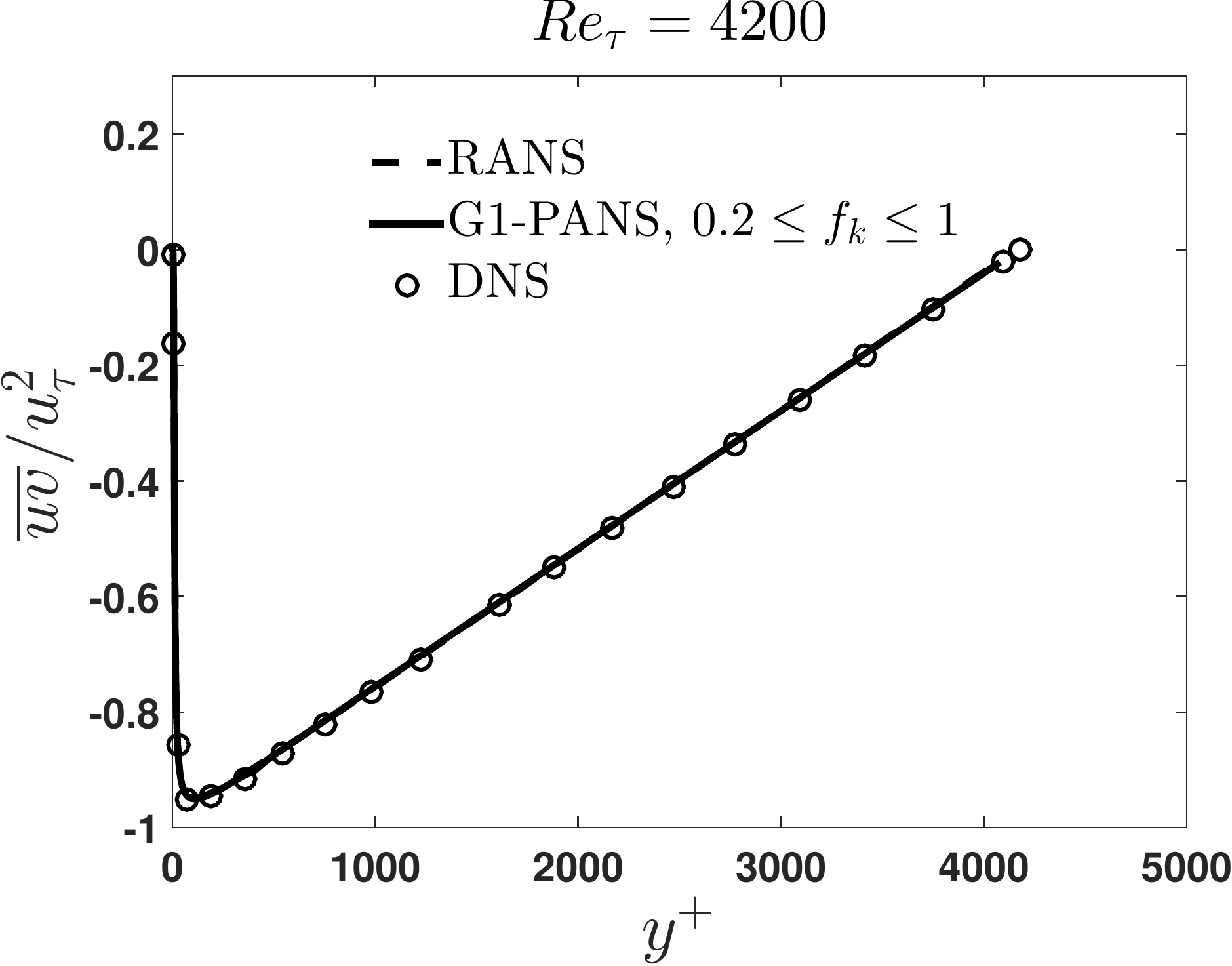}\begin{picture}(0,0)\put(-138,0){(d)}\end{picture}
        \end{subfigure}
\caption{G1.5-PANS simulation of turbulent channel flow for $Re_{\tau}=4200$ (a) Streamwise stress (b) Normal stress (c) Spanwise stress (d) Shear stress}   
\label{varfk_uu}     
\end{figure}
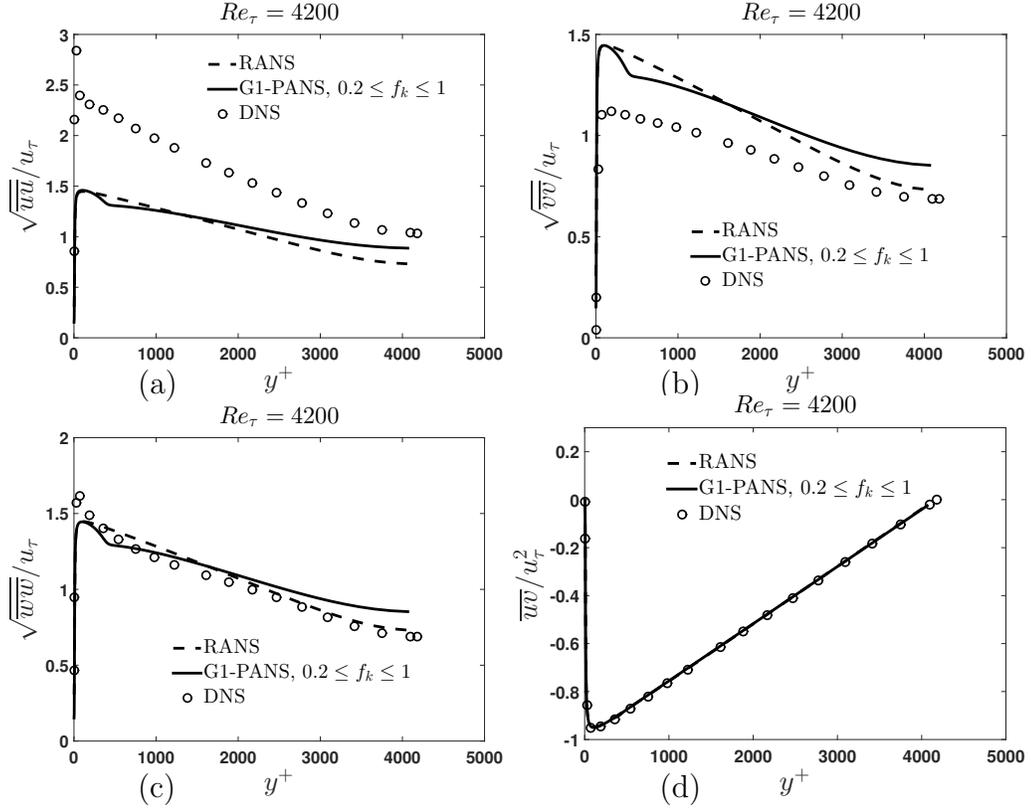

\subsection{G2-PANS simulations of turbulent channel flow}

The G2-PANS simulations for $Re_\tau$= 950-8000 are presented in this section. For each case, the flow statistics are compared with well documented DNS data except for $Re_\tau$=8000 where there is no DNS data available. 

Figure \ref{recovery}a shows the variation of filter parameter at different Reynolds numbers considered in this study. As shown in this figure for the lower Reynolds number cases, $Re_\tau=950$ and $2000$, $f_k$ is one in the laminar sublayer and part of the buffer layer and then is gradually reduced so that in the fully turbulent region it reaches a constant value of $0.2$. For higher Reynolds numbers of $4200$ and $8000$, the switch from RANS solution is purposefully delayed to a further distance from the wall where the laminar sublayer, buffer layer and part of the log layer is fully modeled with RANS. This is because at higher Reynolds numbers, the range of length scales will become wider and more small scale eddies are present in the flow domain. This implies that, with the current grid resolutions, it is not possible to resolve the necessary amount of scales in the buffer layer and early part of the log layer at higher Reynolds number, and therefore they are fully modeled. 

\subsubsection{Eddy Viscosity Recovery}

%
%

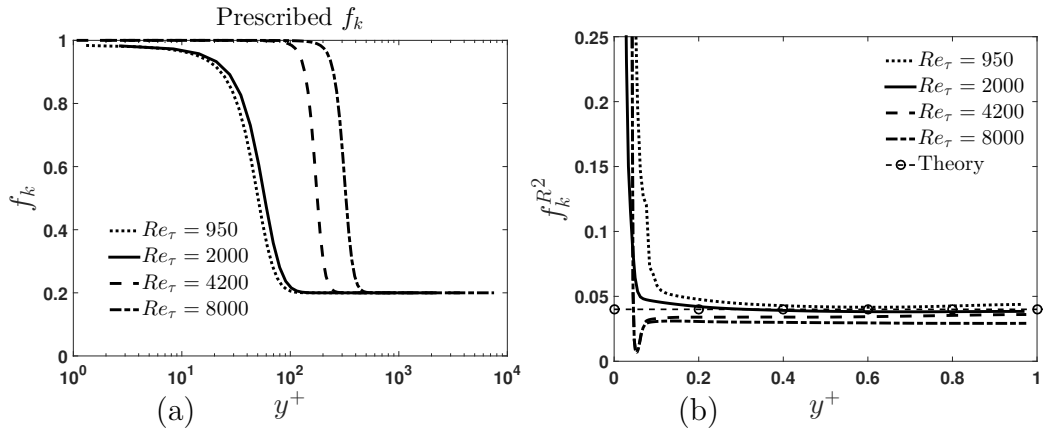
\begin{figure}[H]
        \centering
 		\captionsetup{justification=centering}                                       
 		        \begin{subfigure}[b]{0.45\textwidth}
                \includegraphics[width=\textwidth]{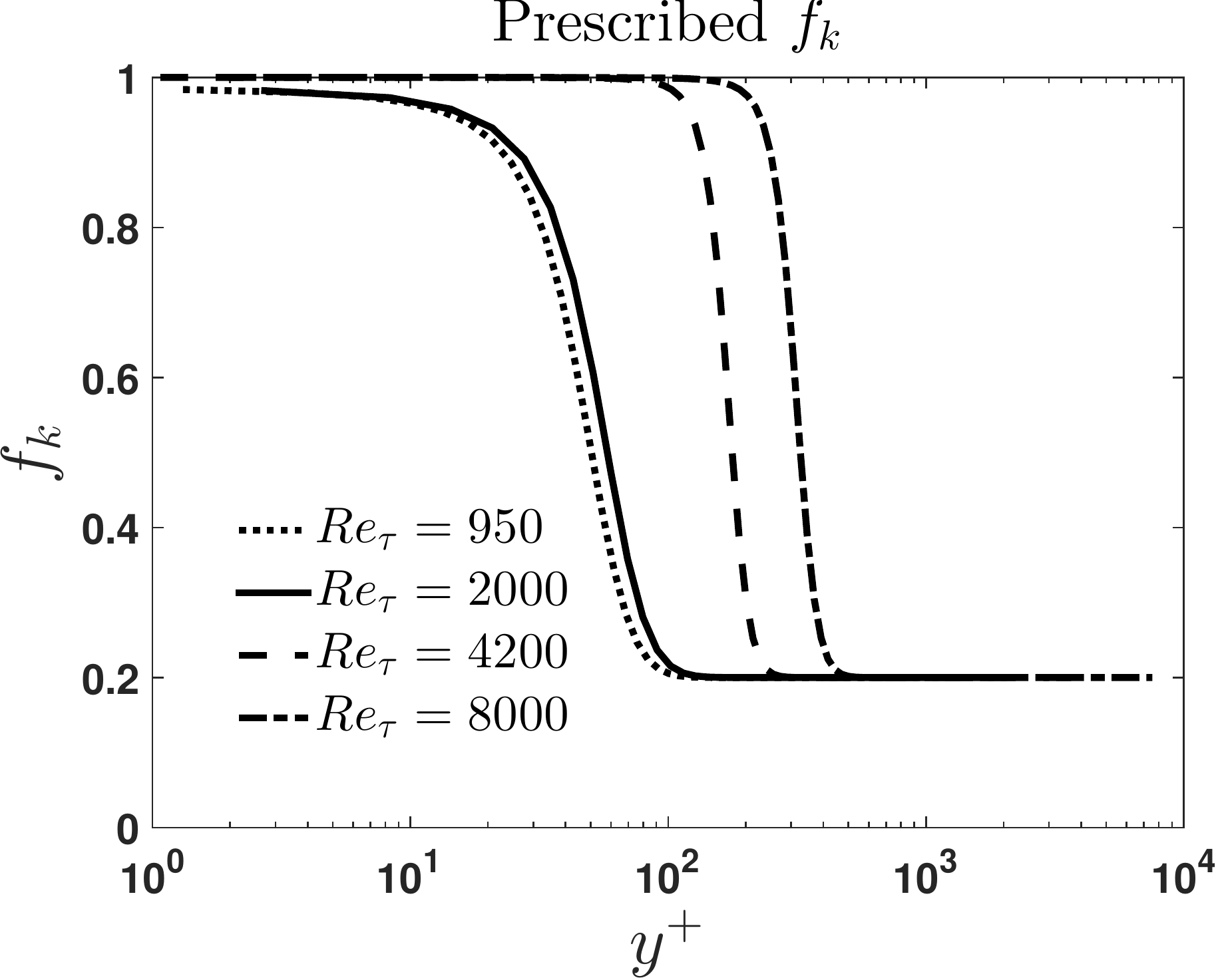}\begin{picture}(0,0)\put(-138,0){(a)}\end{picture}
        \end{subfigure}
                \begin{subfigure}[b]{0.45\textwidth}
                \includegraphics[width=\textwidth]{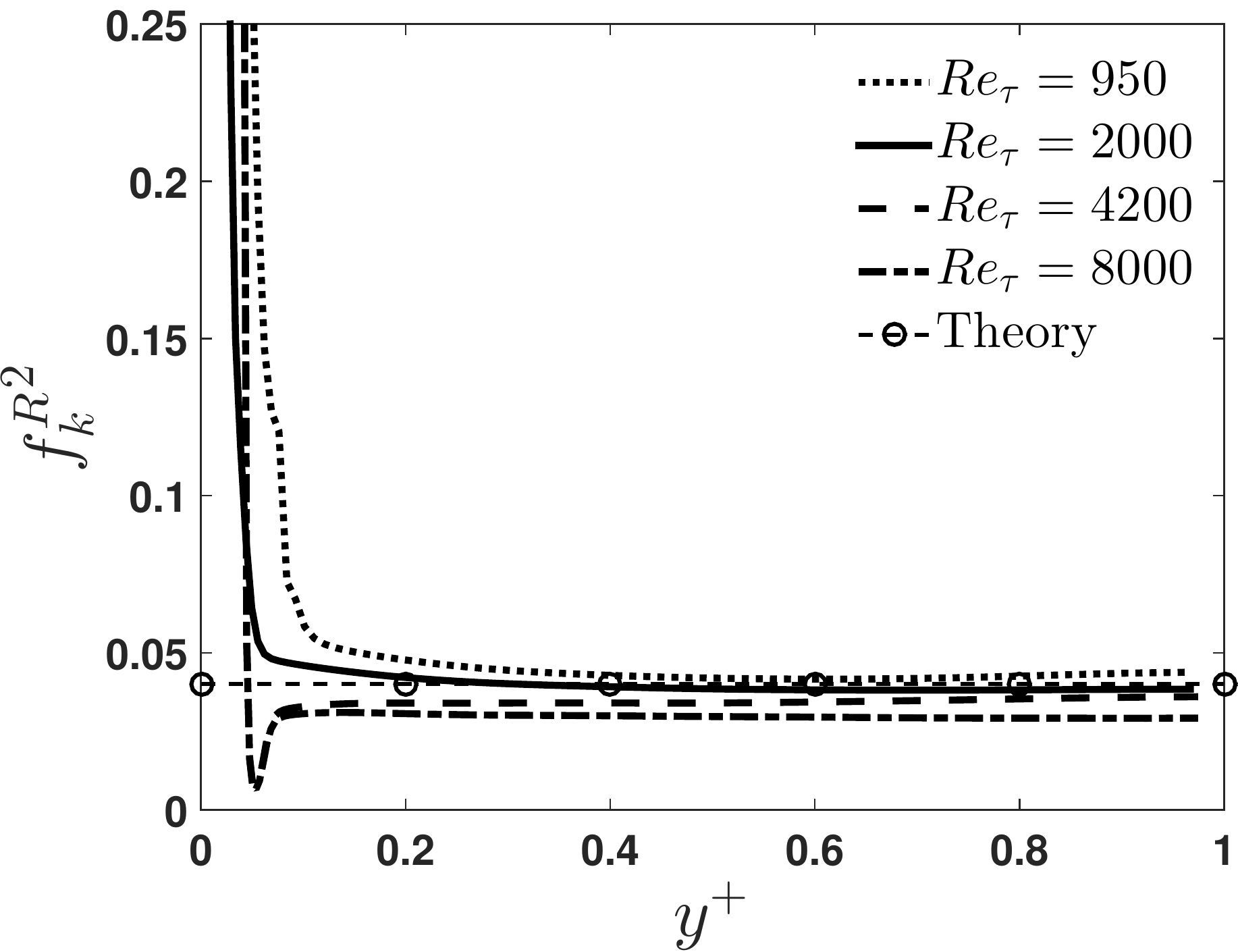}\begin{picture}(0,0)\put(-138,0){(b)}\end{picture}
        \end{subfigure}       
\caption{G2-PANS simulation of turbulent channel flow (a) Prescribed $f_k$ (b) Recovery of $f_k^2$}   
\label{recovery}     
\end{figure}

Variation of $\frac{\nu_u}{\nu}$ with $y^+$ is shown in Fig. \ref{recovery} (b) for the PANS simulations with input $f_k$ values given in Fig. \ref{recovery} (a) at different $Re_\tau$. The eddy viscosity recovery is well observed for almost the entire domain particularly in the middle region away from the walls for the PANS simulations. This ensures that the model performs well, and the right level of eddy viscosity is achieved in the solution domain. 

\subsubsection{Flow statistics}

\begin{figure}[H]
        \centering
 		\captionsetup{justification=centering}                                       
                \includegraphics[width=0.55\textwidth]{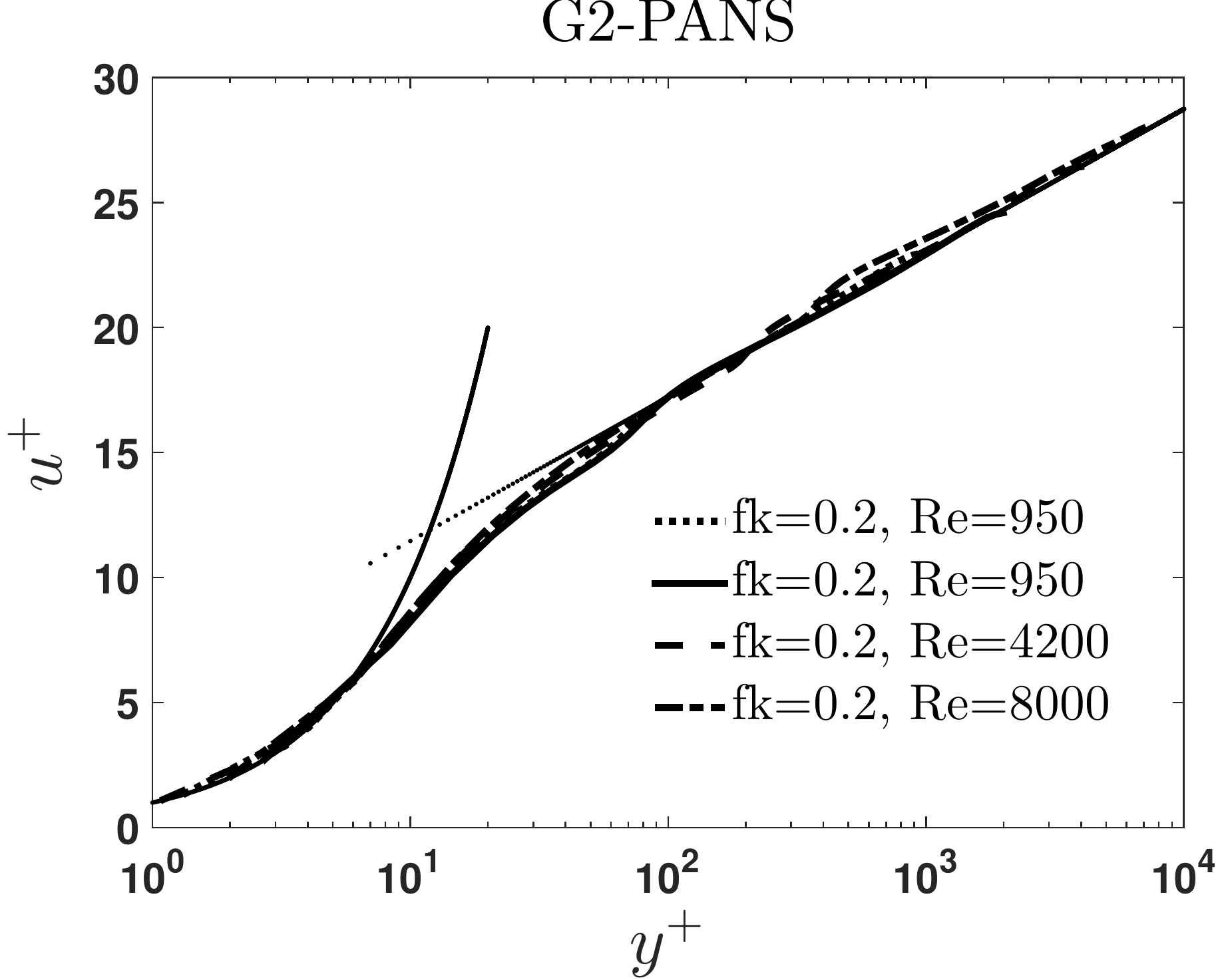}
\caption{G2-PANS simulation of turbulent channel flow: mean velocity at different Reynolds numbers}   
\label{varfk2-u}     
\end{figure}

Figure \ref{varfk2-u} shows the mean velocity profile at different Reynolds numbers for RANS as well as G2-PANS calculations. Remarkably, the log-layer is accurately captured at all Reynolds numbers. It must be pointed out that most hybrid methods exhibit a log-layer mismatch. The accuracy of G2-PANS model is attributed to the right level of energy exchange in the region of resolution variation. 

The numerical transition from steady to unsteady region is noticeable in the velocity profile. The transition occurs at the region of rapid variation in $f_k$. This effect can be alleviated by increasing the number of points in the resolution change region or by prescribing a wider variation of filter parameter. 

\begin{figure}[H]
        \centering
 		\captionsetup{justification=centering}                                       
 		        \begin{subfigure}[b]{0.45\textwidth}
                \includegraphics[width=\textwidth]{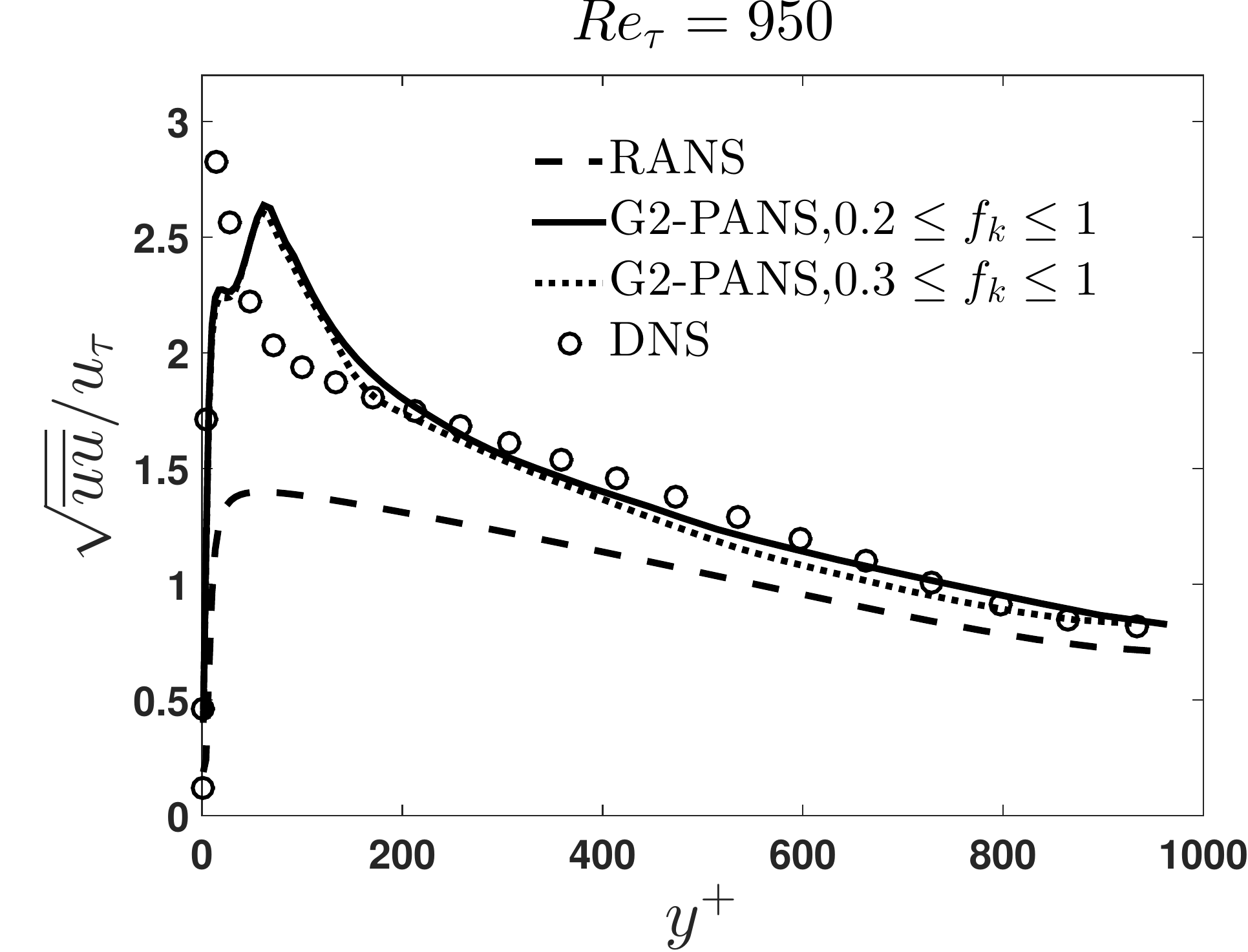}\begin{picture}(0,0)\put(-138,0){(a)}\end{picture}
        \end{subfigure}
        			\begin{subfigure}[b]{0.45\textwidth}
                \includegraphics[width=\textwidth]{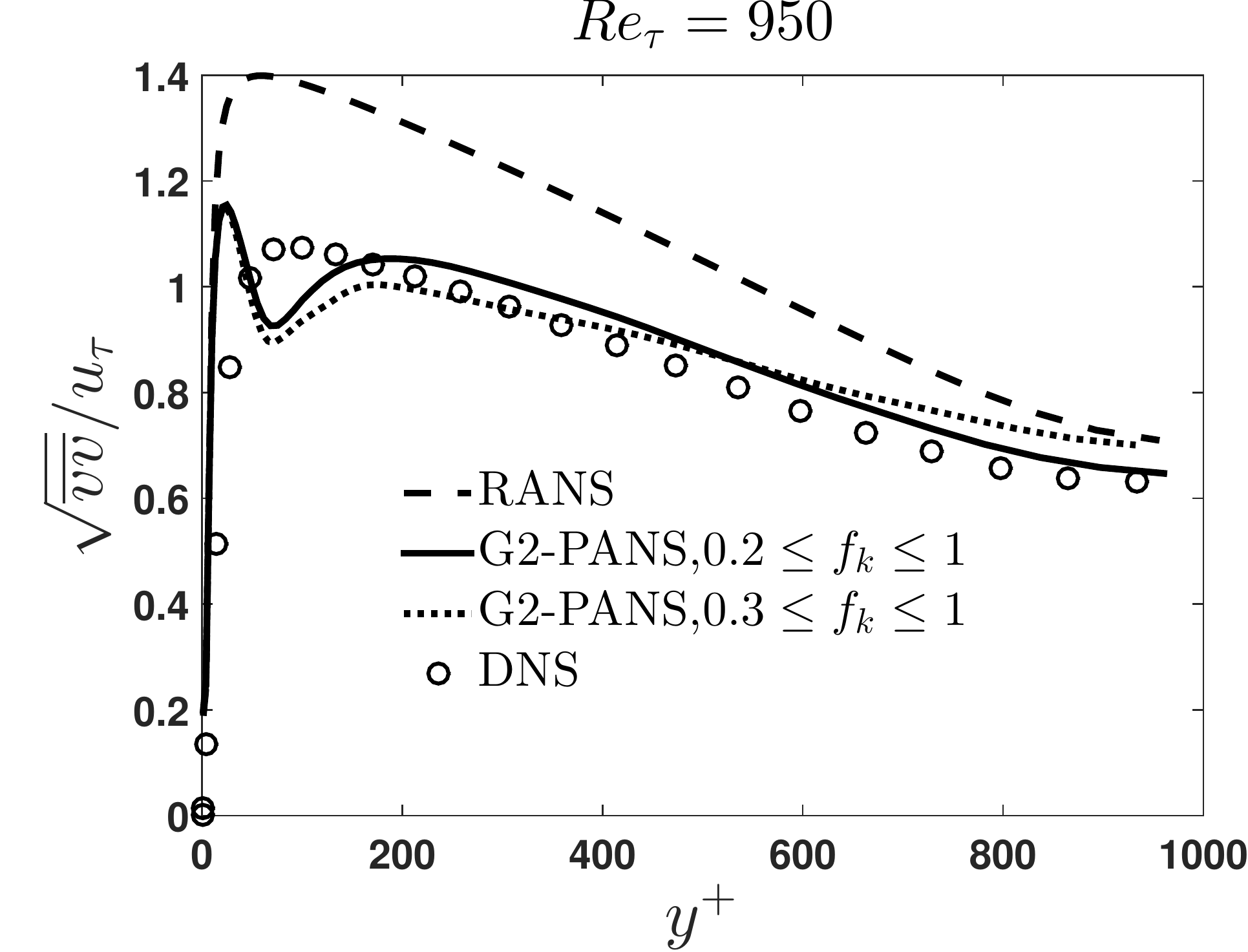}\begin{picture}(0,0)\put(-138,0){(b)}\end{picture}
        \end{subfigure}
                \begin{subfigure}[b]{0.45\textwidth}
                \includegraphics[width=\textwidth]{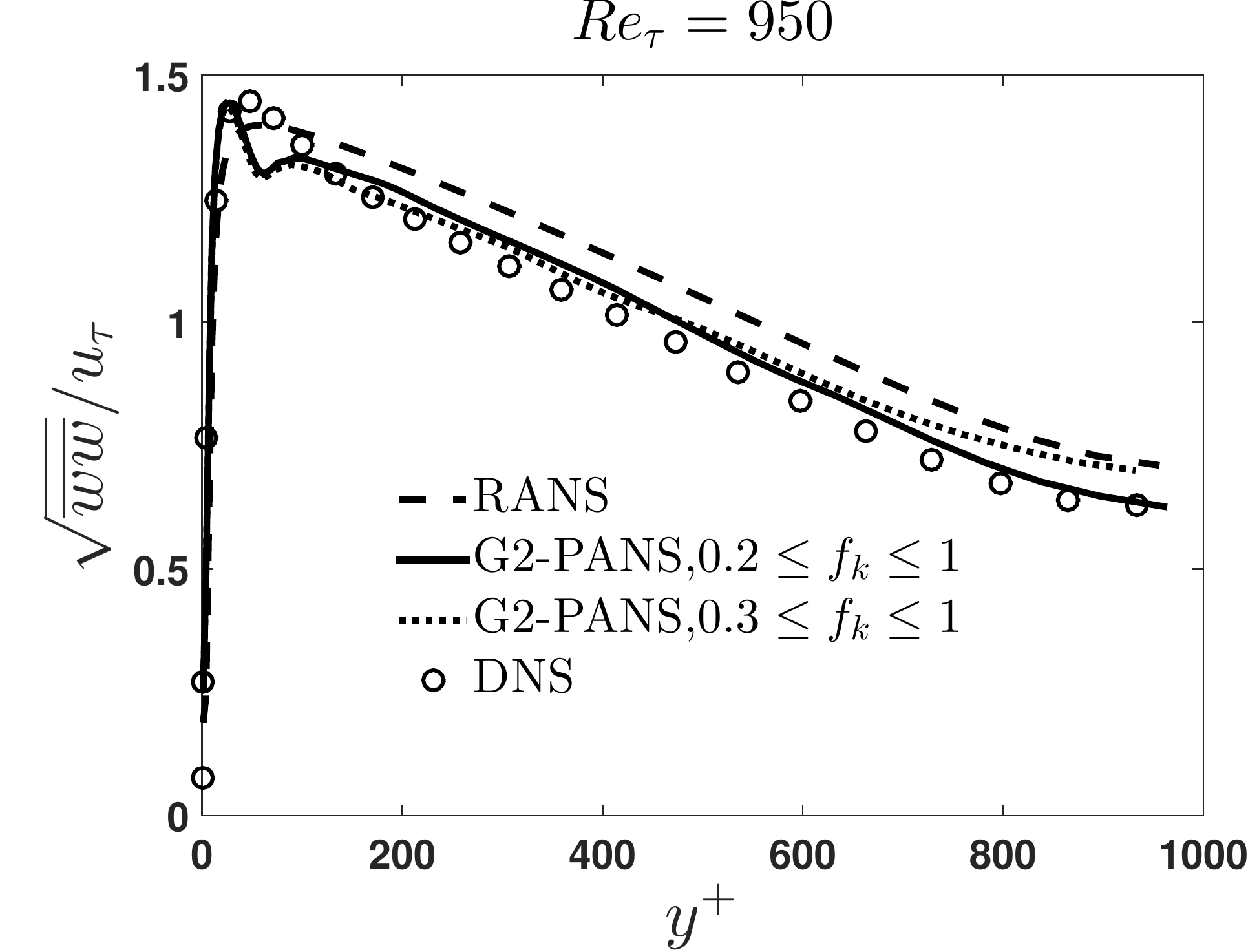}\begin{picture}(0,0)\put(-138,0){(c)}\end{picture}
        \end{subfigure}       
 		        \begin{subfigure}[b]{0.45\textwidth}
                \includegraphics[width=\textwidth]{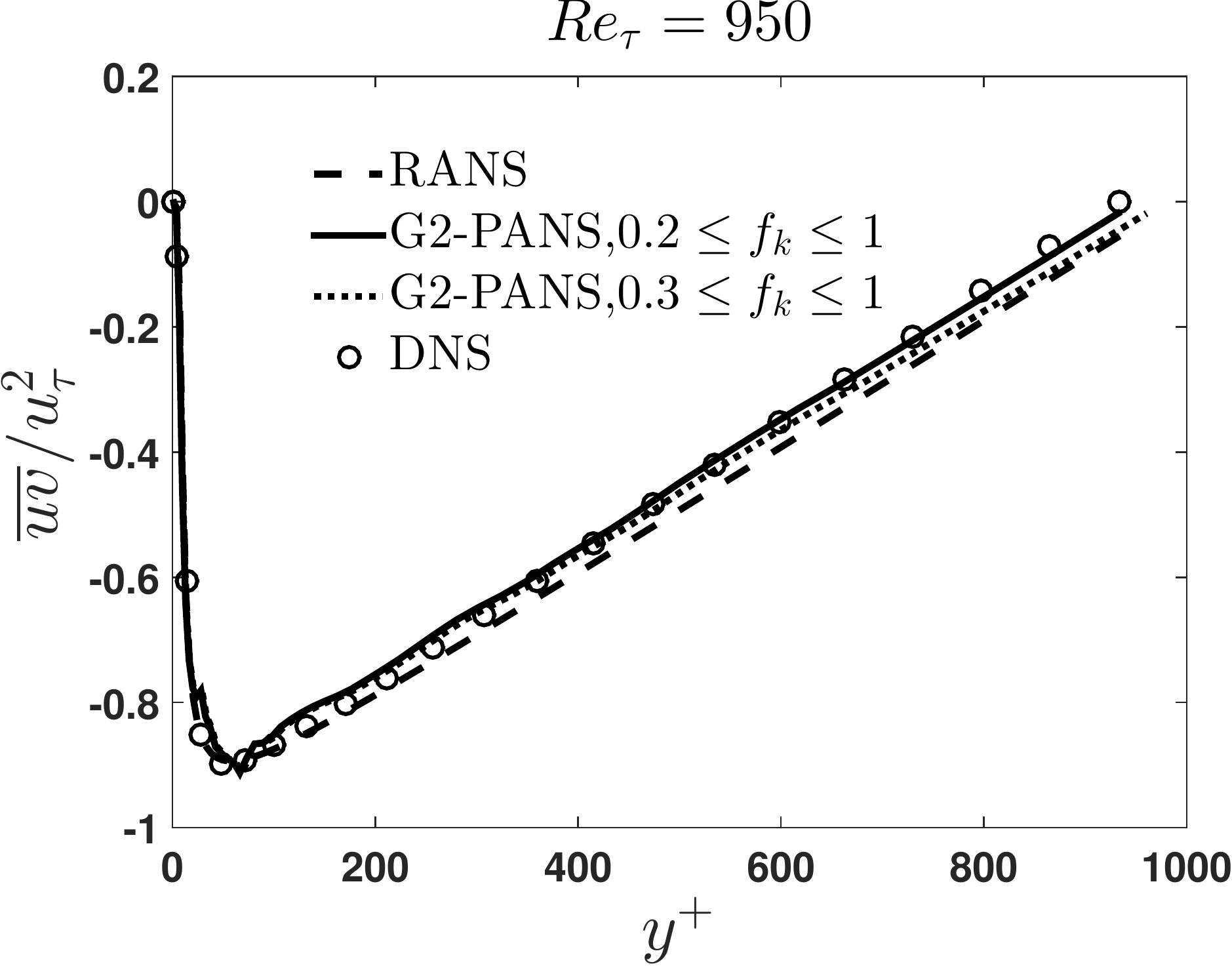}\begin{picture}(0,0)\put(-138,0){(d)}\end{picture}
        \end{subfigure}
\caption{G2-PANS simulation of turbulent channel flow for $Re_{\tau}=950$ (a) Streamwise stress (b) Normal stress (c) Spanwise stress (d) Shear stress}   
\label{varfk950}     
\end{figure}

\begin{figure}[H]
        \centering
 		\captionsetup{justification=centering}                                       
 		        \begin{subfigure}[b]{0.43\textwidth}
                \includegraphics[width=\textwidth]{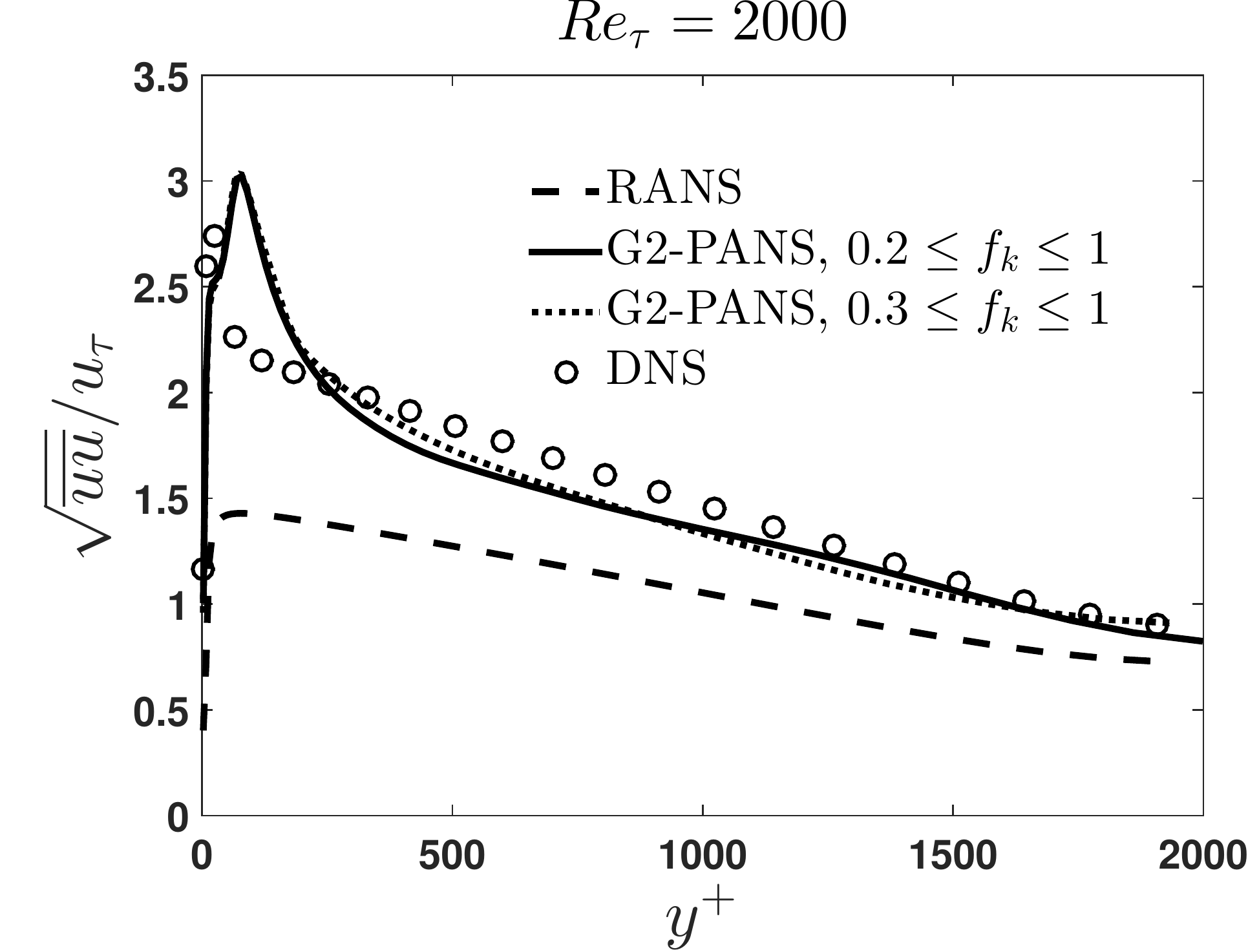}\begin{picture}(0,0)\put(-138,0){(a)}\end{picture}
        \end{subfigure}
        			\begin{subfigure}[b]{0.43\textwidth}
                \includegraphics[width=\textwidth]{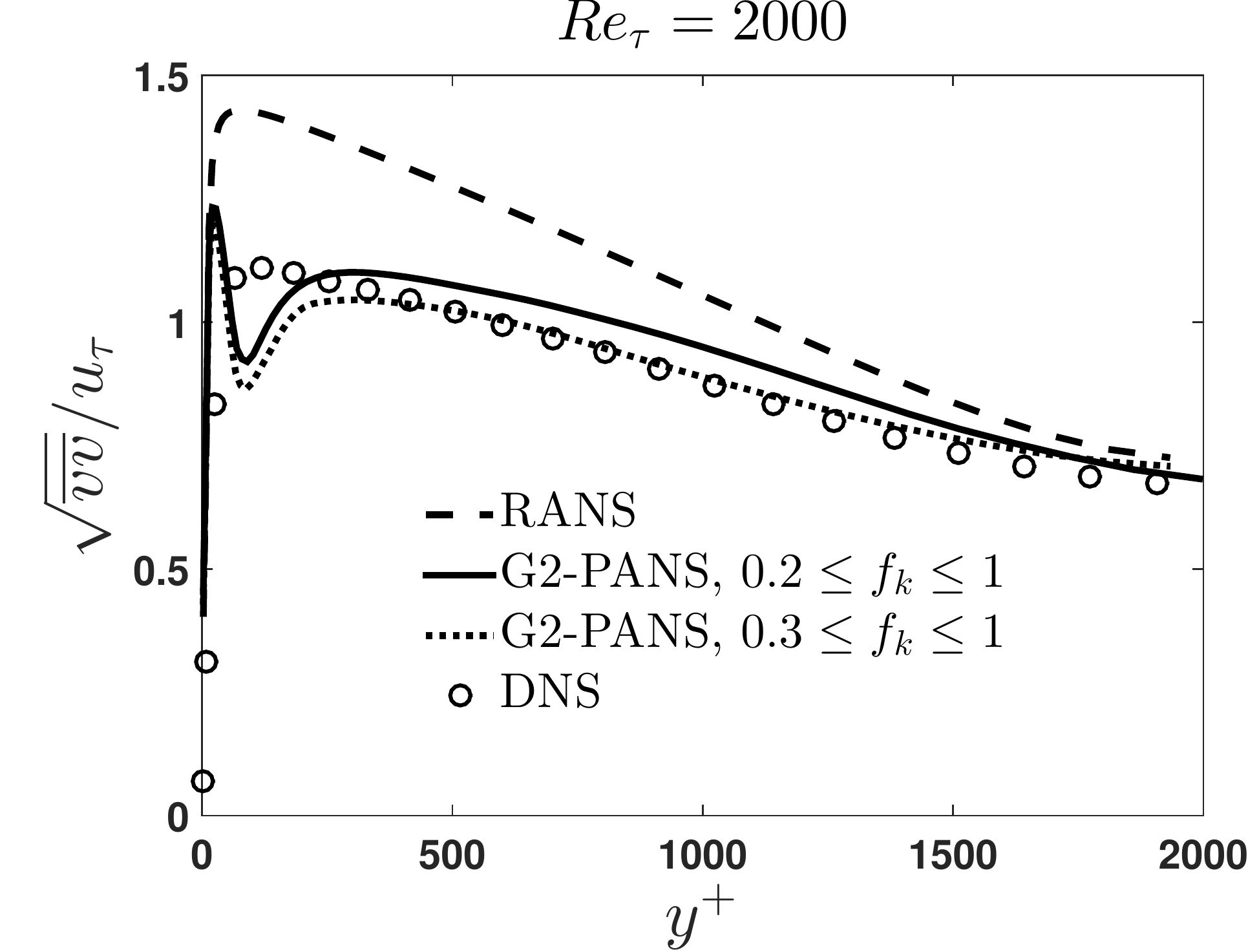}\begin{picture}(0,0)\put(-138,0){(b)}\end{picture}
        \end{subfigure}
                \begin{subfigure}[b]{0.43\textwidth}
                \includegraphics[width=\textwidth]{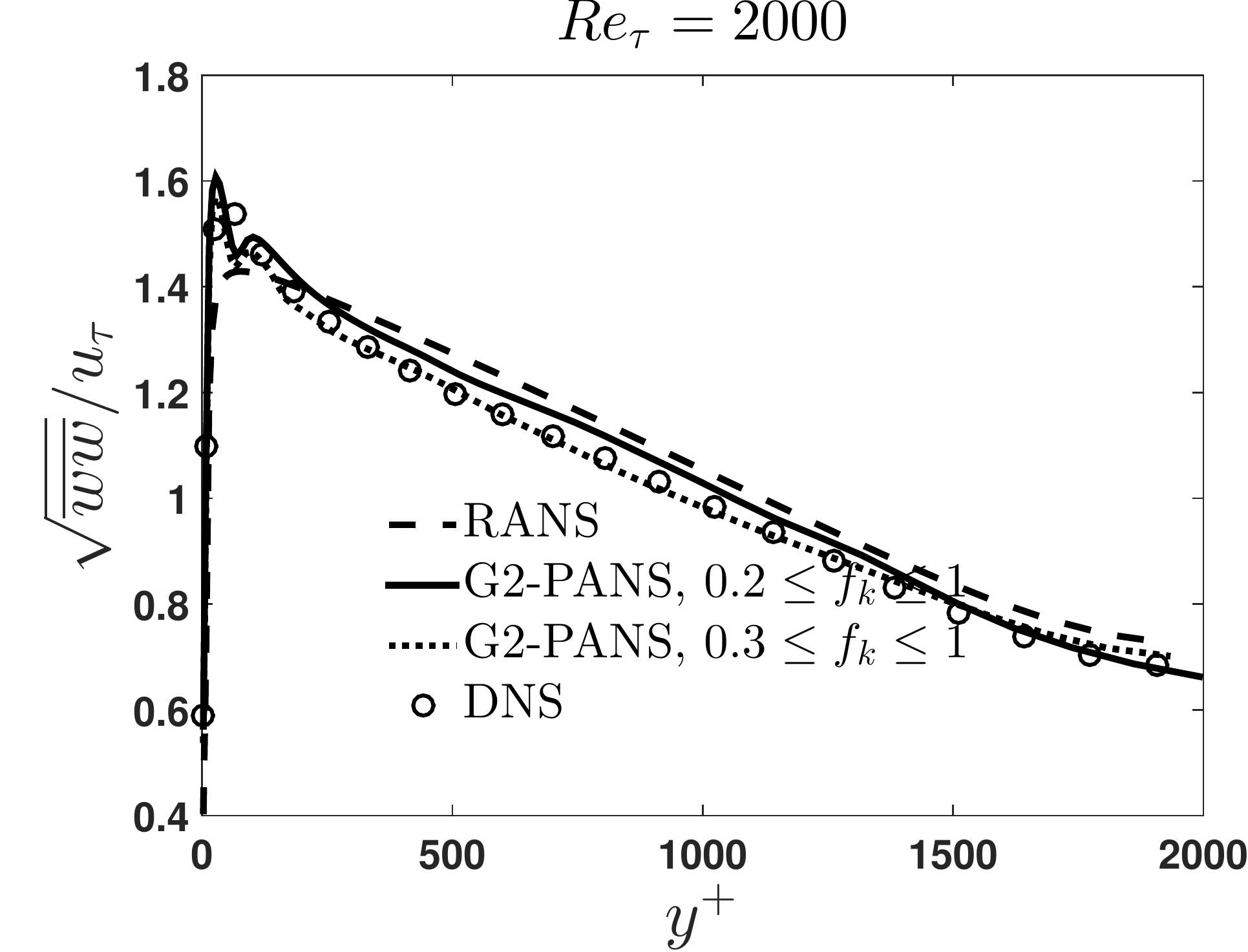}\begin{picture}(0,0)\put(-138,0){(c)}\end{picture}
        \end{subfigure}       
 		        \begin{subfigure}[b]{0.43\textwidth}
                \includegraphics[width=\textwidth]{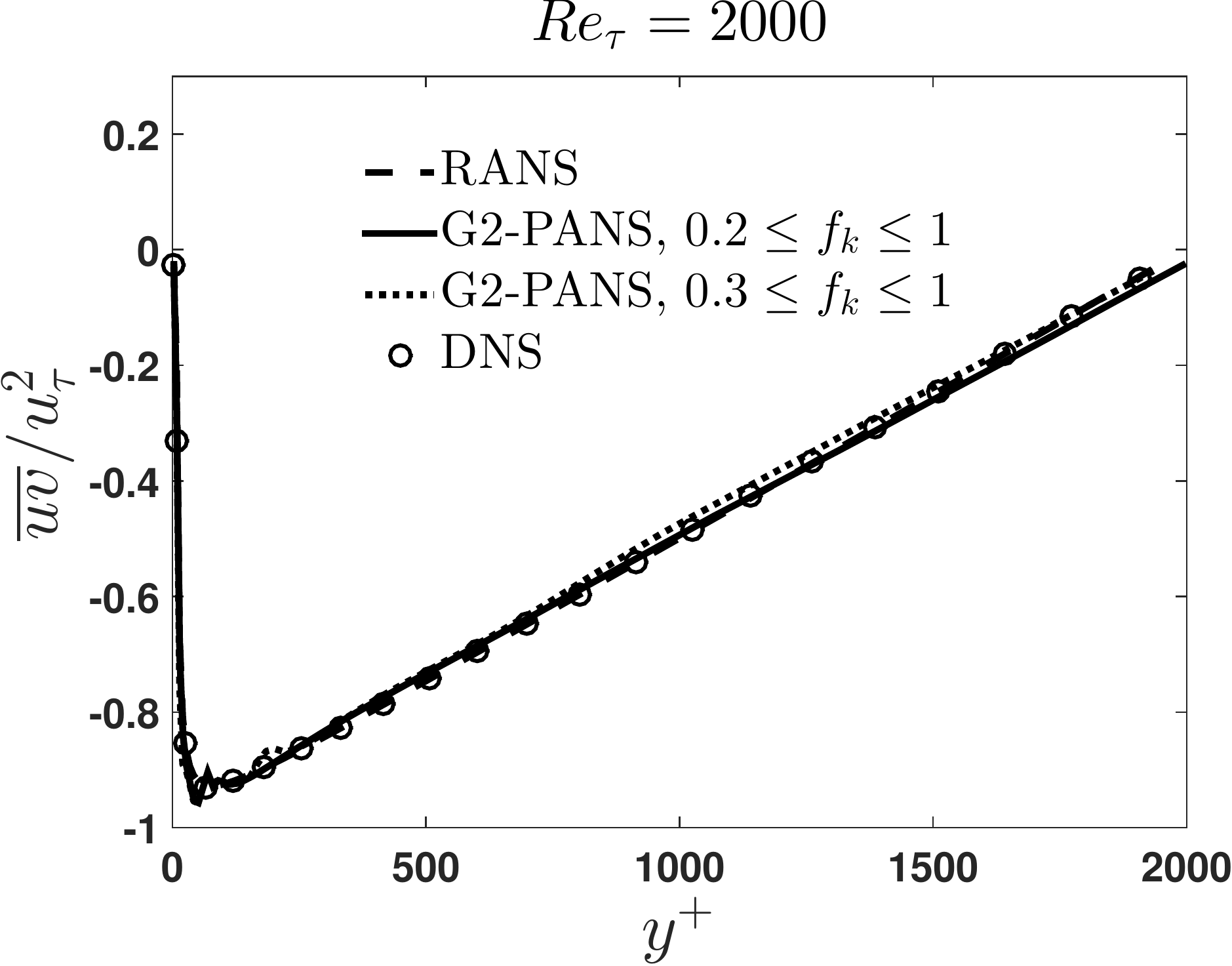}\begin{picture}(0,0)\put(-138,0){(d)}\end{picture}
        \end{subfigure}
\caption{G2-PANS simulation of turbulent channel flow for $Re_{\tau}=2000$ (a) Streamwise stress (b) Normal stress (c) Spanwise stress (d) Shear stress}   
\label{varfk2000}     
\end{figure}
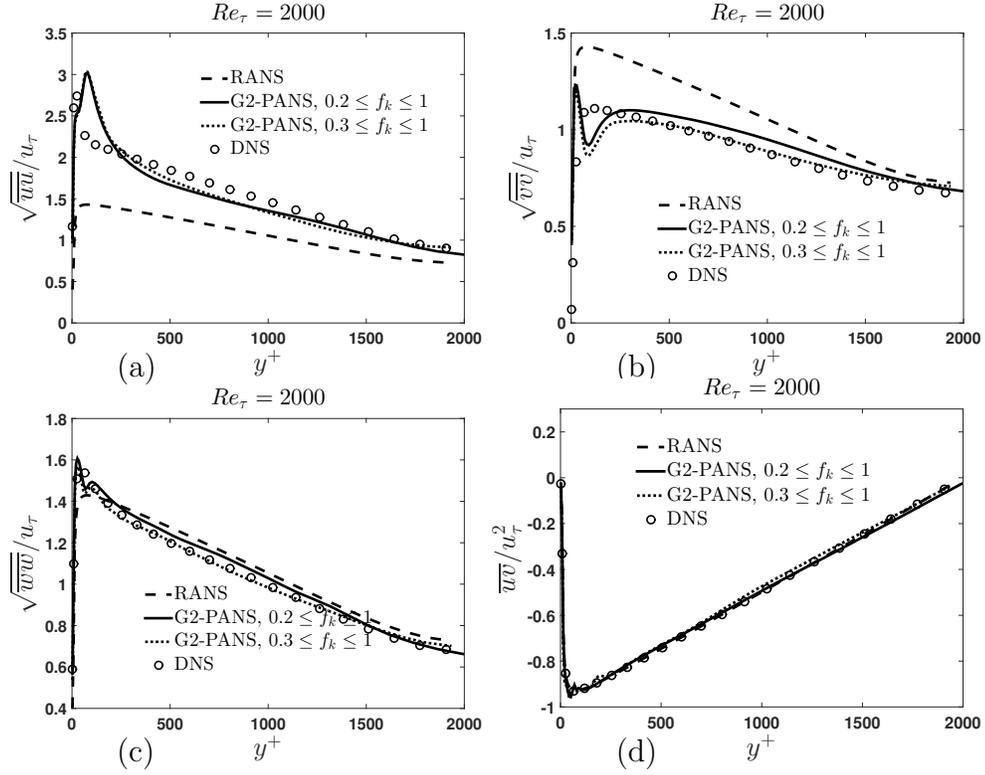

To ensure that the correct results are obtained with the correct underlying physics, several higher-order statistics are compared with corresponding DNS results in figures \ref{varfk950} and \ref{varfk2000} for $Re_\tau$=950 and 2000. All of the individual Reynolds stress components are close to the DNS results outside the laminar-sublayer for both values of $f_k$. Recall, the underlying model is RANS in the sublayer. Most importantly, the simulation undergoes a numerical transition from steady RANS near the wall to unsteady flow-field away from the wall.

Failure of the G1-PANS and the G1.5-PANS models to accurately simulate turbulent channel flow at high Reynolds numbers are associated with lack of grid resolution and improper energy exchange near the wall region, respectively. Evolution of second generation of the PANS model leads to a drastic reduction in computational expense by the means of resolution variation, and results in simulation accuracy by including the energy scale transfer terms. 

\subsubsection{Effect of changing the center of cut-off}

\begin{figure}[H]
        \centering
 		\captionsetup{justification=centering}                                       
                \includegraphics[width=0.45\textwidth]{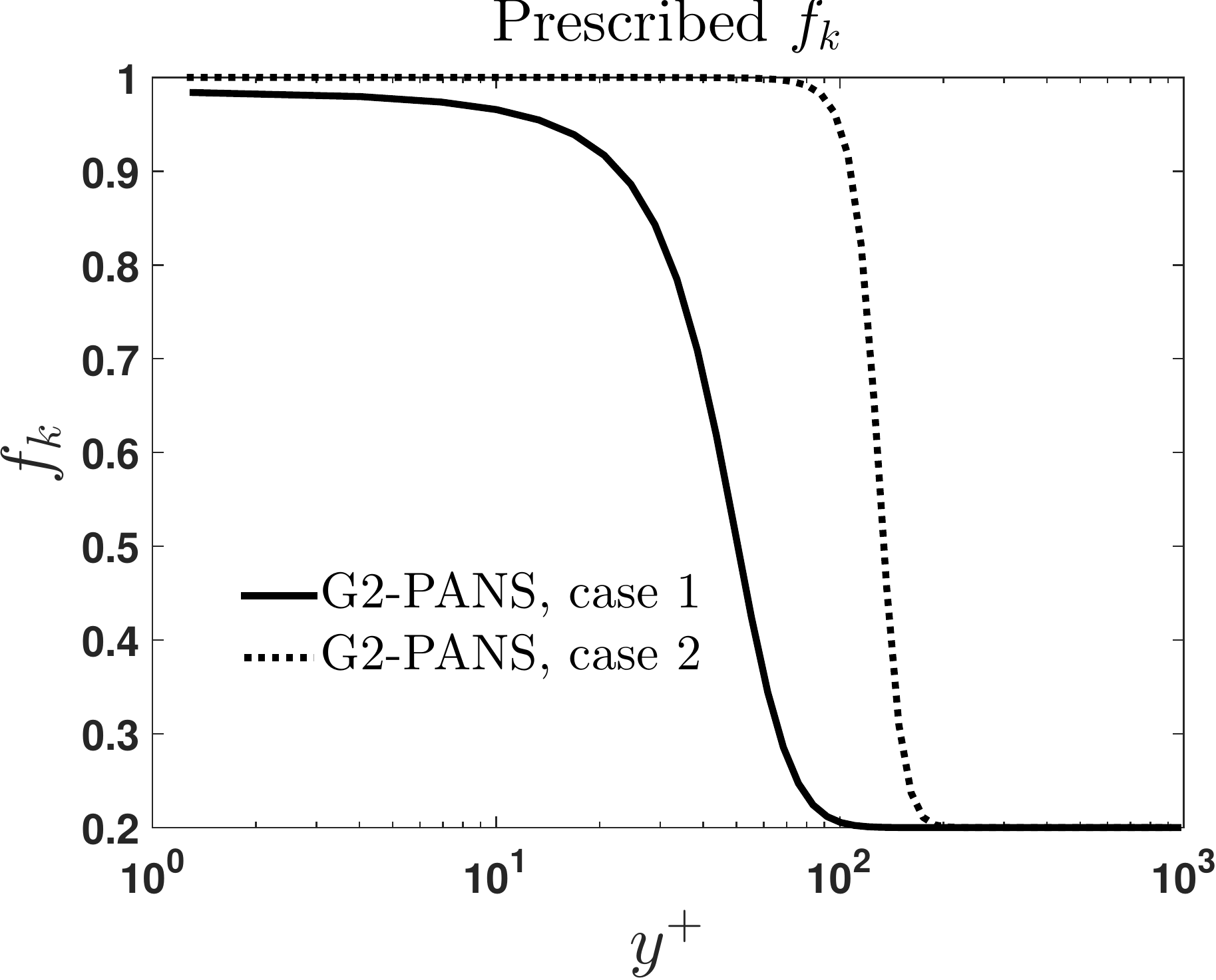}\begin{picture}(0,0)\put(-138,0){}\end{picture}
\caption{Prescribed $f_k$ for different cases at $Re_\tau=950$}   
\label{cutfk}     
\end{figure}
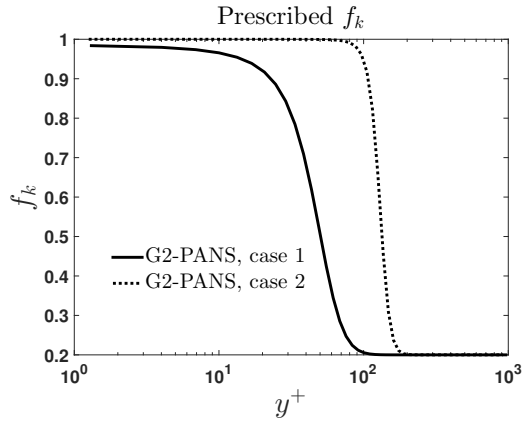

The effect of the location of the RANS-to-PANS transition region is studied in this section. Figure \ref{cutfk} shows the filter variation for two different cases at $Re_\tau$=950. For case 1, resolution of the model changes in the range of $20 < y^+ < 100$ while for the second case it happens within $100 < y^+ < 250$. The second variation implies that all the scales within the laminar sub-layer and buffer layer are modelled, while for the first case, these are partially resolved. 

\begin{figure}[H]
        \centering
 		\captionsetup{justification=centering}                                       
 		        \begin{subfigure}[b]{0.45\textwidth}
                \includegraphics[width=\textwidth]{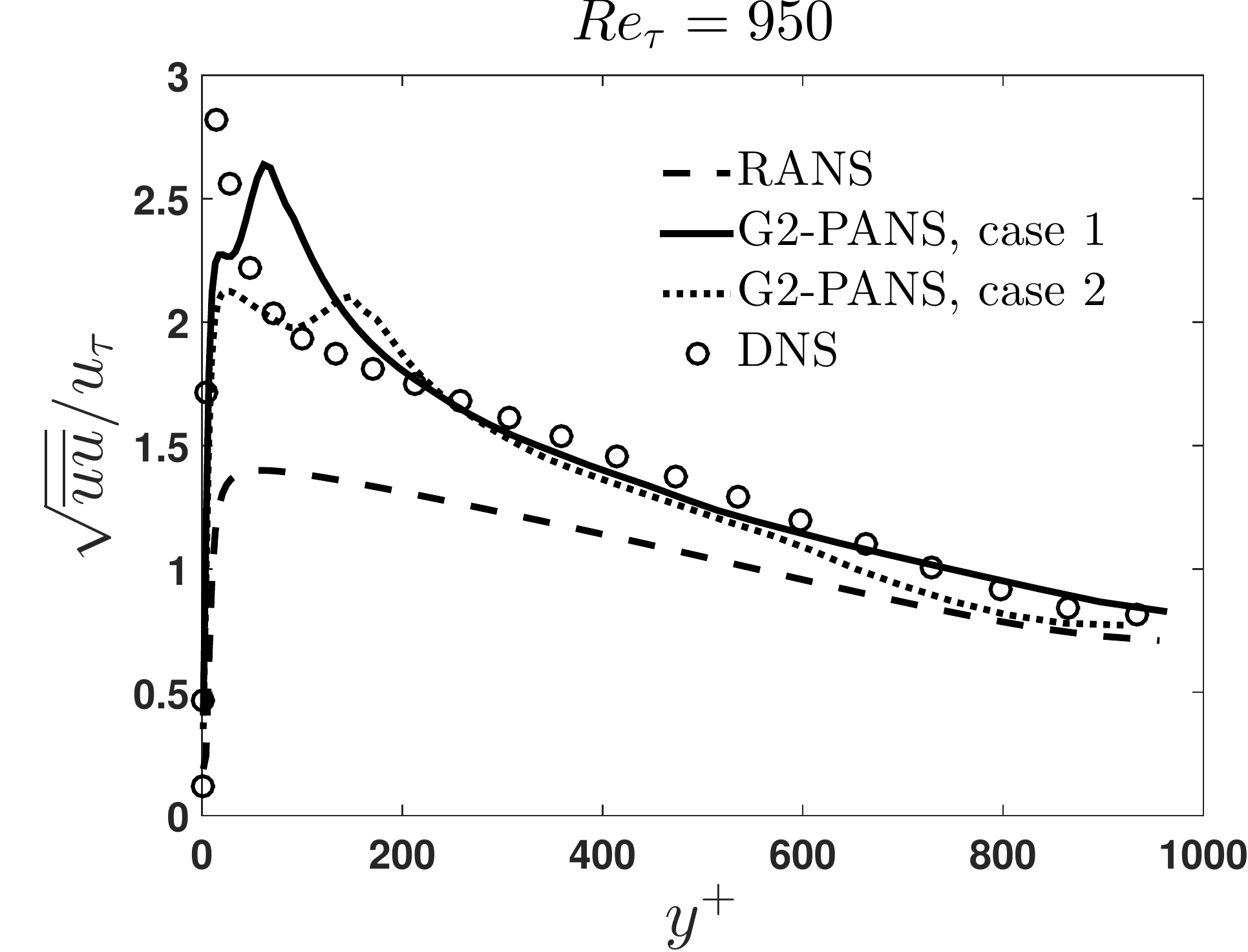}\begin{picture}(0,0)\put(-138,0){(a)}\end{picture}
        \end{subfigure}
        			\begin{subfigure}[b]{0.45\textwidth}
                \includegraphics[width=\textwidth]{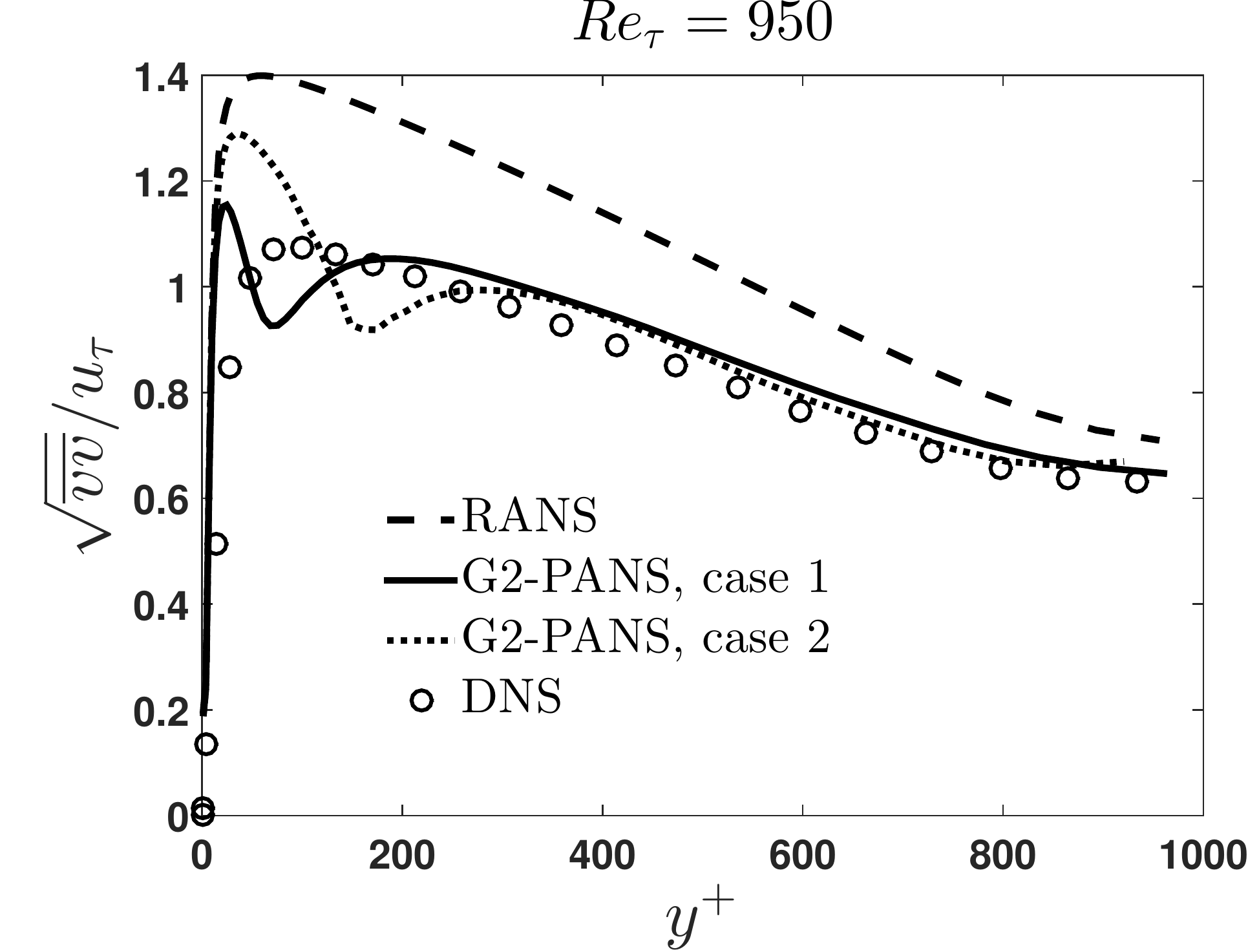}\begin{picture}(0,0)\put(-138,0){(b)}\end{picture}
        \end{subfigure}
                \begin{subfigure}[b]{0.45\textwidth}
                \includegraphics[width=\textwidth]{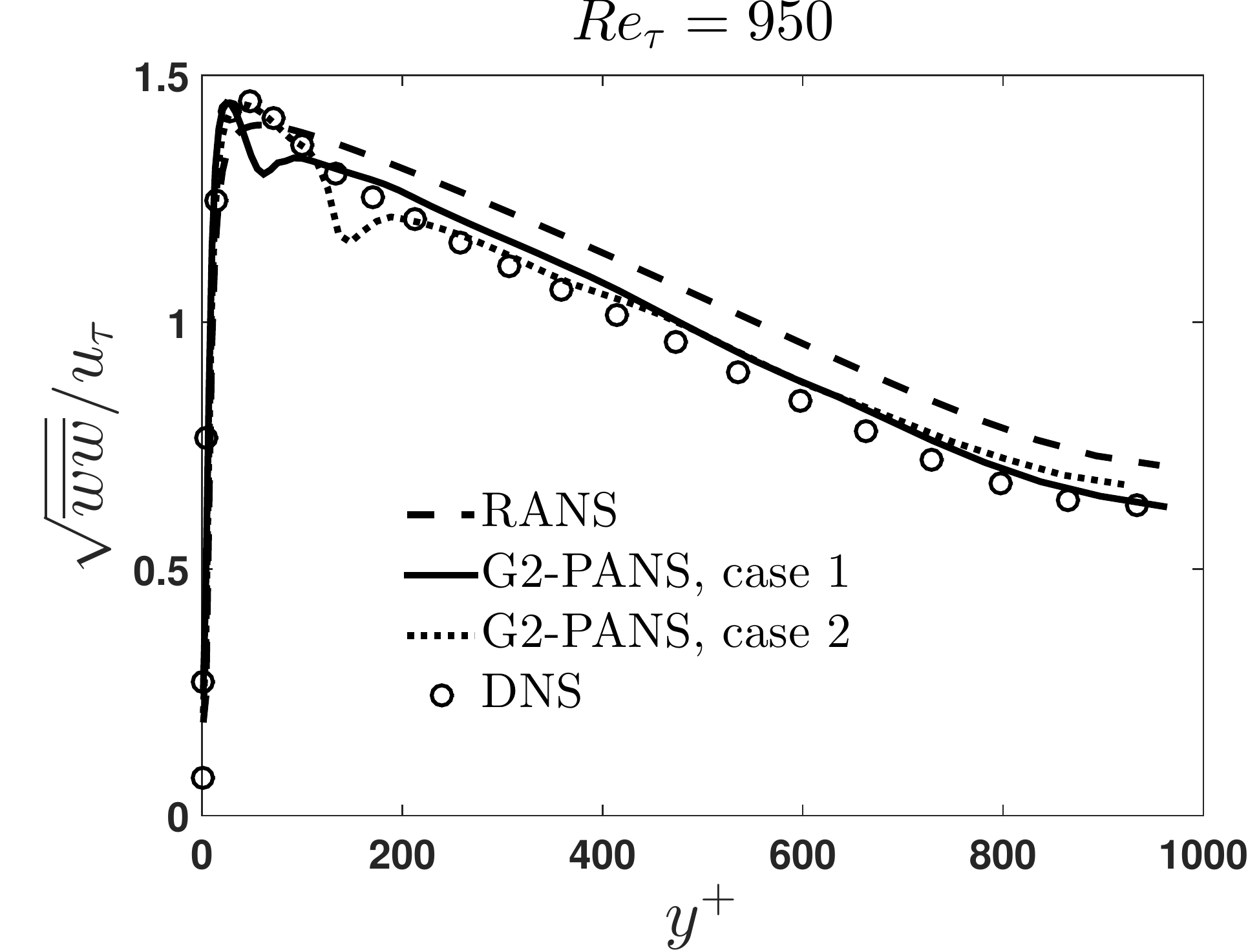}\begin{picture}(0,0)\put(-138,0){(c)}\end{picture}
        \end{subfigure}       
 		        \begin{subfigure}[b]{0.415\textwidth}
                \includegraphics[width=\textwidth]{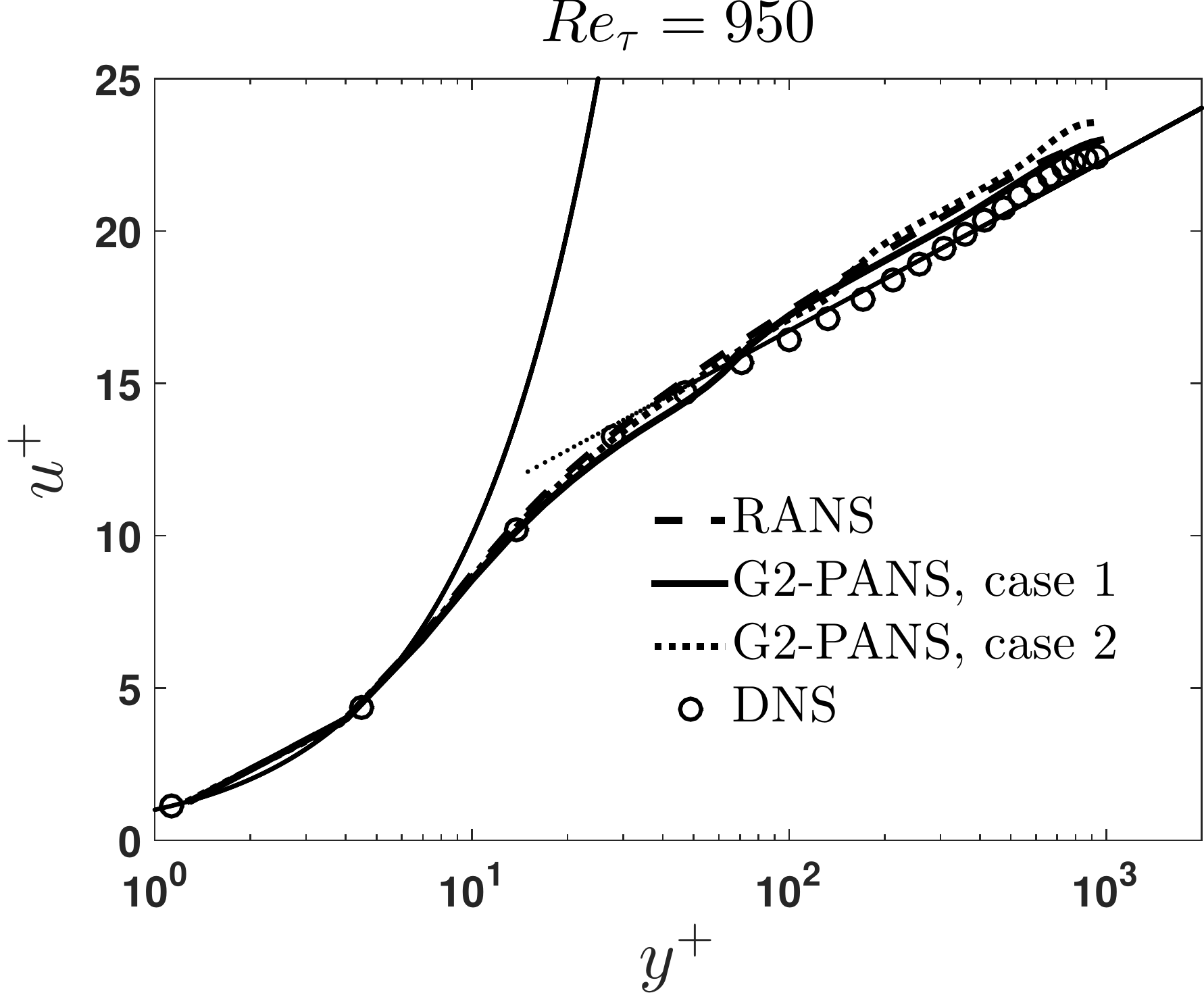}\begin{picture}(0,0)\put(-138,0){(d)}\end{picture}
        \end{subfigure}
\caption{G2-PANS simulation of turbulent channel flow for $Re_{\tau}=950$(a) Streamwise stress (b) Normal stress (c) Spanwise stress (d) Shear stress}   
\label{cutuv}     
\end{figure}

Figure \ref{cutuv} shows the mean velocity and stress profiles for the two cases. This figure infers that by moving the location of transition towards the buffer layer, accuracy of the results are improved. The velocity overshoot seen for the second case is not represented for the first case. Besides, the estimation of normal stresses near the wall and particularly the peak values are closer to DNS for the first case. 

\begin{figure}[H]
        \centering
 		\captionsetup{justification=centering}                                       
 		        \begin{subfigure}[b]{0.45\textwidth}
                \includegraphics[width=\textwidth]{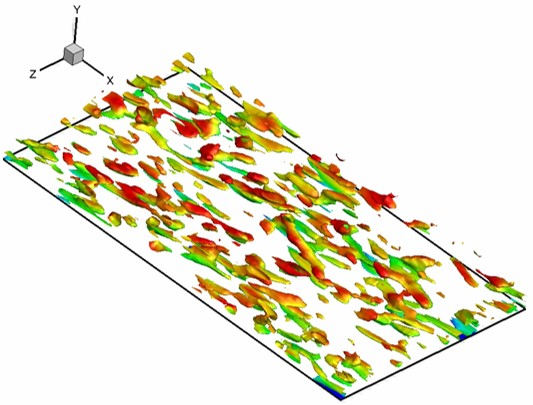}\begin{picture}(0,0)\put(-138,0){(a)}\end{picture}
        \end{subfigure}
        			\begin{subfigure}[b]{0.45\textwidth}
                \includegraphics[width=\textwidth]{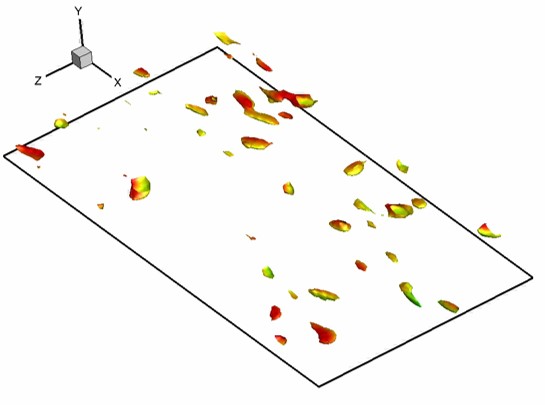}\begin{picture}(0,0)\put(-138,0){(b)}\end{picture}
        \end{subfigure}
          \caption{Q iso-surfaces(a) Case 1 (b) Case 2}   
\label{Q950}     
\end{figure}
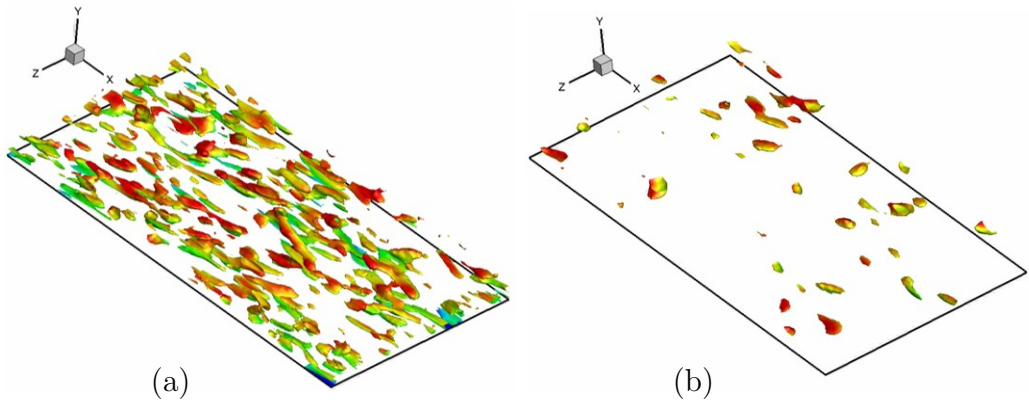

For turbulent channel flow, instabilities stem from the buffer layer and early log-layer region. Less accurate calculation of the second simulation is associated with model failure in resolving the scales in these important regions. Illustration of this fact is seen in Fig. \ref{Q950}. This figure shows Q iso-surfaces coloured by streamwise velocity for the two different cases at $0<y^+<200$. It is clear from this figure that considerably more scales are resolved in the first case where the resolution variation happened closer to the wall. Note that pushing the location of transition closer to the wall may not result in better accuracy as the grid resolution will not suffice for this Reynolds number. 

\subsubsection{Stress components at higher $Re_\tau$}

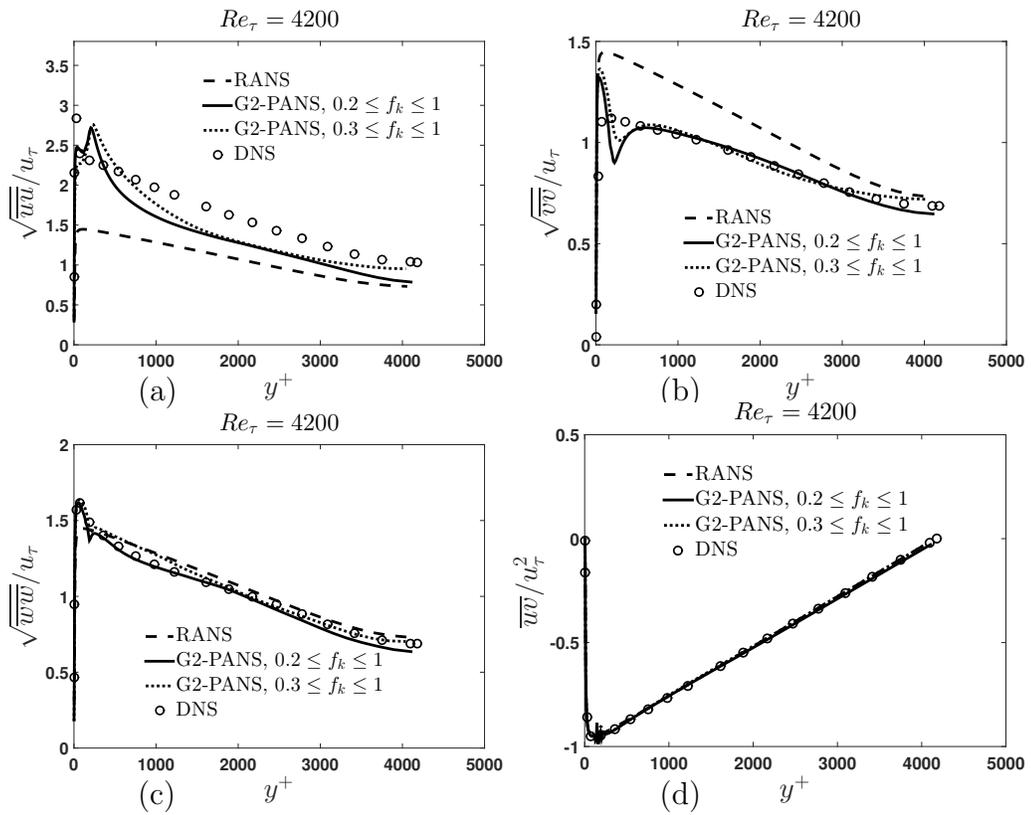
\begin{figure}[H]
        \centering
 		\captionsetup{justification=centering}                                       
 		        \begin{subfigure}[b]{0.45\textwidth}
                \includegraphics[width=\textwidth]{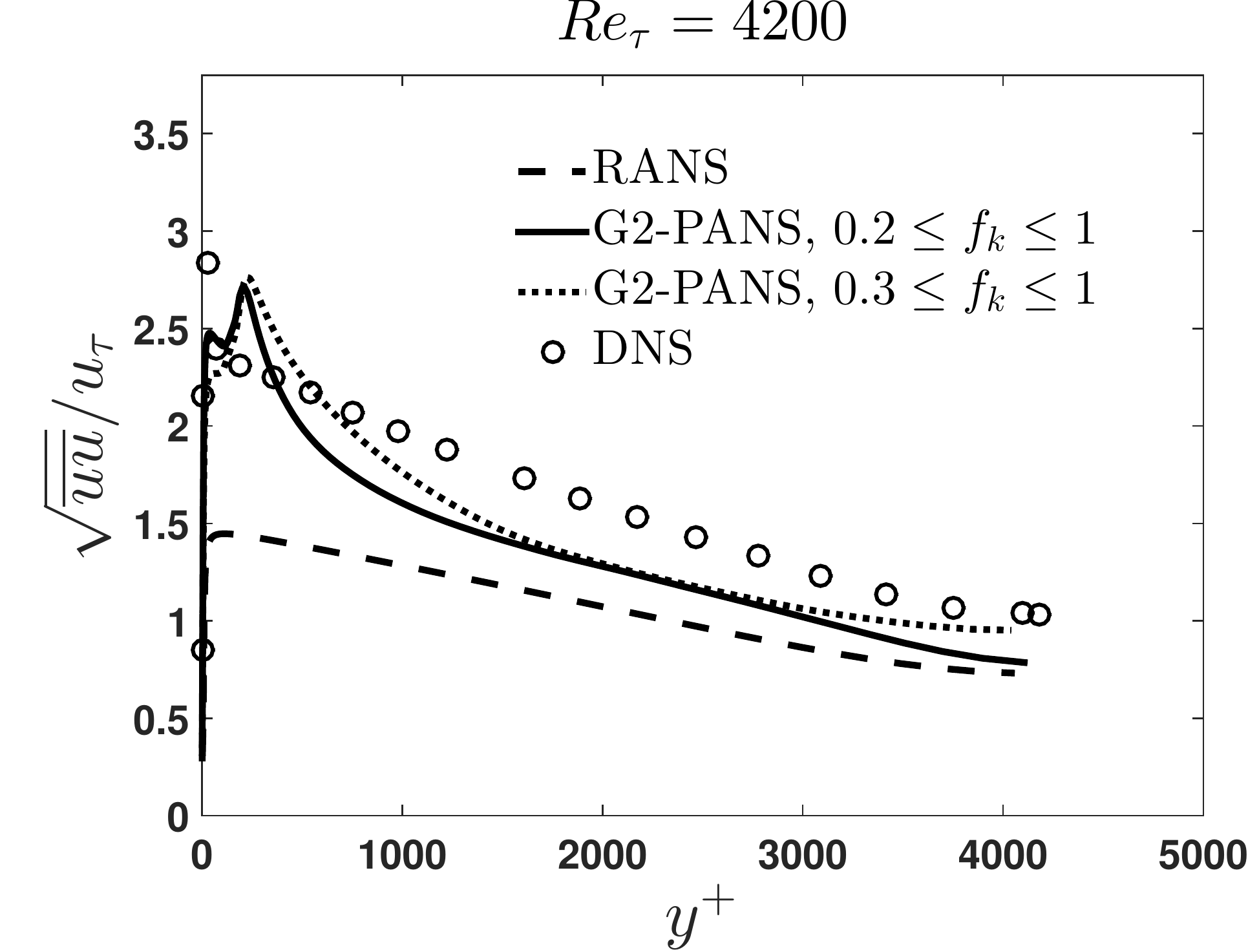}\begin{picture}(0,0)\put(-138,0){(a)}\end{picture}
        \end{subfigure}
        			\begin{subfigure}[b]{0.45\textwidth}
                \includegraphics[width=\textwidth]{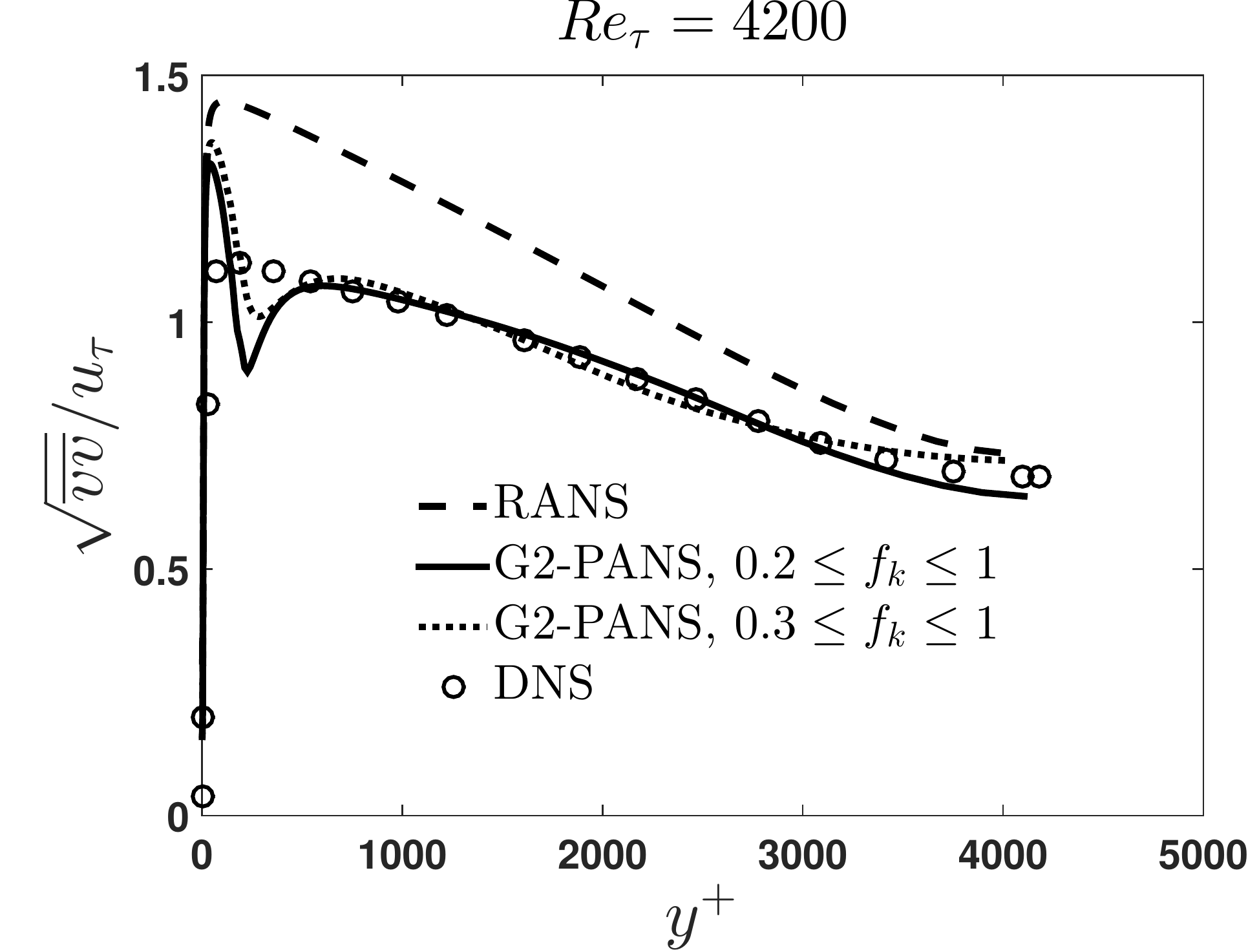}\begin{picture}(0,0)\put(-138,0){(b)}\end{picture}
        \end{subfigure}
                \begin{subfigure}[b]{0.45\textwidth}
                \includegraphics[width=\textwidth]{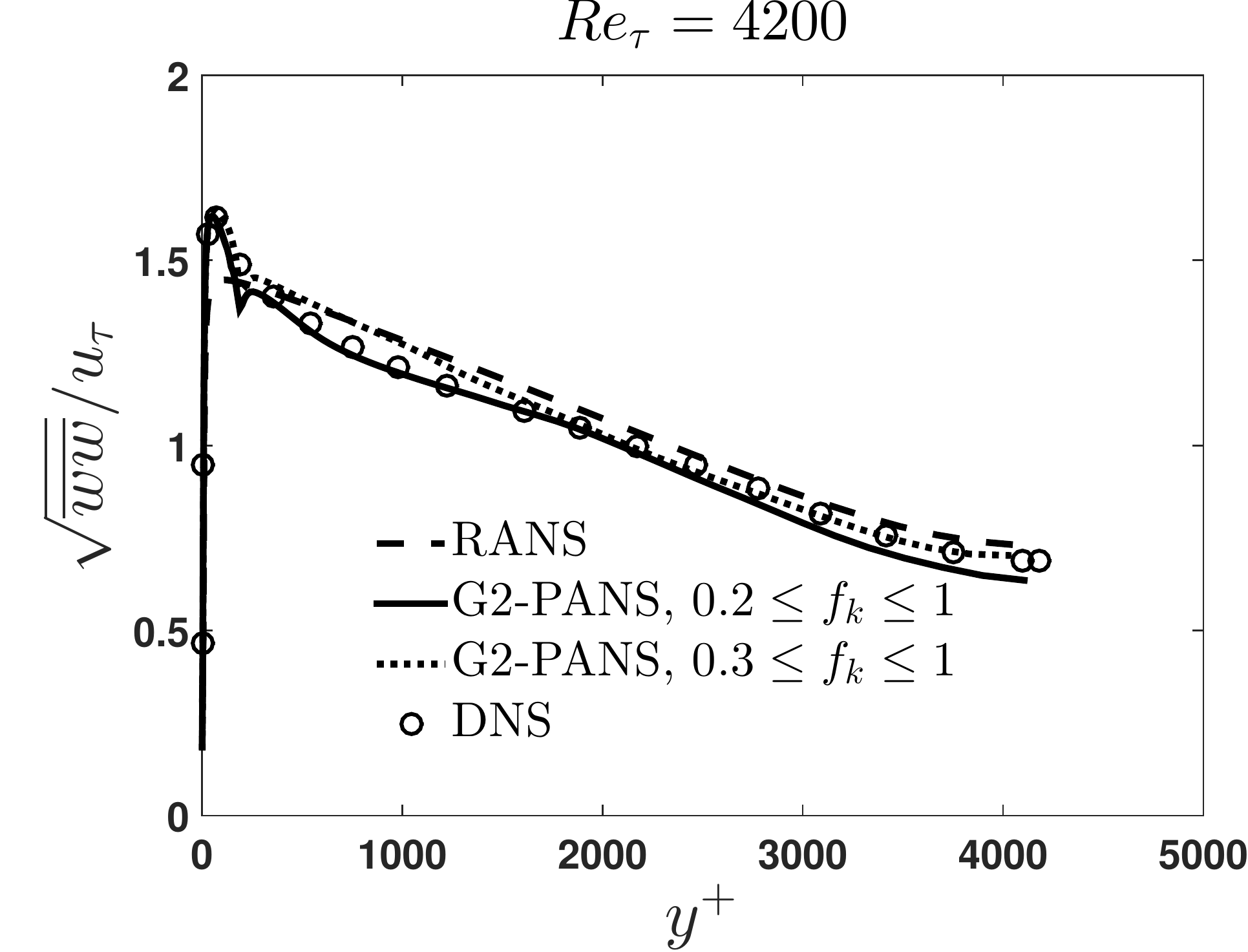}\begin{picture}(0,0)\put(-138,0){(c)}\end{picture}
        \end{subfigure}       
 		        \begin{subfigure}[b]{0.45\textwidth}
                \includegraphics[width=\textwidth]{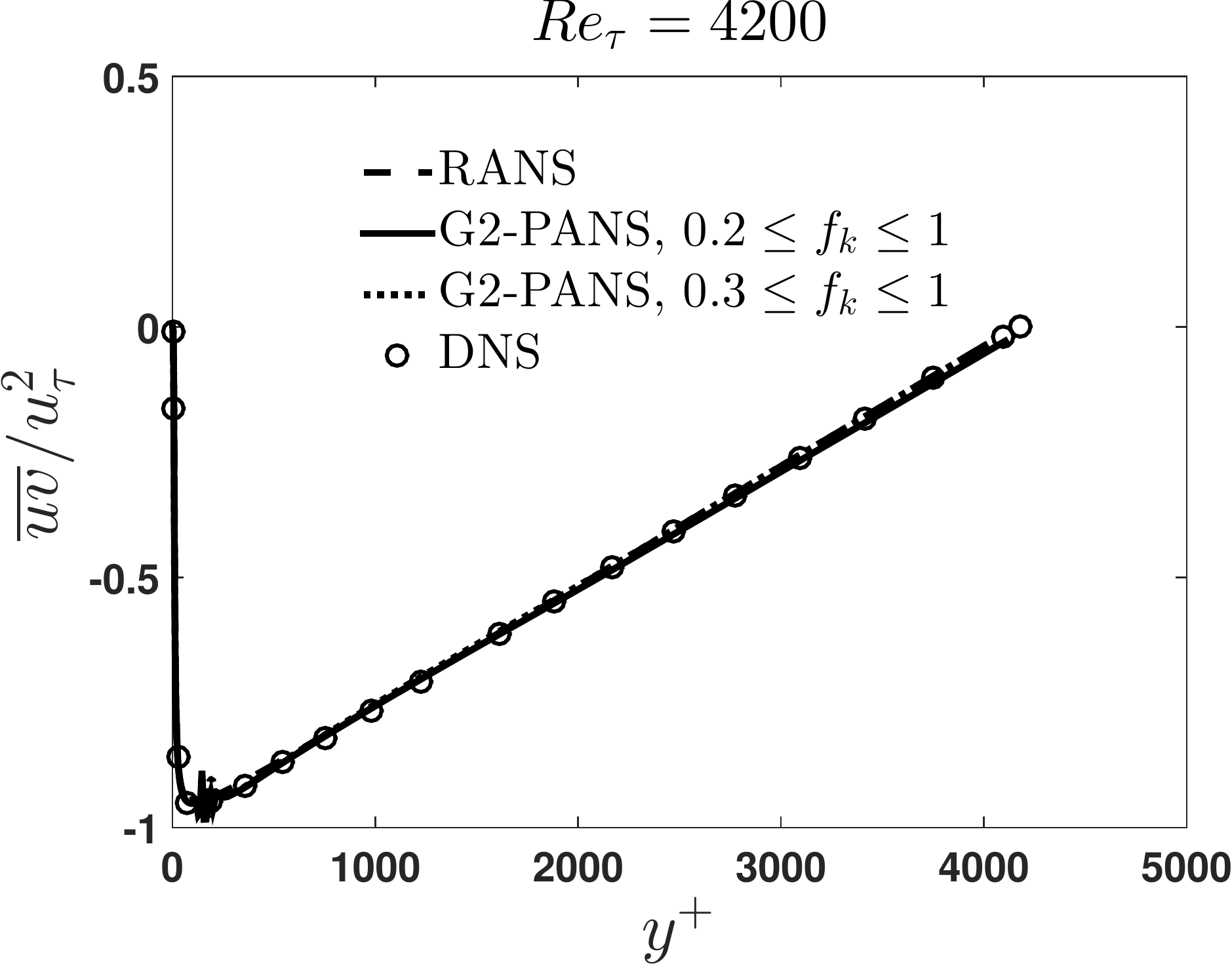}\begin{picture}(0,0)\put(-138,0){(d)}\end{picture}
        \end{subfigure}
\caption{G2-PANS simulation of turbulent channel flow for $Re_{\tau}=2000$ (a) Streamwise stress (b) Normal stress (c) Spanwise stress (d) Shear stress}   
\label{varfk4200}     
\end{figure}

Figure \ref{varfk4200} shows the the stress components for $Re_\tau=4200$. It is illustrative that overall, the agreement of stress profiles with DNS data is fully satisfactory in this case. However, for this case, the deviation of streamwise stresses from DNS data is noticeable unlike the profiles for $Re_\tau$=950 and 2000. Referring to figure \ref{recovery}a, for $Re_\tau$=4200 and 8000, model switches from RANS to PANS after $y^+$ of 100, whereas for $Re_\tau$=950 and 2000, this happened much more earlier. The grid resolution requirement is more critical at higher Reynolds number. With the current grid, resolving the scales in near wall region is impractical, and therefore for higher Reynolds numbers, this region is fully modelled with RANS. As discussed in previous section, by moving the location of RANS-to-PANS transition away from the wall important physical phenomena in near wall region are not captured. This affects the mean flow statistics as shown in Fig. \ref{varfk4200} (a). It is worth reiterating that for $Re\tau$=8000, there were no available DNS data, and therefore the stress profiles are not shown for this case.

%
%
%

\chapter{\uppercase{Simulation of Smooth Surface Separation Using G2-PANS}}
\label{hump}

\section{Introduction}
As it is shown in Sec. \ref{hill}, accurate computation of wall-mounted hump flow using the G1-PANS model with fixed $f_k$ requires very refined grid resolution in near wall region. Therefore, in this study, the flow separation over wall-mounted hump is simulated using G2-PANS $k-\omega$ model with variable resolution near the wall. The simulations are performed for a range of grid resolutions and $f_k$. In Sec. \ref{hump-study1}, the wall-mounted hump configuration, flow conditions and details of the numerical method are explained. The various grids used in this study for different $f_k$ calculations are given in Table \ref{case2}. As shown in this table, the finest grid resolution used for the G2-PANS calculation is 1.5 Million and the coarsest one is around 0.3 Million, whereas this is approximately 9.4 Million for the LES simulation \cite{avdis}. The results are presented for different parameters including the separation and reattachment lengths, first and second order statistics and surface friction coefficient and are compared with experiment and other numerical studies. 


\begin{table}[H]
\centering
\caption{Details of the test cases simulated}
\begin{tabular}{ l l l l l}
\hline\noalign{\smallskip}
\textbf{Study} & \textbf{$f_k$} & \textbf{$f_\epsilon$} & \textbf{Grid} & \textbf{Averaging Period}\\ \hline\noalign{\smallskip}
\textbf{Re=935,892} \\ \hline\noalign{\smallskip}
RANS & 1 & 1 & $201\times60\times20$ & 10T-15T \\ \hline\noalign{\smallskip}
PANS & 0.2 & 1 & $201\times250\times30$ & 10T-15T \\
PANS & 0.2 & 1 & $201\times96\times30$ & 10T-15T \\
PANS & 0.2 & 1 & $201\times70\times30$ & 10T-15T \\
PANS & 0.2 & 1 & $201\times70\times20$ & 10T-15T \\
\hline\noalign{\smallskip}
PANS & 0.25 & 1 & $201\times96\times30$ & 10T-15T \\
\hline\noalign{\smallskip}
PANS & 0.3 & 1 & $201\times96\times30$ & 10T-15T \\
\hline\noalign{\smallskip}
LES & - & - & $768\times96\times128$ & - \\ 
\end{tabular}
\label{case2}
\end{table}

\section{Results}
\label{results}

In this section, the simulation results of G2-PANS are presented for the flow over mounted hump. For G2-PANS simulations, the near wall region is modelled and variable resolution is applied in the wall-normal direction. The resolution change is applied far enough from the wall in order to avoid penetrating too deeply into the boundary layer. If the PANS region would reside too close to the wall due to insufficient resolution, lower viscosity and turbulence levels could be obtained which possibly produces
a premature separation and less accurate flow predictions.

\subsection{Mean Flow Statistics}

\begin{figure}
        \centering
 		\captionsetup{justification=centering}                                               
 		        \begin{subfigure}[b]{0.3\textwidth}
                \includegraphics[width=\textwidth]{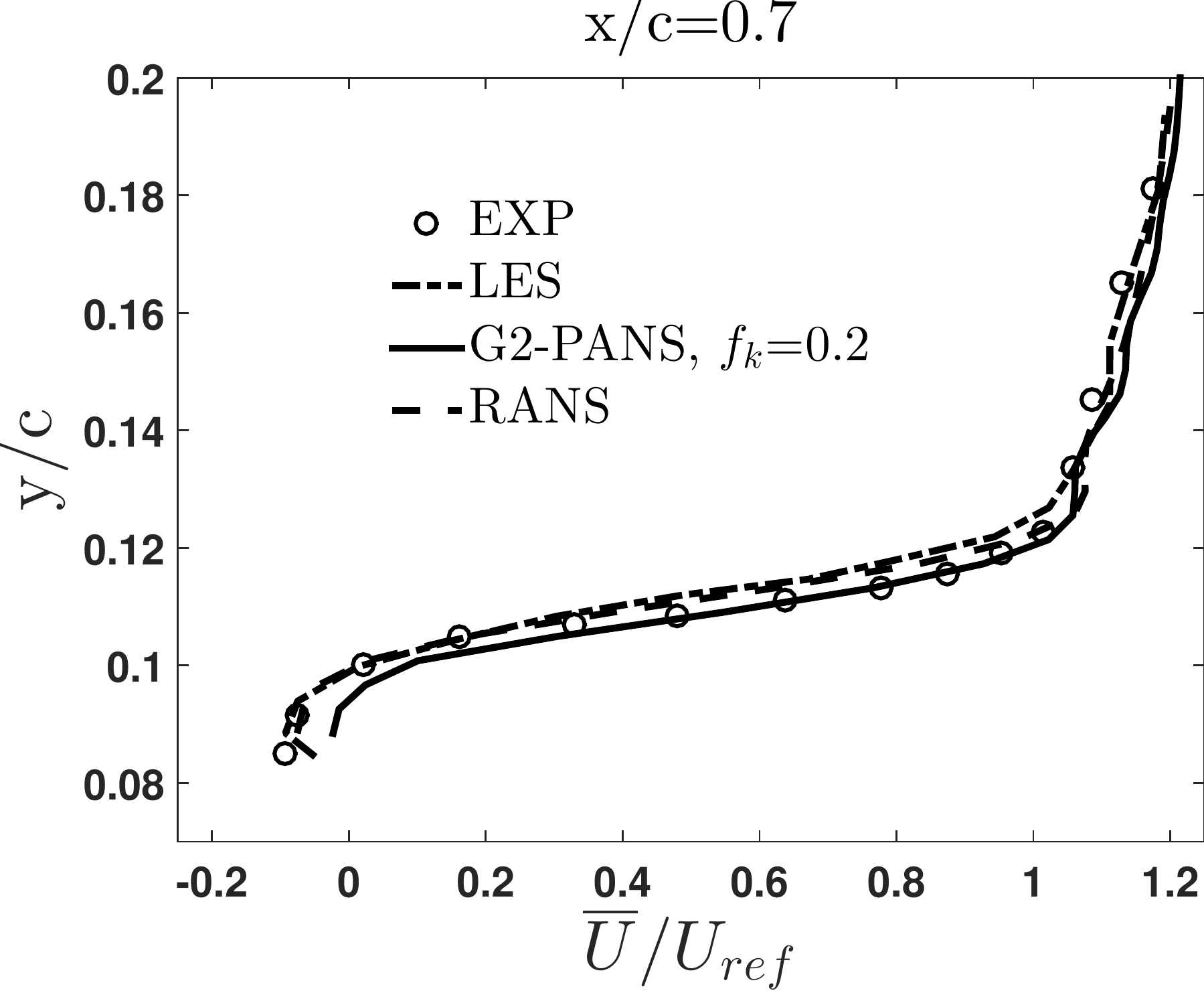}\begin{picture}(0,0)\put(-138,0){(a)}\end{picture}
        \end{subfigure}
        			\begin{subfigure}[b]{0.3\textwidth}
                \includegraphics[width=\textwidth]{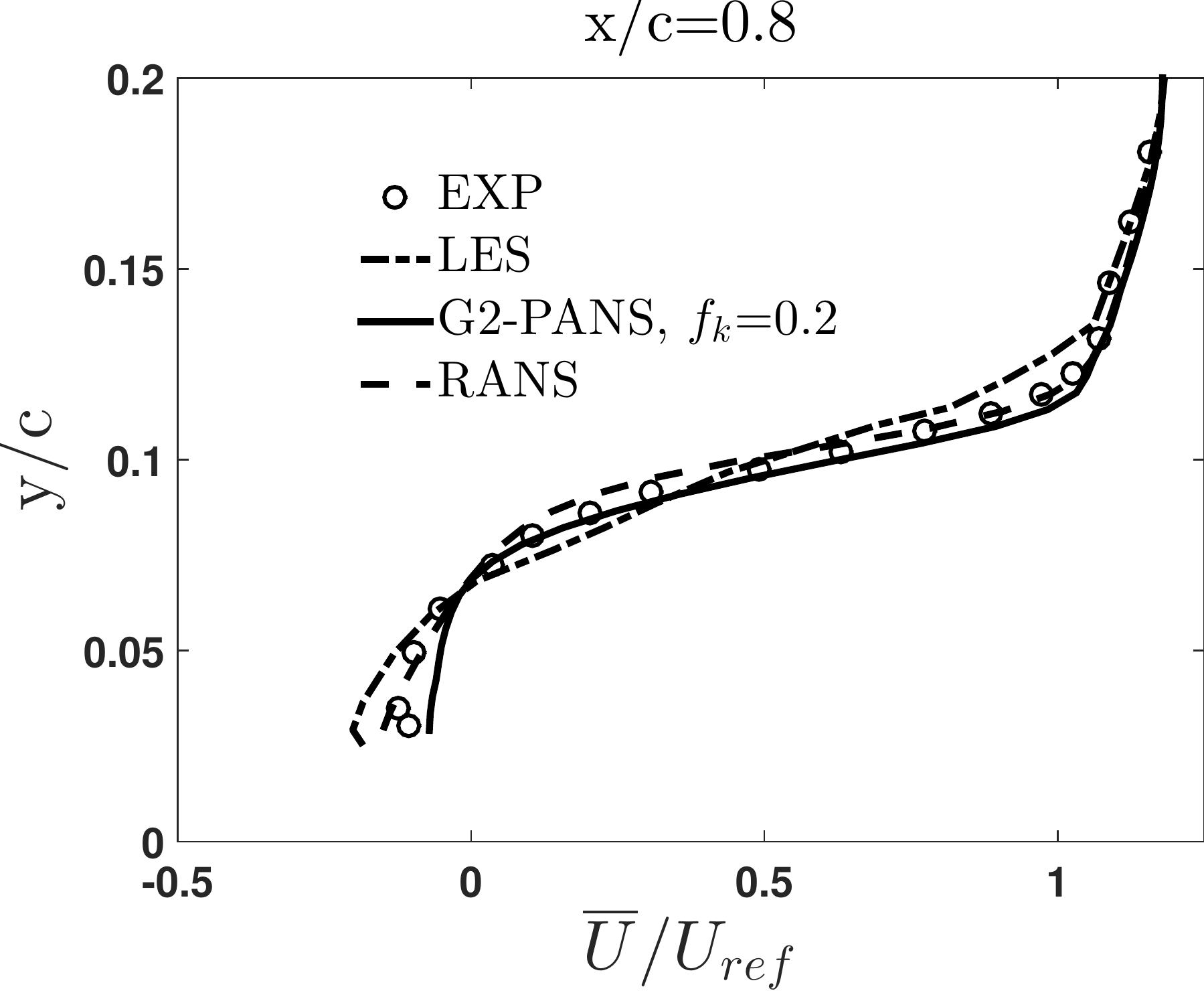}\begin{picture}(0,0)\put(-138,0){(b)}\end{picture}
        \end{subfigure}
                \begin{subfigure}[b]{0.3\textwidth}
                \includegraphics[width=\textwidth]{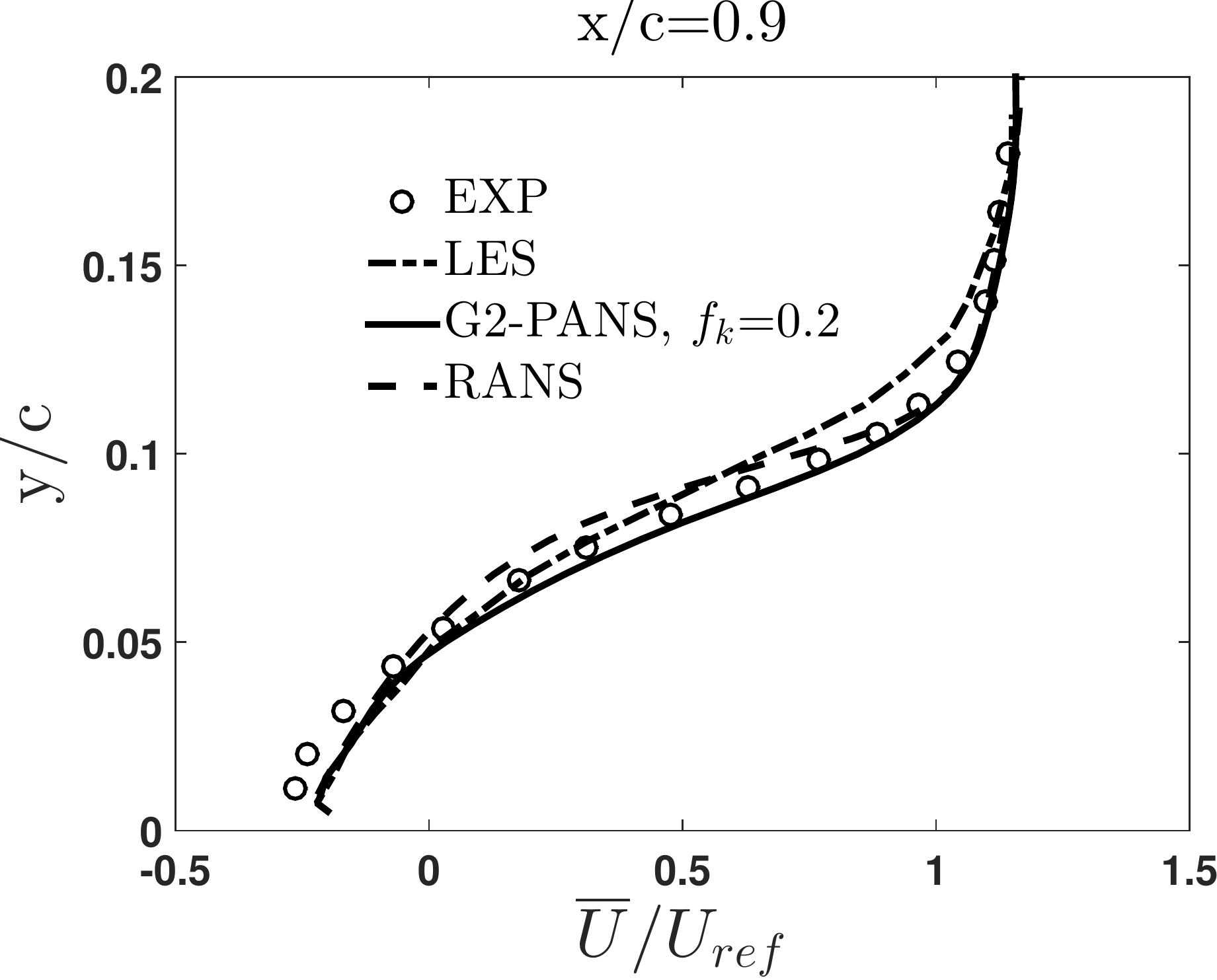}\begin{picture}(0,0)\put(-132,0){(c)}\end{picture}
        \end{subfigure}
        
 		       \begin{subfigure}[b]{0.3\textwidth}
                \includegraphics[width=\textwidth]{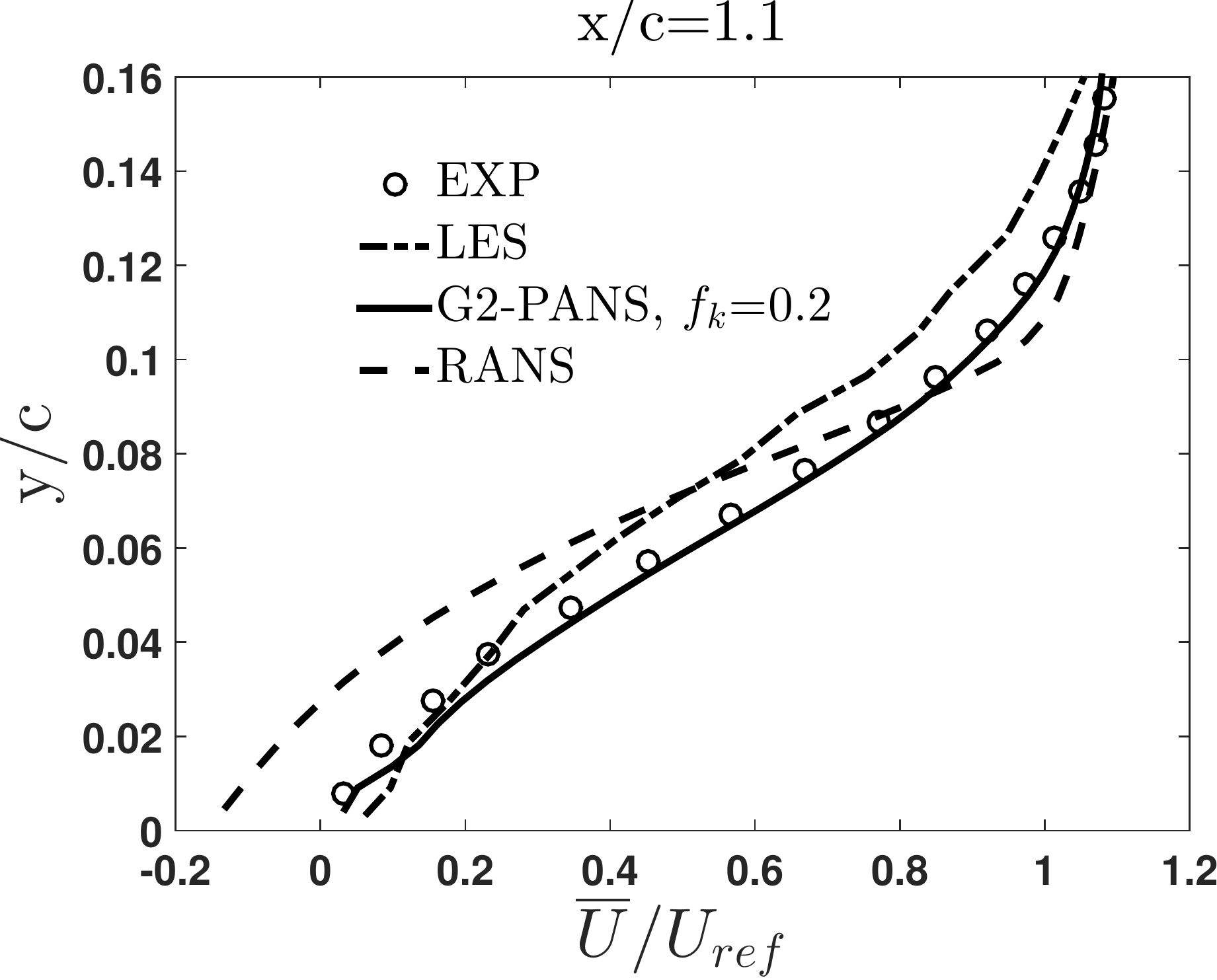}\begin{picture}(0,0)\put(-138,0){(d)}\end{picture}
        \end{subfigure}
        			\begin{subfigure}[b]{0.3\textwidth}
                \includegraphics[width=\textwidth]{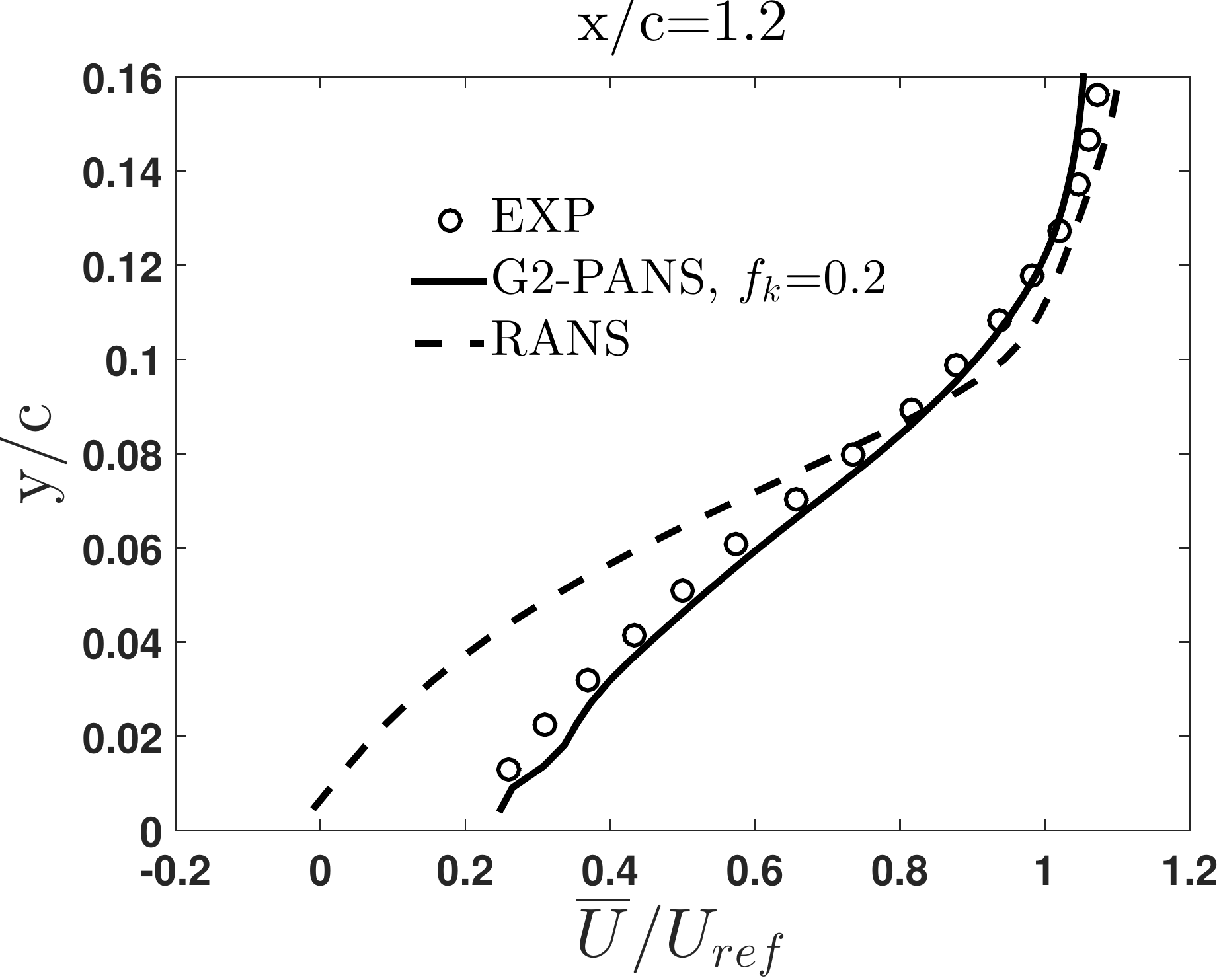}\begin{picture}(0,0)\put(-138,0){(e)}\end{picture}
        \end{subfigure}
                \begin{subfigure}[b]{0.3\textwidth}
                \includegraphics[width=\textwidth]{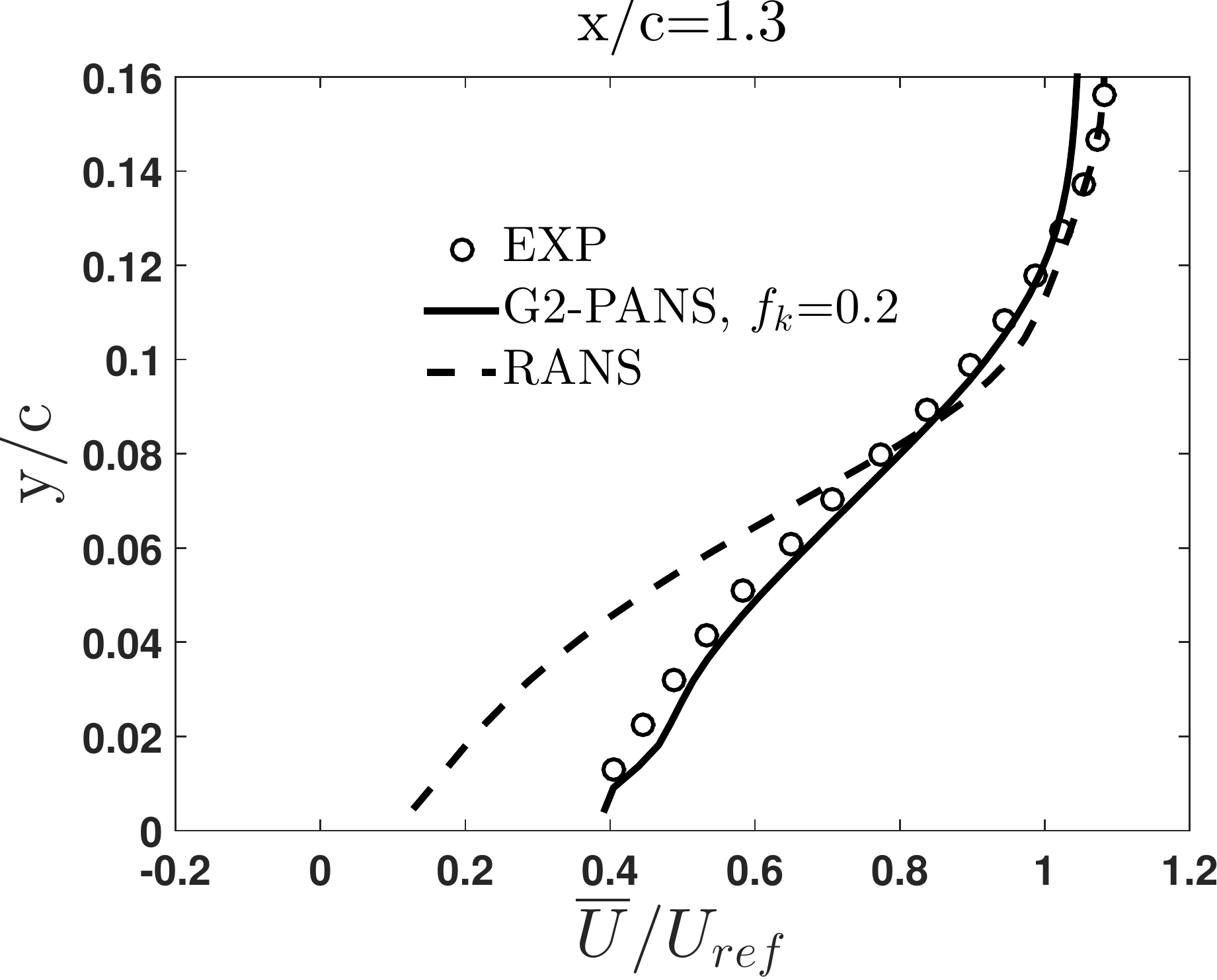}\begin{picture}(0,0)\put(-132,0){(f)}\end{picture}
        \end{subfigure}
\caption{Streamwise velocity (a) $x/c=0.7$ (b) $x/c=0.8$ (c) $x/c=0.9$ (d) $x/c=1.1$ (e) $x/c=1.2$ (f) $x/c=1.3$}   
\label{g2-u}                         
\end{figure}
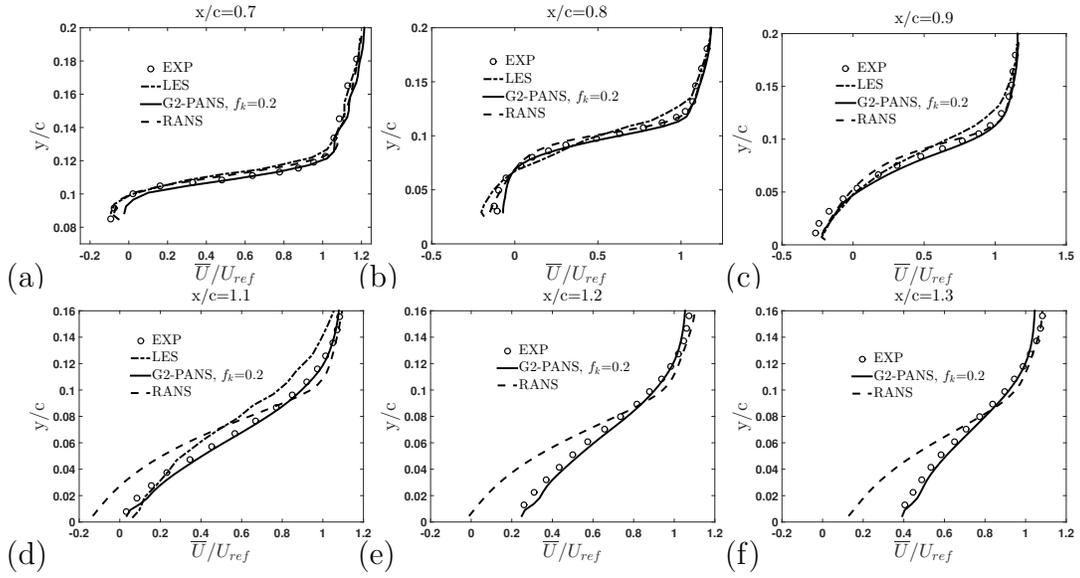

\begin{figure}
        \centering
 		\captionsetup{justification=centering}                                               
 		        \begin{subfigure}[b]{0.3\textwidth}
                \includegraphics[width=\textwidth]{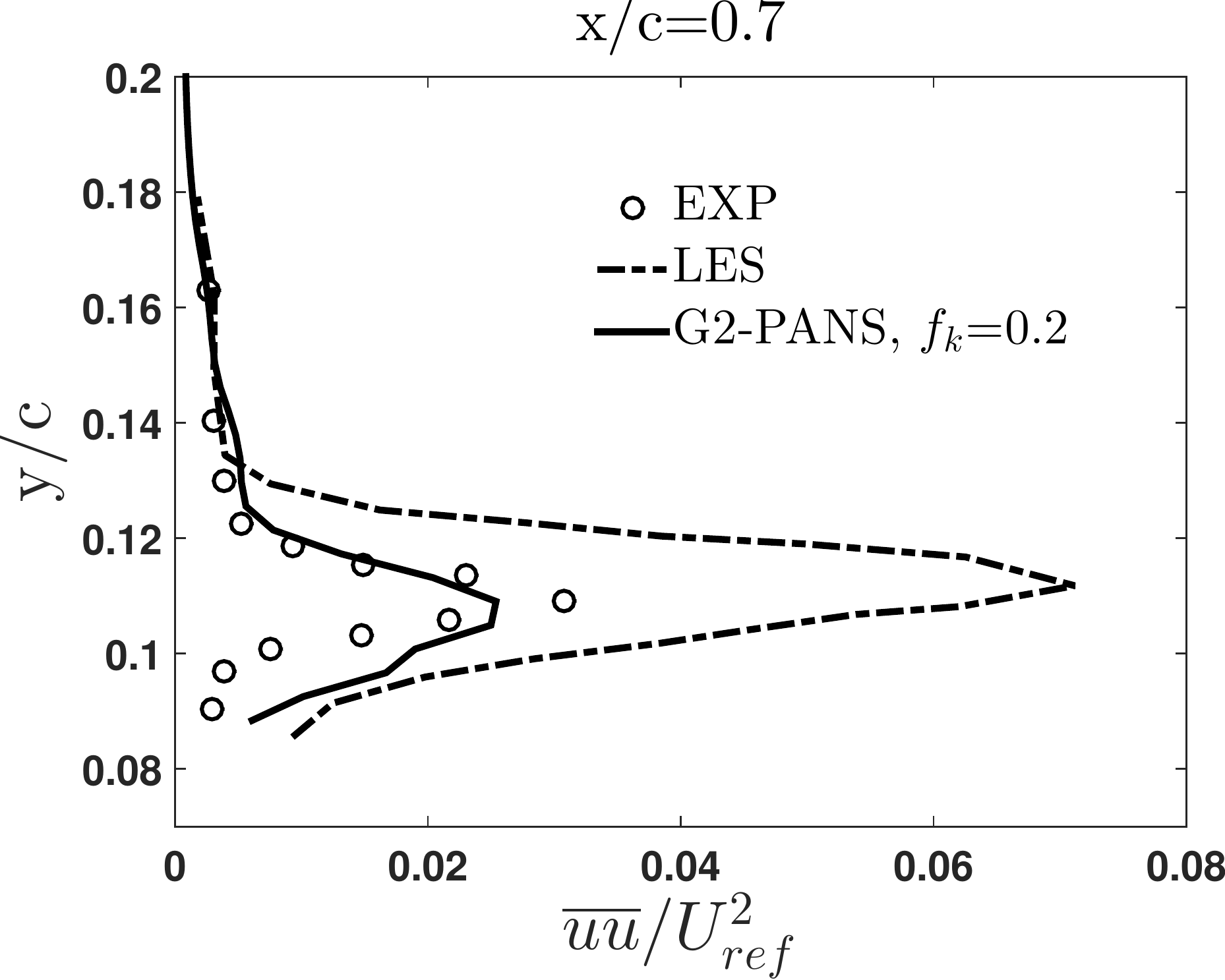}\begin{picture}(0,0)\put(-138,0){(a)}\end{picture}
        \end{subfigure}
        			\begin{subfigure}[b]{0.3\textwidth}
                \includegraphics[width=\textwidth]{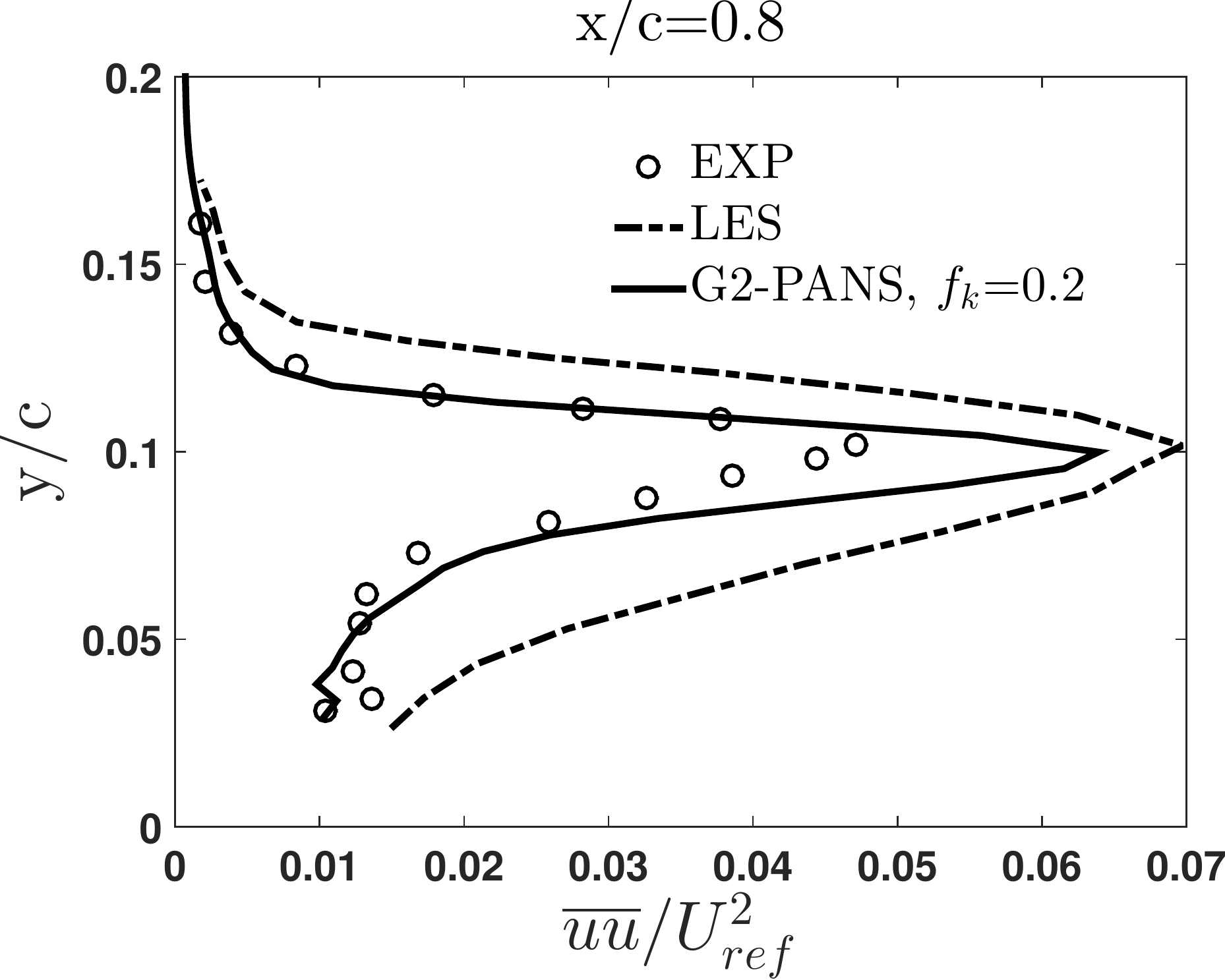}\begin{picture}(0,0)\put(-138,0){(b)}\end{picture}
        \end{subfigure}
                \begin{subfigure}[b]{0.3\textwidth}
                \includegraphics[width=\textwidth]{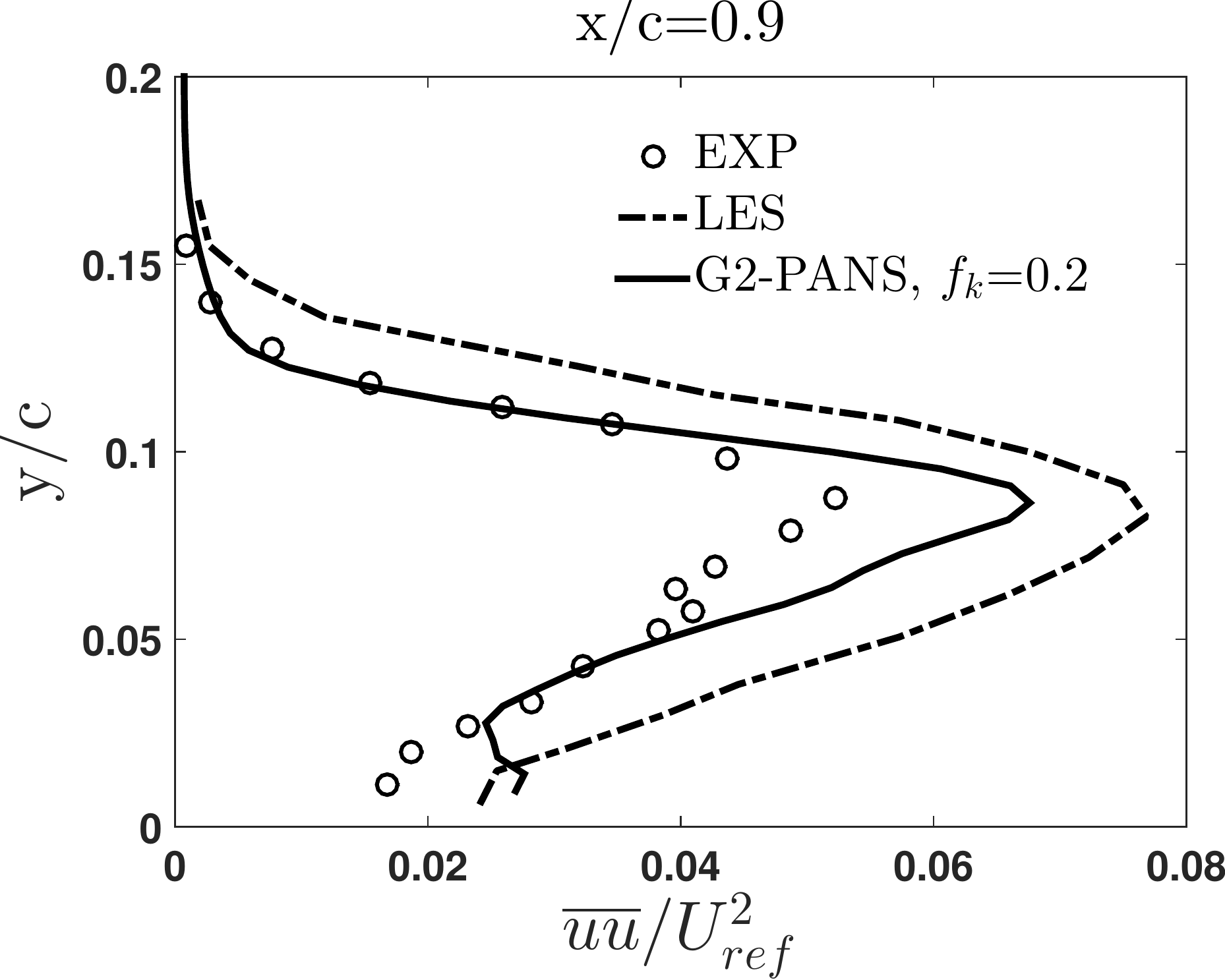}\begin{picture}(0,0)\put(-132,0){(c)}\end{picture}
        \end{subfigure}
        
 		       \begin{subfigure}[b]{0.3\textwidth}
                \includegraphics[width=\textwidth]{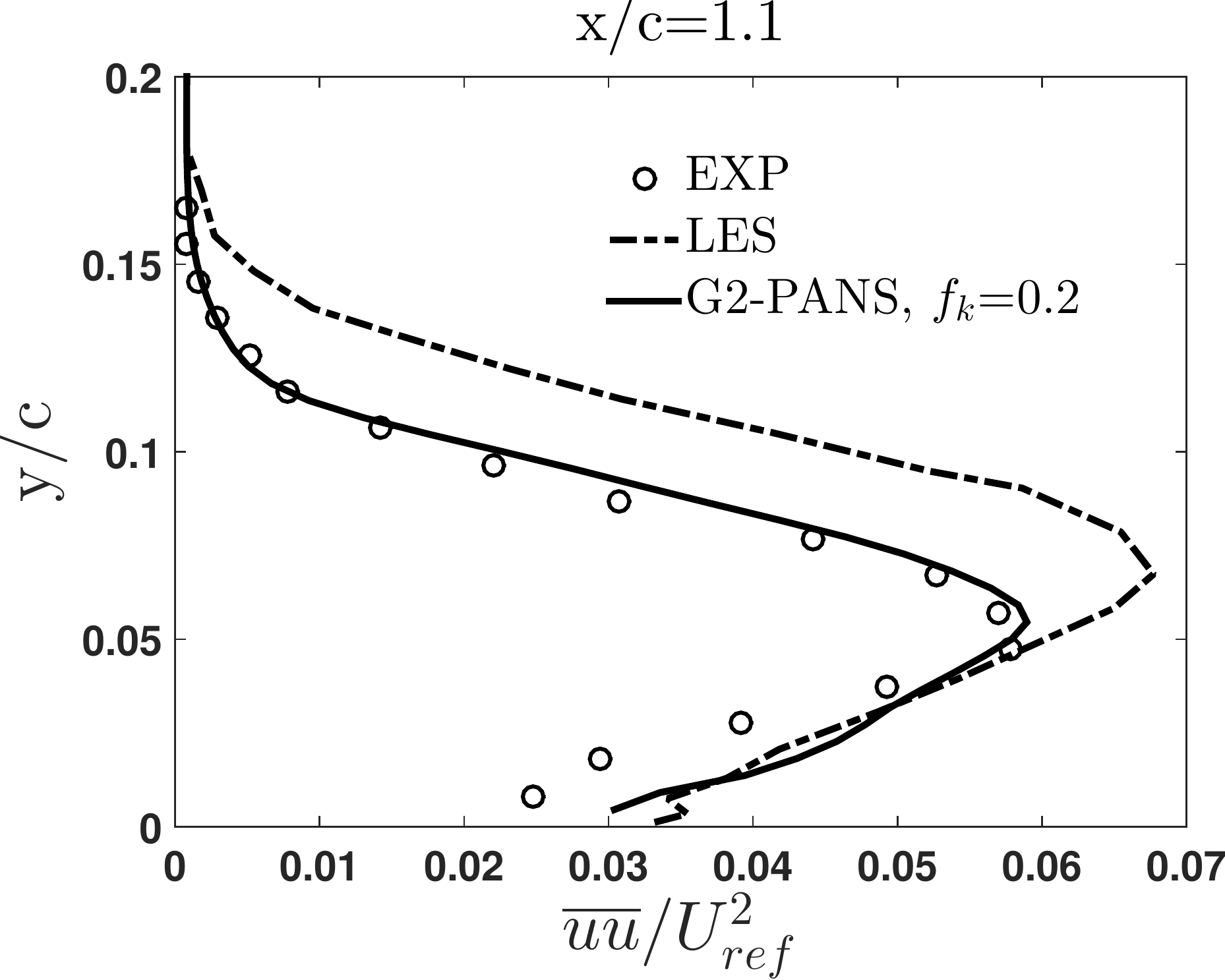}\begin{picture}(0,0)\put(-138,0){(d)}\end{picture}
        \end{subfigure}
        			\begin{subfigure}[b]{0.3\textwidth}
                \includegraphics[width=\textwidth]{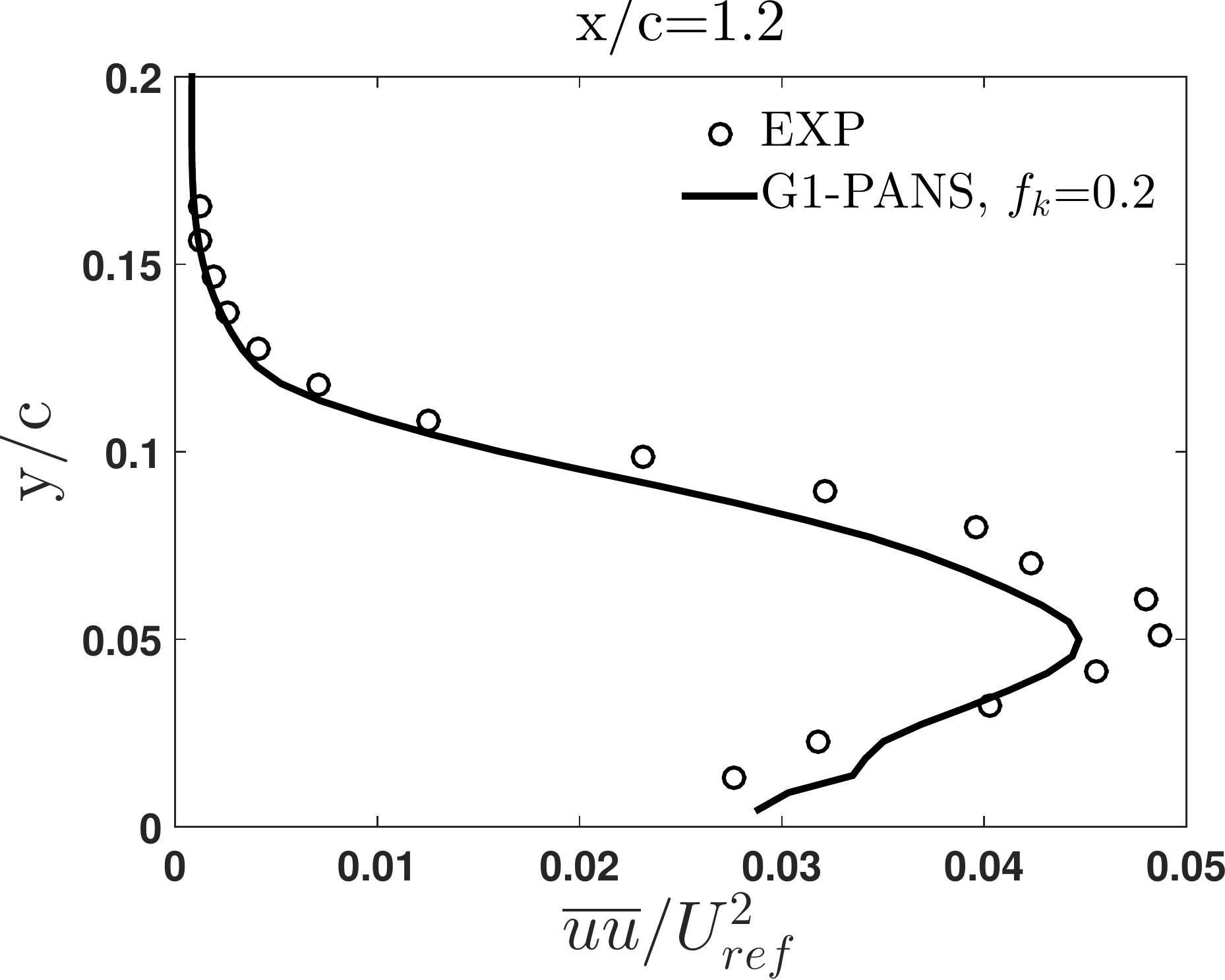}\begin{picture}(0,0)\put(-138,0){(e)}\end{picture}
        \end{subfigure}
                \begin{subfigure}[b]{0.3\textwidth}
                \includegraphics[width=\textwidth]{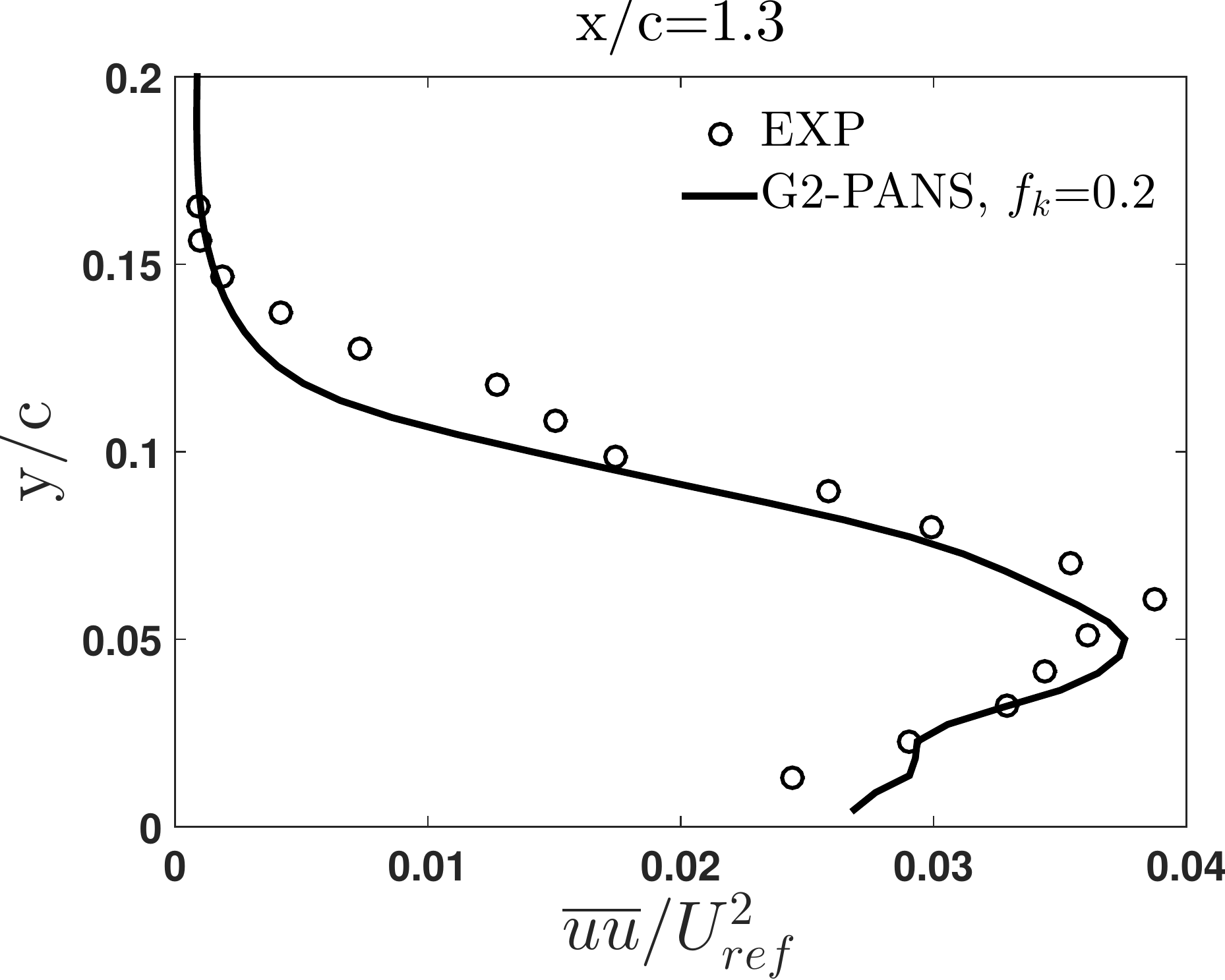}\begin{picture}(0,0)\put(-132,0){(f)}\end{picture}
        \end{subfigure}
\caption{Streamwise stress (a) $x/c=0.7$ (b) $x/c=0.8$ (c) $x/c=0.9$ (d) $x/c=1.1$ (e) $x/c=1.2$ (f) $x/c=1.3$}   
\label{g2-uu}                         
\end{figure}

\begin{figure}
        \centering
 		\captionsetup{justification=centering}                                               
 		        \begin{subfigure}[b]{0.3\textwidth}
                \includegraphics[width=\textwidth]{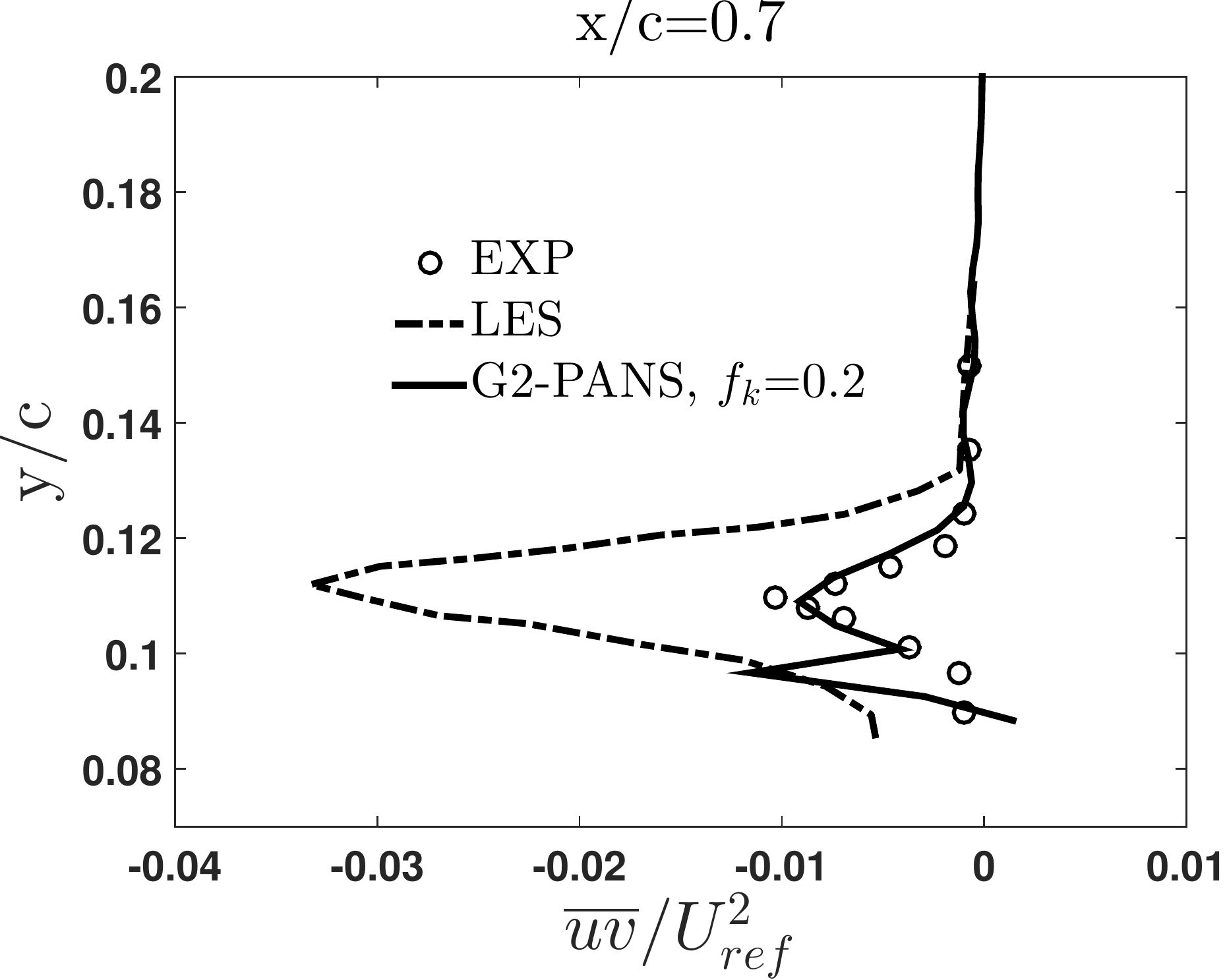}\begin{picture}(0,0)\put(-138,0){(a)}\end{picture}
        \end{subfigure}
        			\begin{subfigure}[b]{0.3\textwidth}
                \includegraphics[width=\textwidth]{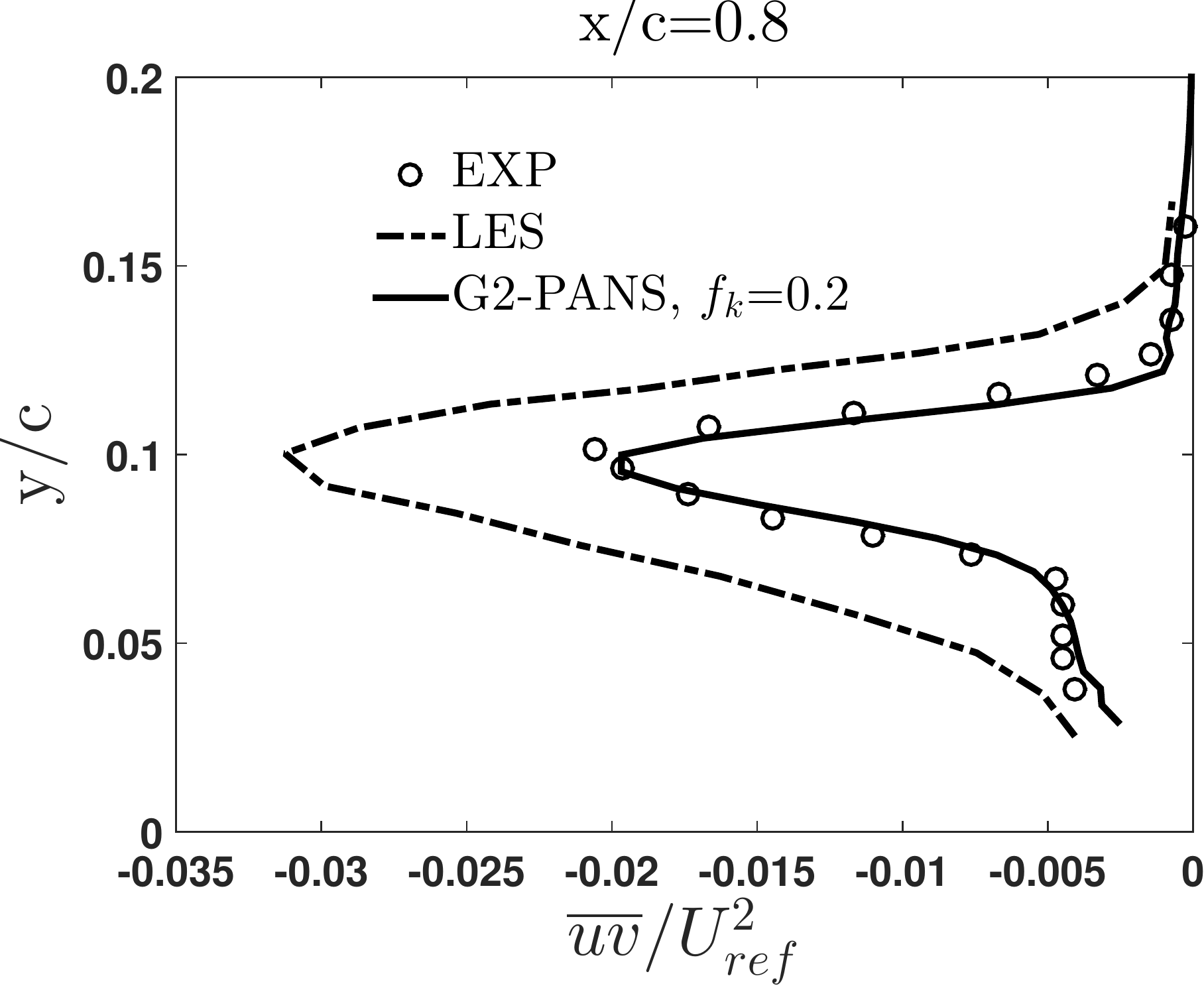}\begin{picture}(0,0)\put(-138,0){(b)}\end{picture}
        \end{subfigure}
                \begin{subfigure}[b]{0.3\textwidth}
                \includegraphics[width=\textwidth]{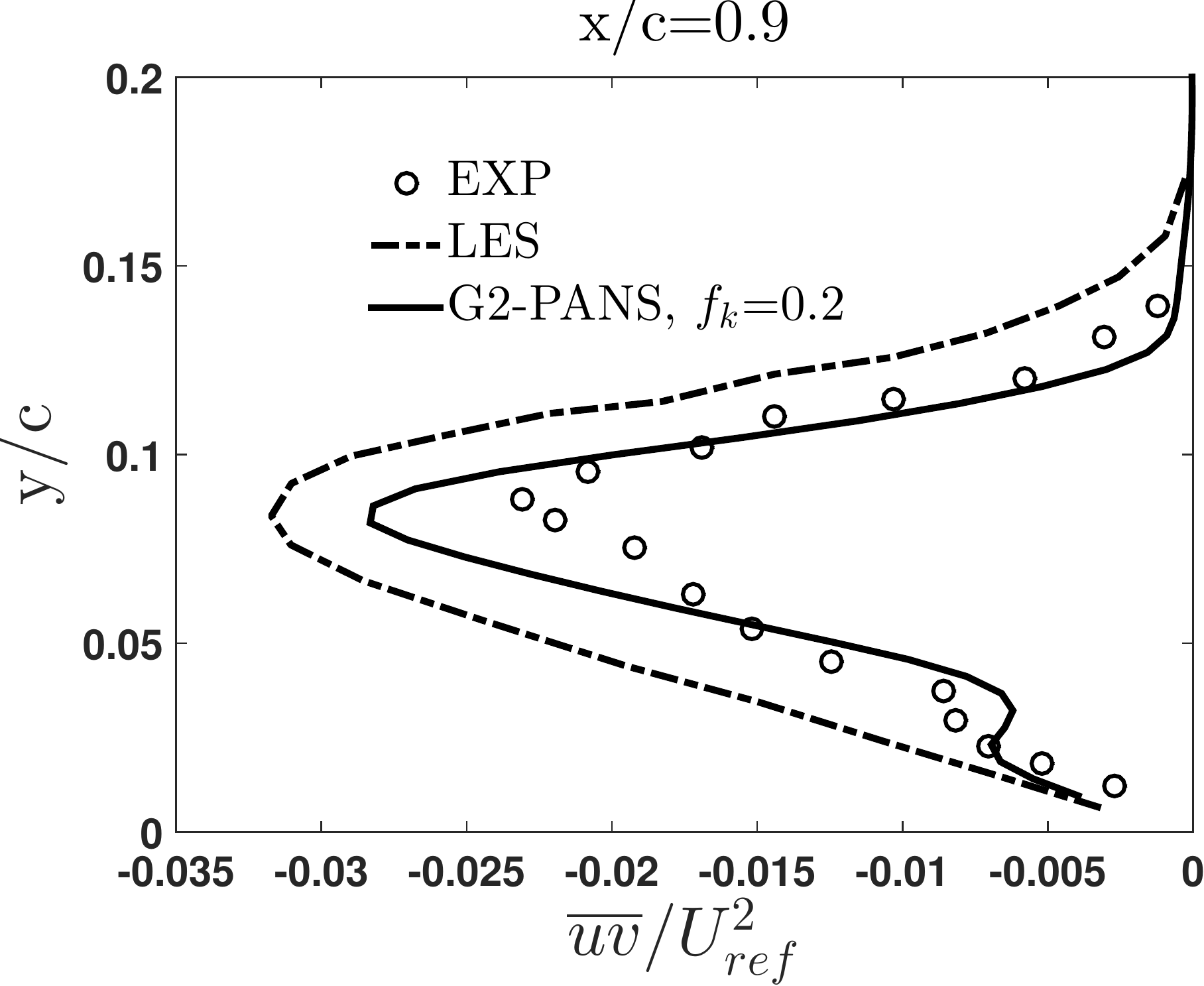}\begin{picture}(0,0)\put(-132,0){(c)}\end{picture}
        \end{subfigure}
        
 		       \begin{subfigure}[b]{0.3\textwidth}
                \includegraphics[width=\textwidth]{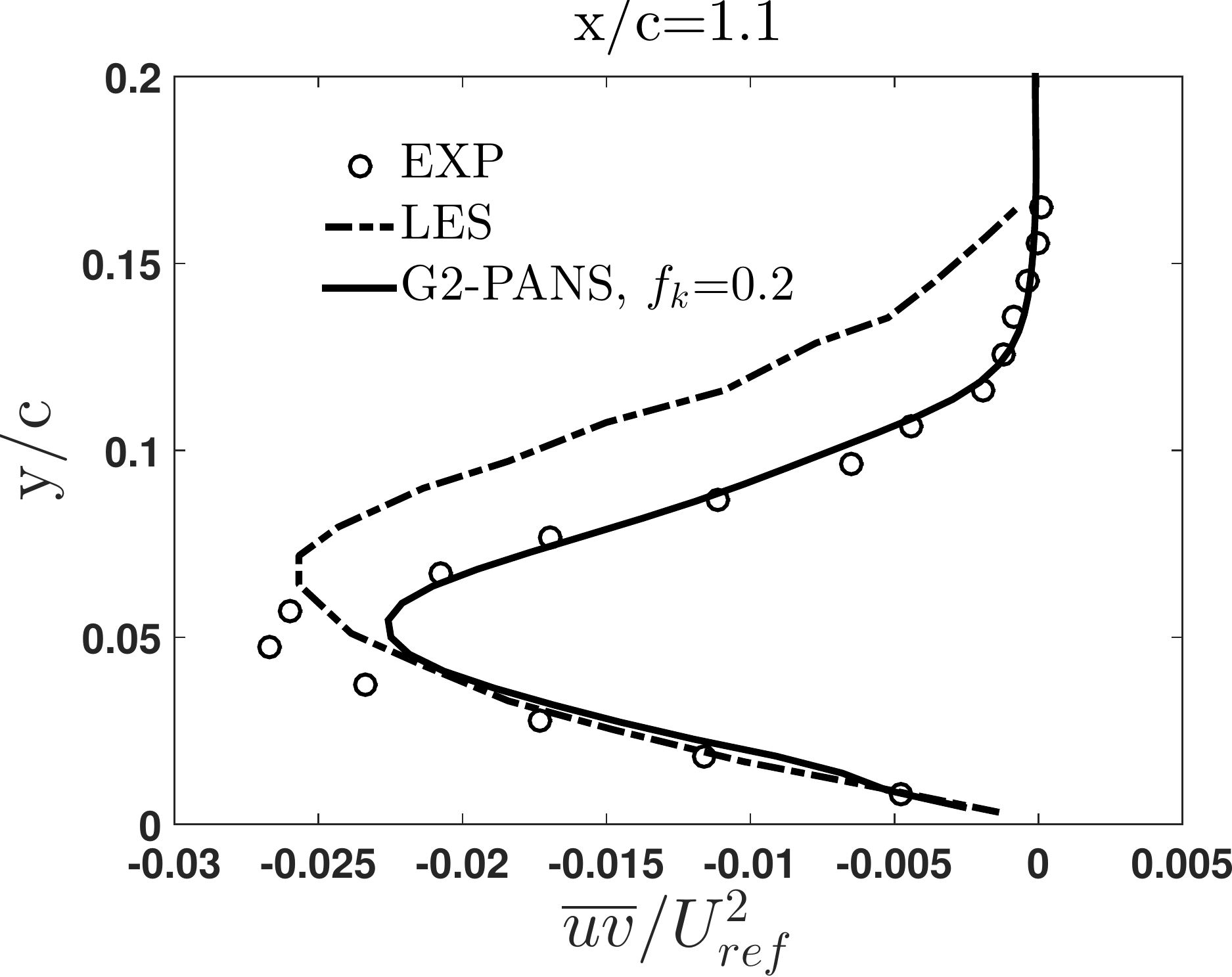}\begin{picture}(0,0)\put(-138,0){(d)}\end{picture}
        \end{subfigure}
        			\begin{subfigure}[b]{0.3\textwidth}
                \includegraphics[width=\textwidth]{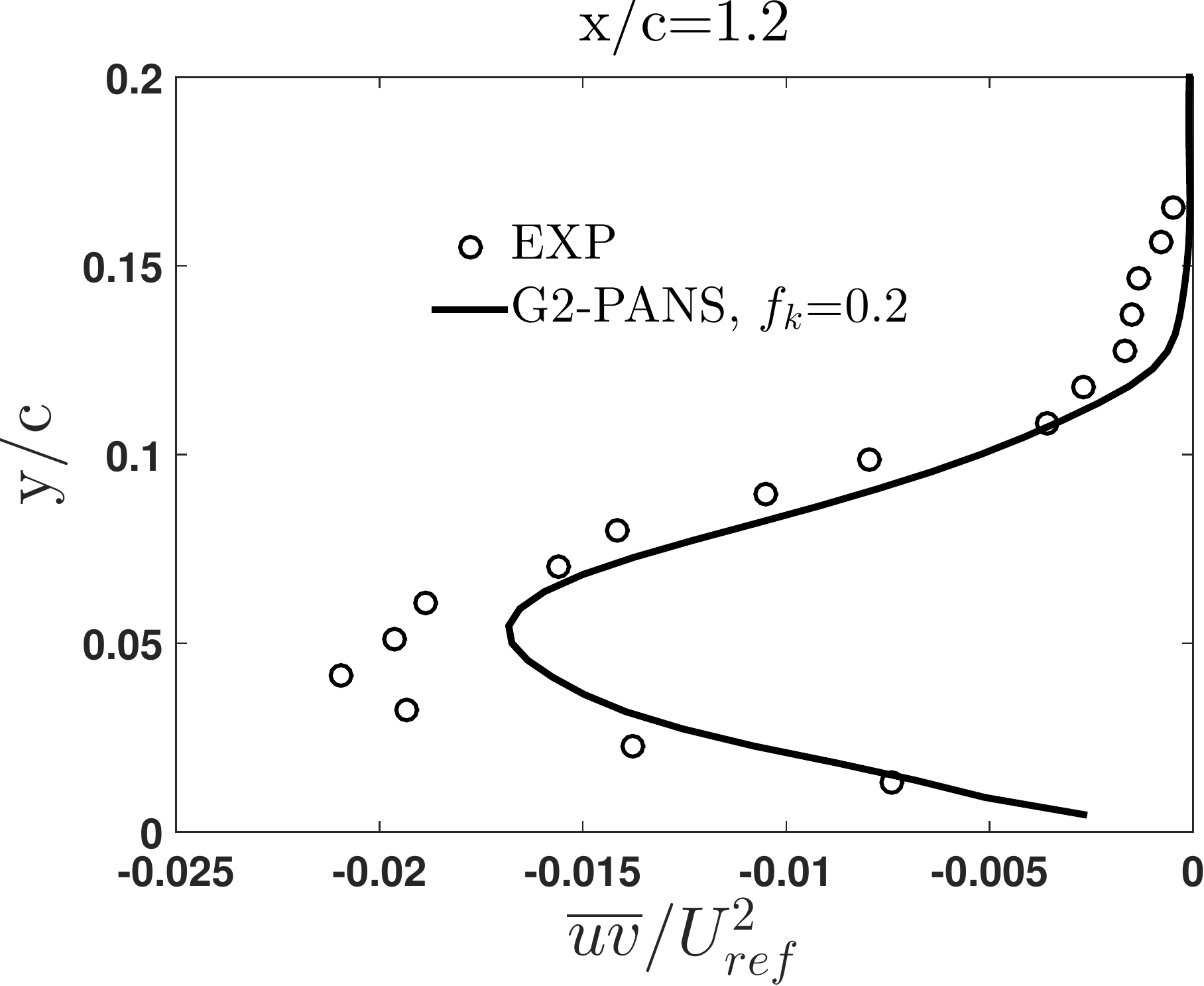}\begin{picture}(0,0)\put(-138,0){(e)}\end{picture}
        \end{subfigure}
                \begin{subfigure}[b]{0.3\textwidth}
                \includegraphics[width=\textwidth]{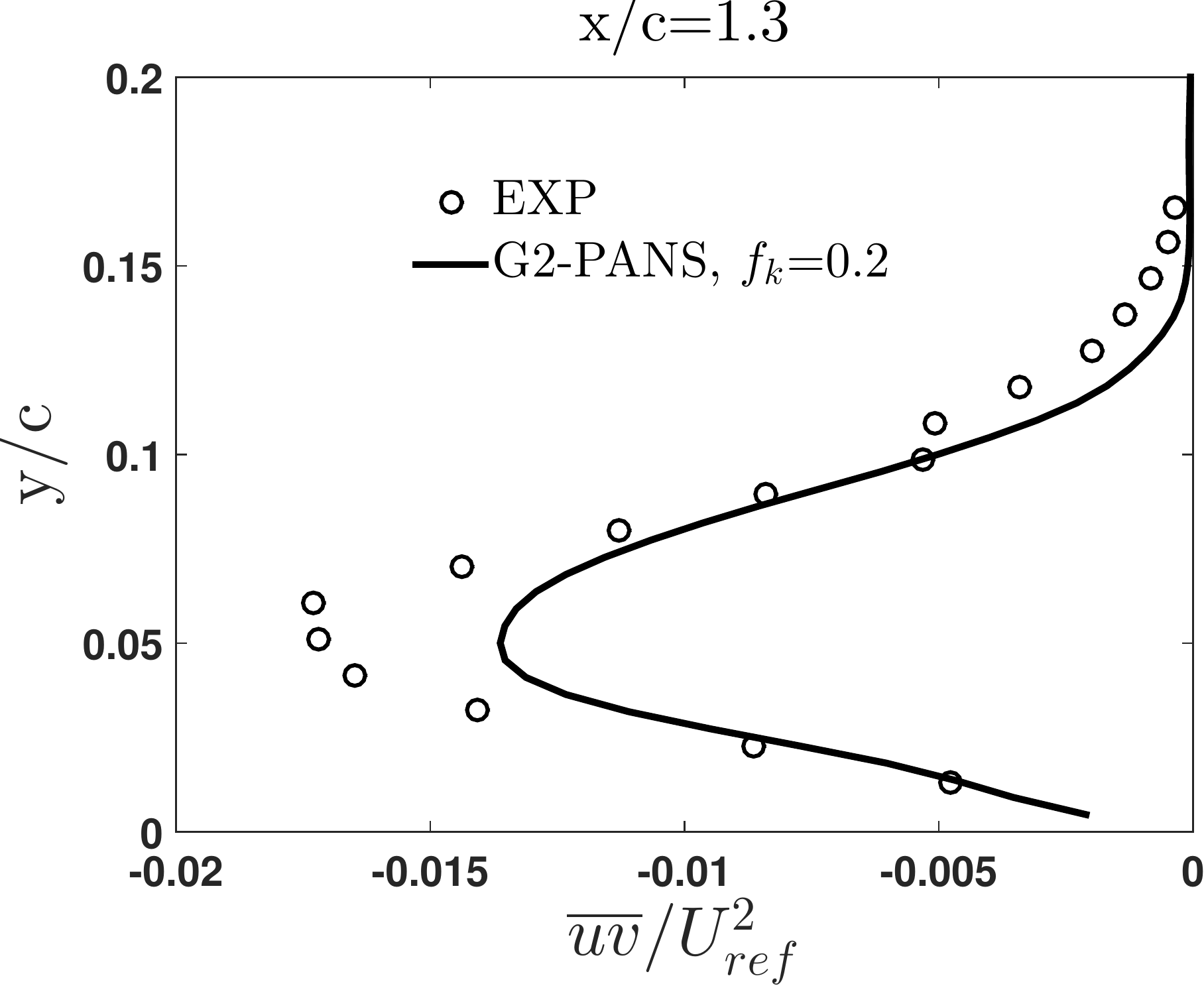}\begin{picture}(0,0)\put(-132,0){(f)}\end{picture}
        \end{subfigure}
\caption{Shear stress  (a) $x/c=0.7$ (b) $x/c=0.8$ (c) $x/c=0.9$ (d) $x/c=1.1$ (e) $x/c=1.2$ (f) $x/c=1.3$}   
\label{g2-uv}                         
\end{figure}

Figures \ref{g2-u}-\ref{g2-uv} show the non-dimensional streamwise velocity $\overline{U}$, streamwise stress component $\overline{uu}$, and the shear stress $\overline{uv}$ at six streamwise locations. These locations include both the separation and post-reattachment regions. G2-PANS calculations on 0.6 million grid cells are compared to a LES simulation with 9.4 million grid nodes \cite{avdis}, and experimental data \cite{greenblatt2006}. RANS simulations are also included for the comparison purposes. It should be noted that the LES study did not provide the profiles for the two post-reattachment locations, $x/c=1.2$ and $x/c=1.3$, and the G2-PANS results are compared only with experiment at these locations.

Figure \ref{g2-u} shows good agreement between the experimental and simulated profiles at $x/c=0.7-0.9$ which locate within the separation bubble. However, close to the reattachment point at $x/c=1.1$, noticeable differences are observed for various simulations. While the G2-PANS results are in very good agreement with experiment, RANS simulation predicts a much stronger reverse flow and LES model simulates early reattached flow. 

The differences between the RANS and G2-PANS simulation become more apparent at post-reattachment region, where the RANS results deviates significantly from the data. Also, it can be inferred from Fig. \ref{g2-u} that the LES simulation predicts larger shear layer thickness than the measured data. This is specifically visible by looking at the upper part of the velocity profile where the estimated velocity by LES is remarkably low. Consequently, the stresses are over-predicted by the LES simulation as shown in Figs. \ref{g2-uu} and \ref{g2-uv}. However, for the G2-PANS simulations, size of the recirculation zone and thickness of the shear layer is very well anticipated which are accompanied by a good prediction of stress components. Thus, the close agreement of G2-PANS (0.6M cells) results with experiment represent an improvement over LES (9.4M cells) simulation.     

The variation of skin-friction coefficient over a region that encompasses pre and post-reattachment is shown in Fig. \ref{cf} for the G2-PANS ($f_k=0.2)$ and LES calculations. As seen in this plot, G2-PANS prediction of $C_f$ is closer to the experimental values than the LES simulation. Prior to separation, friction coefficient increases as a result of flow acceleration in that region. The LES simulation over-predicts the friction coefficient before separation and under-predicts that in most part of the separation region. Specially, LES simulation is not able to estimate the local minimum of $C_f$ plot in the core of separation bubble. Although G2-PANS simulation is able to estimate the friction coefficient reasonably well before the hump leading edge, once separation occurs, dramatic improvement for the G2-PANS $C_f$ values is achieved. 

The locations where the value of $C_f$ goes to zero corresponds to the separation point and reattachment location for each computation. Fig. \ref{cf} further illustrates that the LES simulation predicts early reattachment, while the G2-PANS model reattachment is very close to experiment. The results presented in Figs. \ref{g2-u}-\ref{cf} indicate that an accurate PANS solution can be obtained even with a grid resolution which is around 15 times lower than the corresponding LES simulation. It is also important to prove that with increasing grid resolution, we are able to reproduce/improve the simulation results. Therefore, a mesh-independence study was conducted to identify such a plateau in terms of the separation and reattachment point for this configuration.

\begin{figure}
        \centering
 		\captionsetup{justification=centering}                                               
                \includegraphics[width=0.55\textwidth]{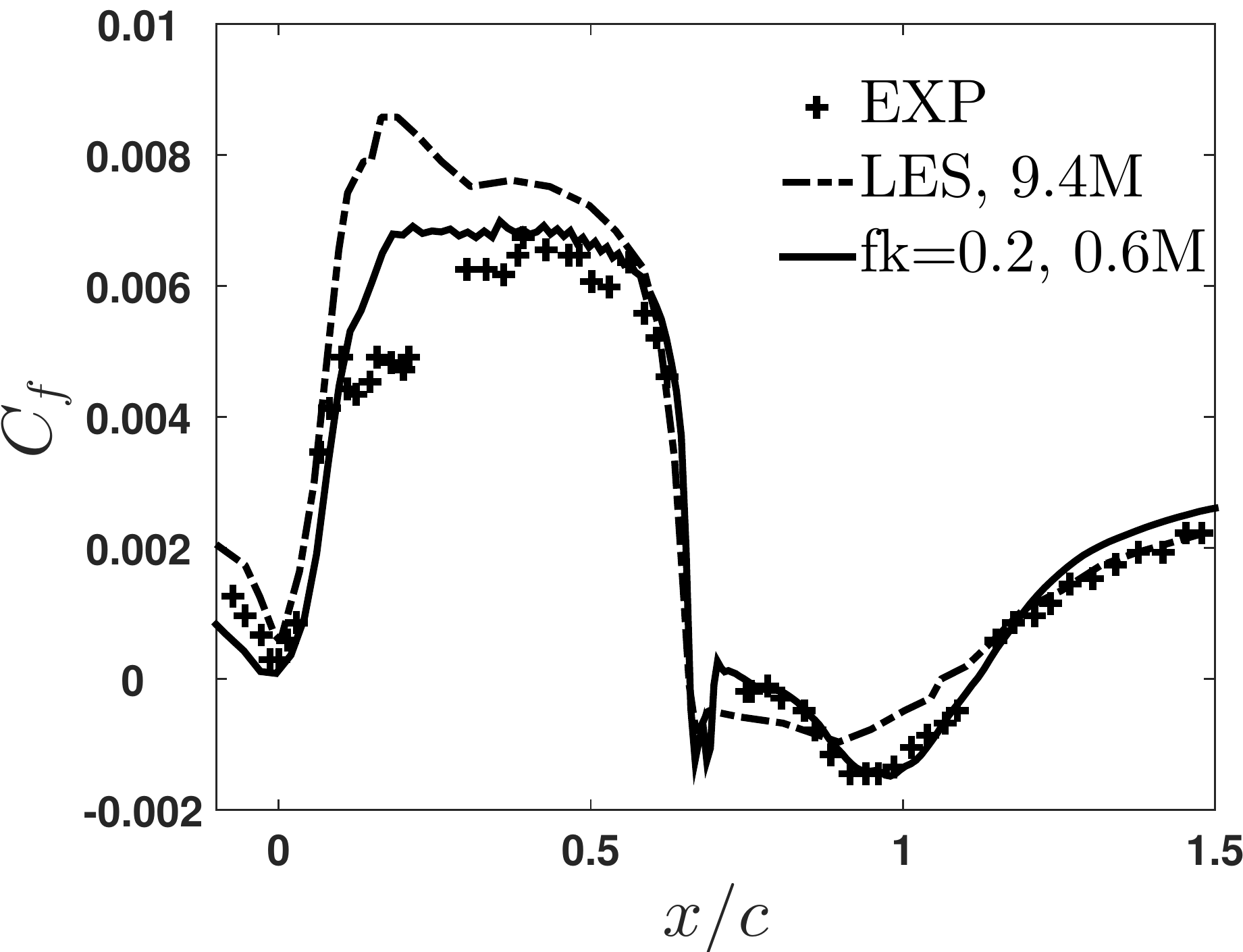}\begin{picture}(0,0)\put(-138,0){}\end{picture}
\caption{Friction coefficient}   
\label{cf}                         
\end{figure}
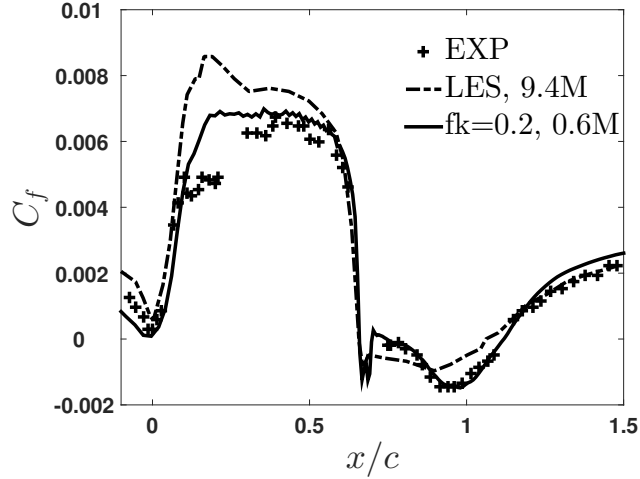

Figure \ref{sep} shows the result of this study at four different grid resolutions in the range of 0.3-1.5 million grid nodes for G2-PANS simulations with $f_k$=0.2. Change of grid resolution is particularly applied in the normal and spanwise directions as indicated in table \ref{setup}. The experimental reattachment and separation points are at $x/c=1.11 \pm 0.003$ and $x/c=0.665$, respectively. It is observed from this figure that the separation point is estimated accurately at all the grid resolutions, and the reattachment point prediction improves by increasing the grid resolution from 0.3 to 0.4 million nodes. No substantial improvement for the two parameters is seen for higher grid resolution than 0.4 million. It can be further confirmed from this plot that grid resolution of 0.6 million grid nodes is adequate for the G2-PANS simulation with the specified $f_k$.
 
\begin{figure}
        \centering
 		\captionsetup{justification=centering}                                               
                \includegraphics[width=0.55\textwidth]{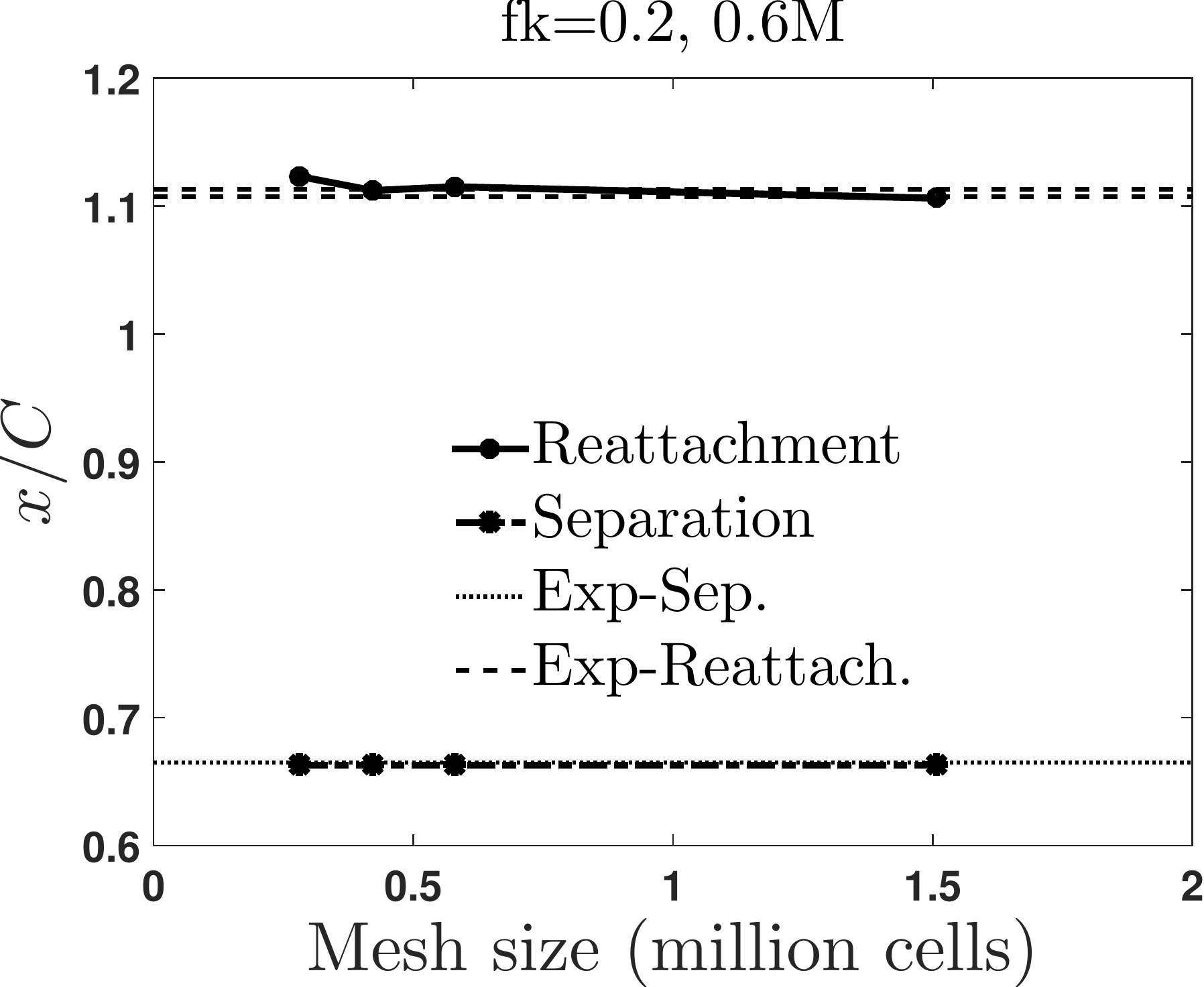}\begin{picture}(0,0)\put(-132,-10){}\end{picture}  
\caption{Reattachment and separation location for G2-PANS simulations at different grid resolutions}   
\label{sep}                         
\end{figure}
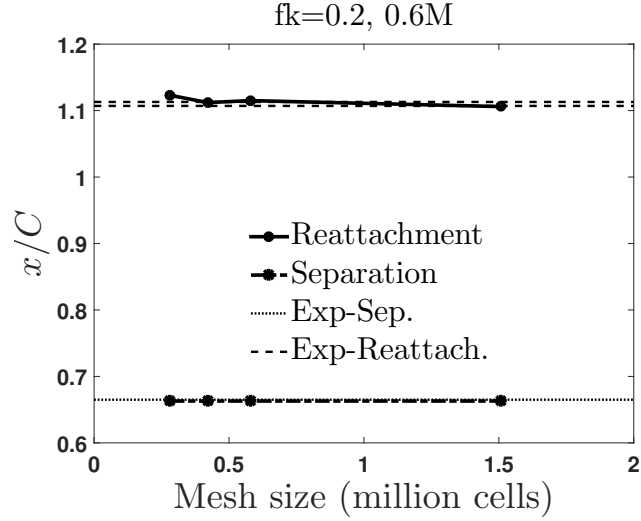

\subsection{$f_k$ study}

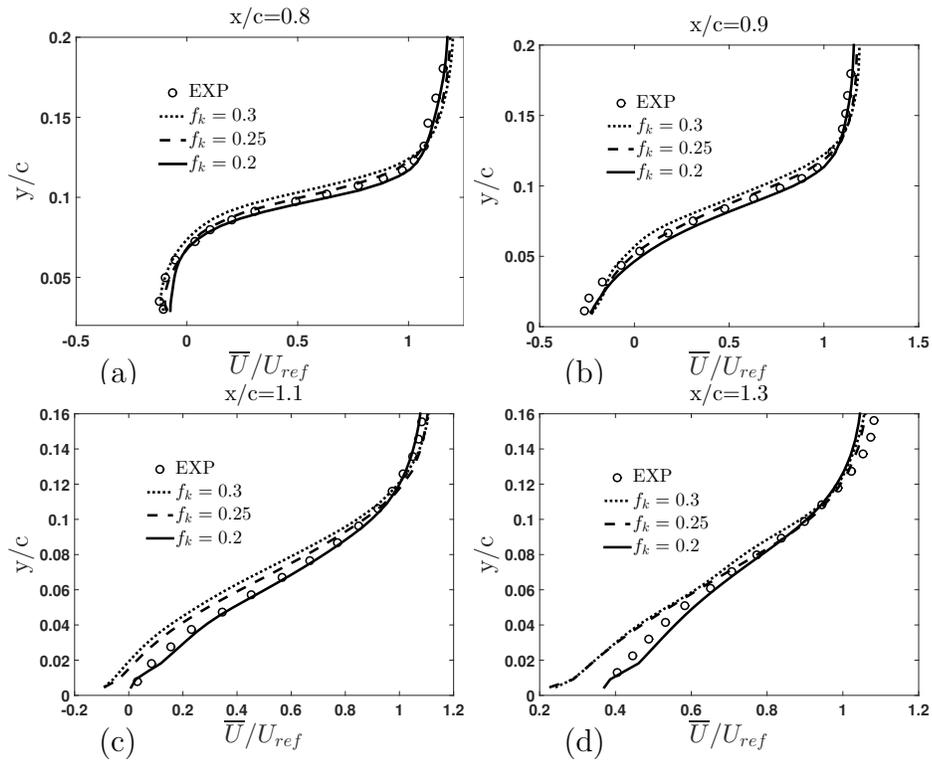
\begin{figure}
        \centering
 		\captionsetup{justification=centering}                                               
 		        \begin{subfigure}[b]{0.4\textwidth}
                \includegraphics[width=\textwidth]{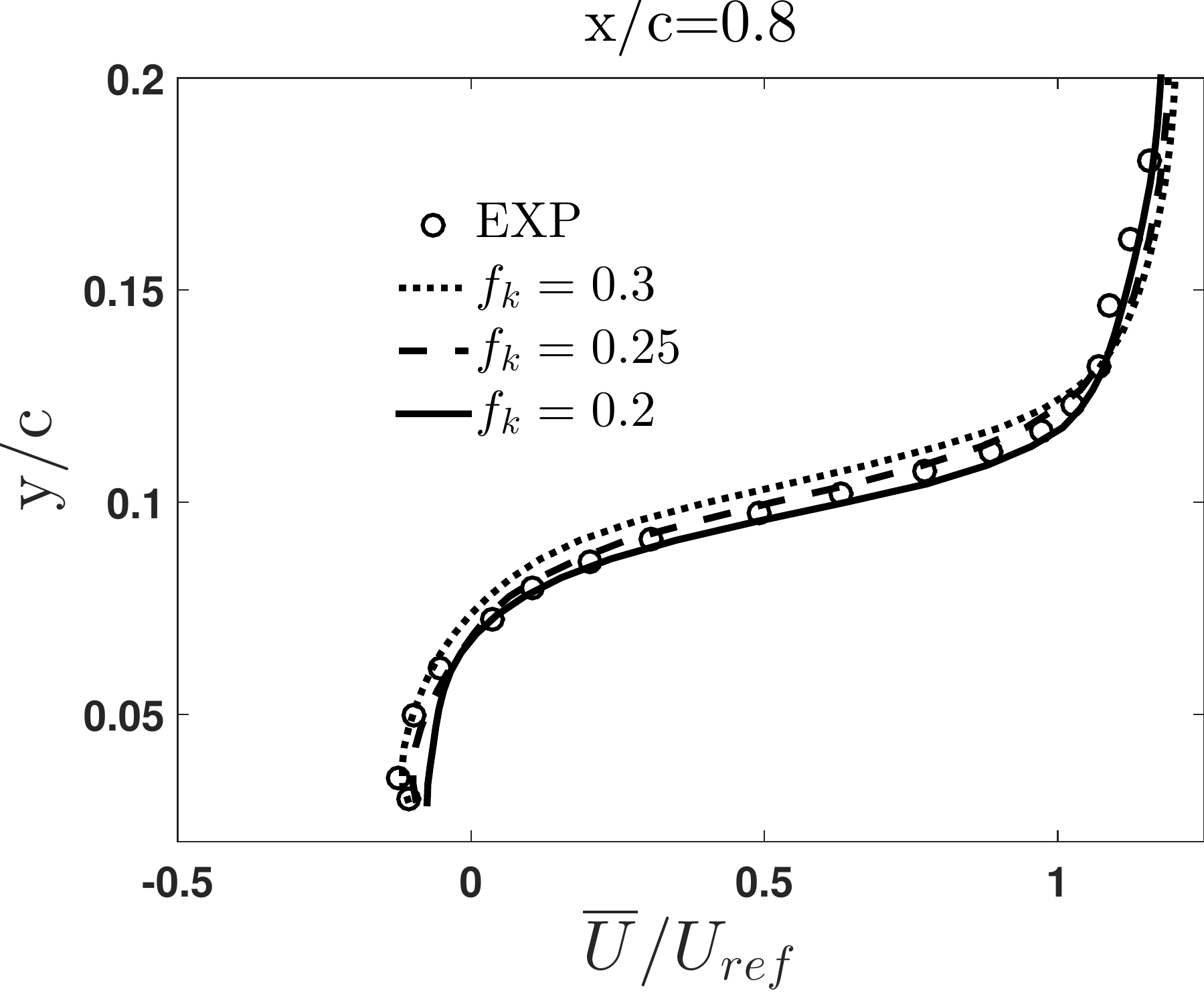}\begin{picture}(0,0)\put(-138,0){(a)}\end{picture}
        \end{subfigure}
        			\begin{subfigure}[b]{0.4\textwidth}
                \includegraphics[width=\textwidth]{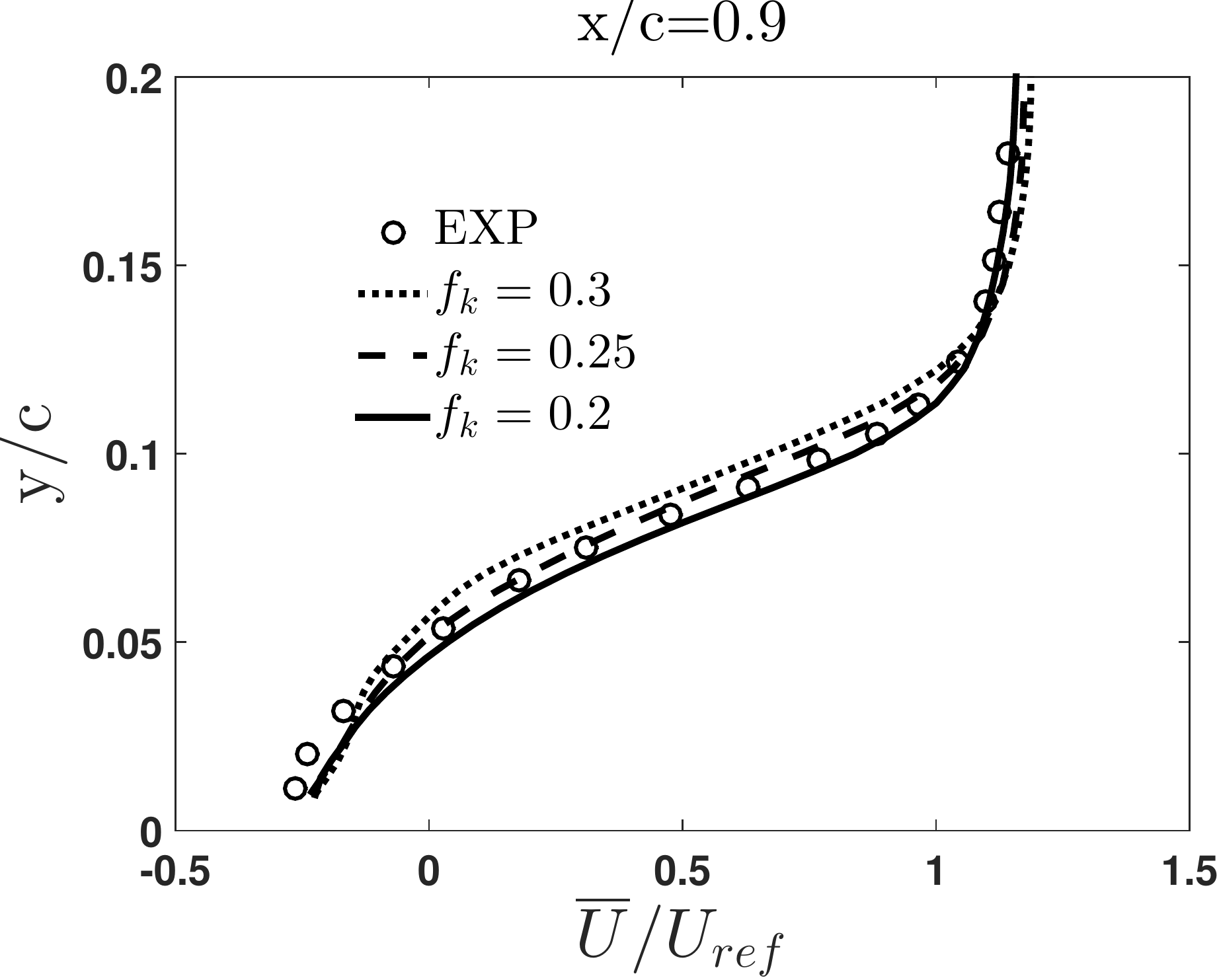}\begin{picture}(0,0)\put(-138,0){(b)}\end{picture}
        \end{subfigure}
                \begin{subfigure}[b]{0.4\textwidth}
                \includegraphics[width=\textwidth]{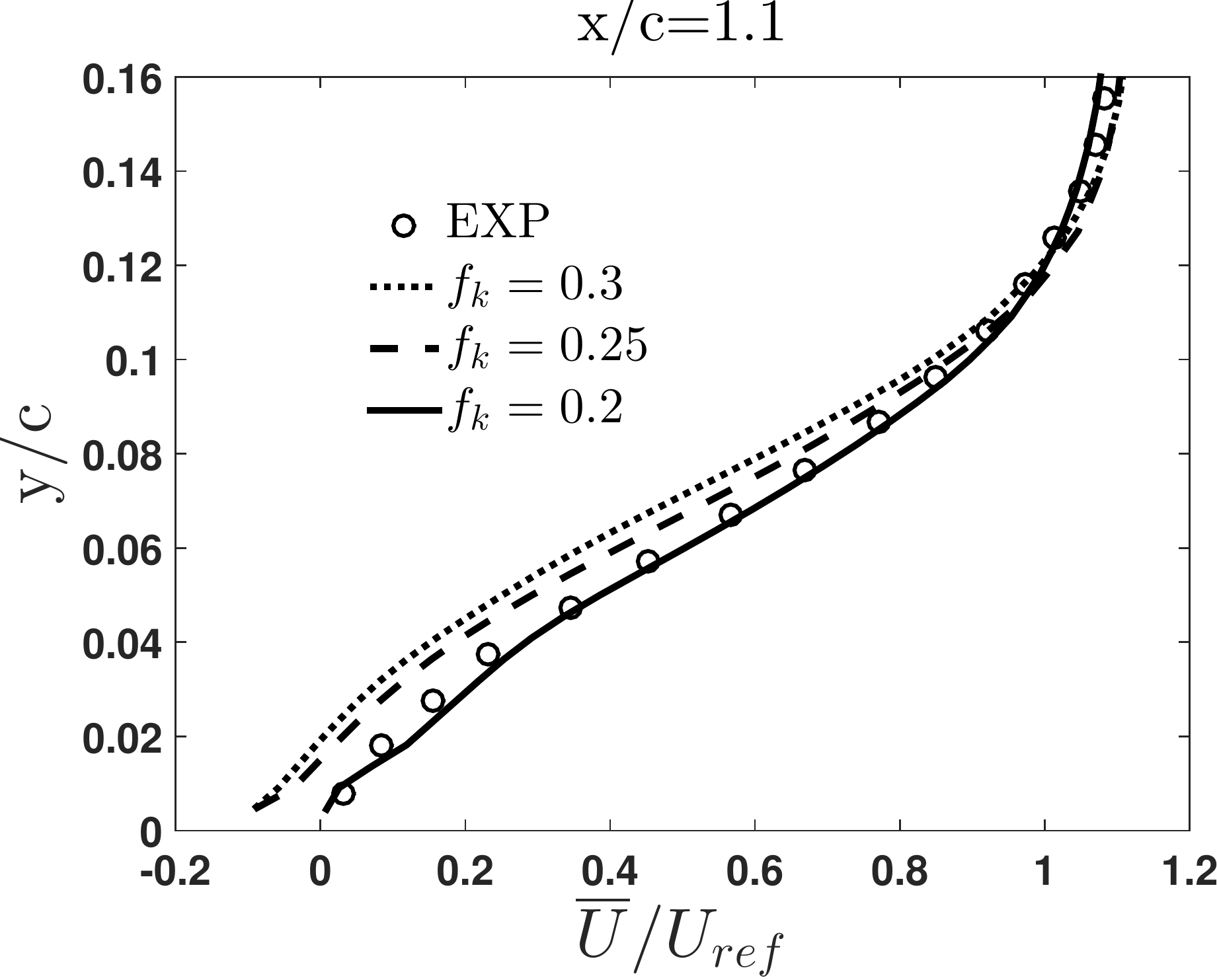}\begin{picture}(0,0)\put(-138,0){(c)}\end{picture}
        \end{subfigure}
 		       \begin{subfigure}[b]{0.4\textwidth}
                \includegraphics[width=\textwidth]{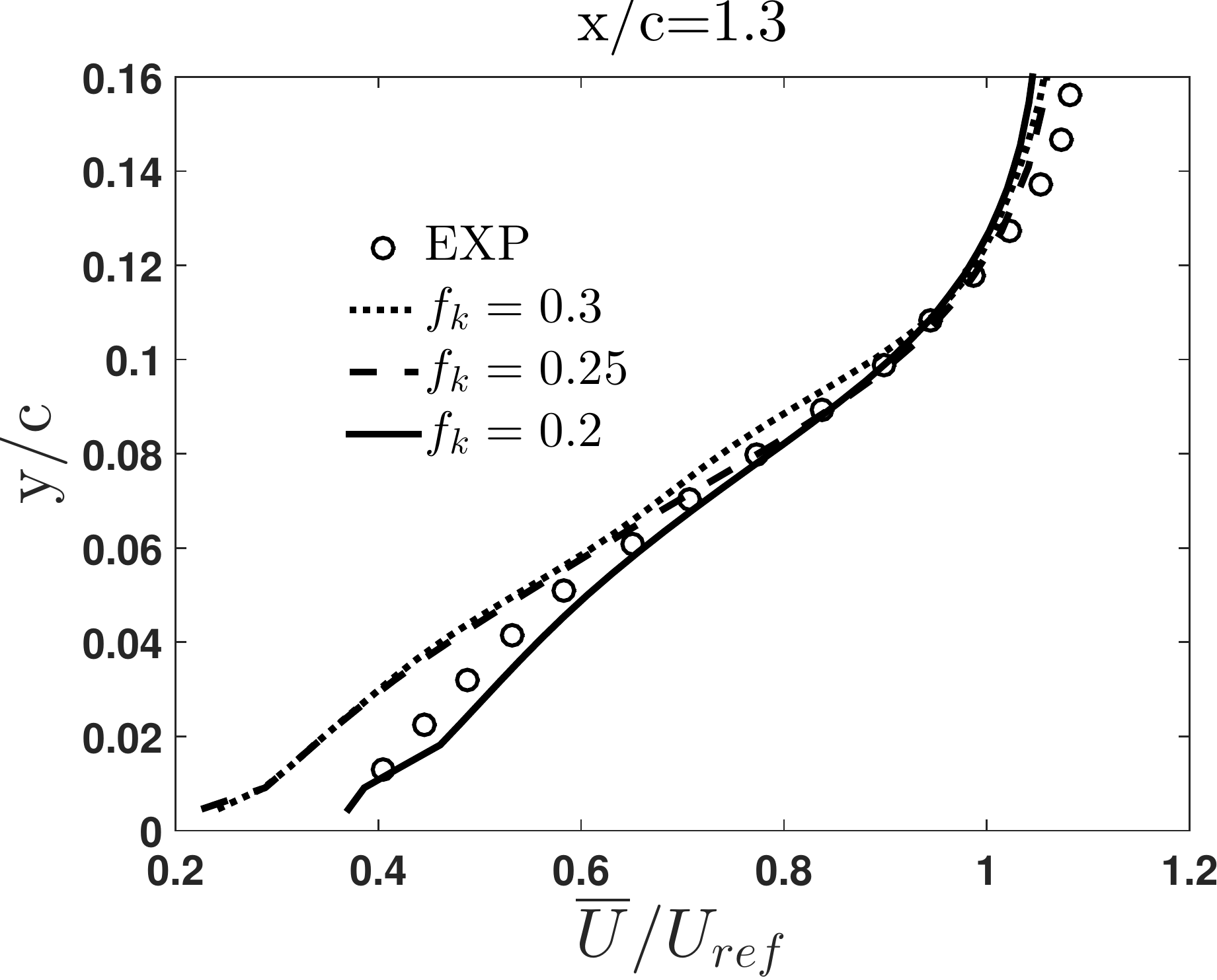}\begin{picture}(0,0)\put(-138,0){(d)}\end{picture}
        \end{subfigure}

\caption{Streamwise velocity (a) $x/c=0.7$ (b) $x/c=0.8$ (c) $x/c=0.9$ (d) $x/c=1.1$ (e) $x/c=1.2$ (f) $x/c=1.3$}   
\label{fk-u}                         
\end{figure}

\begin{figure}
        \centering
 		\captionsetup{justification=centering}                                               
 		        \begin{subfigure}[b]{0.4\textwidth}
                \includegraphics[width=\textwidth]{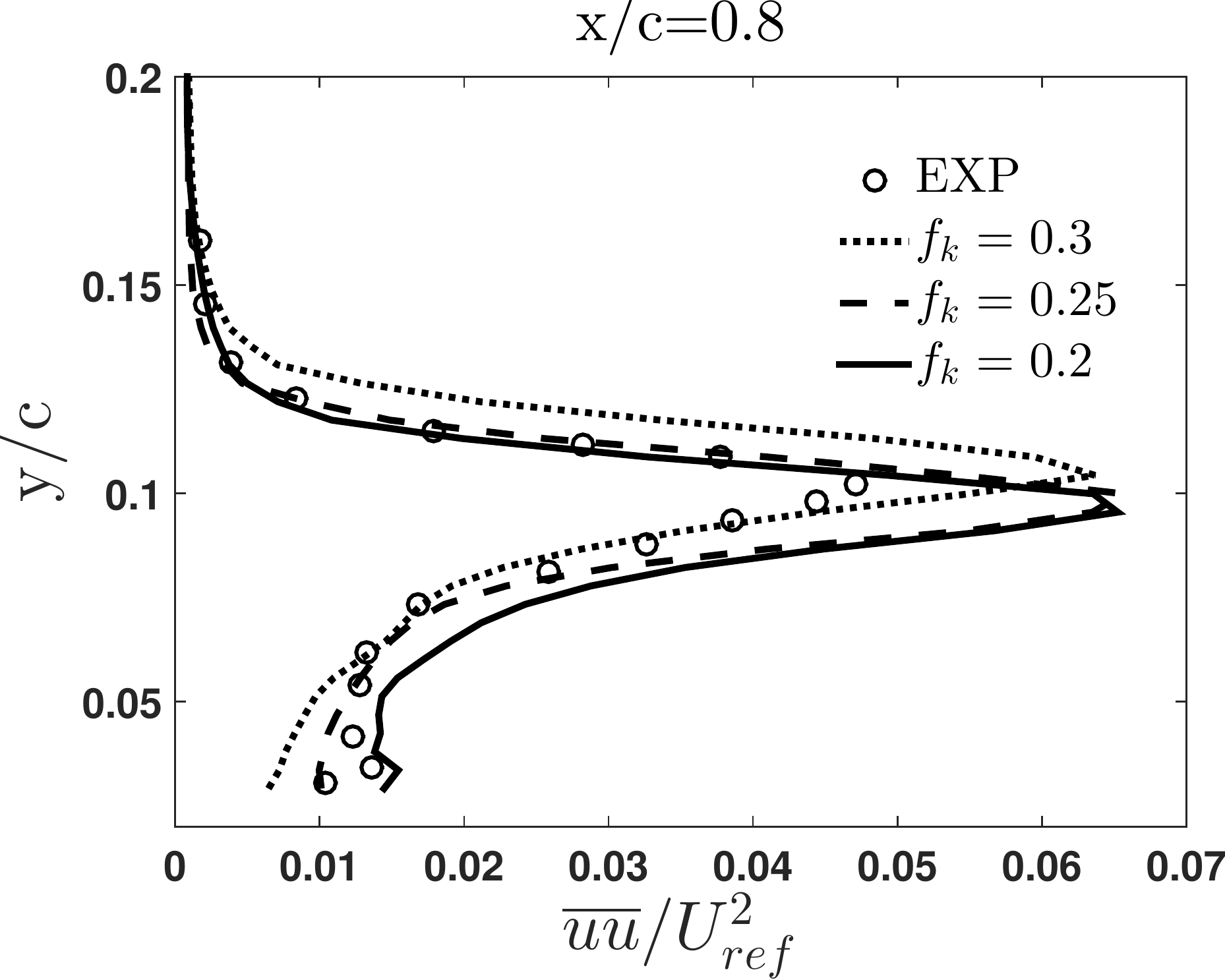}\begin{picture}(0,0)\put(-138,0){(a)}\end{picture}
        \end{subfigure}
        			\begin{subfigure}[b]{0.4\textwidth}
                \includegraphics[width=\textwidth]{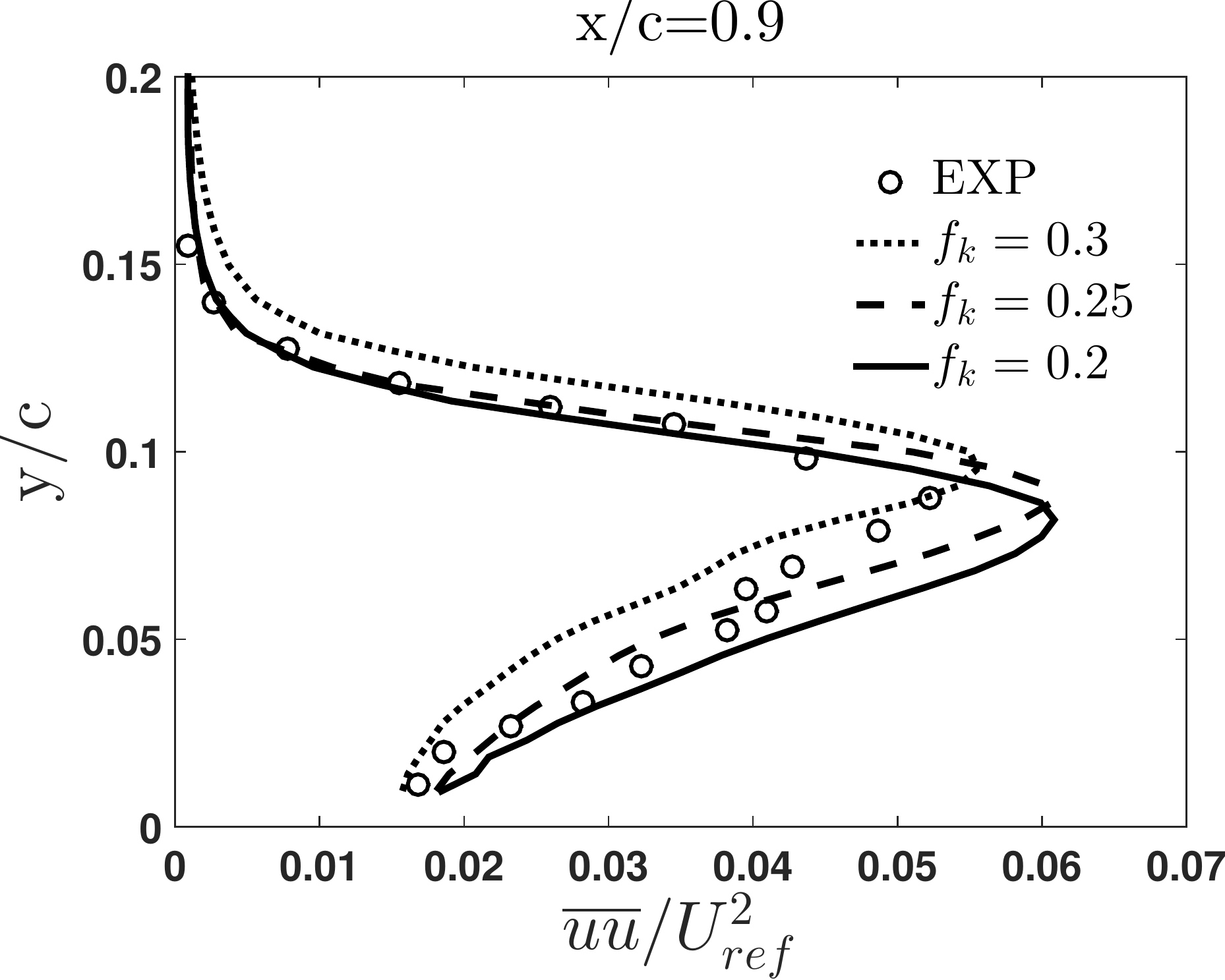}\begin{picture}(0,0)\put(-138,0){(b)}\end{picture}
        \end{subfigure}
                \begin{subfigure}[b]{0.4\textwidth}
                \includegraphics[width=\textwidth]{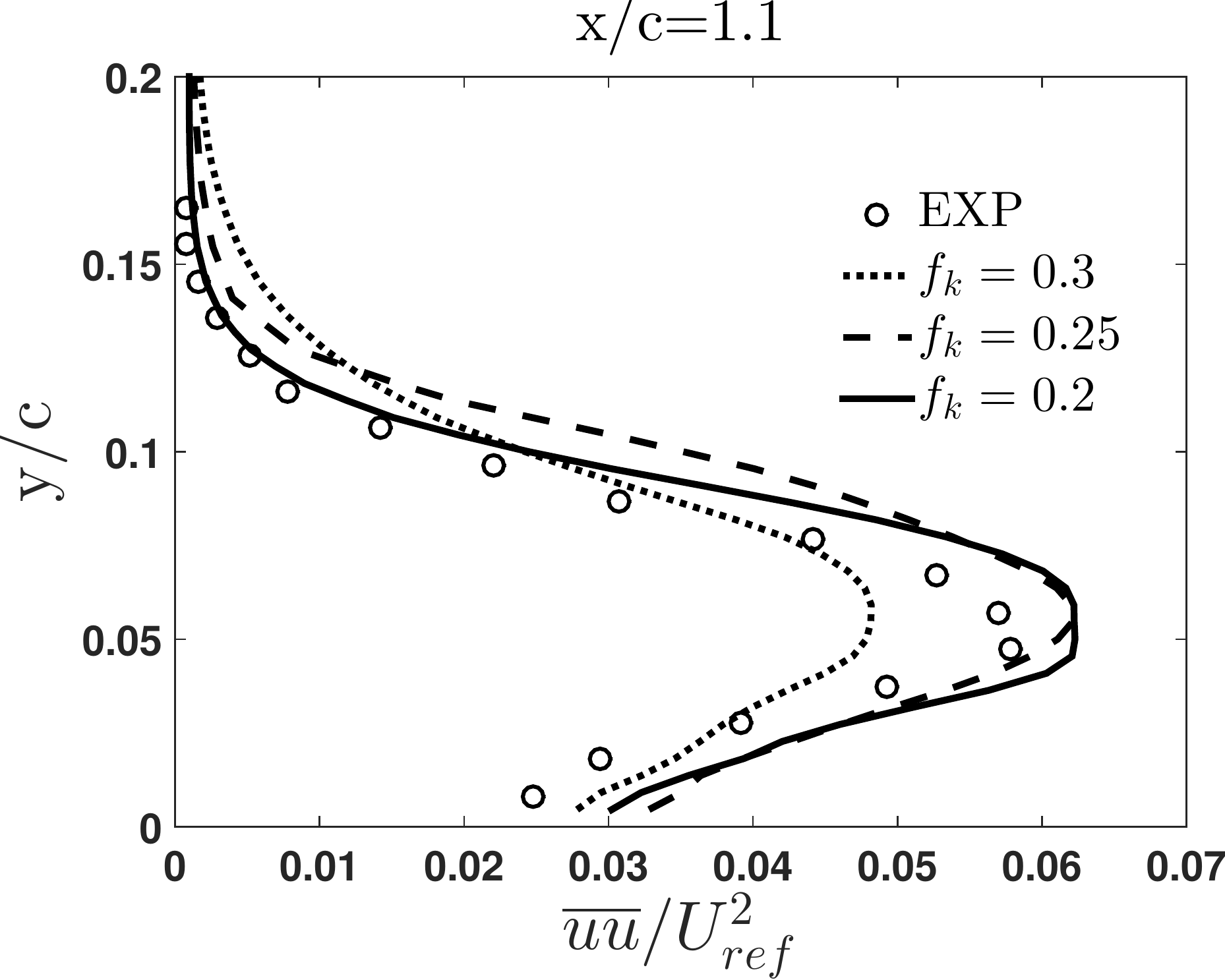}\begin{picture}(0,0)\put(-138,0){(c)}\end{picture}
        \end{subfigure}
 		       \begin{subfigure}[b]{0.4\textwidth}
                \includegraphics[width=\textwidth]{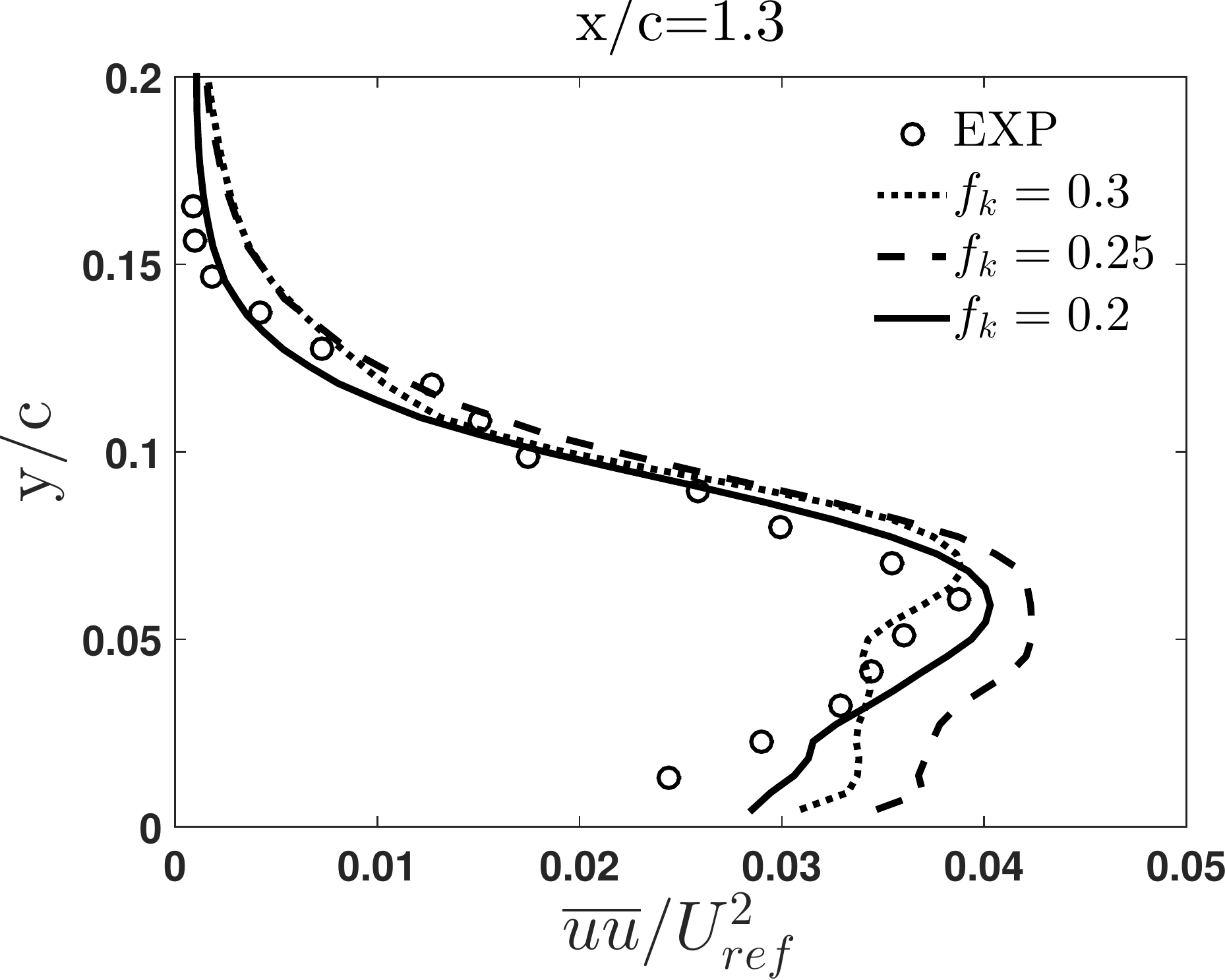}\begin{picture}(0,0)\put(-138,0){(d)}\end{picture}
        \end{subfigure}

\caption{Streamwise stress (a) $x/c=0.7$ (b) $x/c=0.8$ (c) $x/c=0.9$ (d) $x/c=1.1$ (e) $x/c=1.2$ (f) $x/c=1.3$}   
\label{fk-uu}                         
\end{figure}

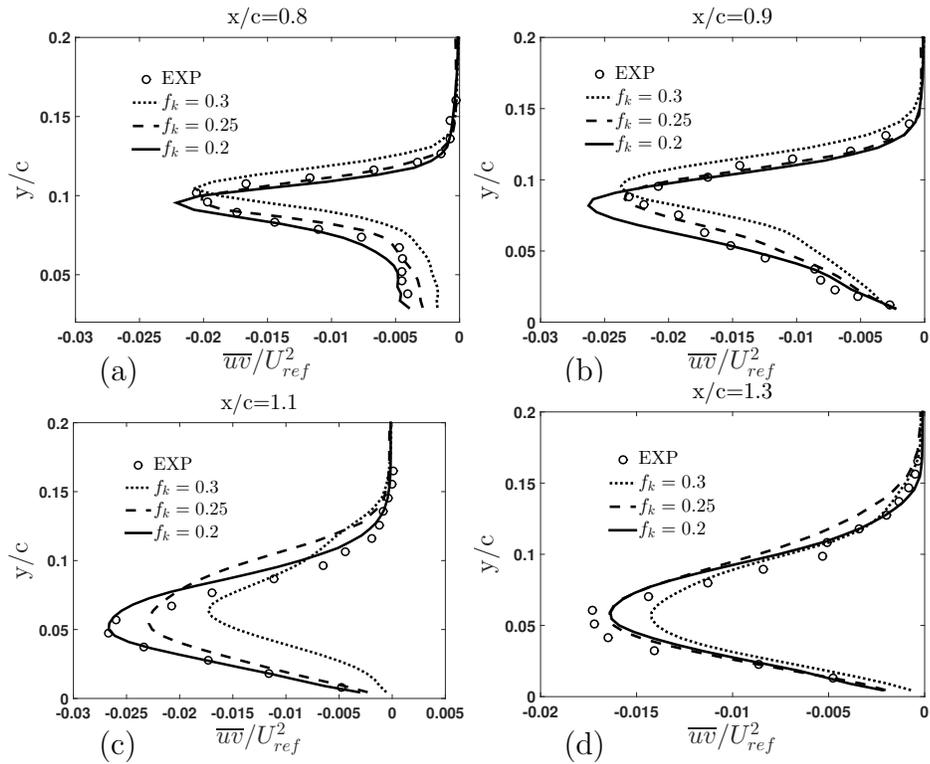
\begin{figure}
        \centering
 		\captionsetup{justification=centering}                                               
 		        \begin{subfigure}[b]{0.4\textwidth}
                \includegraphics[width=\textwidth]{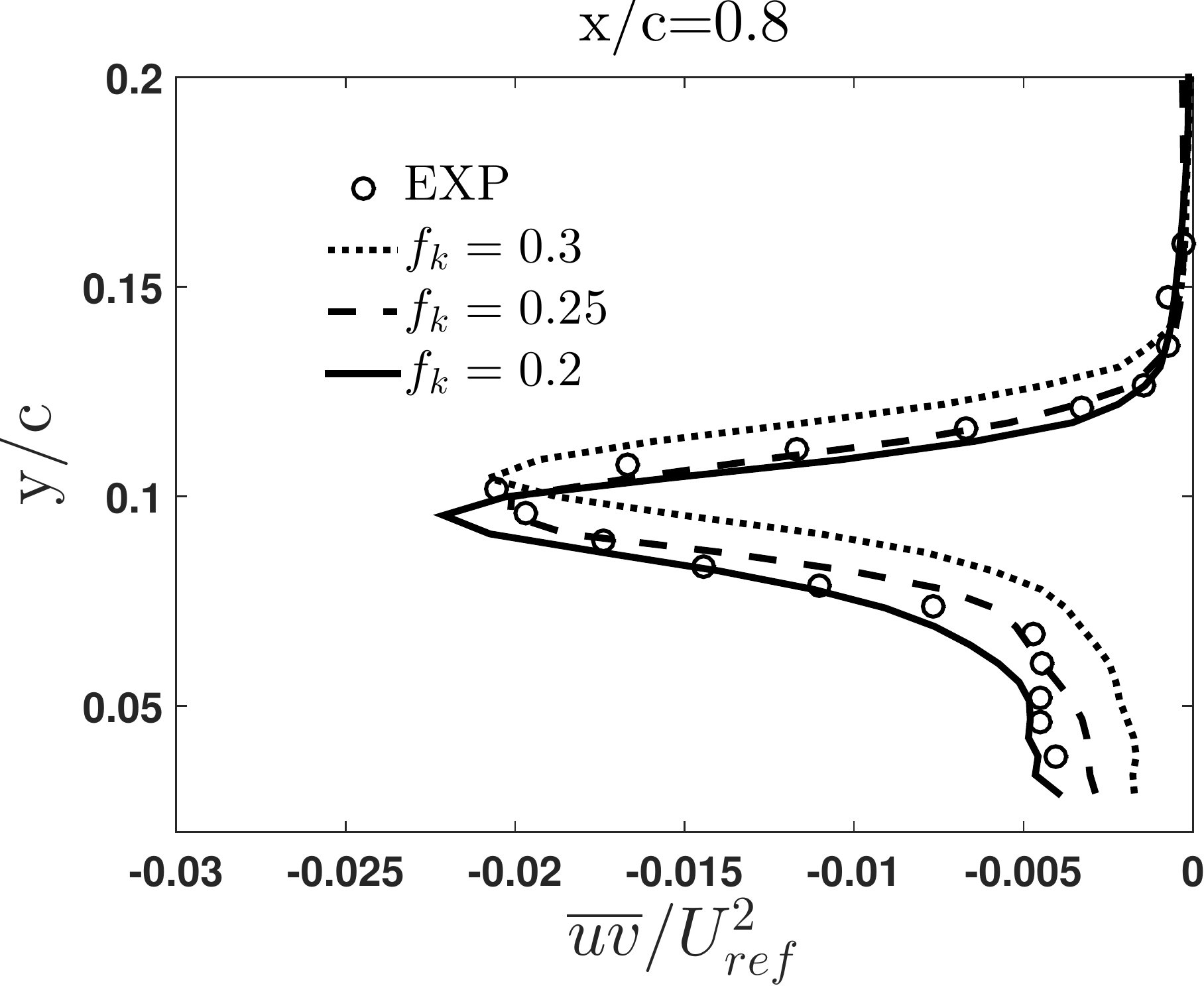}\begin{picture}(0,0)\put(-138,0){(a)}\end{picture}
        \end{subfigure}
        			\begin{subfigure}[b]{0.4\textwidth}
                \includegraphics[width=\textwidth]{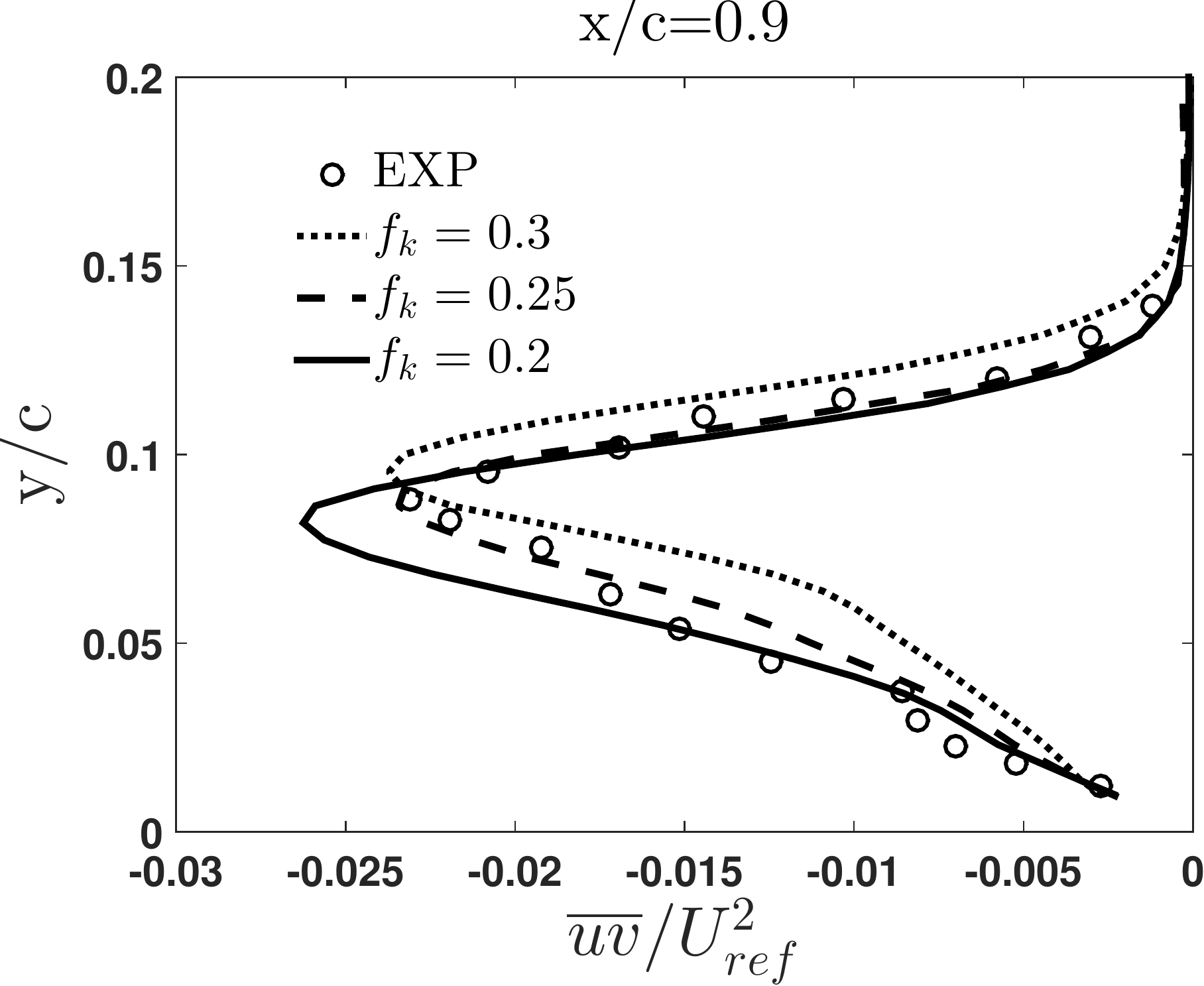}\begin{picture}(0,0)\put(-138,0){(b)}\end{picture}
        \end{subfigure}
                \begin{subfigure}[b]{0.4\textwidth}
                \includegraphics[width=\textwidth]{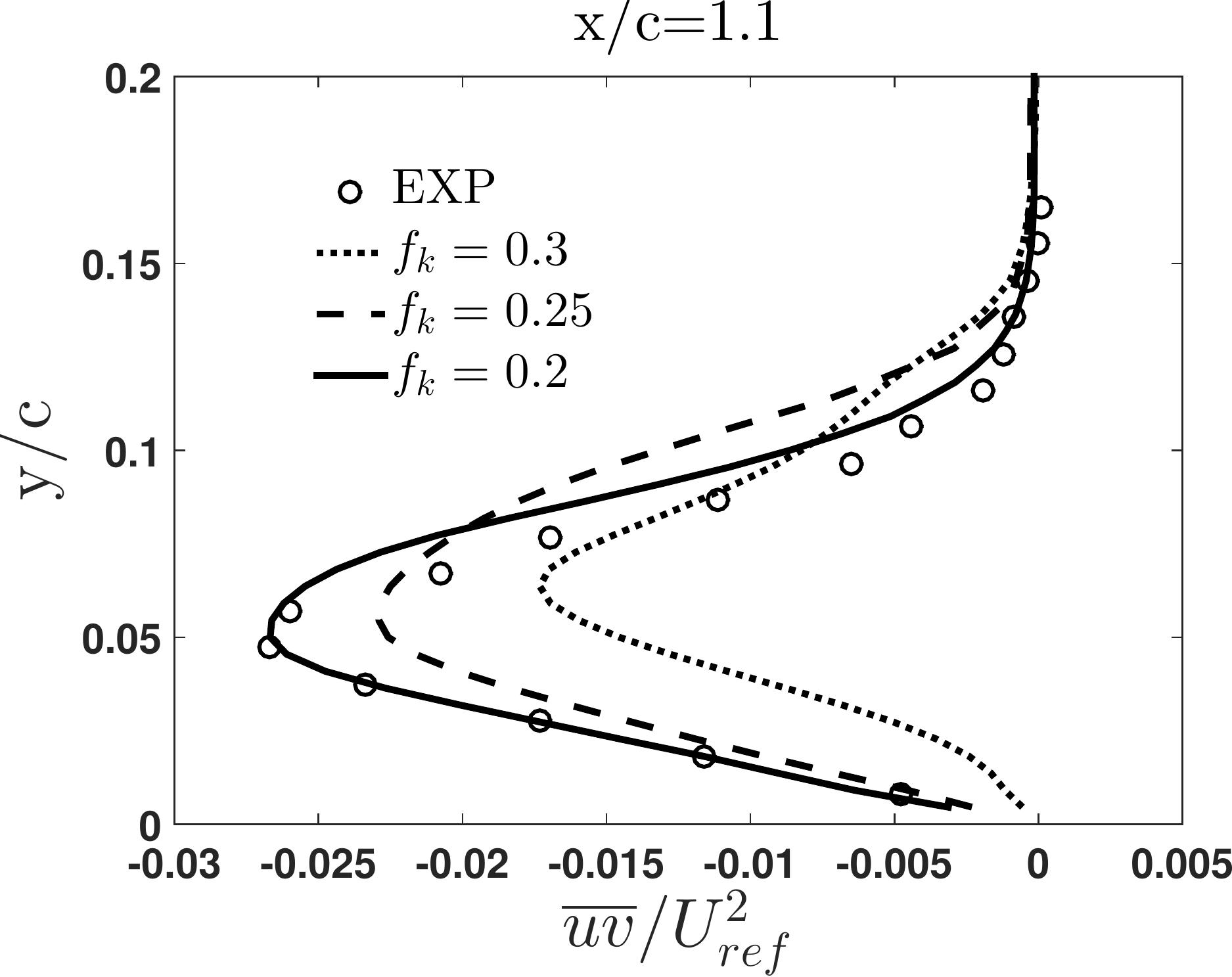}\begin{picture}(0,0)\put(-138,0){(c)}\end{picture}
        \end{subfigure}
 		       \begin{subfigure}[b]{0.4\textwidth}
                \includegraphics[width=\textwidth]{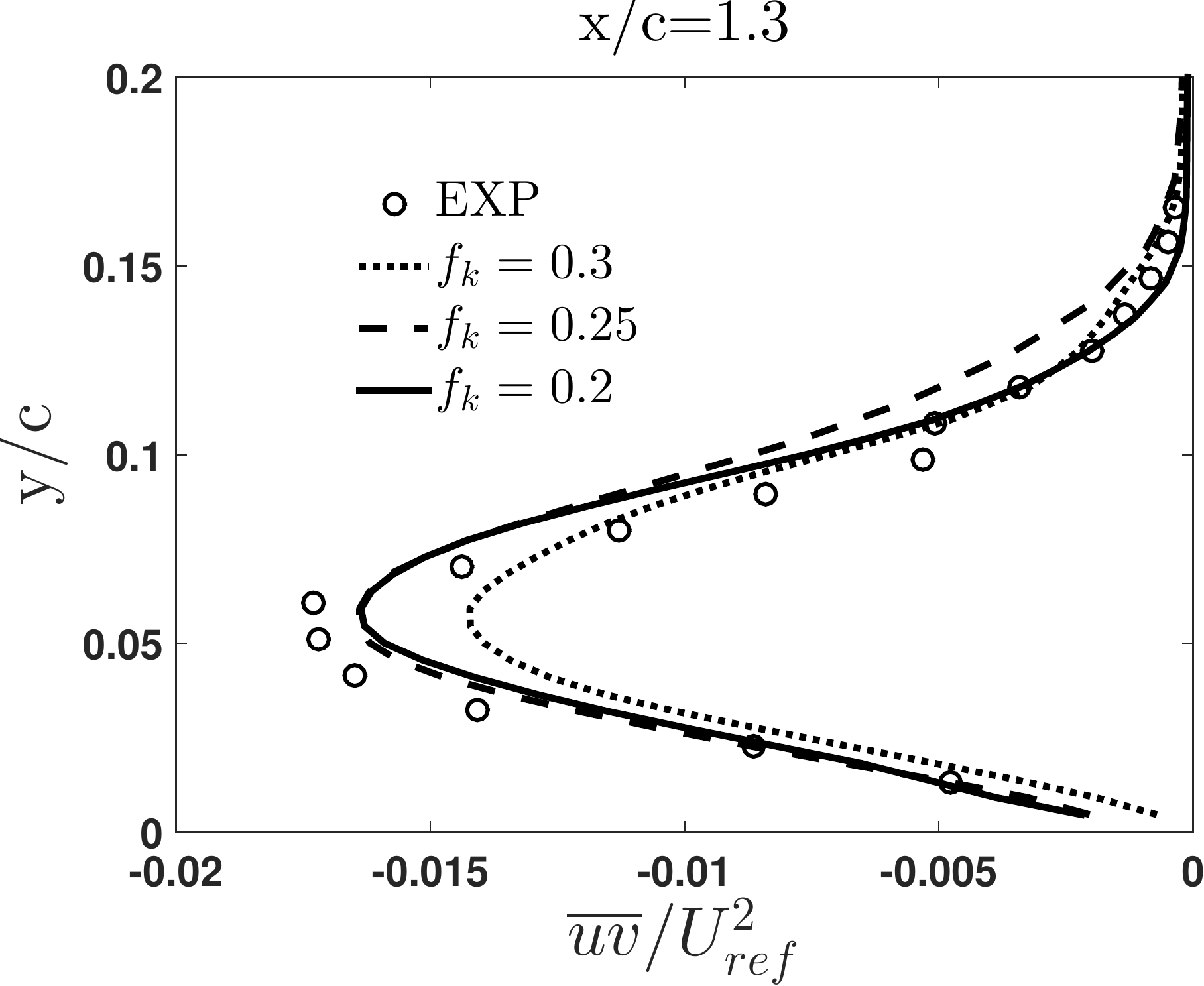}\begin{picture}(0,0)\put(-138,0){(d)}\end{picture}
        \end{subfigure}

\caption{Shear stress (a) $x/c=0.7$ (b) $x/c=0.8$ (c) $x/c=0.9$ (d) $x/c=1.1$ (e) $x/c=1.2$ (f) $x/c=1.3$}   
\label{fk-uv}                         
\end{figure}

Mean flow statistics at selected streamwise locations for different cut-off values are depicted in Figs. \ref{fk-u}-\ref{fk-uv}. All the calculations are performed on the grid with 0.6 million cells. The mean velocity profile shown in Fig. \ref{fk-u} clearly reveals that a better estimation of the flow behavior is obtained by lowering the cut-off parameter. This is particularly visible in regions close to reattachment and post-reattachment. Delayed reattachment is seen for the G2-PANS simulations with $f_k=0.25, 3$. The delayed reattachment can be related to the under-prediction of the second order statistics as seen in Figs. \ref{fk-uu} and \ref{fk-uv}. As shown in these figures, the peak values of stress components are well captured by $f_k=0.2$ simulation in most of the streamwise locations. However, still the results for all the $f_k$ values are in good agreement with experiment specially for $f_k=0.2$. 

\subsection{Flow structure}

\begin{figure}
        \centering
 		\captionsetup{justification=centering}                                               
 		        \begin{subfigure}[b]{0.4\textwidth}
                \includegraphics[trim=1cm 0.8cm 0.75cm 15.2cm, clip=true, scale=0.3]{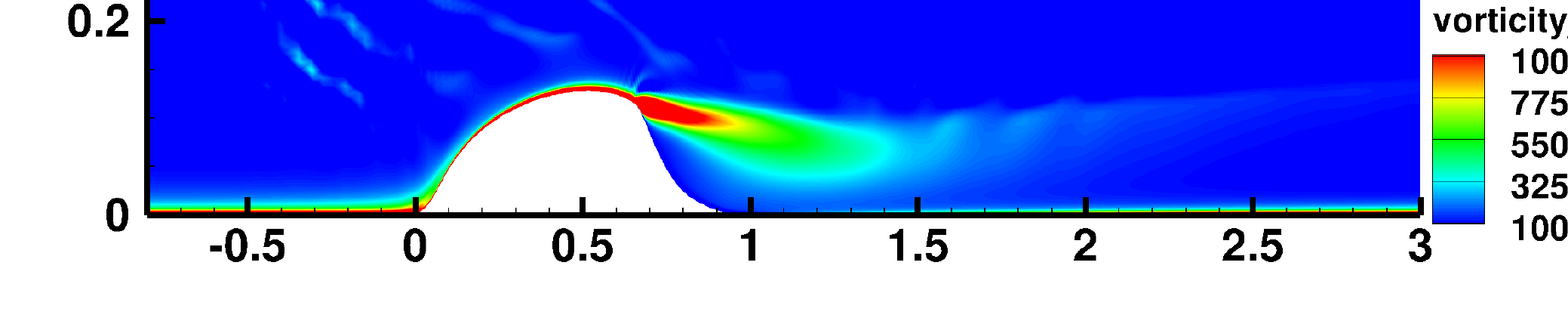}\begin{picture}(0,0)\put(-138,0){(a)}\end{picture}
        \end{subfigure}
        			\begin{subfigure}[b]{0.4\textwidth}
                \includegraphics[trim=1cm 0.8cm 0.75cm 15.2cm, clip=true, scale=0.3]{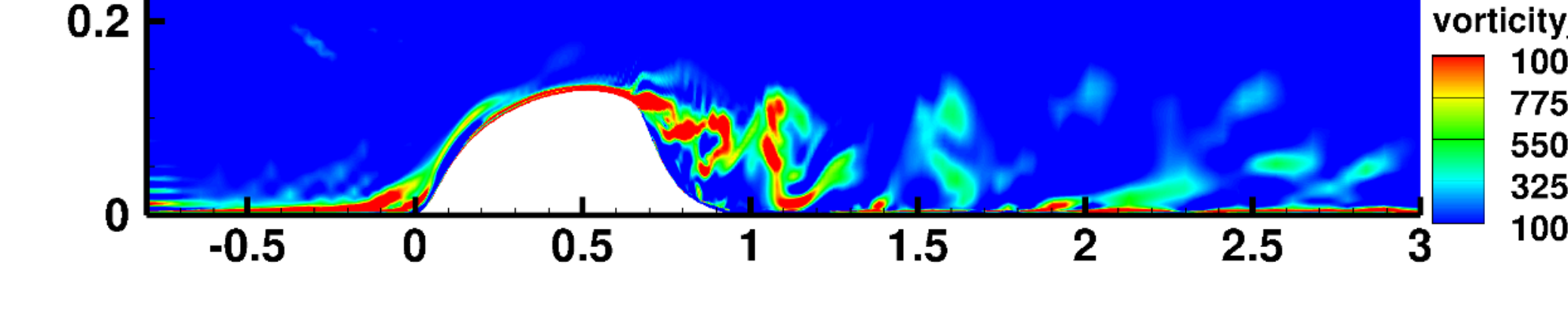}\begin{picture}(0,0)\put(-138,0){(b)}\end{picture}
        \end{subfigure}

       \begin{subfigure}[b]{0.4\textwidth}
                \includegraphics[trim=1cm 0.8cm 0.75cm 15.2cm, clip=true, scale=0.3]{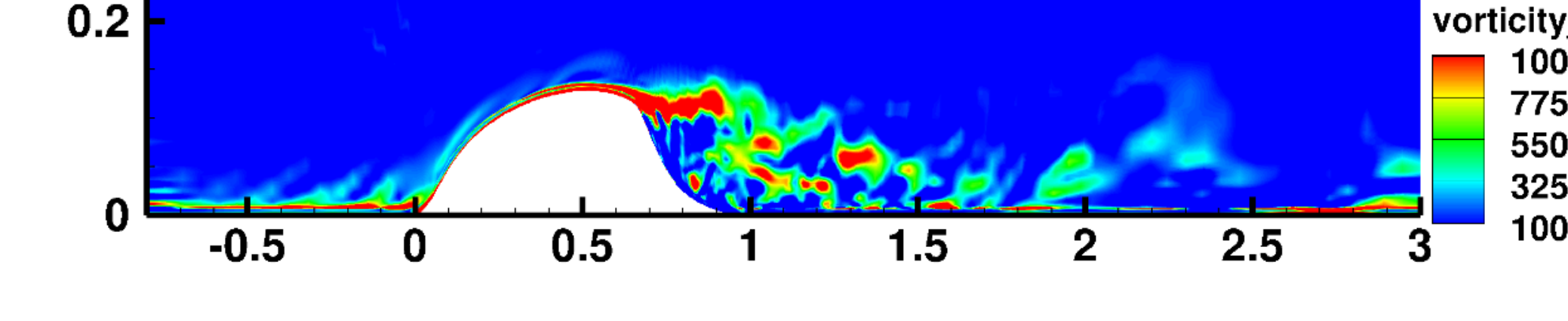}\begin{picture}(0,0)\put(-138,0){(c)}\end{picture}
        \end{subfigure}
 		       \begin{subfigure}[b]{0.4\textwidth}
                \includegraphics[trim=1cm 0.8cm 0.75cm 15.2cm, clip=true, scale=0.3]{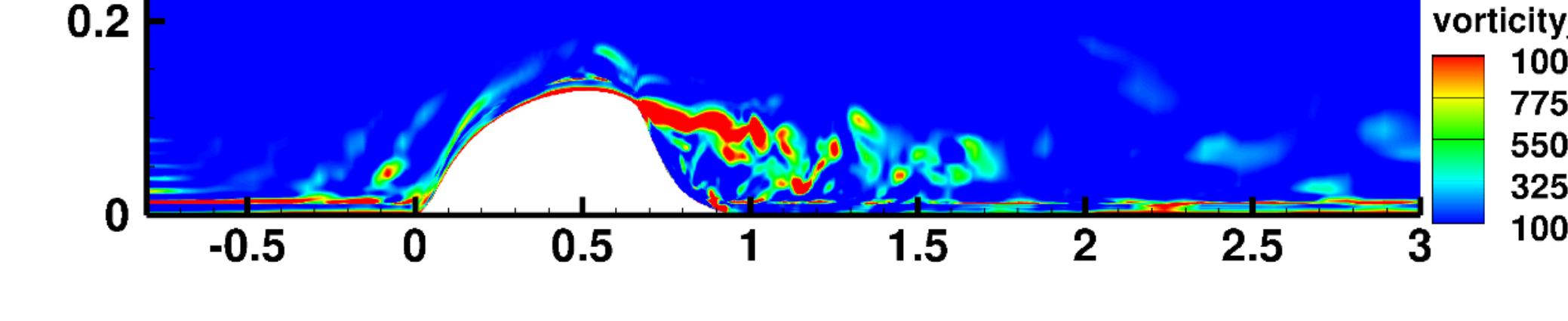}\begin{picture}(0,0)\put(-138,0){(d)}\end{picture}
        \end{subfigure}

\caption{Vorticity contours (a) RANS (b) $f_k=0.3$ (c) $f_k=0.25$ (d) $f_k=0.2$}   
\label{vorticity}                         
\end{figure}
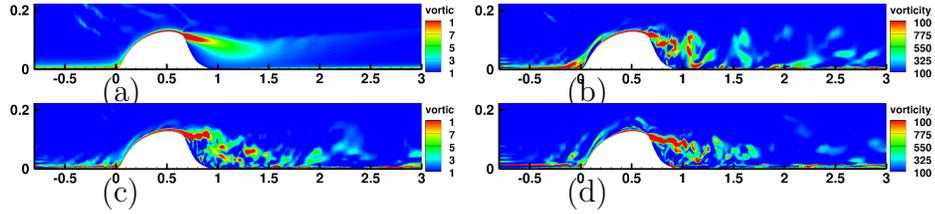

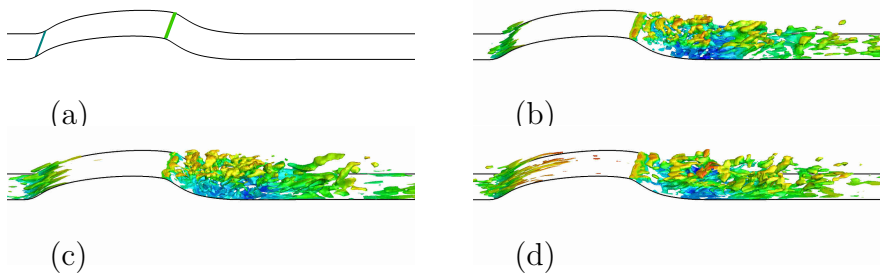
\begin{figure}
        \centering
 		\captionsetup{justification=centering}                                               
 		        \begin{subfigure}[b]{0.4\textwidth}
                \includegraphics[trim=0.5cm 10cm 0.75cm 3cm, clip=true, scale=0.25]{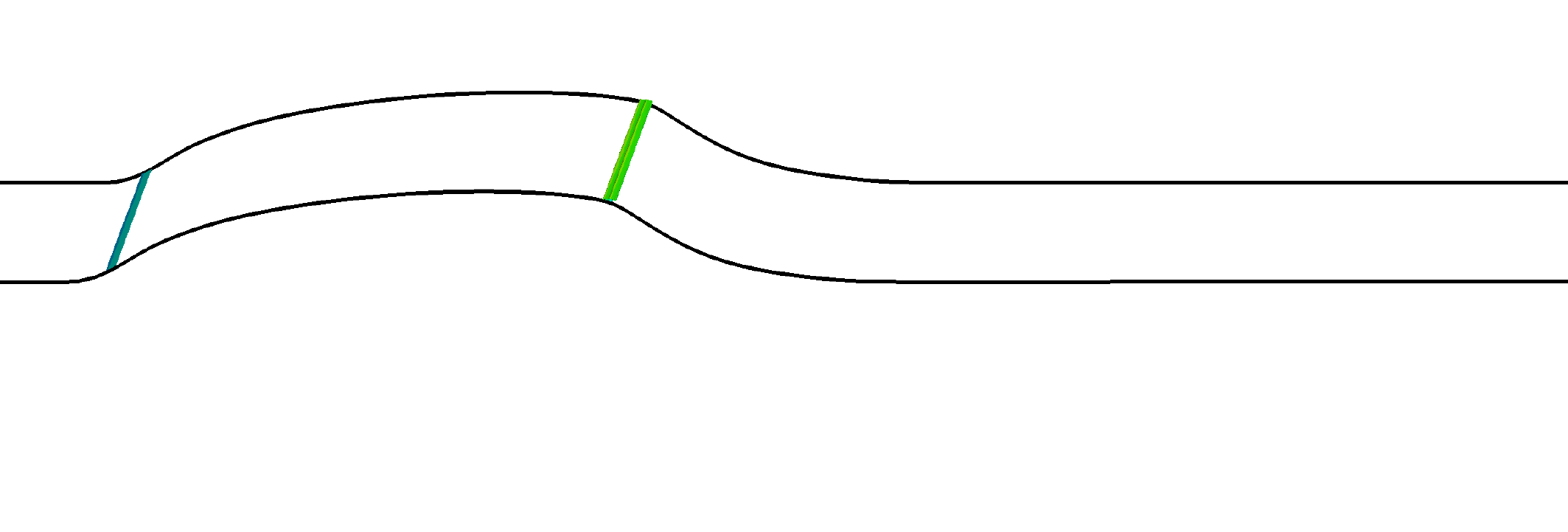}\begin{picture}(0,0)\put(-138,0){(a)}\end{picture}
        \end{subfigure}
        			\begin{subfigure}[b]{0.4\textwidth}
                \includegraphics[trim=0.5cm 10cm 0.75cm 3cm, clip=true, scale=0.25]{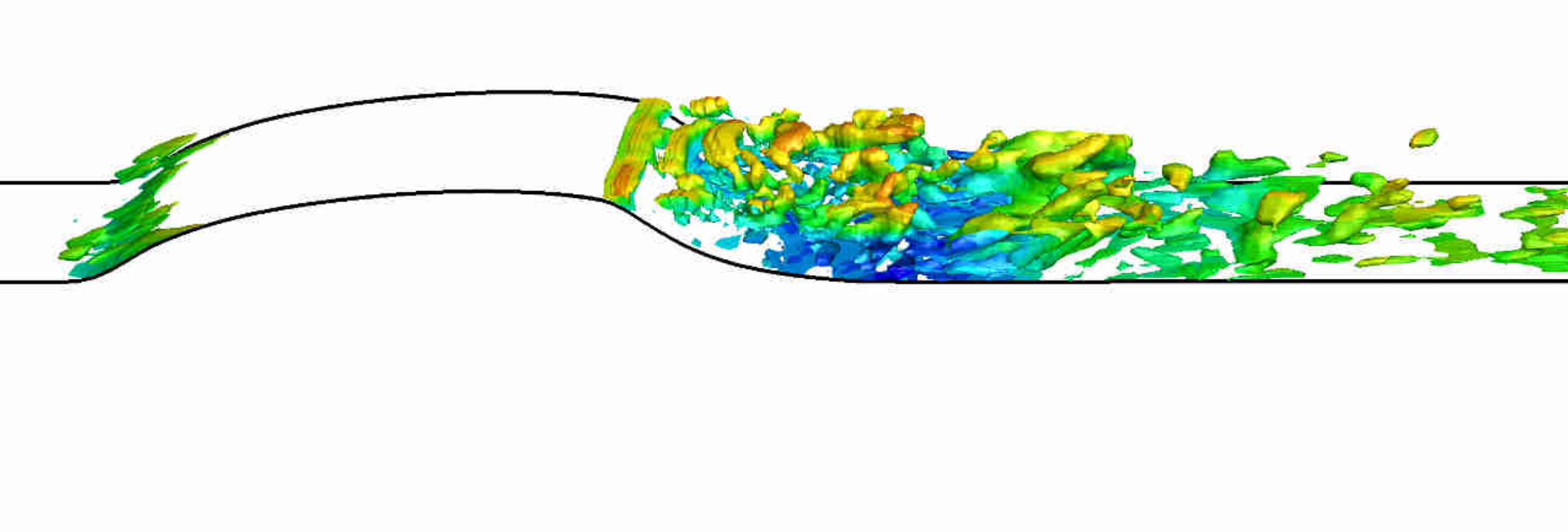}\begin{picture}(0,0)\put(-138,0){(b)}\end{picture}
        \end{subfigure}
                \begin{subfigure}[b]{0.4\textwidth}
                \includegraphics[trim=0.5cm 10cm 0.75cm 3cm, clip=true, scale=0.25]{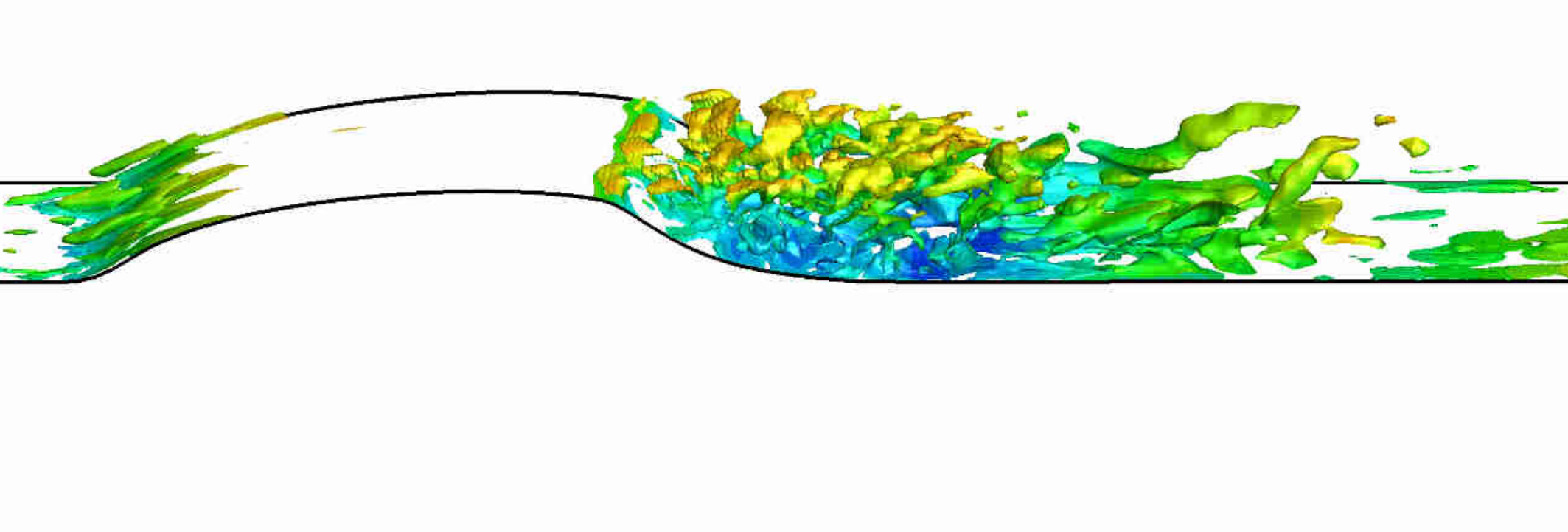}\begin{picture}(0,0)\put(-138,0){(c)}\end{picture}
        \end{subfigure}
 		       \begin{subfigure}[b]{0.4\textwidth}
                \includegraphics[trim=0.5cm 10cm 0.75cm 3cm, clip=true, scale=0.25]{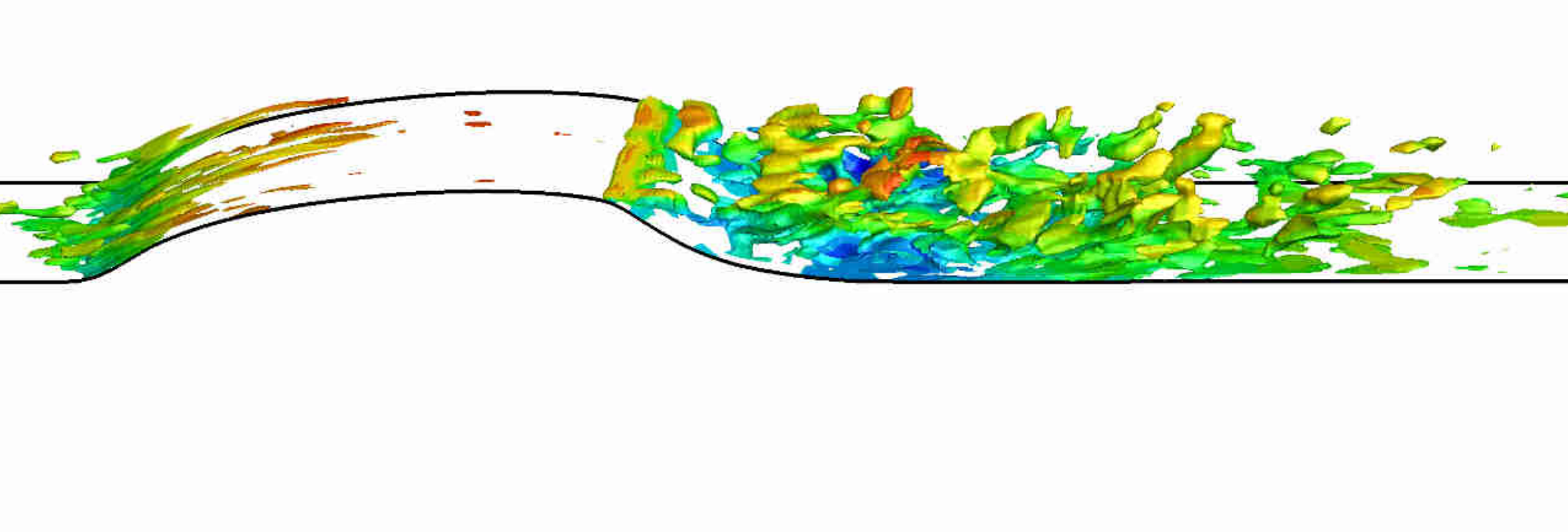}\begin{picture}(0,0)\put(-138,0){(d)}\end{picture}
        \end{subfigure}

\caption{Q iso-surfaces (a) RANS (b) $f_k=0.3$ (c) $f_k=0.25$ (d) $f_k=0.2$}   
\label{Q}                         
\end{figure}

Instantaneous snapshots of the hump flow is given in Figs. \ref{vorticity} and \ref{Q}. The fields are displayed through iso-surfaces of the second invariant of the velocity gradient tensor, the $Q$ criterion and the vorticity contours. The iso-surfaces of Q are coloured with contours of the streamwise velocity. These figures reveal the existence of flow structures with wide range of scales in the separated shear-layer and at the hump leading edge. As discussed in previous sections, the level of kinetic energy and turbulent stresses of the flow at the leading edge is critical in predicting the right behavior of the flow after the separation. Better estimation of the mean flow statistics by lowering the cut-off value can be explained by looking at Fig. \ref{Q}. This figure illustrates that more scales of the flow motion at the hump leading edge are resolved by reducing $f_k$ which consequently results in superior estimation of the separation bubble size.

\chapter{\uppercase{Conclusions}}
\label{conclusion}

The overarching purpose of this work is to advance the scale-resolving PANS approach to a frontline computational fluid dynamics (CFD) tool that can meet the accuracy demand of complex engineering flows and afford the computational expenses. This dissertation addresses the fidelity of model with fixed and variable scale resolution for simulating smooth surface separation, shear flow and combination of both at different Reynolds numbers. The main conclusions of each of the three studies included in this document are summarized in this section.

i) Simulations of flow over periodic hill are performed using G1-PANS implemented in OpenFoam for a fixed value of $f_k$. The results for the flow Reynolds number of 10590 are compared against an experimental study and LES. Overall, very good agreement of the PANS simulations with literature data is obtained for the mean flow statistics. It was shown that at this Reynolds number, the results for the G1-PANS simulation with $f_k$=0.15 is closely following LES data and notably provide a better estimation of the separation bubble on a much coarser grid. Constructing the invariant map for the second and third anisotropy invariants showed that all states lie within Lumley's realizability constraints. The flow near the lower wall is found to traverse to the two-component state along axisymmetric contraction for the separated and early reattachment region which is consistent with LES findings. 

Additionally, internal consistency criteria is set as a sanity check for all the PANS calculations. For this purpose, the ratio of modeled to total eddy viscosity within the simulation is calculated and is investigated to recover the input value of $f_{k}^2$. The consistency of the recovered filter parameter with the input value for all the PANS calculations presented in this study is seen at all the streamwise locations.  

For higher Reynolds number of 37000, two studies are performed to investigate the effect of reducing cut-off ratio, $f_k$ and grid resolution on the accuracy of the PANS results. Results from PITM method and an experimental study are also included for the sake of comparison. For the finest grid investigated in this work, reducing $f_k$ value results in improved agreement with experimental data regarding the size of separation bubble and flow statistics. For the two grid resolutions studied here, PANS method did not show remarkable sensitivity to grid resolution. However, PITM method failed to predict the size of separation bubble and mean quantities accurately for the coarse grid. 

Observation of flow structure for the PANS simulations divulge eddies with wide range of scales which demonstrates the three dimensionality nature and unsteadiness of the flow field. None of these phenomena is resolved by the RANS calculation. The results presented in this study further illustrates that PANS method can be useful in terms of accuracy and computational cost for simulating smooth curvature separation.      

ii) Two transport models proposed for partially-resolved flow simulations known as zero transport model (ZTM) and the maximum transport model (MTM) were examined for turbulent boundary layer. It is shown that the ZTM model is more consistent with log-layer behavior, and it estimates the second order statistics better than the MTM model. Besides, more scales of flow motion is resolved by ZTM in near wall region. 

In addition, it was concluded that the G1-PANS model is only able to recover the boundary layer physics if adequate grid resolution is provided. Since higher Reynolds number flow calculations demand extremely large computational domain even for an intermediate scale resolution, employment of variable resolution in near wall region is investigated in this work. In order to accurately simulate variable resolution calculations, the energy transfer from the unresolved to resolved scales in case of decreasing resolution and vice versa in case of increasing resolution must be accounted for. Girimaji and Wallin \cite{girimaji2013} adopted commutation error terms in the PANS closure to appropriately model temporal variation in resolution. In this work, similar approach is taken to identify and develop the commutation terms for both temporal and spatial variations in $f_k$. This modeling effort resulted in emergence of the second generation of the PANS model (G2-PANS). 

Multiple simulations of turbulent channel flow are performed in the range of $Re_\tau$ = 150-8000 using the G1-PANS and G2-PANS models. The near-wall resolution for G2-PANS calculations is varied from $f_k$ = 1 (at the wall) to $f_k$ = 0.2 or 0.3 as required near the center-line. The results showed that the log-layer is accurately captured at all Reynolds numbers for the G2-PANS calculations. It must be pointed out that the corresponding simulations with G1-PANS model on the same grids exhibited log-layer mismatch at high Reynolds number. Besides, the individual stress components obtained from G2-PANS simulations are close to DNS data specially in the region outside the RANS subdomain. 

Finally, the effect of resolution variation location is studied for G2-PANS calculations. It is shown that the mean flow statistics and resolved scales in near wall region is dependent on the location of resolution variation. More accurate results are obtained if the location of resolution change is closer to the wall providing the right amount of grid resolution. For high Reynolds number calculations of 4200 and 8000, this effect becomes visible due to the computational limitations of near wall region where the $f_k$ variation is applied farther from the wall compared to $Re_\tau$=950 and 2000. 

iii) G2-PANS model is validated for flow separation over wall-mounted hump. This flow geometry is a challenge test case  as accurate simulation of separation process is strongly dependent on precise development of turbulent boundary layer upstream of the hump. As shown in Sec. \ref{wall-model}, computational efforts to simulate turbulent boundary layer can be substantially relaxed by applying G2-PANS model which provides proper modeling of unresolved-to-resolved energy transfer. Therefore for the hump simulation, the near wall region is modelled and variable resolution is applied in the wall-normal direction.  

The G2-PANS results are compared against RANS, LES (9.4 M grid nodes) and experiments. The results indicate that the separation and reattachment locations for several $f_k$ calculations and computational grid nodes in the range of 0.3M to 1.5M are closely approximated by the PANS simulation. The flow structures illustrated that more scales of the flow motion at the hump leading edge are resolved by reducing $f_k$ which consequently results in superior estimation of the separation bubble size. 
 
While RANS performs very poorly, even the coarsest PANS simulation agrees well with data. The variation of skin-friction coefficient and mean flow statistics for the G2-PANS calculations reveals an improvement over LES. In this study, the ability of G2-PANS to generate LES-quality predictions at a much reduced cost is exhibited.

\section{Future work}

The first and second version of the PANS model is tested for the smooth and sharp-edge separation, secondary flow and complex separated flows. Although these validation studies demonstrated potential advantages of the PANS model, the model needs to be examined for a broad range of aviation flows including mixing layers in all speed regimes, curved flows and flows with strong secondary motion. 

Besides, In the first two generations of PANS simulations, a precursor RANS simulation is performed to estimate $k$ and
$\epsilon$, following which $f_k$ and $f_\epsilon$ fields are specified. This involves user intervention which is not very desirable. For automated specification of $f_k$ and $f_\epsilon$ fields, further investigation is required. 

In addition, the PANS methodology needs to be extended for the compressible flow regime, and re-derived to account for compressibility corrections. Besides, most of the validation studies are in the incompressible flow limit and the model should be examined for compressible test cases.

\let\oldbibitem\bibitem
\renewcommand{\bibitem}{\setlength{\itemsep}{0pt}\oldbibitem}
%
%
%


\phantomsection
\addcontentsline{toc}{chapter}{REFERENCES}

\renewcommand{\bibname}{{\normalsize\rm REFERENCES}}

\bibliographystyle{abbrv}
\bibliography{references}

\end{document}